\pdfoutput=1

\documentclass[preprint,fleqn,5p,numbers,sort&compress]{elsarticle}

\usepackage{graphicx}
\usepackage{amssymb,amsmath}
\usepackage{natbib}
\usepackage{hyperref}
\usepackage{gensymb}
\hypersetup{pdftitle = The title of my PDF, pdfauthor = My name, pdfsubject= The subject, pdfkeywords = keyword1 keyword2 keyword3}
\hypersetup{colorlinks = true, linkcolor = blue, anchorcolor = red, citecolor = blue, filecolor = red, pagecolor = red, urlcolor = blue}

\journal{International Journal of Non-Linear Mechanics}

\begin{document}

\begin{frontmatter}

\title{On the Newton-Raphson basins of convergence associated with the libration points
in the axisymmetric five-body problem: the concave configuration}

\author[mss]{Md Sanam Suraj\corref{cor1}}
\ead{mdsanamsuraj@gmail.com}

\author[ps]{Prachi Sachan}
%\eed{$prachi.sachan80@gmail.com$}

\author[eez]{Euaggelos E. Zotos}
%\ead{evzotos@physics.auth.gr}

\author[am]{Amit Mittal}
%\ead{cto.amitmittal@gmail.com}

\author[ra]{Rajiv Aggarwal}
%\ead{rajiv_agg1973@yahoo.com}

\cortext[cor1]{Corresponding author}

\address[mss]{Department of Mathematics,
Sri Aurobindo College, University of Delhi, 
Malvia nagar, Delhi-110017, India}

\address[ps]{Department of Mathematics,
University of Delhi, Delhi-110007, India}

\address[eez]{Department of Physics, School of Science,
Aristotle University of Thessaloniki, GR-541 24, Thessaloniki, Greece}

\address[am]{Department of Mathematics, ARSD College,
University of Delhi, Dhaula Kuan, Delhi-110021, India}

\address[ra]{Department of Mathematics, Deshbandhu College,
University of Delhi, Kalkaji, Delhi-110019, India}

\begin{abstract}
The axisymmetric five-body problem with the concave configuration has been studied numerically to reveal the basins of convergence, by exploring the Newton-Raphson iterative scheme, corresponding to the coplanar libration points (which act as attractors). In addition, four primaries are set in axisymmetric central configurations introduced by \'{E}rdi and Czirj\'{a}k \cite{erd16} and the motion is governed by mutual gravitational attraction only. The evolution of the positions of libration points is illustrated, as a function of the value of angle parameters. A systematic and rigorous investigation is performed in an effort to unveil how the angle parameters affect the topology of the basins of convergence. In addition, the relation of the domain of basins of convergence with required number of iterations and the corresponding probability distributions are illustrated.
\end{abstract}

\end{frontmatter}

\section{Introduction}
\label{intro}
In celestial mechanics and dynamical system, the problem of $N$-bodies still remains, undoubtedly, one of the most interesting and challenging topics.  In particular, a plethora of research papers are available on the restricted problem of $N$-bodies for $N=3, 4$ in classical as well as when the effective potential of the classical version of restricted problem of three and four bodies has been modified by including various type of perturbing terms due to considering the additional forces (e.g., \cite{AM14a, AM14b, AM15a, AM15b,
AM15c, AG16, PE18}, \cite{EAK16}, \cite{A18}). These research topics have various practical applications in the field of planetary physics and in galactic dynamics. Therefore, the general extension of $N-$body problem is the study of problem of $N$-bodies for $N>4$ (e.g., \cite{L18}). In  few decades, the restricted problem of $N=5$ bodies has fascinated many researchers and scientists. The restricted five-body problem deals with the motion of the test particle (with an infinitesimal mass relative to the  primaries), under the combined effect of gravitational pull of four primary bodies.

A profusion of research papers are available on the topic of restricted five-body problem where the motion of the test particle is investigated under various  perturbing forces in addition to gravitational force. The restricted problem of five bodies is to study the motion of the test particle under the gravitational field of four primary when three primary bodies are placed on the vertices of equilateral triangle while one primary is placed on the center of mass of the system (e.g., \cite{oll88}). Moreover, the extra perturbations are included in the gravitation potential by many authors, viz., the effect of radiation of primaries in the context of five-body problem (e.g., \cite{pap07}), the effect of variable mass  in the five-body problem (e.g.,\cite{sur18e}). The existence of the libration points in the axisymmetric five-body problem has been studied in \cite{gao17} by considering the configuration of  \cite{erd16}.
In the study of $N$-body problem, knowing the positions of the libration points is an important issue as the position of the test particle remains unperturbed relative to the primary bodies. Unfortunately, except the restricted three-body problem, we do not have explicit formulae to determine the positions of the libration points for $N>3$-body problem. However, the only possible way is to use one of the available numerical method to determine the positions of the libration points in these type of systems, i.e., we require to use multivariate iterative method to solve the non-linear system of equations. The multivariate version of the Newton-Raphson iterative scheme is one of the well known method to solve these equations. The well known fact is that every iterative scheme depends strongly on the choice of the initial conditions (i.e., the starting value for an iterative method) used.  Moreover, it is also necessary to note that the iterative scheme converges quickly for some of the initial conditions, where as for other initial conditions it may take a huge number of iterations to converge or some of the initial conditions do not converge to any of the libration points (which act as attractors) or for some initial conditions, the iteration scheme may enter into an endless cycle in a periodic or aperiodic manner or may diverge to infinity. In available literature concerning, the various iterative schemes unveil that the initial conditions which fall into the domain of the basins of convergence, and located in the fractal regions require a considerable number of iterations to converge at one of the initial conditions. These facts provide considerable amount of reasons to justify the study of the basins of convergence associated with the libration points of any dynamical system.  Therefore, we can select easily those initial conditions for which the iterative scheme converges to predefined attractors with lowest number of iterations.

Thus, the Newton-Raphson iterative scheme to determine the basins of convergence associated with the libration points, provides intrinsic properties of the dynamical system. A series of literature is available which deals with the basins of convergence associated with the libration points in various type of dynamical system such as  the restricted three-body problem (e.g., \cite{zot16}, \cite{sur18c}), the restricted four-body problem with and without various perturbations (e.g.,\cite{amt18}, \cite{agg18}, \cite{asi16}, \cite{sur17a, sur17b}, \cite{zot17a, zot17b}, \cite{sur18a}), the restricted five-body problem (e.g., \cite{zot18a}), the Hill's problem (e.g., \cite{dou10}, \cite{zot17d}), pseudo-Newtonian three and four body problem (e.g., \cite{zot17c}, \cite{Sur18d}), the Copenhagen problem (e.g., \cite{zot18b}, \cite{sur18b}),  or even  the Sitnikov Problem in three and four body problem (e.g., \cite{dou12}, \cite{zot18c, zot18d}).

The present paper is described with following structure: the basic properties of the model of axisymmetric five-body problem are presented in
 Sect.\ref{Properties of the dynamical system}. In Sect.  \ref{The libration points of the system}, we illustrate the parametric evolution of the libration points of the system.
  The following section deals with the Newton-Raphson basins of convergence where the numerical outcomes are presented. The paper ends with Sect. \ref{Concluding remarks} where the outcomes of the research are provided.
%%%%%%
\begin{figure*}[!t]
\centering
(a)\includegraphics[scale=.6]{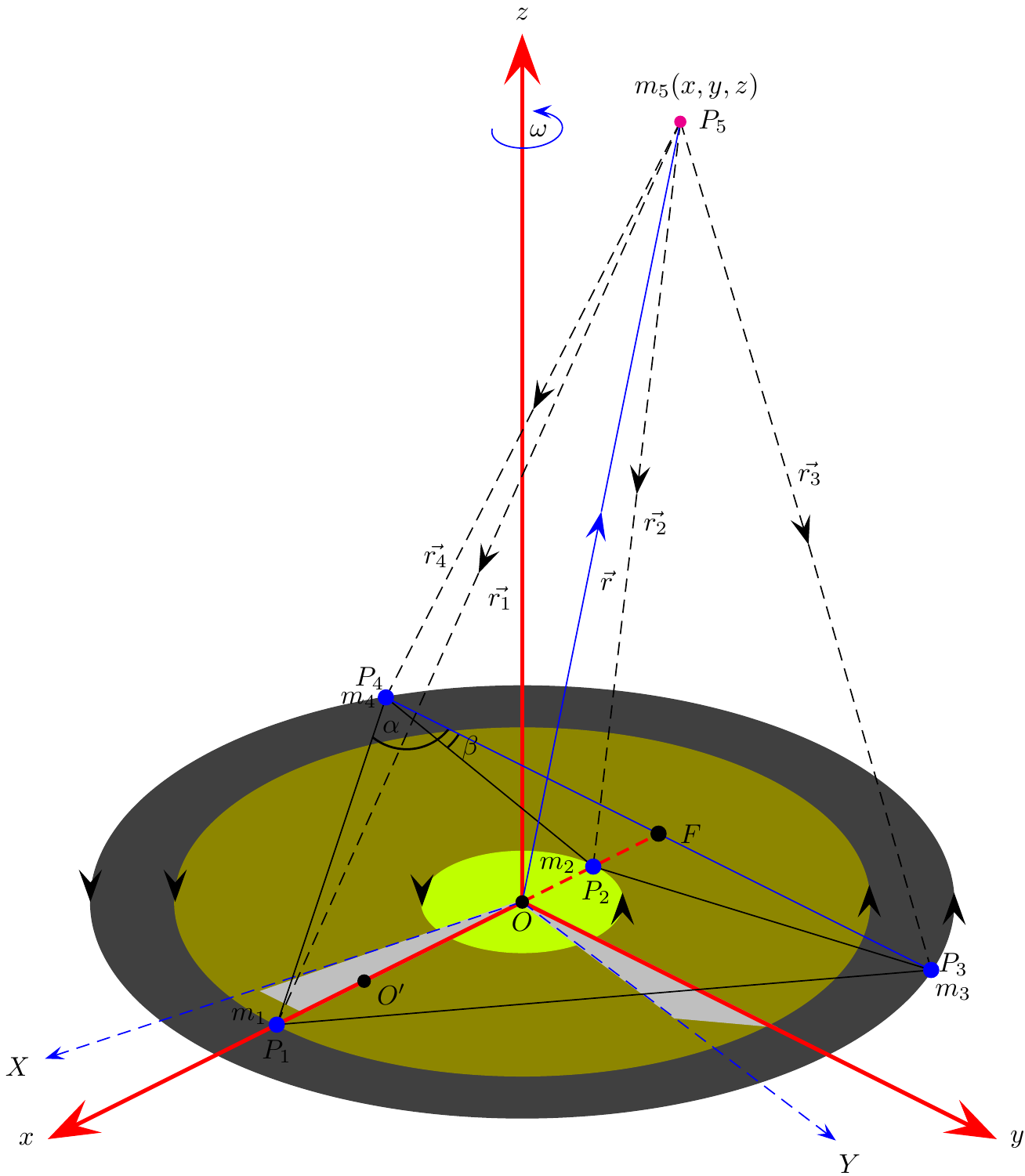}
(b)\includegraphics[scale=.6]{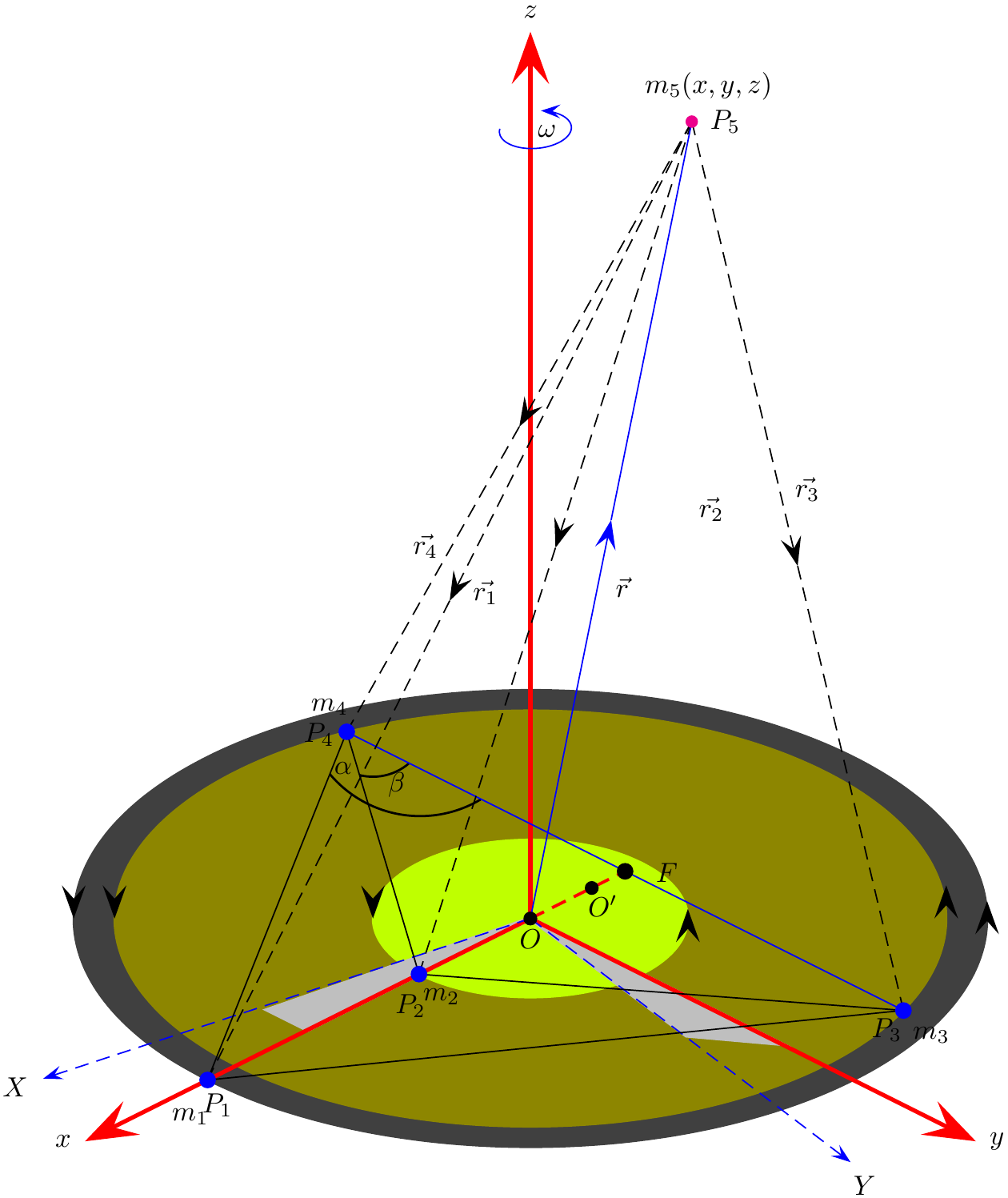}
\caption{The configurations of the axisymmetric five-body problem: (a) first concave case, (b) second concave case, with the synodic coordinate system $Oxyz$ and the inertial frame $OXYZ$. More details are given in the text. (Color figure online).}
\label{Fig:1}
\end{figure*}
%%%%
\section{Properties of the dynamical system}
\label{Properties of the dynamical system}
%%%%%
%\begin{figure}[!tH]
%\centering
%\resizebox{\hsize}{!}{\includegraphics{graph_lines}}
%\caption{ (Color figure online).}
%\label{Fig:2}
%\end{figure}
%%%%%
The dynamical system investigated in the paper mainly contains four primary bodies $P_i$, $i=1, 2, ..., 4$, which move, in circular orbits around their common center of mass. A fifth body with infinitesimal mass in comparison of mass of the primaries, always referred as test particle, is moving under the gravitational pull of these primaries whereas its motion does not influence the orbits of any primaries.
 To study the motion of infinitesimal mass is always referred as the restricted five-body problem.
%%%%%
\begin{figure*}[!t]
\centering
\resizebox{\hsize}{!}{\includegraphics{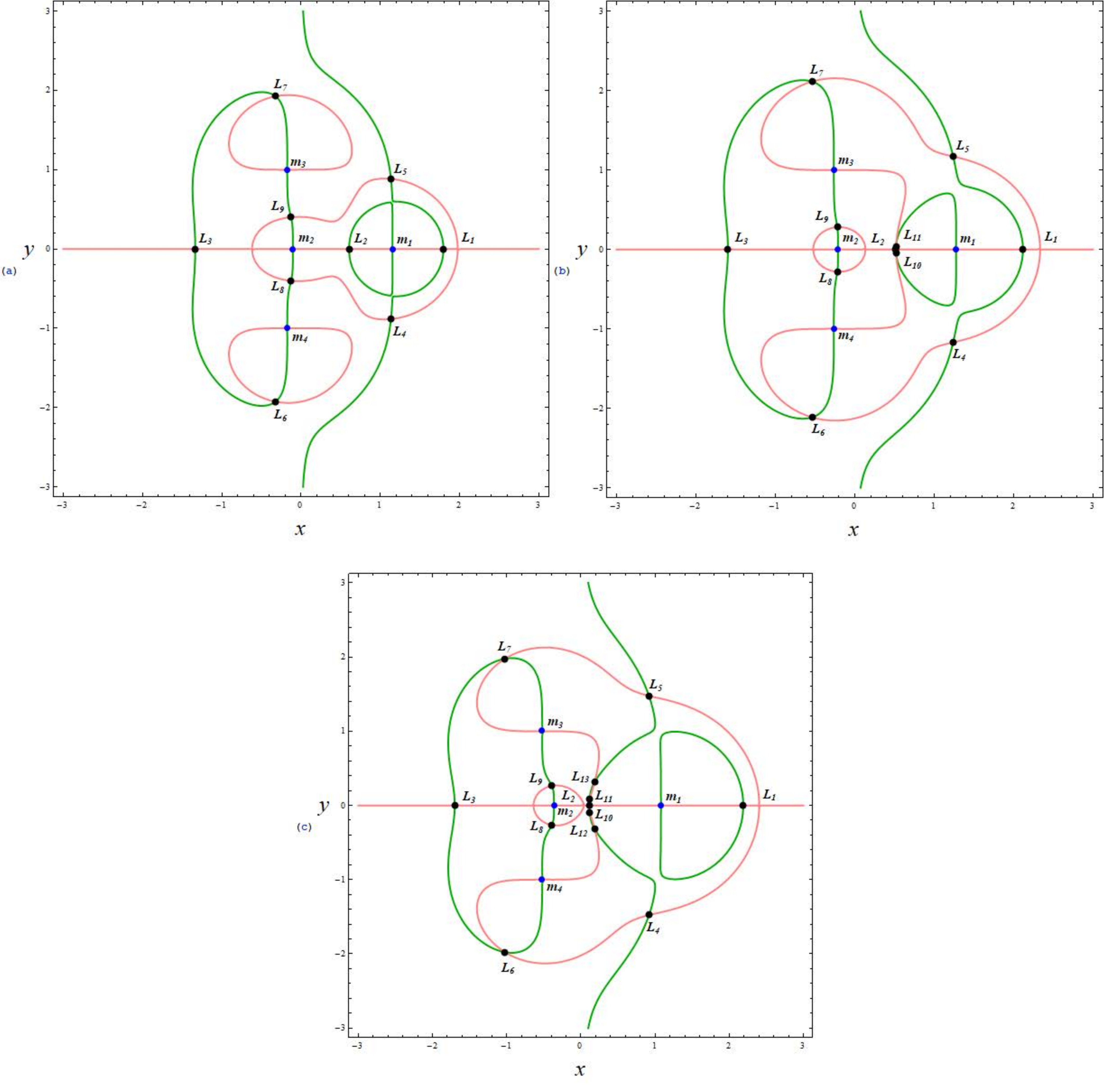}}%\includegraphics{Fig3Comb.pdf}
\caption{Positions (black dots) and numbering of the libration points ($L_i; i = 1,2,...9$ or $11$ or $13$) in first concave case through the intersections of $\Omega_x= 0$ (green) and $\Omega_y= 0$ (pink), when (a-first row, left): $\alpha=53\degree, \beta=4\degree$ (nine libration points), (b-first row, right): $\alpha=57\degree, \beta=3\degree$ (eleven libration points), and (c-second row) $\alpha=58\degree, \beta=9\degree$ (thirteen libration points). The blue dots denote the centers $(P_i, i = 1, 2, 3, 4)$ of the primaries. (Color figure online).}
\label{Fig:3}
\end{figure*}
%%%%%%%%%%
\begin{figure*}[!t]
\centering
\resizebox{\hsize}{!}{\includegraphics{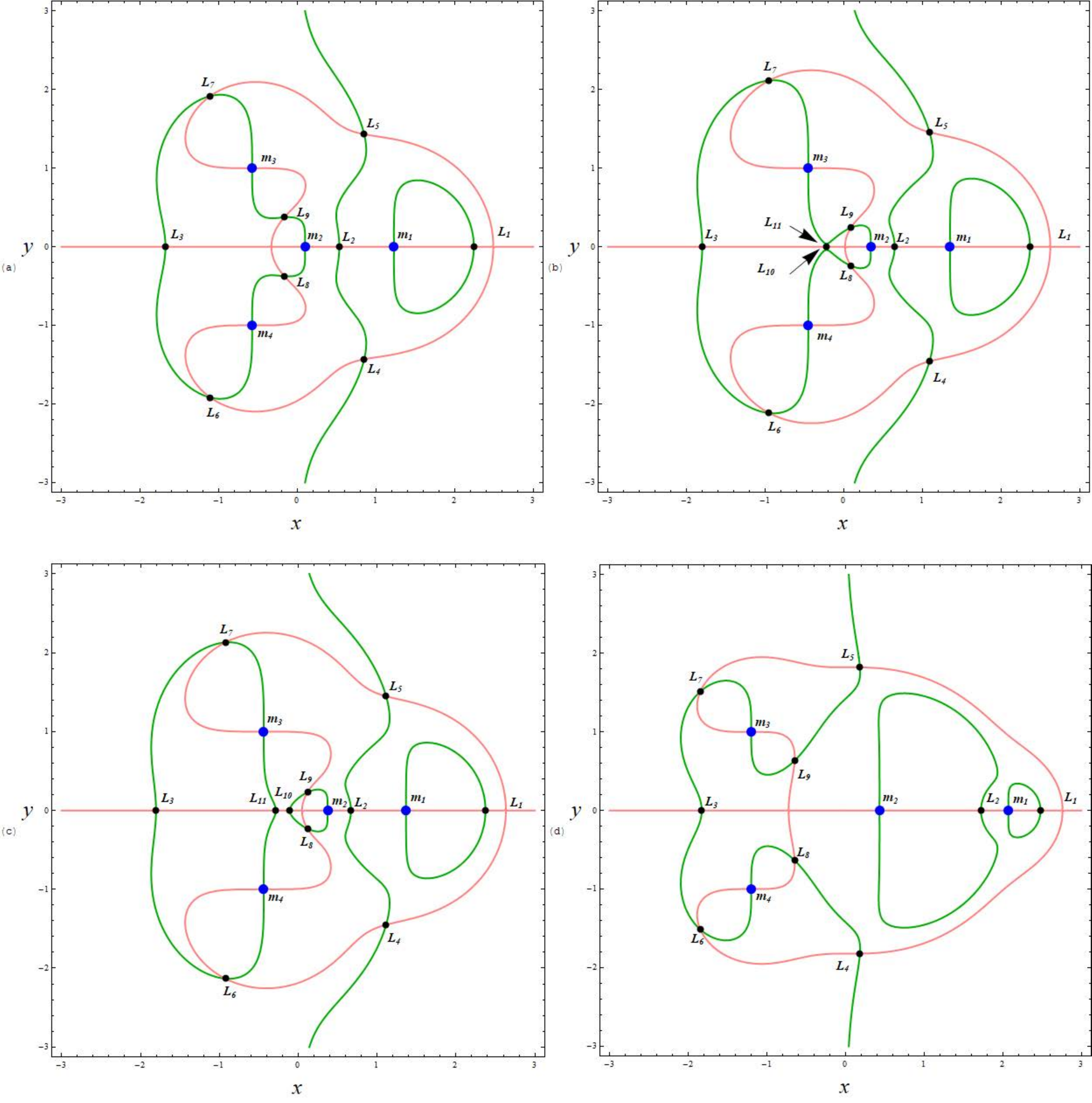}}%\includegraphics{Fig3Comb.pdf}
\caption{Positions (black dots) and numbering of the libration points ($L_i; i = 1,2,...9$ or $11$) in second concave case through the intersections of $\Omega_x= 0$ (green) and $\Omega_y= 0$ (pink), when (a-first row, left): $\alpha=61\degree, \beta=34\degree$ (nine libration points), (b-first row, right): $\alpha=61\degree, \beta=38.5\degree$, (c-second row, left): $\alpha=61\degree, \beta=39\degree$   (eleven libration points), and (d-second row, right) $\alpha=73\degree, \beta=58.5\degree$ (nine libration points). The blue dots denote the centers $(P_i, i = 1, 2, 3, 4)$ of the primaries. (Color figure online).}
\label{Fig:4}
\end{figure*}
%%%%
The primaries $P_i$ with masses $m_i$ and coordinates $(x_i, y_i, z_i)$ are taken in planar reference plane where the line joining the center of the primaries $m_1$ and $m_2$ is
 $Ox$-axis while the line perpendicular to $Ox$-axis and passes through the center of mass of the four-body system with co-ordinate $(0, g, 0)$ is considered as $y-$axis. In particular, when $g=0, y_{3,4}\neq 0$, the center of mass coincides with the origin of the coordinate system and the $Ox-$axis remains the axis of symmetry.
 Therefore, we obtain a symmetric configuration where two equal masses $m_{3,4}$ are placed opposite to each other and are symmetrical about the $Ox$-axis.  In the present scenario, it is assumed that $(m_2 \geq m_1)$, whereas $(m_1 \geq m_2)$ is the mirror configuration.
%%%

According to \cite{erd16}, there are three possible configurations of four bodies with an axis of symmetry passing through the center of two bodies ($P_1$ and $P_2$), whereas the two other bodies of equal masses ($P_3$ and $P_4$) are situated symmetrically with respect to this axis. Furthermore, the point $F$ located on the axis of symmetry, is the middle point of line joining the center of primaries $P_3$ and $P_4$. In addition, let us take $O'$ the center of mass of $P_1, P_3$ and $P_4$ and $O$ is the center of mass of the entire system. On the basis of the position of the points $F$, $O$, and $O'$, the following configurations are possible: (a) the convex configuration: when the point $F$ lies between the primaries $P_1$  and $P_2$. (b) The concave configuration: (i) first concave configuration: the primary $P_2$ lies between $F$ and $O' $, (see Fig. \ref{Fig:1}a) and (ii) second concave configuration: the primary $P_2$ lies between $O'$ and $P_1$ (see Fig. \ref{Fig:1}b).

In the present paper, we have considered only the second case, i.e., the concave configuration. According to \cite{gao17}, the time-independent effective potential of the axisymmetric five-body problem, in a synodic coordinates system $Oxyz$ is represented by following expression:
\begin{equation}\label{Eq:1}
\Omega(x, y, z)=\frac{1}{\Delta}\sum_{i=1}^{4}\frac{m_i}{r_i}+\frac{1}{2}\big(x^{2}+y^{2}\big),
\end{equation}
where $(x,y,z)$ are the coordinates of the test particle, while $r_i$ are the distances of the test particle from the primaries $P_i$, respectively and
\begin{align*}
r_{i}&=\sqrt{(x-x_i)^{2}+(y-y_i)^{2}+(z-z_i)^{2}},\\
x_1&=a, x_2=b, x_3=x_4=c, \\
y_1&=y_2=0, y_3=-y_4=1,\\
z_i&=0, (i=1,2,...,4).
\end{align*}
and the values of the remaining parameters are presented in \ref{Appendix}. Since, we have considered the concave case only, the angle coordinates $\alpha$ and $\beta$ must satisfy the following inequalities:
\begin{itemize}
\item First concave case:
\begin{align*}
2\alpha -\beta &>90\degree, \  \alpha < 60\degree, \  \alpha >0\degree.
\end{align*}
\item Second concave case:
\begin{align*}
2\alpha -\beta &<90\degree, \  \alpha > 60\degree, \  \beta <60\degree.
\end{align*}
\end{itemize}
Applying the transformation from the inertial to the synodic coordinate system, while scaling the physical quantities, the motion of the test particle in the rotating frame of reference is governed by the following equations:
\begin{subequations}
\begin{eqnarray}
\label{Eq:2a}
\ddot{x}-2\dot{y}&=&\Omega_x,\\
\label{Eq:2b}
\ddot{y}+2\dot{x}&=&\Omega_y,\\
\label{Eq:2c}
\ddot{z}&=&\Omega_z.
\end{eqnarray}
\end{subequations}
The dynamical system where the motion is governed by the equations (\ref{Eq:2a}-\ref{Eq:2c}) admits only one integral of motion (also called  Jacobi integral). The corresponding Hamiltonian is
\begin{equation}\label{Eq:3}
 J(x,y,z,\dot{x}, \dot{y}, \dot{z})=2\Omega(x,y,z)-(\dot{x}^2+\dot{y}^2+\dot{z}^2)=C,
\end{equation}
where $\dot{x}, \dot{y}$, and $\dot{z}$ are of course the velocities, whereas the numerical value of Jacobian constant is represented by $C$  and it remains conserved.
\section{The libration points of the system}
\label{The libration points of the system}
In the following section, we will determine how the angle parameters influence the dynamical attributes of the coplanar libration points.

The necessary and sufficient conditions,  which must be satisfied to determine and depict the positions of libration points,  are:
\begin{equation}\label{Eq:4}
  \dot{x}=\dot{y}=\dot{z}=0, \ddot{x}=\ddot{y}=\ddot{z}=0.
\end{equation}
The required  coordinates of the libration points can be evaluated by solving the following system of the first order partial  derivatives, numerically:
\begin{equation}\label{Eq:5}
  \Omega_x=0, \Omega_y=0, \Omega_z=0,
\end{equation}
where
\begin{subequations}
\begin{eqnarray}
 % \nonumber % Remove numbering (before each equation)
\label{Eq:6a}
\Omega_x&=&x-\frac{1}{\Delta}\sum _{i=1}^{4}\frac{m_i\tilde{x_i}}{r_i^{3}},\\
\label{Eq:6b}
\Omega_y&=&y-\frac{1}{\Delta}\sum _{i=1}^{4}\frac{m_i\tilde{y_i}}{r_i^{3}},\\
\label{Eq:6c}
\Omega_z&=&-\frac{1}{\Delta}\sum _{i=1}^{4}\frac{m_i\tilde{z_i}}{r_i^{3}},
\end{eqnarray}
\end{subequations}
$\tilde{x_i}=x-x_i, \tilde{y_i}=y-y_i$, and $\tilde{z_i}=z$ with $i=1,2,...,4$.

The intersections of the nonlinear equations $\Omega_x=0$,  and $\Omega_y=0$ illustrate the positions of the libration points on the $(x, y)$  plane (see Fig. \ref{Fig:3}, \ref{Fig:4}).
\subsection{The first concave case}
We start our analysis with the first concave case where we have considered three set of values of the angle parameter $\alpha$  and accordingly the values of the angle parameter $\beta$ are as follows:
\begin{enumerate}
  \item When $\alpha$ is $53\degree$,
  \begin{itemize}
    \item [-]$\beta \in (0\degree, 16\degree)$, there exist 9 libration points, in which 3 are collinear and 6 are non-collinear.
  \end{itemize}
  \item When $\alpha$ is $57\degree$,
  \begin{itemize}
    \item [-]$\beta \in (0\degree, 2.909\degree]$, there exist 9 libration points, in which 3 are collinear and 6 are non-collinear.
    \item [-]$\beta \in [2.910\degree, 4.491\degree]$, there exist 11 libration points, in which 3 are collinear and 8 are non-collinear.
    \item [-]$\beta \in [4.492\degree, 24\degree)$, there exist 9 libration points, in which 3 are collinear and 6 are non-collinear.
  \end{itemize}
  \item When $\alpha$ is $58\degree$,
  \begin{itemize}
    \item [-]$\beta \in (0\degree, 1.654\degree]$, there exist 9 libration points, in which 3 are collinear and 6 are non-collinear.
    \item [-]$\beta \in [1.655\degree, 8.740\degree]$, there exist 11 libration points, in which 3 are collinear and 8 are non-collinear.
    \item [-]$\beta \in [8.741\degree, 10.001\degree]$, there exist 13 libration points, in which 3 are collinear and 10 are non-collinear.
    \item [-]$\beta \in [10.002\degree, 26\degree)$, there exist 9 libration points, in which 3 are collinear and 6 are non-collinear.
  \end{itemize}
\end{enumerate}

\begin{figure*}[!t]
\centering
\resizebox{\hsize}{!}{\includegraphics{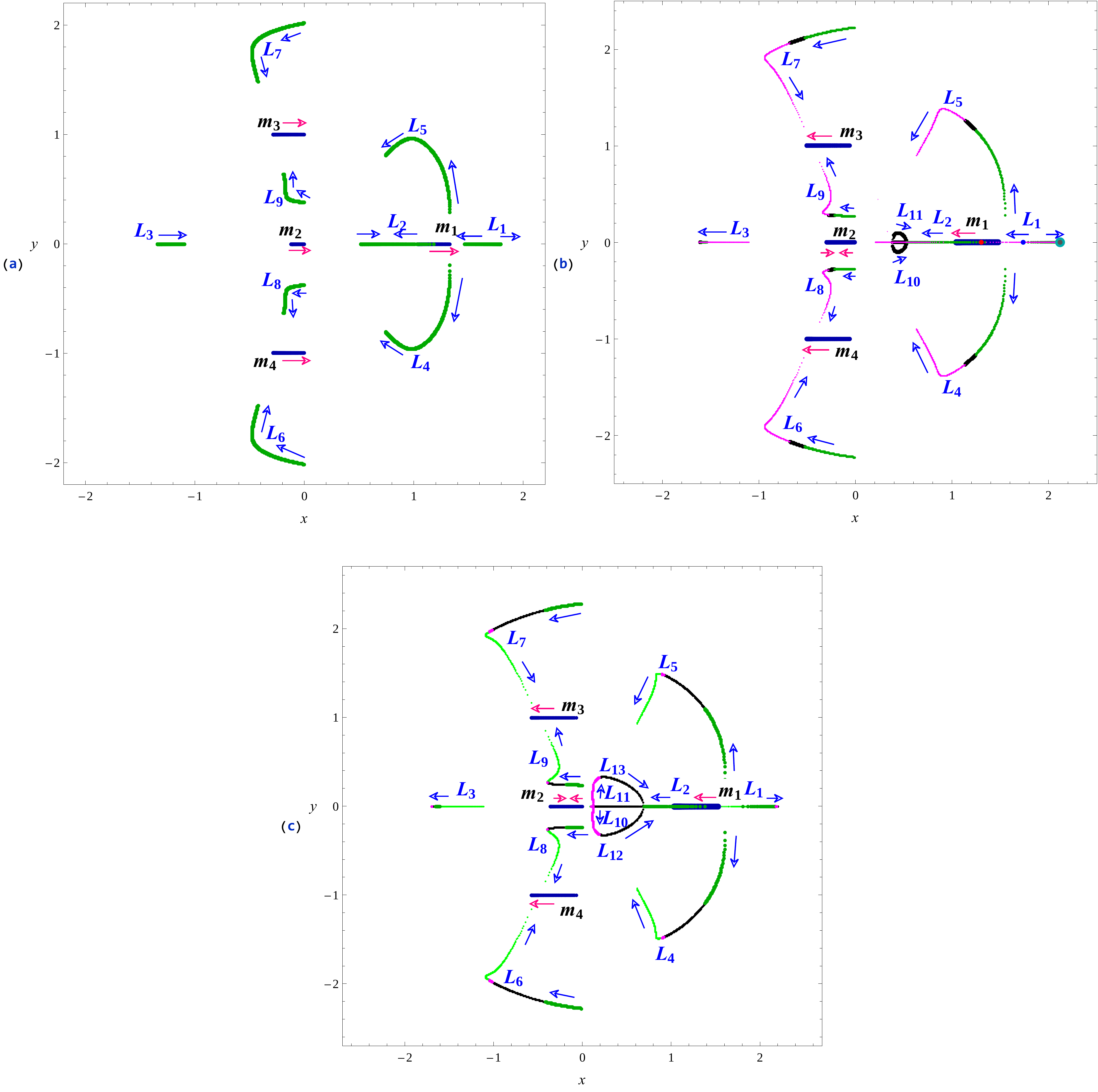}}%\includegraphics{Fig3Comb.pdf}
\caption{The parametric evolution of the positions of the libration points, $L_i, i = 1, 2,..., 9$ or $11$ or $13$, in the axisymmetric five-body problem: first concave case, (a): for $\alpha=53\degree$, and $\beta\in(0\degree, 16\degree)$,  (nine libration points, dark green), (b): for $\alpha=57\degree$, with $\beta\in(0\degree, 2.909\degree]$ (nine libration points, dark green), $\beta\in[2.910\degree, 4.491\degree]$ (eleven libration points, black), and $\beta\in[4.492\degree, 24\degree)$ (nine libration points, magenta), and (c): for $\alpha=58\degree$, with $\beta\in(0\degree, 1.654\degree]$ (nine libration points, dark green), $\beta\in[1.655\degree, 8.740\degree]$ (eleven libration points, black), $\beta\in[8.741\degree, 10.001\degree)$ (thirteen libration points, magenta) and $\beta\in[10.002\degree, 26\degree)$ (nine libration points, green). The blue dot, gray dot, teal colour big dot, and red dots show the positions of libration point $L_1$ for $\beta=0.033\degree, 2.909\degree, 4.492\degree$ and $23.966\degree$ respectively. The blue and magenta arrows indicate the movement direction of the libration points and primaries, respectively, as the value of the angle parameter $\beta$ increases. The navy blue straight line pinpoints the fixed centers of the primaries for particular value of the angle parameter $\beta$. (Color figure online).}
\label{Fig:5}
\end{figure*}

The evolution of the positions of the libration points in the first concave case are depicted in Fig. \ref{Fig:5}, where three set of the angle parameter $\alpha$ are considered for it. In Fig. \ref{Fig:5}a, the parametric evolution for $\alpha=53\degree$  is presented for the permissible value of $\beta\in (0\degree, 16\degree)$ in which nine libration points exist. It is observed that four libration points namely $L_{1,2,4,5}$ emerged in the vicinity of the primary $m_1$ where as the libration points $L_{6,8}$ and $L_{7,9}$ exist in the vicinity of primaries $m_4$ and $m_3$ respectively, and move towards them. The position of the libration point $L_3$ move along $x-$axis towards $m_2$. Moreover, the position of libration points $L_{1,2}$ first move from left to right and then it start to move from right to left. It is also observed that the positions of the primaries are not fixed on the contrary these positions vary with the increasing value of angle parameter $\beta$. The positions of the primaries move from left to right along the straight line.

In Fig. \ref{Fig:5}b, the parametric evolution of the positions of libration points are illustrated for $\alpha=57\degree$, where the number of  libration points are either nine or eleven in different intervals of the angle parameter $\beta$. The value $\beta^*= 2.909\degree$, and $4.491\degree$ are the critical values of the angle parameter $\beta$, since they demarcate the point where the number of libration points changes. The movement of the positions of nine libration points shown in dark green line, for $\beta \in (0\degree, 2.909\degree]$ are same as in previous panel except $L_3$ which get reversed in direction whereas for $\beta \in [2.910\degree, 4.491\degree]$ when eleven libration points exist, shown in black line, the nine libration points among them continue their movement and two new points $L_{10,11}$ originate in the vicinity of the libration point $L_2$ and move along the circular arc which further annihilate in $L_2$ for $\beta=4.492\degree$. For $\beta \in [4.492\degree, 24\degree]$, nine libration points exist, shown in magenta line maintain their movement. It is interesting to note that the movement of the positions of the primaries are reversed i.e., now they move from right to left. But the movement of the position of the primary $m_2$  again changes and it move from left to right for $\beta=8.405\degree$. The movement of the position of the libration point $L_1$ is left to right when $\beta \in(0\degree, 4.491\degree]$ whereas its movement reversed in direction when  $\beta \in[4.492\degree, 24\degree)$.

The movement of the positions of libration point are depicted in Fig. \ref{Fig:5}c, for $\alpha=58\degree$  where the permissible range of  $\beta \in (0\degree, 26\degree)$. In this interval of $\beta$, the number of libration points are either 9, 11 or 13. The movement of the libration points $L_{2,...,11}$ are same as in previous panel whereas the direction of libration point $L_1$ is left to right. Two new libration points $L_{12, 13}$ exist for very small interval and the direction of the movement of these libration points are same as $L_{10, 11}$ of the previous panel respectively.
\subsection{The second concave case}
\begin{figure*}[!t]
\centering
\resizebox{\hsize}{!}{\includegraphics{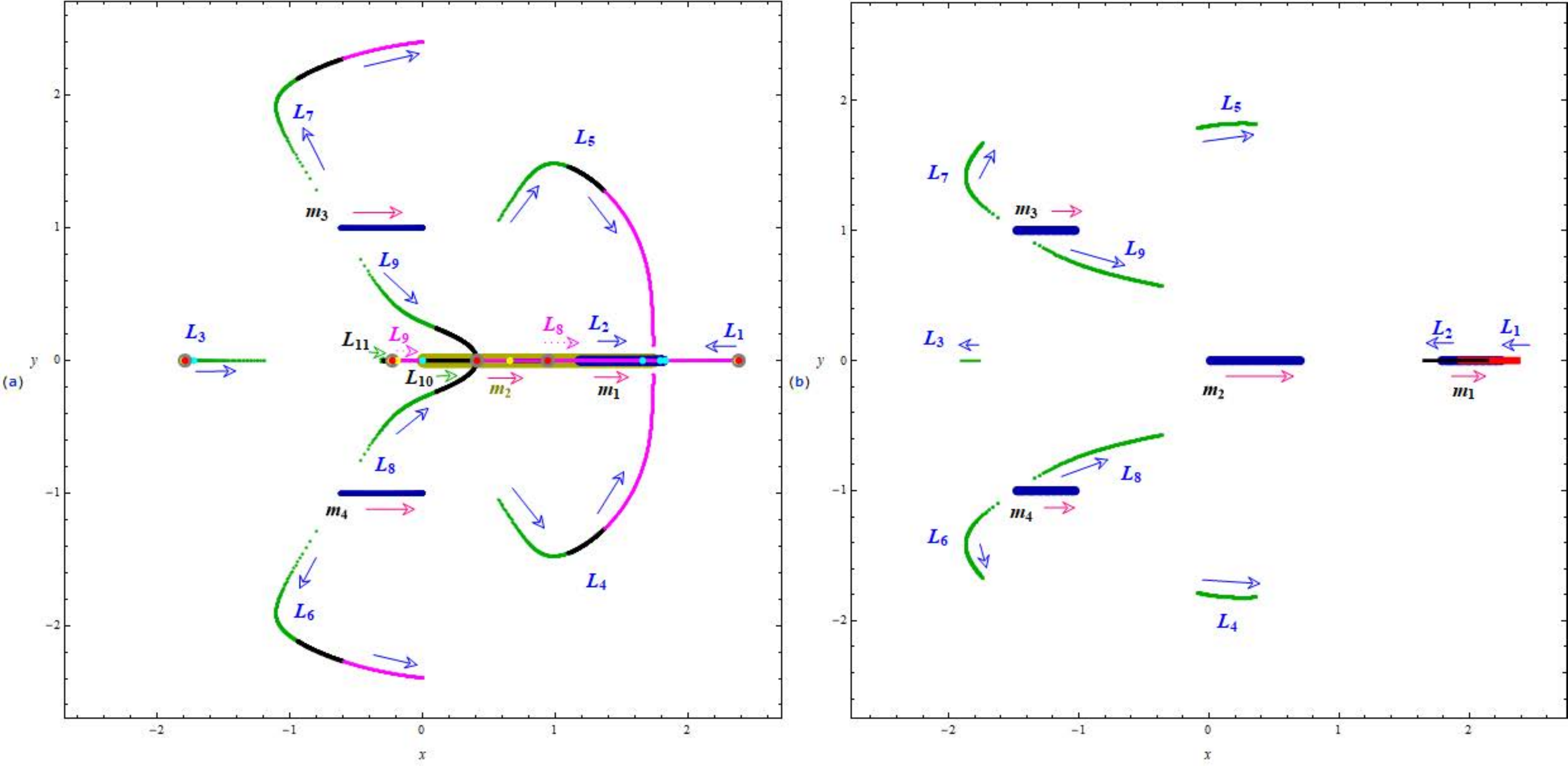}}
\caption{The parametric evolution of the positions of the libration points, $L_i, i = 1, 2,..., 9$ or $11$, in the axisymmetric five-body problem: second concave case, (a): for $\alpha=61, \degree$, with $\beta\in(32\degree, 38.568\degree]$ (nine libration points, dark green), $\beta\in[38.569\degree, 44.402\degree]$ (eleven libration points, \emph{black}), and $\beta\in[44.403\degree, 60\degree)$ (nine libration points, \emph{magenta}). The  \emph{yellow, red, gray} and \emph{cyan} dots show the five collinear libration points for $\beta= 38.569\degree, 44.402\degree, 44.403\degree$, and $59.966\degree$ respectively where as the thick "\emph{Olive}" line shows the position of primary $m_2$ and (b): for $\alpha=73\degree$, and $\beta\in(56\degree, 60\degree)$,  (nine libration points, \emph{dark green}), the positions of $L_{1,2}$ are shown in red and black line respectively. The \emph{blue} and \emph{magenta} arrows indicate the movement direction of the libration points and primaries, respectively, as the value of the angle parameter $\beta$ increases. The \emph{navy blue} straight line pinpoint the fixed centers of the primaries for particular value of the angle parameter $\beta$. (Color figure online).}
\label{Fig:6}
\end{figure*}

We continue our analysis with the second concave case for two set of values of the angle parameter $\alpha$  and corresponding value of the angle parameter $\beta$ are as follows:
\begin{enumerate}
  \item When $\alpha$ is $61\degree$,
\begin{itemize}
    \item [-]$\beta \in (32\degree, 38.568\degree]$, there exist 9 libration points, in which 3 are collinear and 6 are non-collinear.
    \item [-]$\beta \in [38.569\degree, 44.402\degree]$, there exist 11 libration points, in which 5 are collinear and 6 are non-collinear.
    \item [-]$\beta \in [44.403\degree, 60\degree)$, there exist 9 libration points, in which 5 are collinear and 4 are non-collinear.
  \end{itemize}
  \item When $\alpha$ is $73\degree$,
  \begin{itemize}
    \item [-]$\beta \in (56\degree, 60\degree)$, there exist 9 libration points, in which 3 are collinear and 6 are non-collinear.
  \end{itemize}
\end{enumerate}
The movement of the positions of the libration points in the second concave case is illustrated in Fig. (\ref{Fig:6}). The Fig. (\ref{Fig:6})a is depicted for the case when $\alpha=61\degree$ where the permissible range of $\beta \in(32\degree, 60\degree)$ in which 9 or 11 libration points exist.

The value $\beta^*\approx 38.5685 \degree$, and  $44.4025\degree$ are the critical values of the angle parameter $\beta$, since at these values the number of the libration points changes from $9$ to $11$, and again from $11$ to $9$ respectively. The movement of the positions of nine libration points shown in dark green line, for $\beta \in (32\degree, 38.568\degree]$. It is observed that the positions of libration points $L_{2, 3}$ move from left to right while the position of libration point $L_1$ moves right to left. Moreover, the libration points $L_{7, 9}$ and $L_{6, 8}$ move away from the primaries $m_3$ and $m_4$ respectively while the libration points $L_{4,5}$ move towards the primary $m_1$. For the angle parameter $\beta \in[38.569\degree, 44.402\degree]$, there exist eleven libration points shown in black line. The movement of the positions of libration points $L_{1,2,...,9}$ are same as in previous interval of the angle parameter $\beta$. In addition, two new collinear libration points namely $L_{10, 11}$ exist in either side of the primary $m_2$ and move with it from left to right. Moreover, the libration points $L_{8, 9, 10}$ annihilate for $\beta=44.4025\degree$ and further continue as $L_8$ in new interval of angle parameter $\beta \in [44.403\degree, 60\degree)$ which moves from left to right while the libration point $L_{11}$ continues as $L_9$ in it.

The Fig. (\ref{Fig:6})b is depicted for the case when $\alpha=73\degree$ and the permissible range of $\beta \in(56\degree, 60\degree)$ in which 9 libration points exist. We have observed that three libration points $L_{1,2,3}$ move from right to left along the $x-$axis. Moreover, the positions of the primaries are not fixed rather they move from left to right as the value of angle parameter $\beta$ increases. We have also noticed that the libration points $L_{5, 7, 9}$ and $L_{4, 6, 8}$ move away from the primaries $m_{3, 4}$ respectively opposite to the first concave case.

%%%%%%
%%%%
%%%%
\section{The basins of attraction}
\label{The basins of attraction}
Undoubtedly, the Newton-Raphson iterative method is one of the well known method to solve the system of nonlinear equations. Applying the philosophy and technique used in the recent study by \cite{sur17b}, \cite{sur18a}, \cite{zot16, zot17a, zot17b, zot17c, zot18a}, the well known Newton-Raphson basins of convergence or  basins of  attraction or even attracting domains are composed of  sets of the initial conditions which lead to the same attractor are analyzed.
\begin{figure*}[!t]
\centering
(a)\includegraphics[scale=.35]{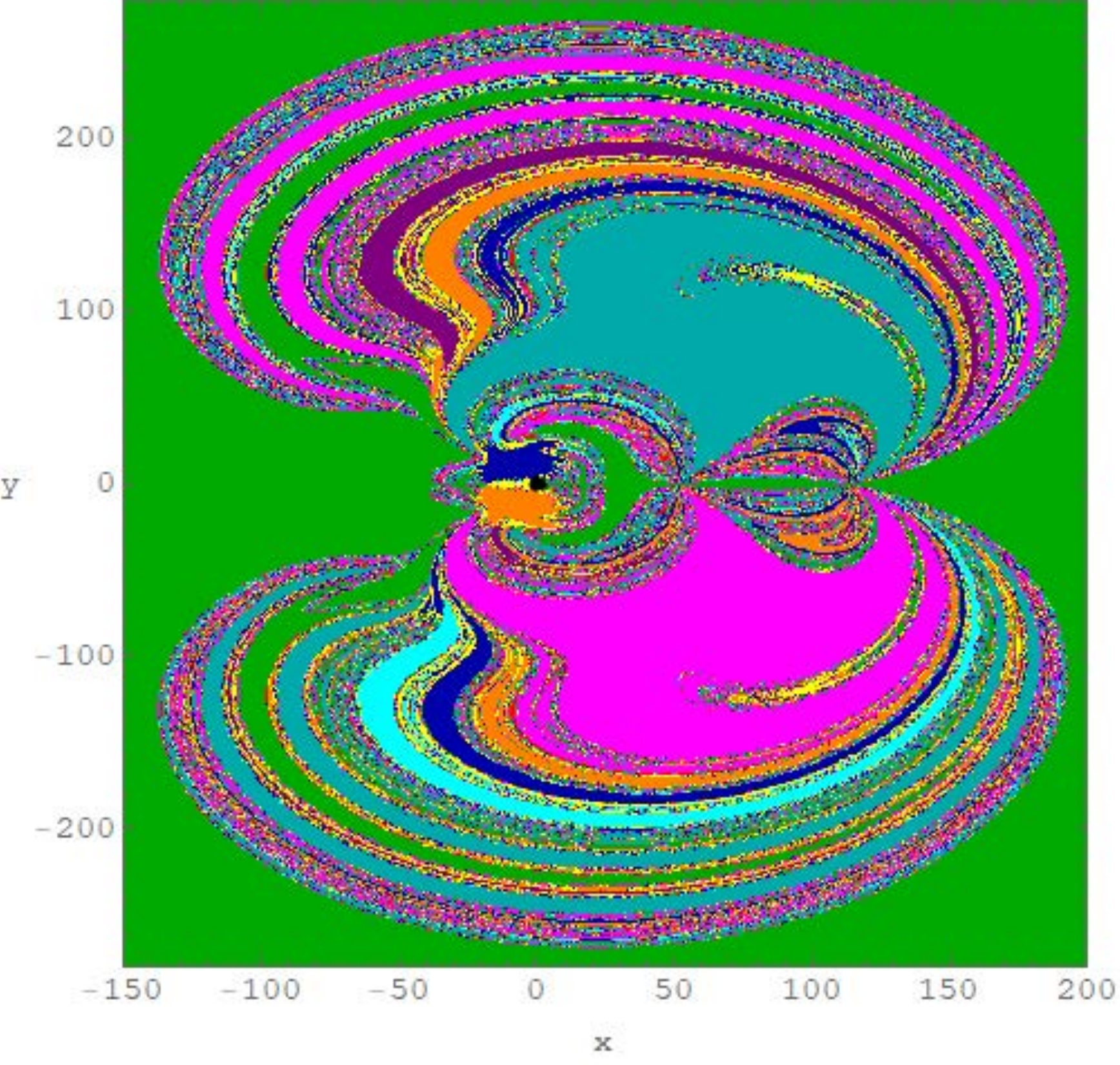}
(b)\includegraphics[scale=.35]{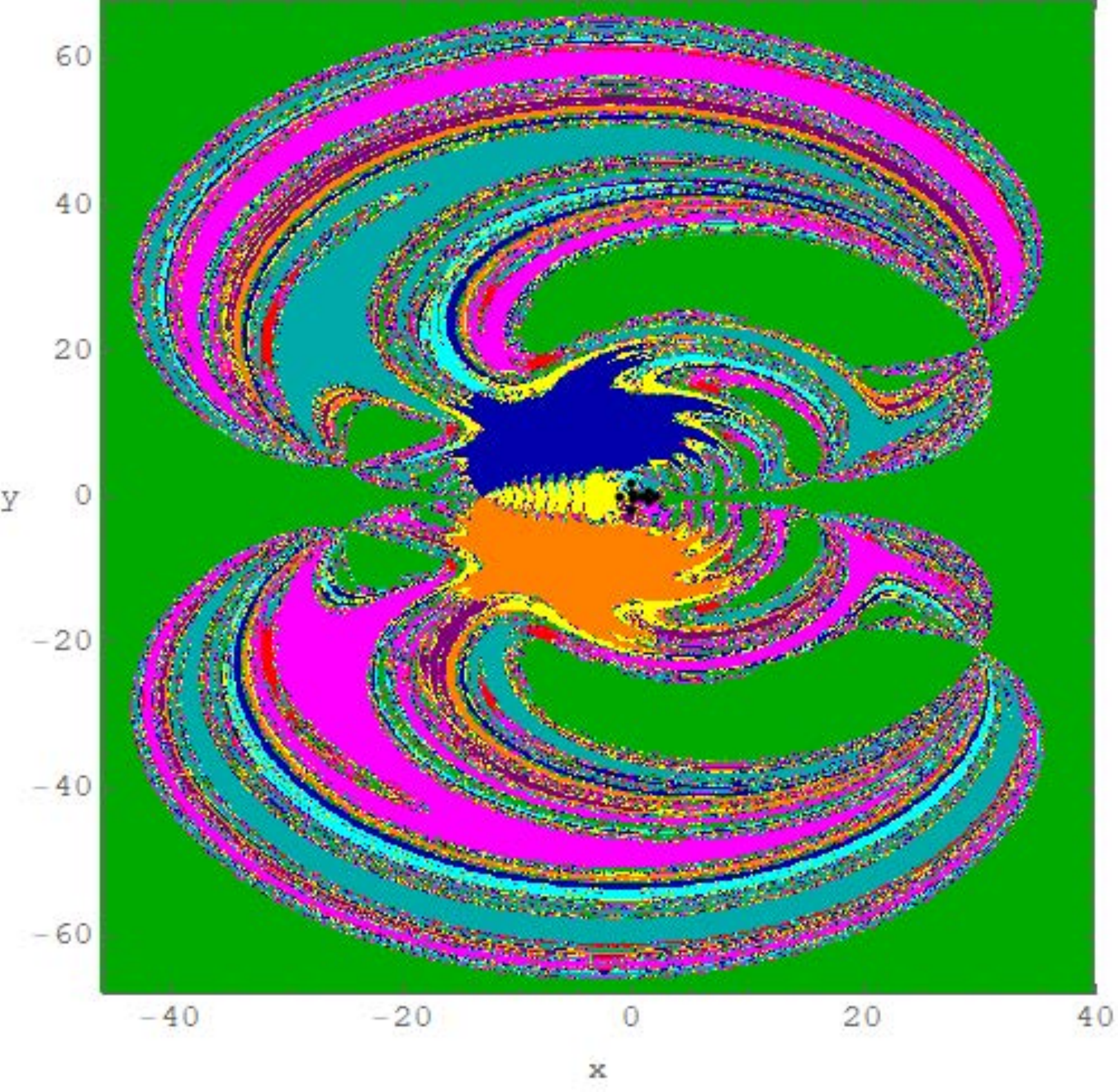}\\
(c)\includegraphics[scale=.35]{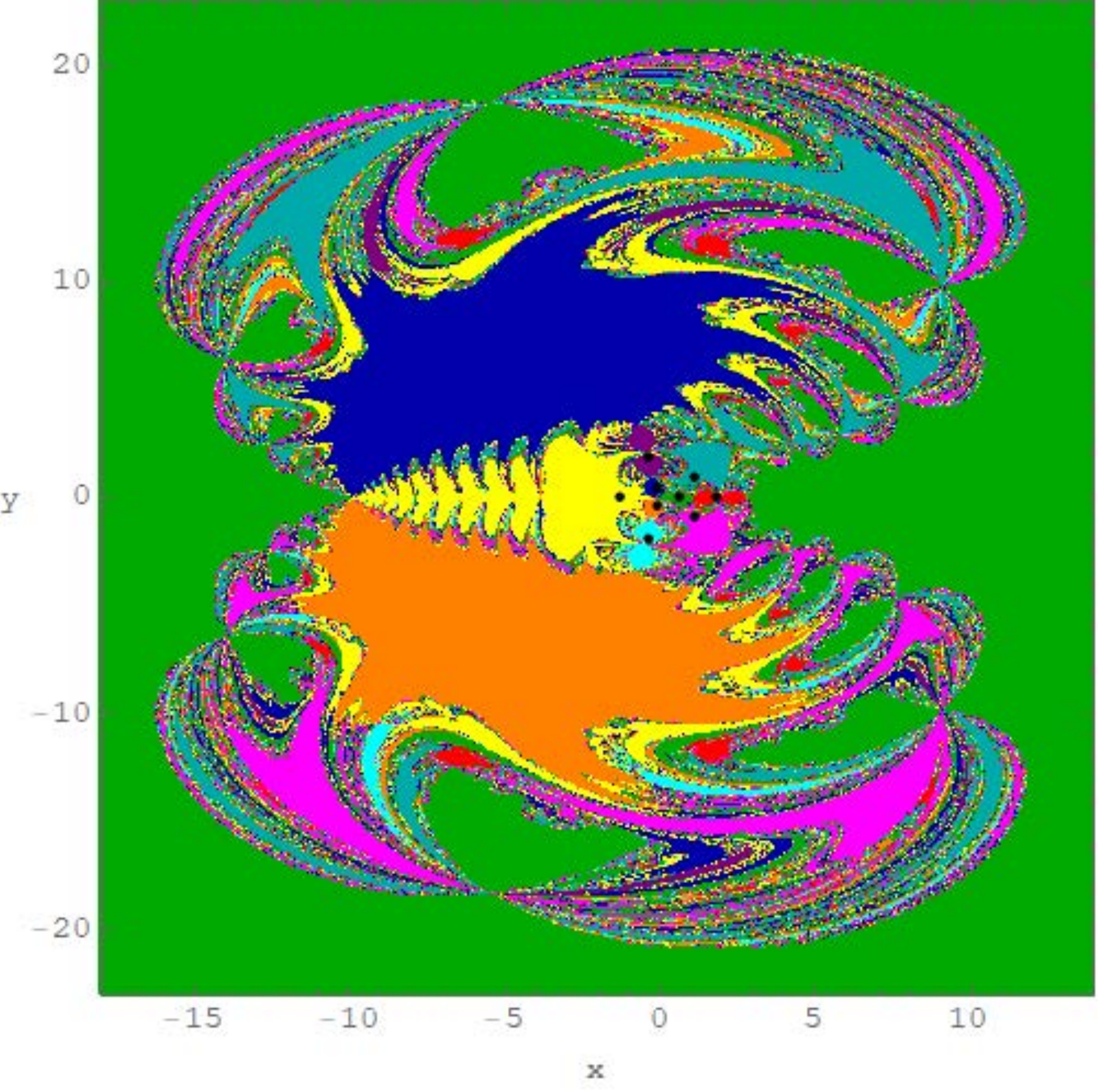}
(d)\includegraphics[scale=.35]{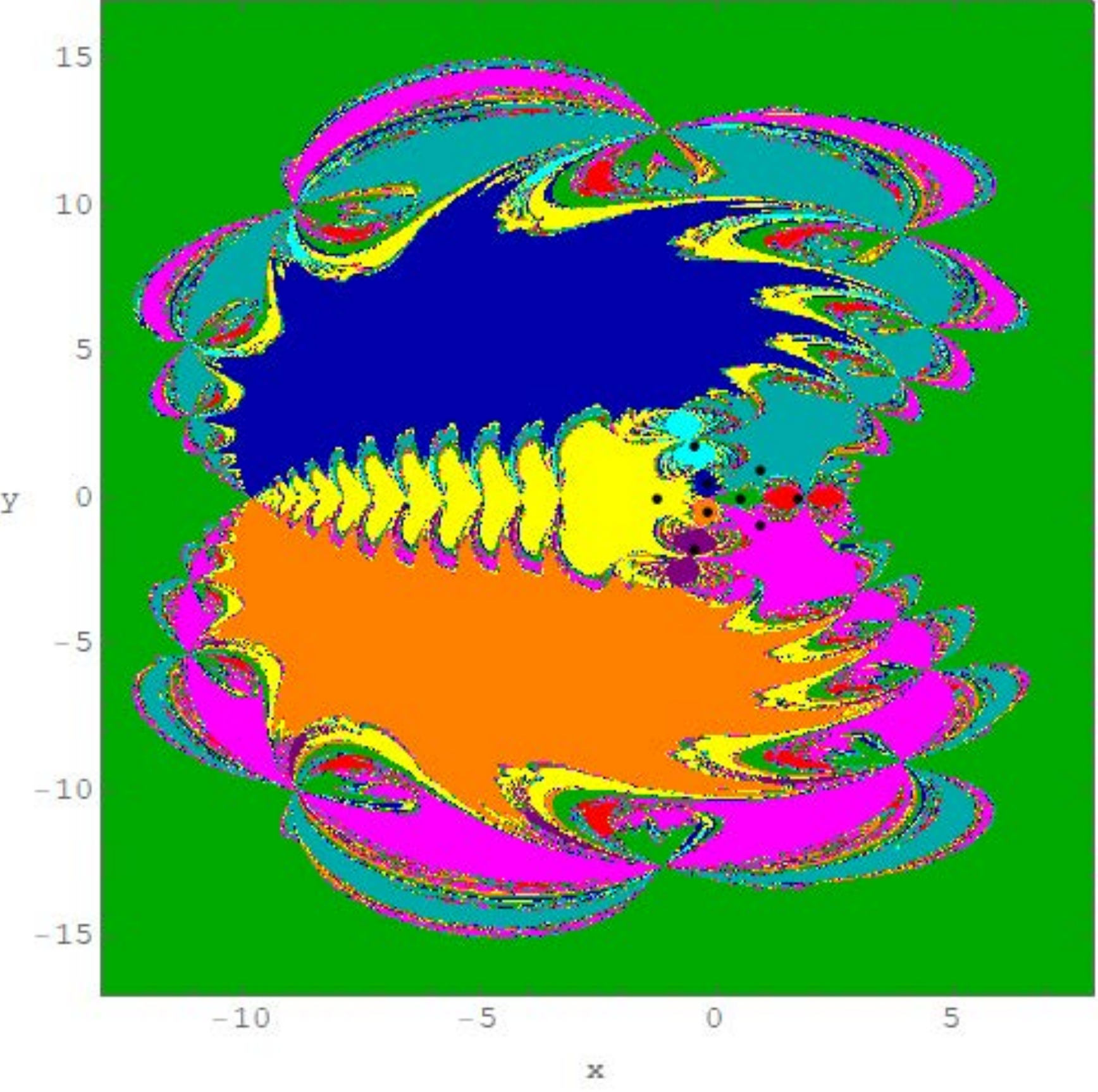}
(e)\includegraphics[scale=.35]{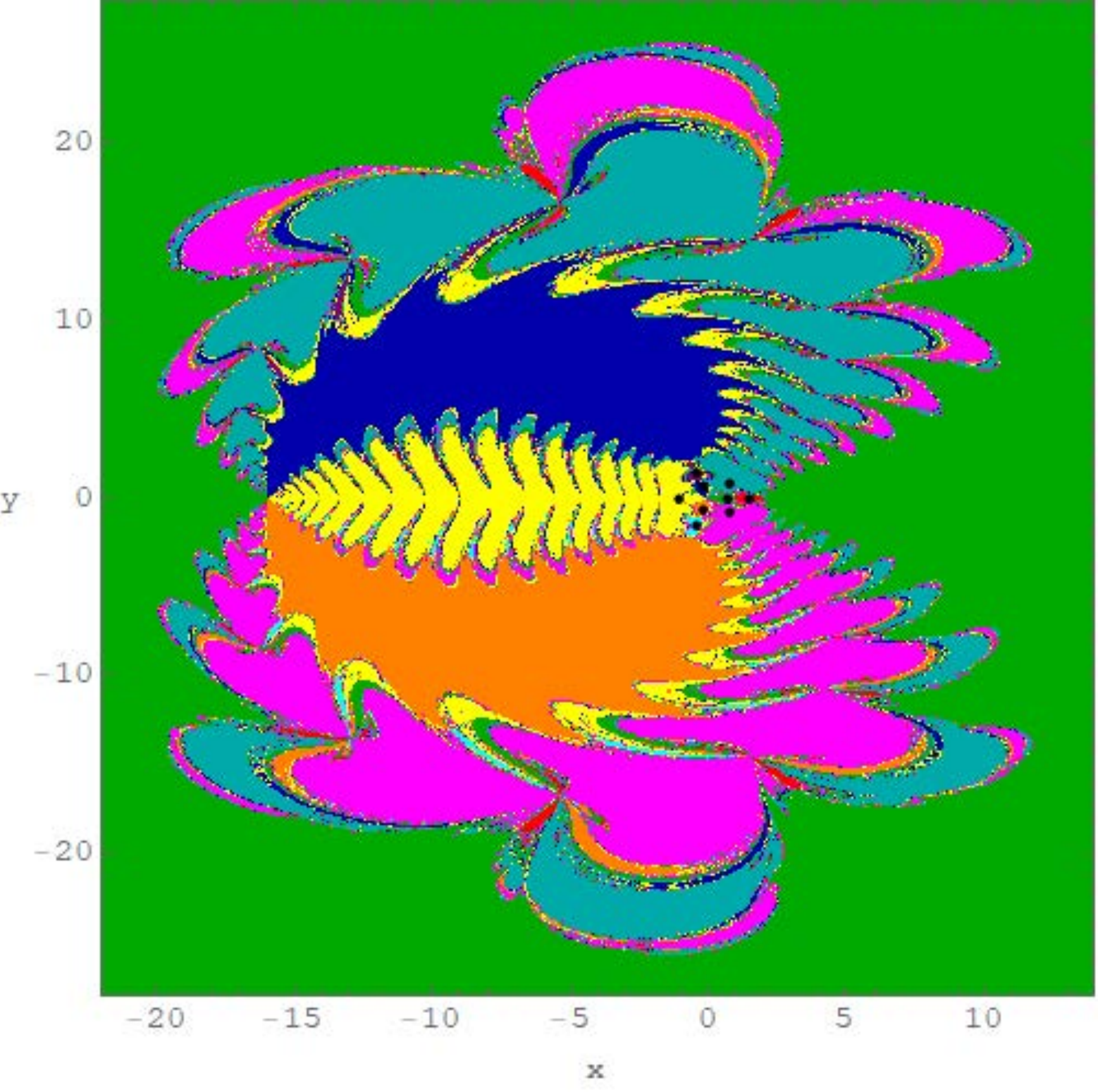}
\caption{The Newton-Raphson basins of attraction on the $xy$-plane for the
case when nine libration points exist for  fixed value of
$\alpha=53 \degree$  and for:
(a) $\beta=1 \degree$; (b) $\beta=2\degree$;
(c) $\beta=5 \degree$; (d) $\beta=10\degree$; (e) $\beta=15\degree$. The color code denoting the attractors is as follows:
$L_1\emph{(red)}$; $L_2\emph{(darker green)}$; $L_3\emph{ (yellow)}$; $L_4\emph{ (magenta)}$;
$L_5\emph{(crimson)}$; $L_6\emph{(purple)}$; $L_7\emph{(cyan)}$; $L_8\emph{(orange)}$; $L_9
\emph{(blue)}$; and non-converging points (white).
 (Color figure online).}
\label{NR_Fig_1}
\end{figure*}
%%%%
\begin{figure*}[!t]
\centering
(a)\includegraphics[scale=.35]{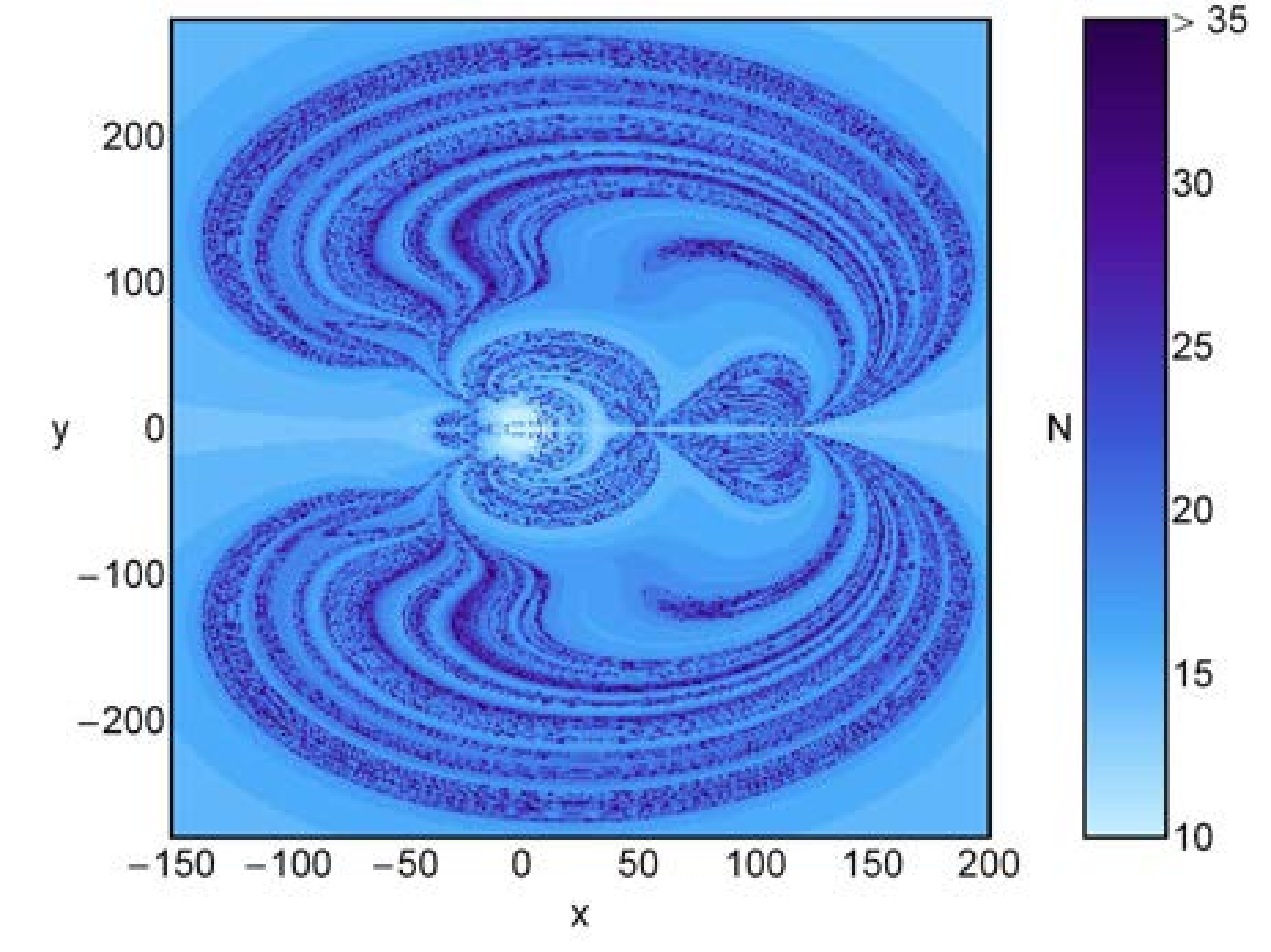}
(b)\includegraphics[scale=.35]{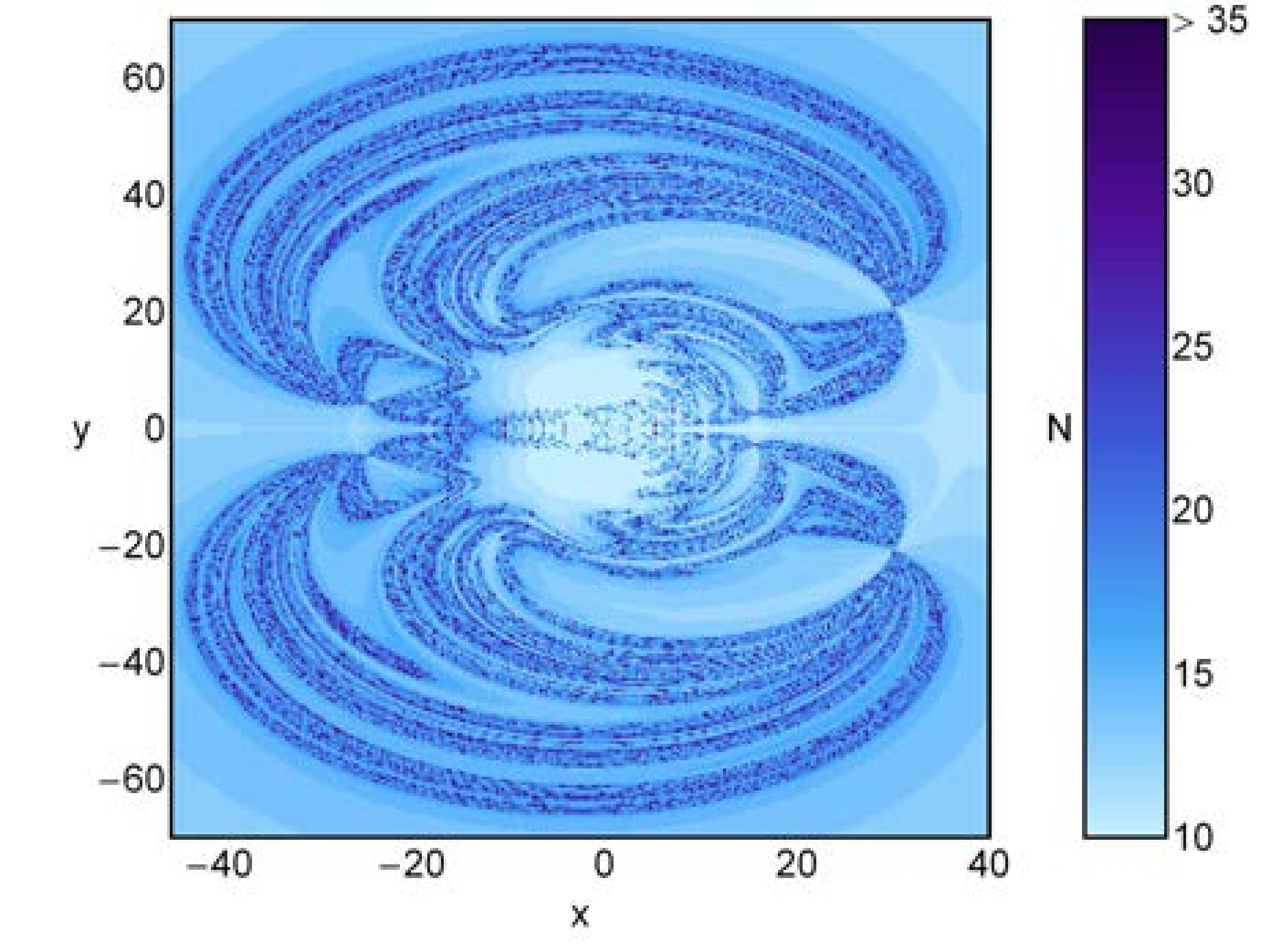}\\
(c)\includegraphics[scale=.35]{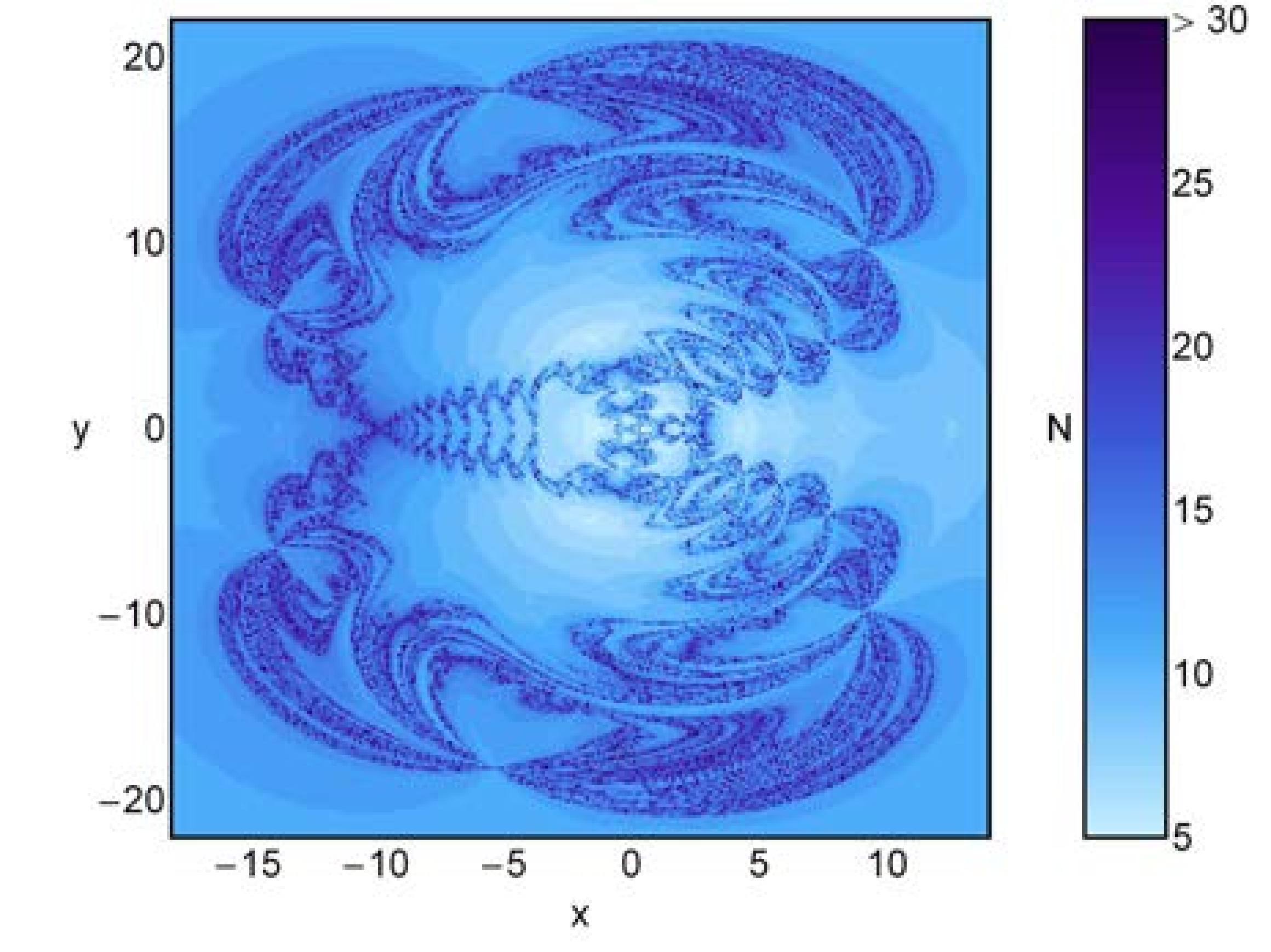}
(d)\includegraphics[scale=.35]{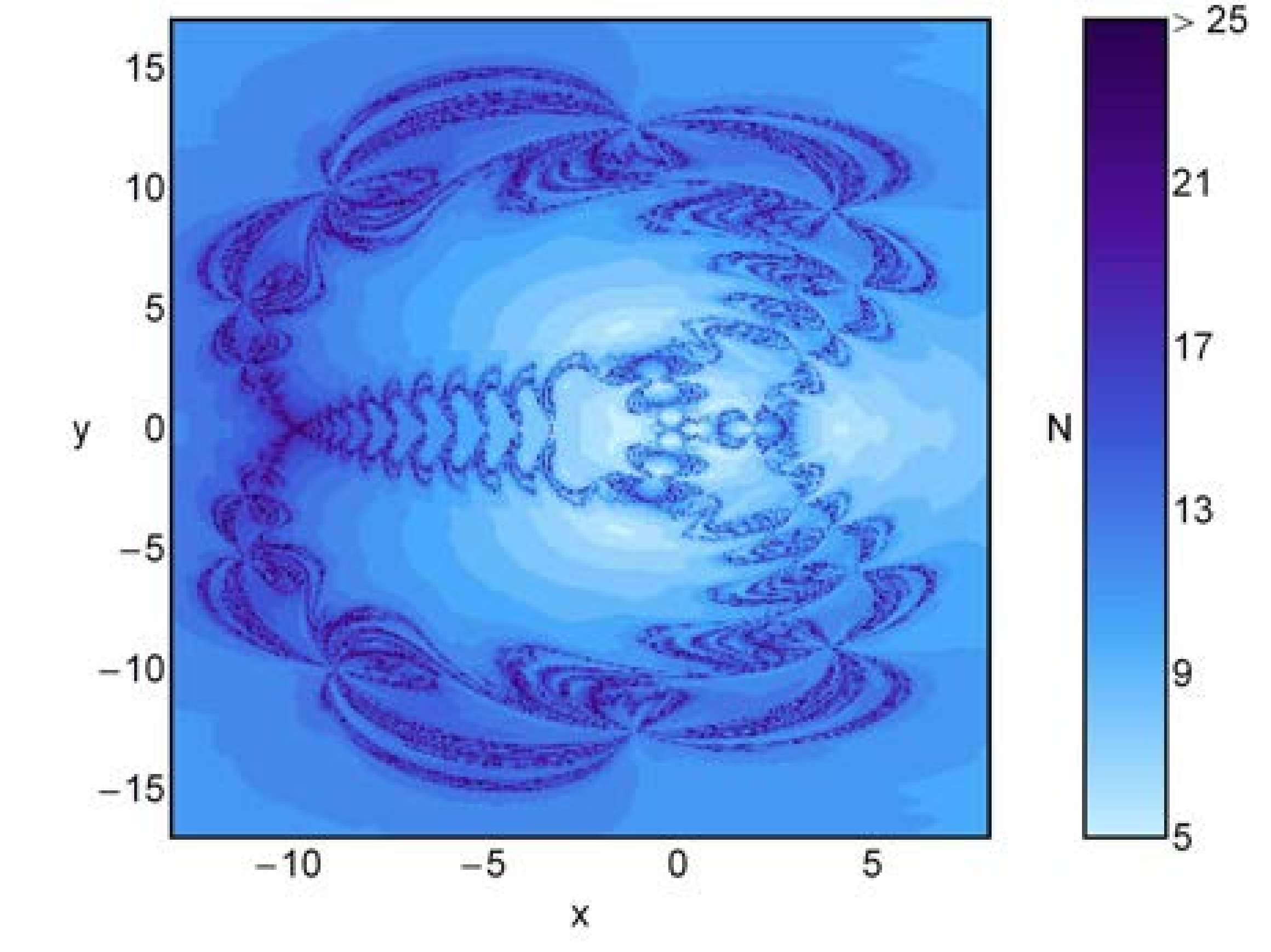}\\
(e)\includegraphics[scale=.35]{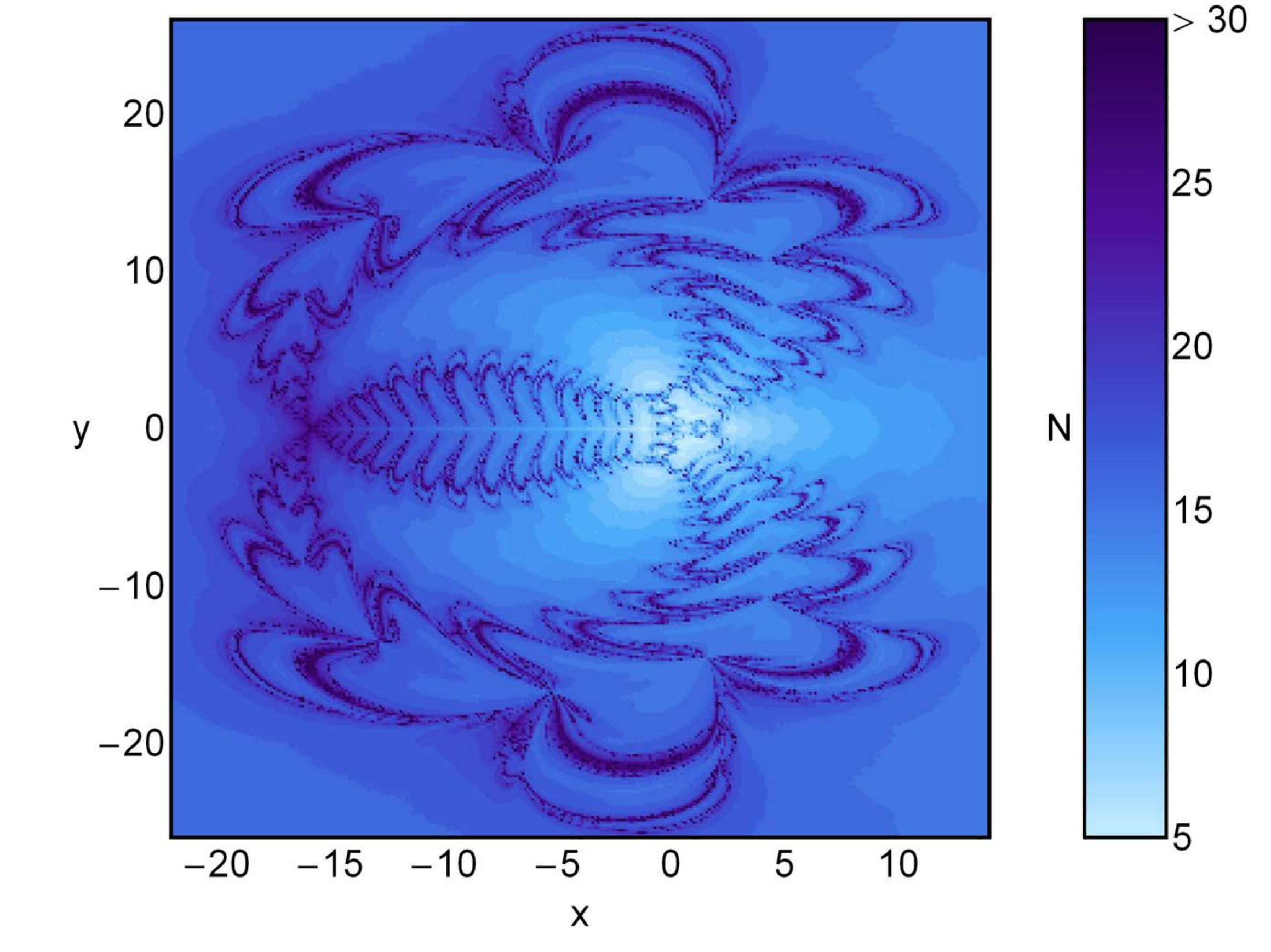}
\caption{The corresponding distributions of number $N$ of the required iterations for obtaining the Newton-Raphson basins of convergence,
shown in Fig. \ref{NR_Fig_1}(a-e). (Color figure online). }
\label{NR_Fig_1a}
\end{figure*}
%%%%
\begin{figure*}[!t]
\centering
(a)\includegraphics[scale=.35]{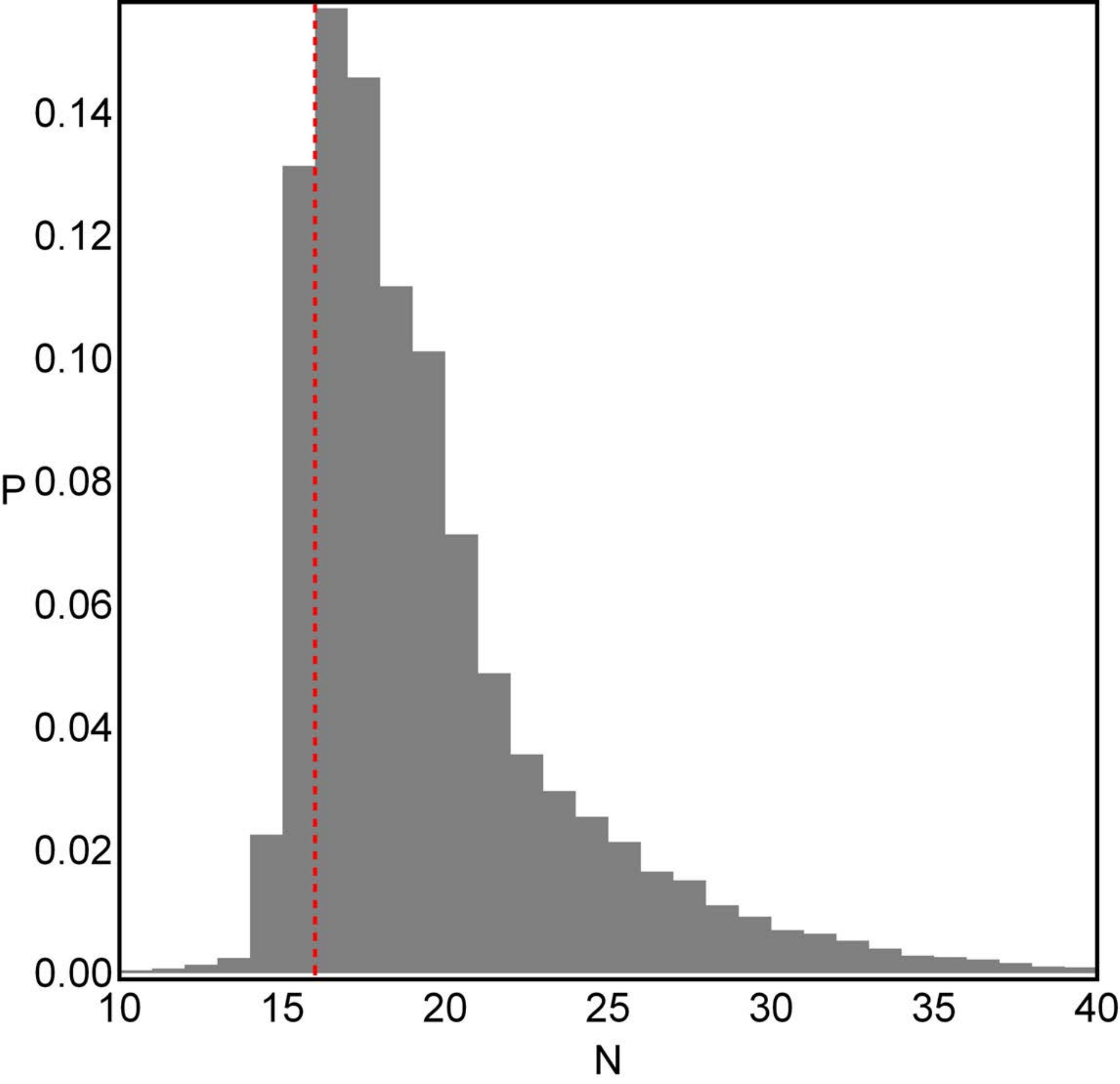}
(b)\includegraphics[scale=.35]{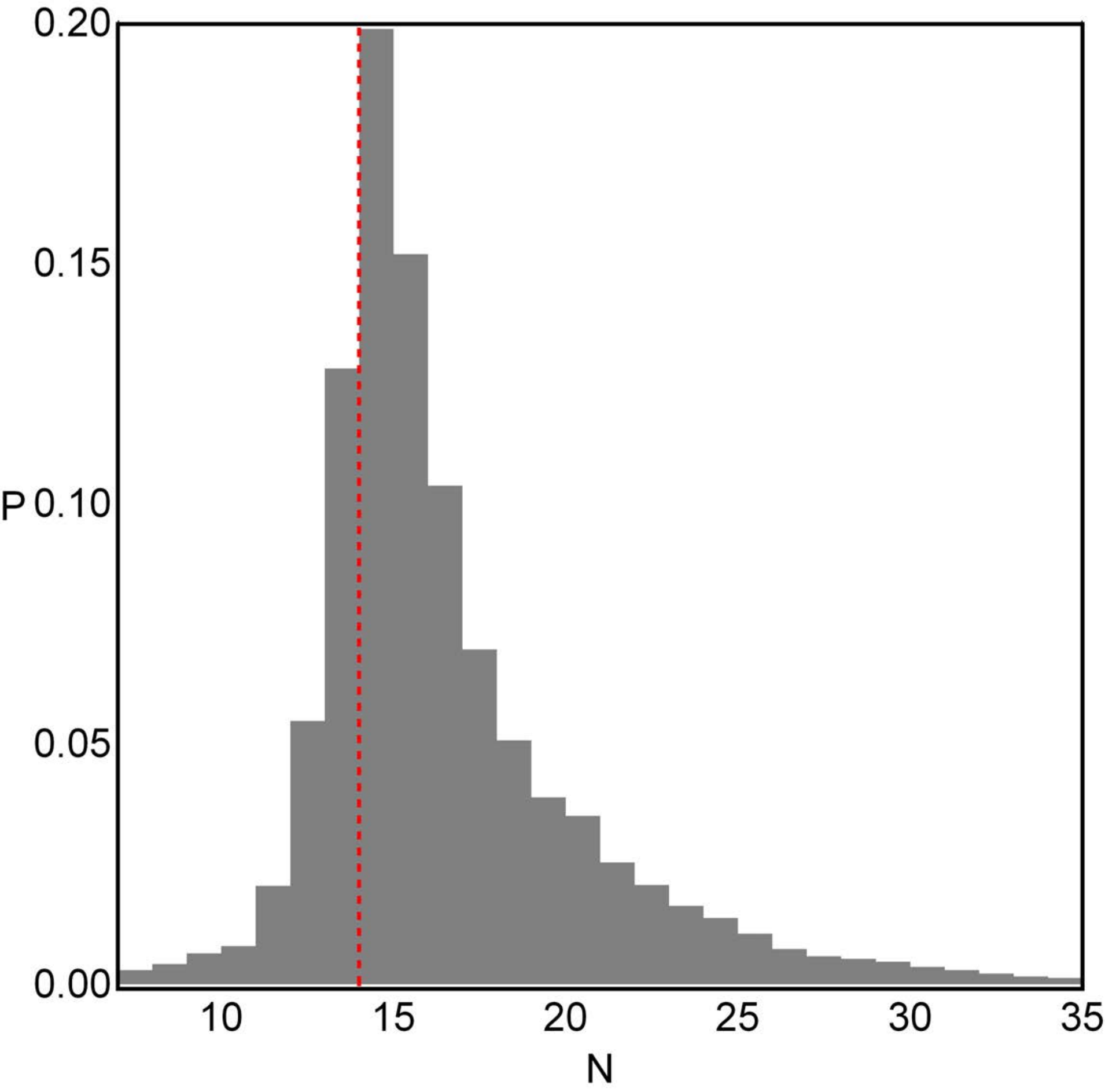}\\
(c)\includegraphics[scale=.35]{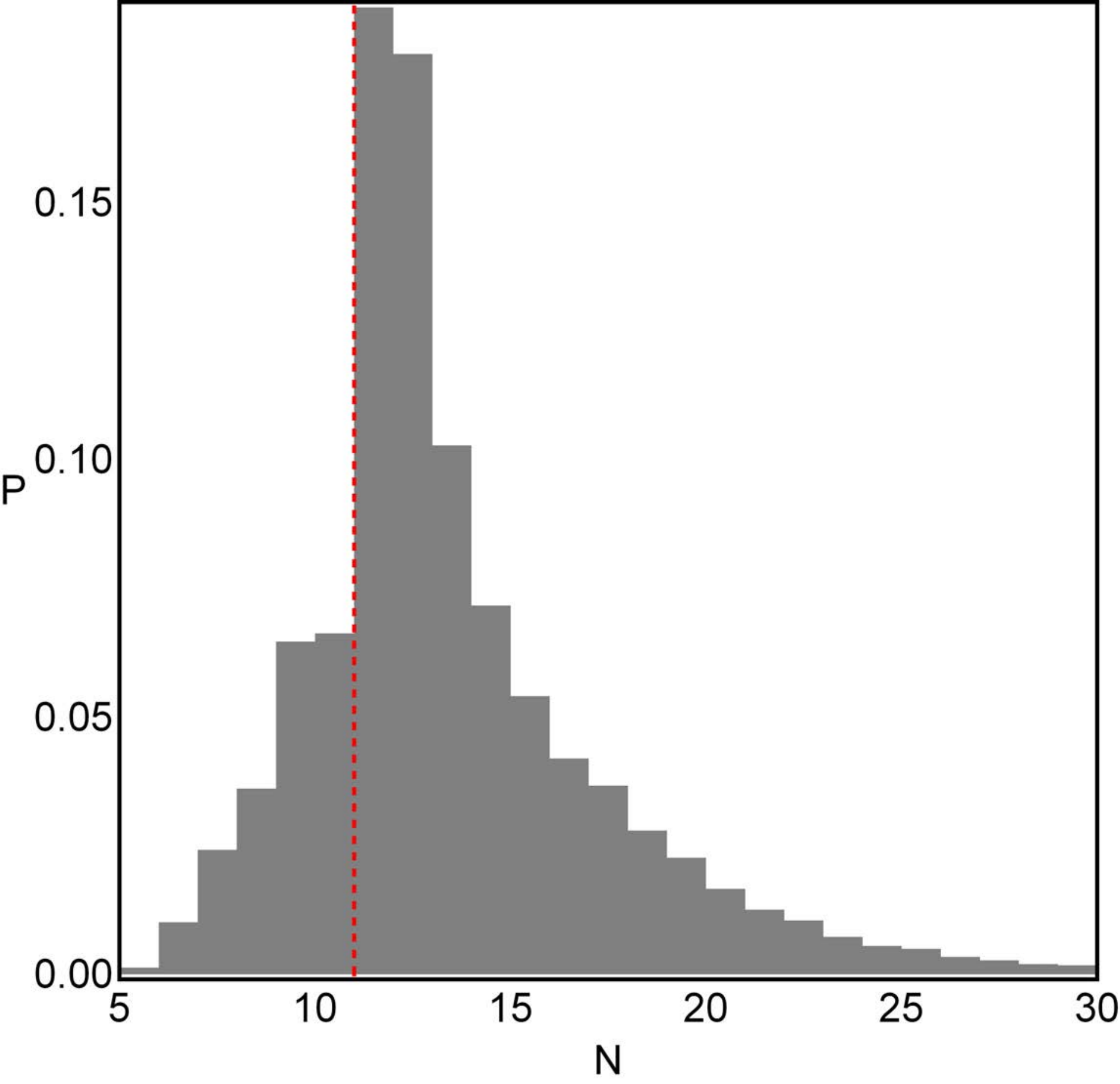}
(d)\includegraphics[scale=.35]{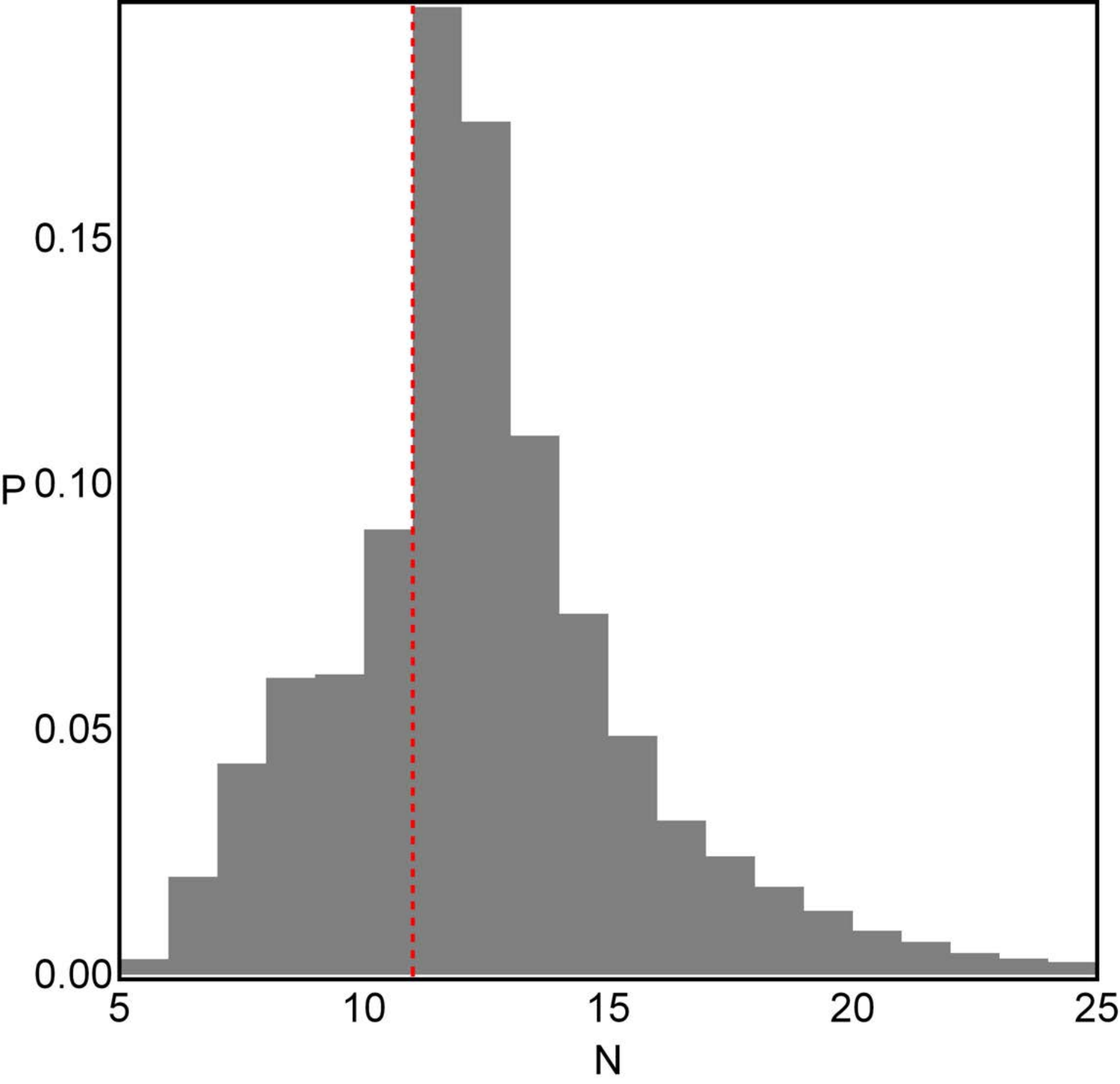}
(e)\includegraphics[scale=.35]{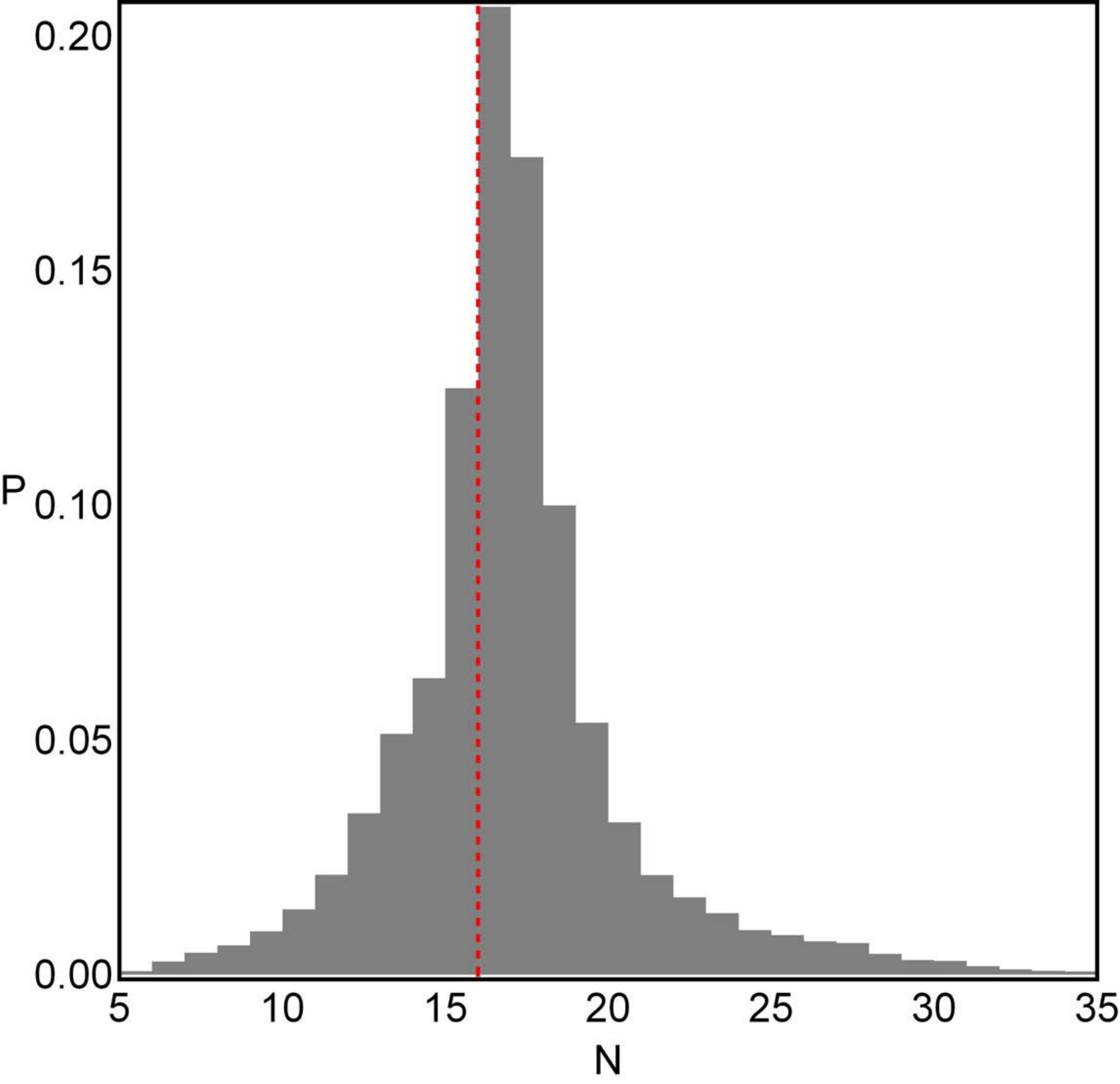}
\caption{The corresponding probability distributions of required number of iterations for obtaining the Newton-Raphson basins of convergence, shown in Fig. \ref{NR_Fig_1}(a-e). The vertical dashed red line indicates, in each case, the most probable number $N^*$ of iterations. (Color figure online)}
\label{NR_Fig_1b}
\end{figure*}
For the configuration $(x, y)$ plane, the iterative formulae for each coordinate are read as:
\begin{subequations}
\begin{eqnarray}
\label{Eq:403a}
x_{n+1} &= x_n - \left( \frac{\Omega_x \Omega_{yy} - \Omega_y \Omega_{xy}}{\Omega_{yy} \Omega_{xx} - \Omega^2_{xy}} \right)_{(x,y) = (x_n,y_n)}, \\
\label{Eq:403b}
y_{n+1} &= y_n - \left( \frac{\Omega_y \Omega_{xx}-\Omega_x \Omega_{yx}}{\Omega_{yy} \Omega_{xx} - \Omega^2_{xy}} \right)_{(x,y) = (x_n,y_n)},
\end{eqnarray}
\end{subequations}
where $\Omega_x, \Omega_y$ are given by Eqs. \ref{Eq:6a}, \ref{Eq:6b} respectively and
\begin{subequations}
\begin{eqnarray}
 % \nonumber % Remove numbering (before each equation)
\label{Eq:8a}
\Omega_{xx}&=&1-\frac{1}{\Delta}\sum _{i=1}^{4}m_i\Big(\frac{1}{r_i^3}-\frac{3\tilde{x_i}^2}{r_i^{5}}\Big),\\
\label{Eq:8a}
\Omega_{yy}&=&1-\frac{1}{\Delta}\sum _{i=1}^{4}m_i\Big(\frac{1}{r_i^3}-\frac{3\tilde{y_i}^2}{r_i^{5}}\Big),\\
\label{Eq:8a}
\Omega_{xy}&=&\frac{1}{\Delta}\sum _{i=1}^{4}\frac{3m_i\tilde{x_i}\tilde{y_i}}{r_i^{5}}=\Omega_{yx}.
\end{eqnarray}
\end{subequations}
%%%%
In the following subsections, we explore how the angle parameters $\alpha$ and $\beta$ influence the shape and geometry of the Netwon-Raphson basins of convergence in axisymmetric five-body problem, by considering two cases regarding the setup of the primaries, i.e., the first concave case and second concave case. In each subsection, the basins of convergence associated with the libration points are further analyzed on the basis of total number of libration points exist. The color coded diagrams are used to assign a different color for each nodes on the configuration $(x, y)$ plane, according to the final state of corresponding initial condition. Moreover, it is necessary to note  that the size of each color codded diagram (i.e., the minimum and the maximum values of $x$ and $y$) is taken, in each panel, in such a manner so as to have a complete or zoomed view of the shape and geometry of the basins of convergence.
\subsection{Case:I The first concave case}
\label{Case:I The first concave case}
We start our analysis of the basins of convergence associated with the libration points by taking the first concave case, for three different values of the angle parameter $\alpha=53\degree, 57\degree, 58\degree$ and corresponding permissible range of $\beta$. In this case, there exist either 9, 11 or 13 libration points depending upon the various combination of the angle parameters. Therefore, we have considered three different subcases to reveal the effect of these angle parameters on the geometry of the basins of convergence.
\subsubsection{when nine libration points exist}
\label{when nine libration points exists}
The analysis of the basins of convergence in this subcase start with those value of angle parameter for which the dynamical system possesses nine libration points.

For $ \alpha=53\degree$ and permissible range of $\beta$, there is only nine libration points and therefore the basins of convergence associated with these libration points are depicted in Fig.(\ref{NR_Fig_1}). It is seen that in all the considered value of angle parameter $\beta$, the configuration $(x, y)$ plane contains various well-defined basins of convergence associated with the libration points, which are highly sensitive with the change in the value of angle parameter $\beta$. The extent of the basin of convergence associated to libration point $L_2$ is infinite whereas for the remaining libration points the domain of corresponding basins of convergence are finite. Moreover, the basins boundaries are composed of highly chaotic mixture of the initial conditions and the domain of the basins of convergence associated with the libration point $L_2$ increases as the angle parameter $\beta$ increases. The domain of the basins of convergence associated with the libration points $L_{1, 6, 7}$ looks like exotic bugs with many legs and antenna where as for libration points $L_{4, 5}$ (\emph{magenta } and \emph{crimson}) (see panel-\ref{NR_Fig_1}e) look like multiple butterfly wings which was very irregular in previous panels.

In Fig. \ref{NR_Fig_1a}, using tones of blue, we have illustrated the corresponding number $N$ of iterations to obtain the predefined accuracy. It is further observed that the initial conditions which lie in the domain of the basins of attraction have relatively fast $(N < 15)$ convergence, whereas the initial conditions which lie in the neighborhood of the basins boundaries converge slowly $(N > 15)$. Moreover,  in Fig. \ref{NR_Fig_1b}, the corresponding probability distribution of the required iterations is depicted. In Fig. \ref{NR_Fig_1b}, the tail of the histograms
decreases so as to cover $95\%$ of the corresponding distribution of iterations. In addition, the \emph{red} dashed line which shows the most probable number $N^*$ of iteration neither increases nor decreases constantly with the increase in $\beta$. Therefore, it is not possible to predict the most probable number of iteration for any predefined value of $\beta$. For this analysis, the probability is defined as $P=N_0/N_t$, if $N_0$ is the number of initial conditions $(x_0, y_0)$ converge, after $N$ iterations, to one of the libration points and $N_t$ defined as the total number of nodes in every color codded diagram.

\begin{figure*}[!t]
\centering
(a)\includegraphics[scale=.27]{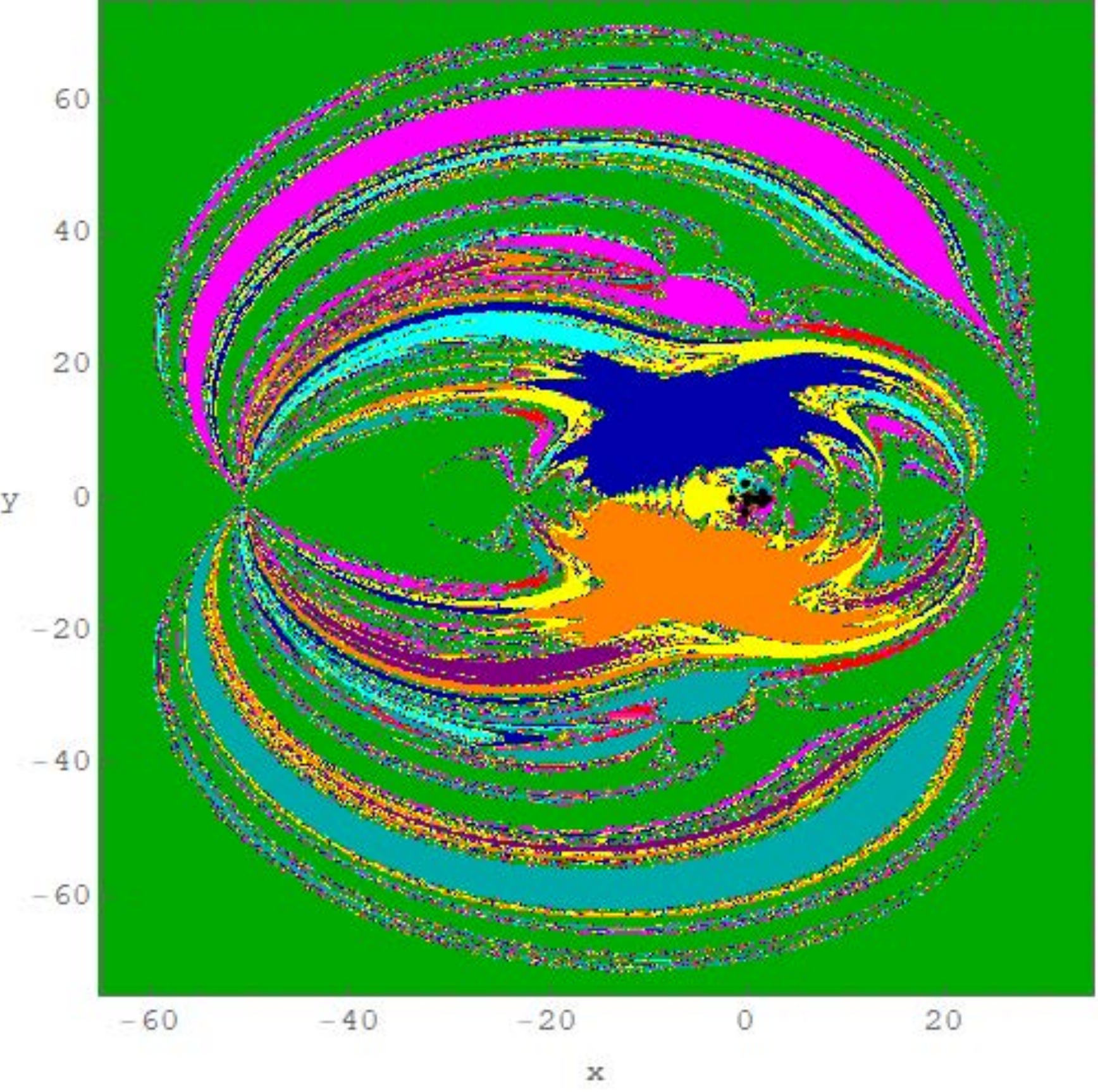}
(b)\includegraphics[scale=.27]{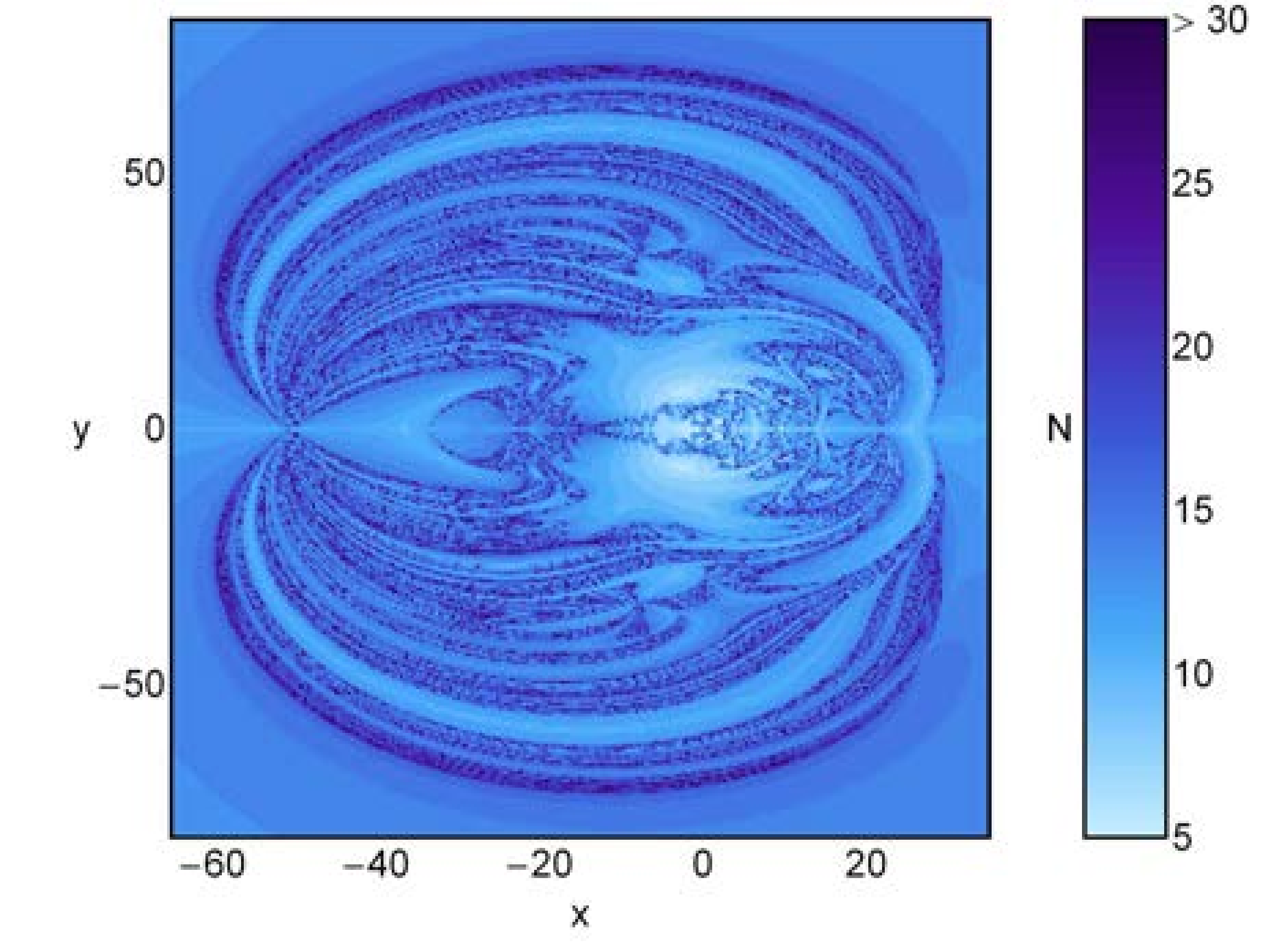}
(c)\includegraphics[scale=.25]{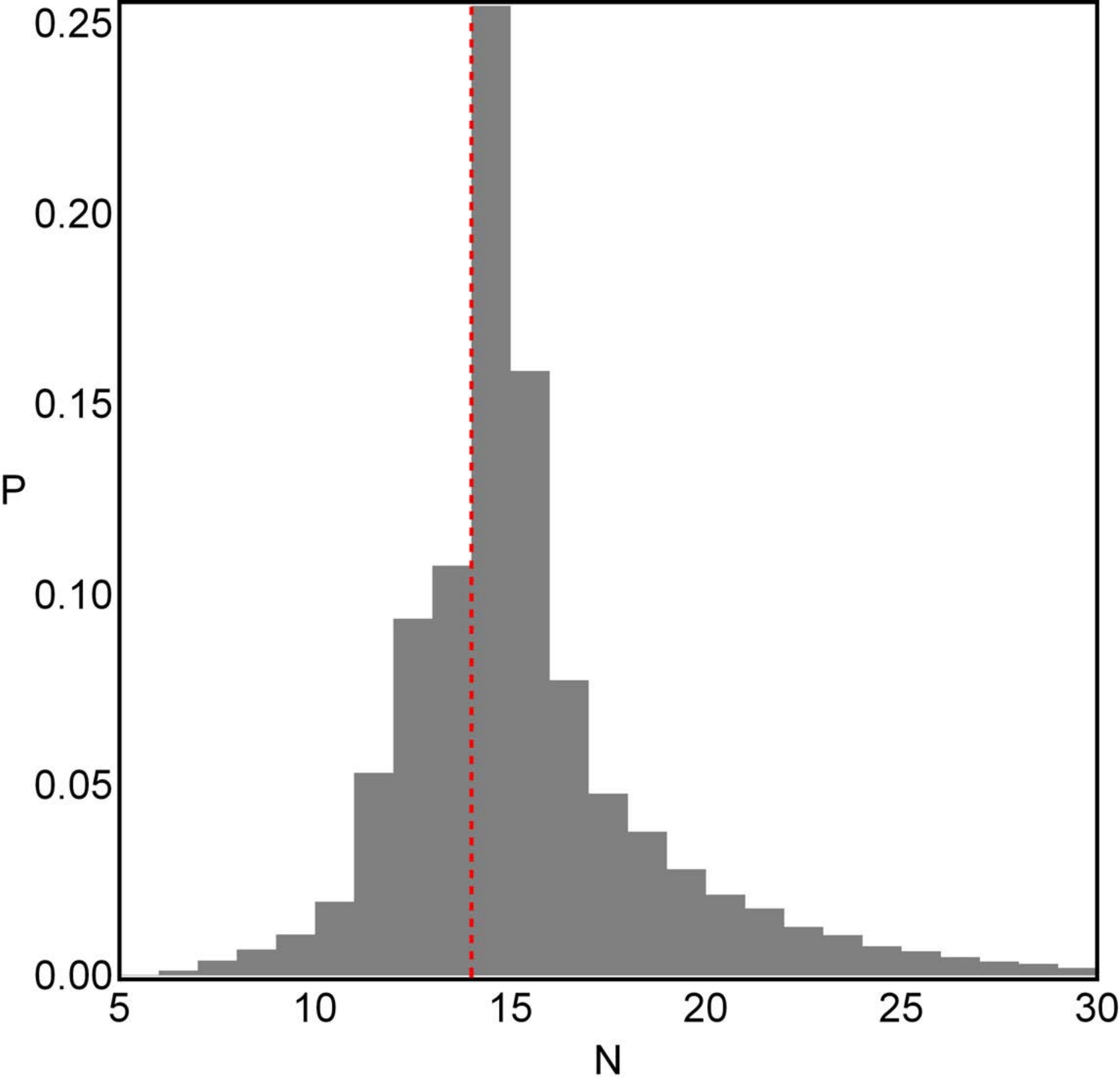}\\
(d)\includegraphics[scale=.27]{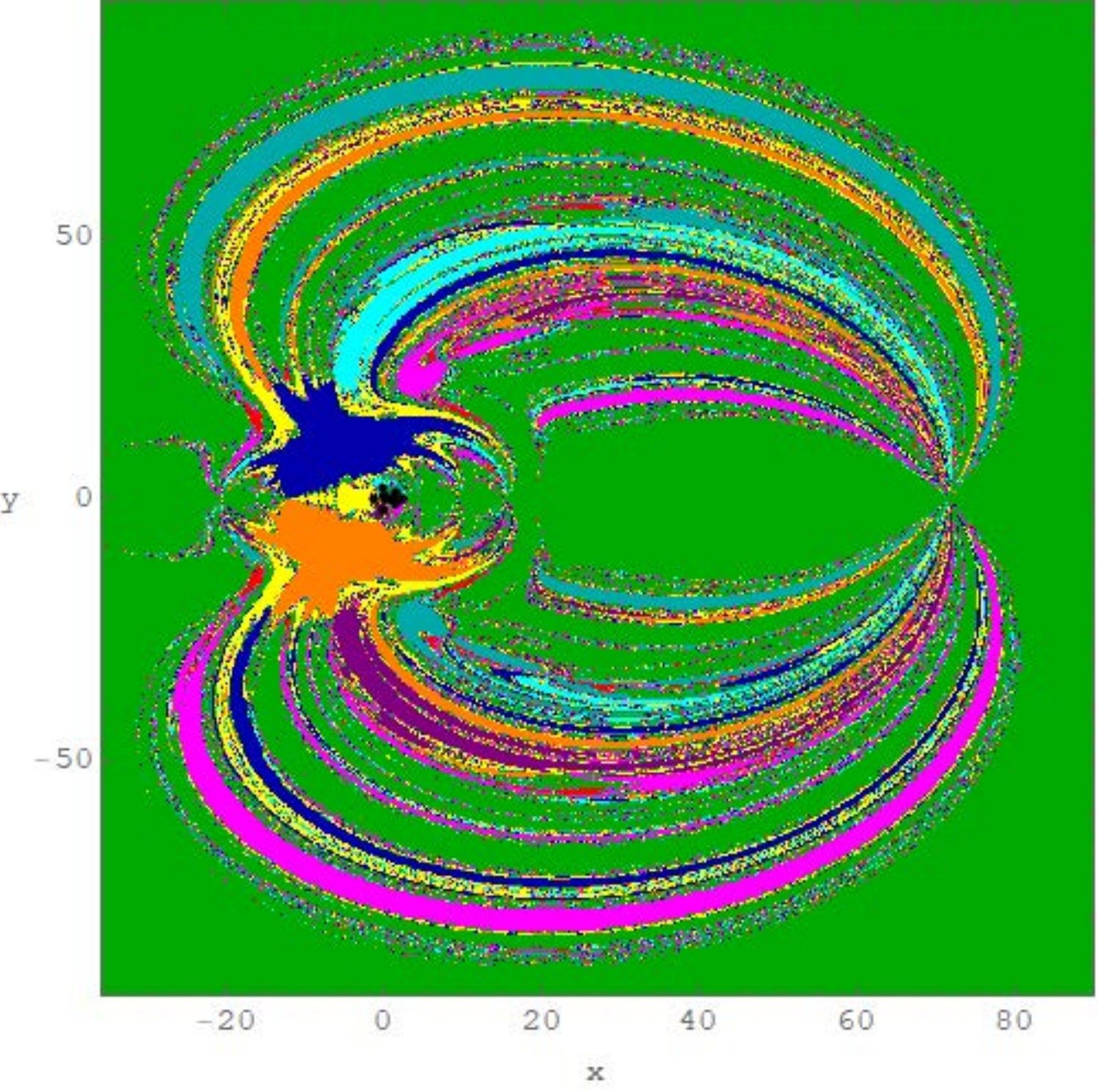}
(e)\includegraphics[scale=.27]{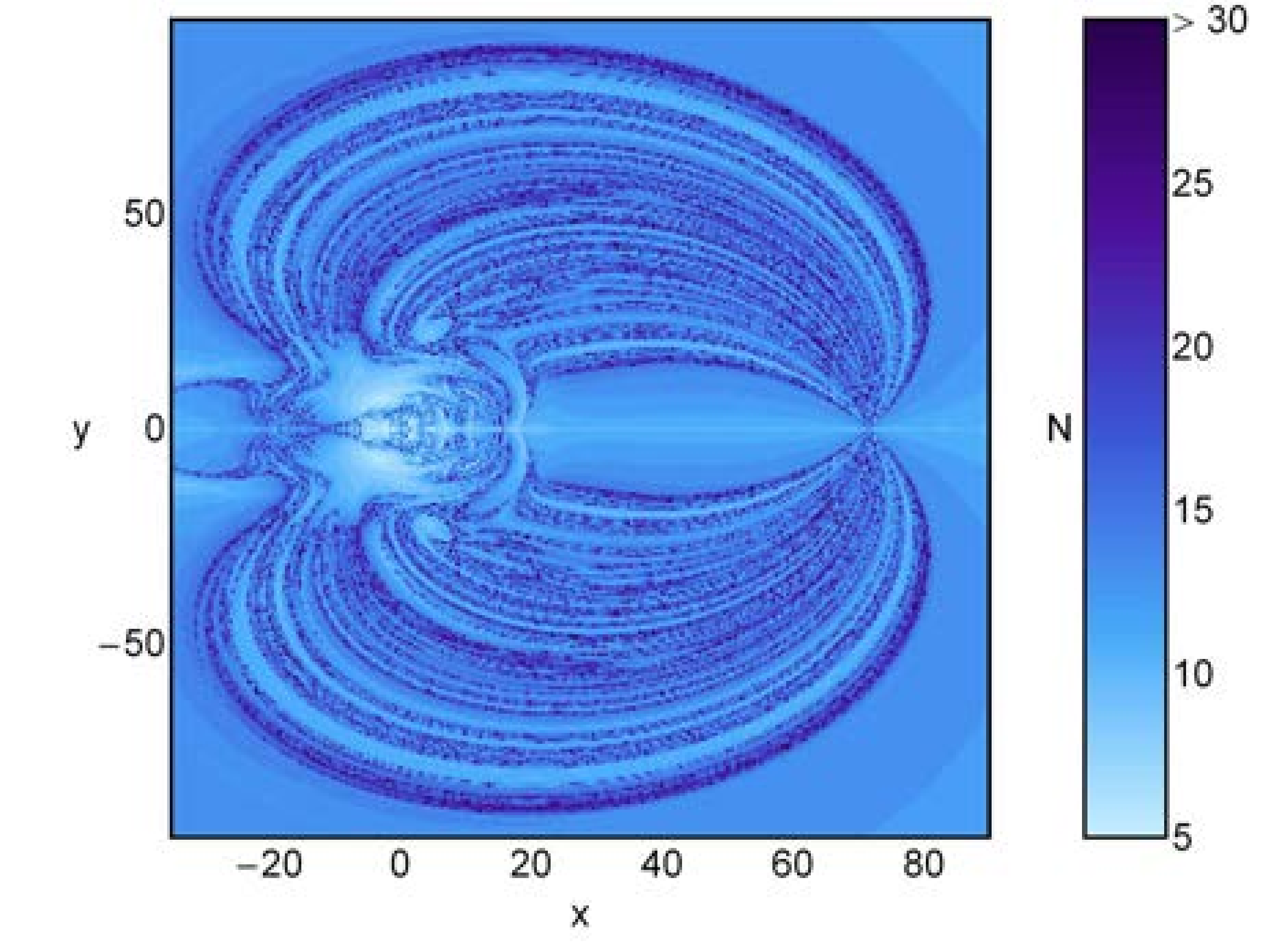}
(f)\includegraphics[scale=.25]{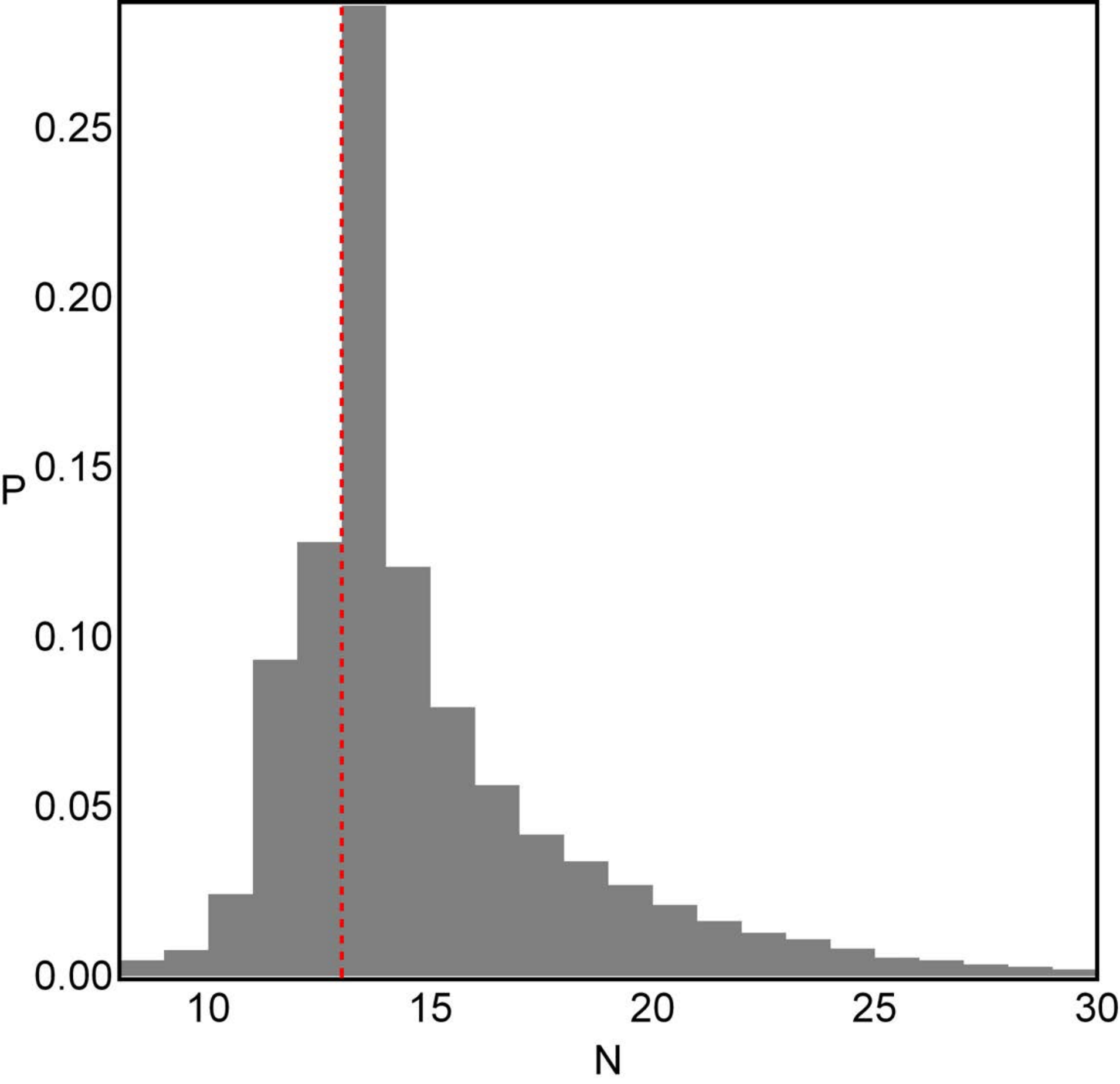}
(g)\includegraphics[scale=.27]{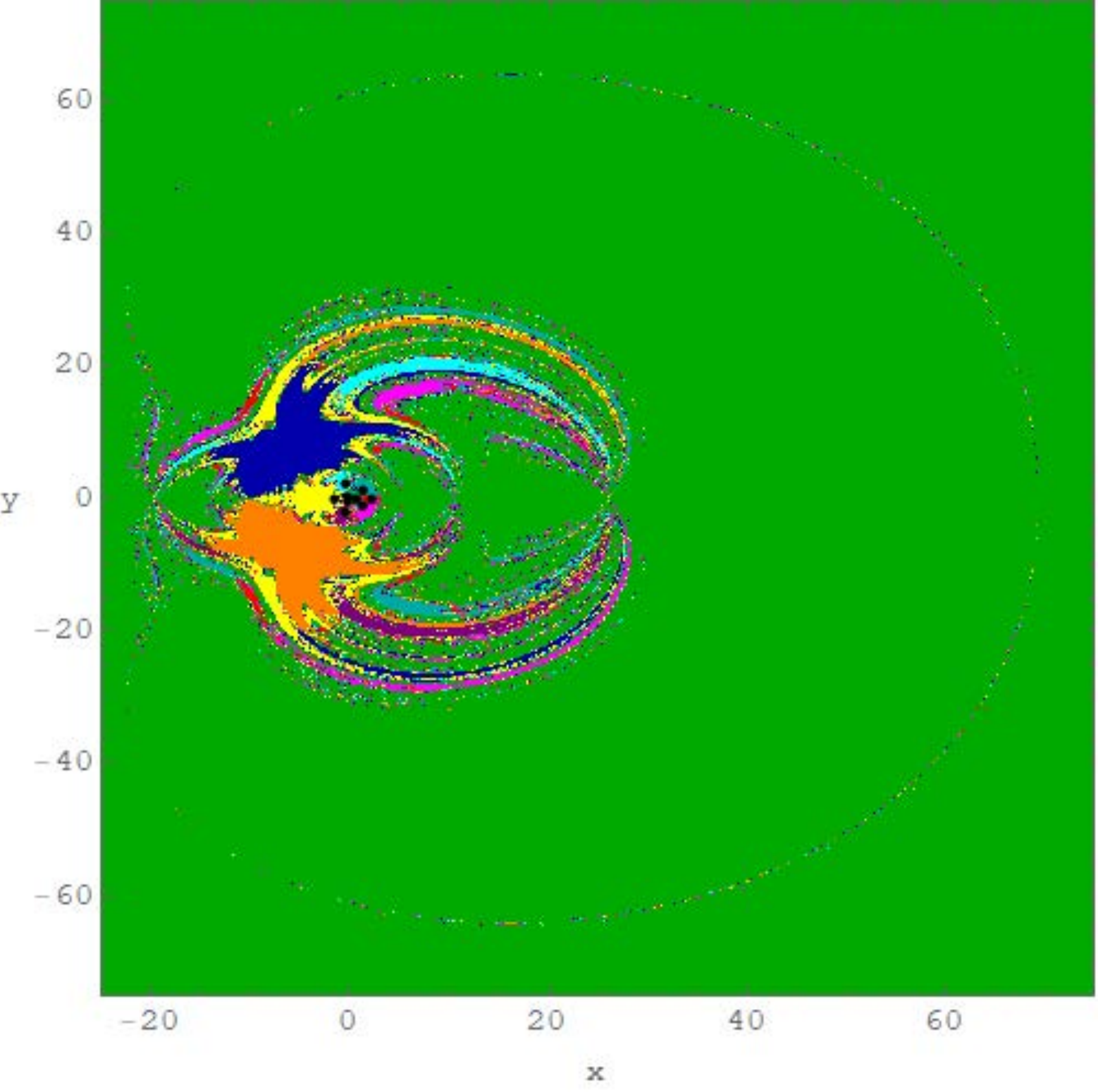}
(h)\includegraphics[scale=.27]{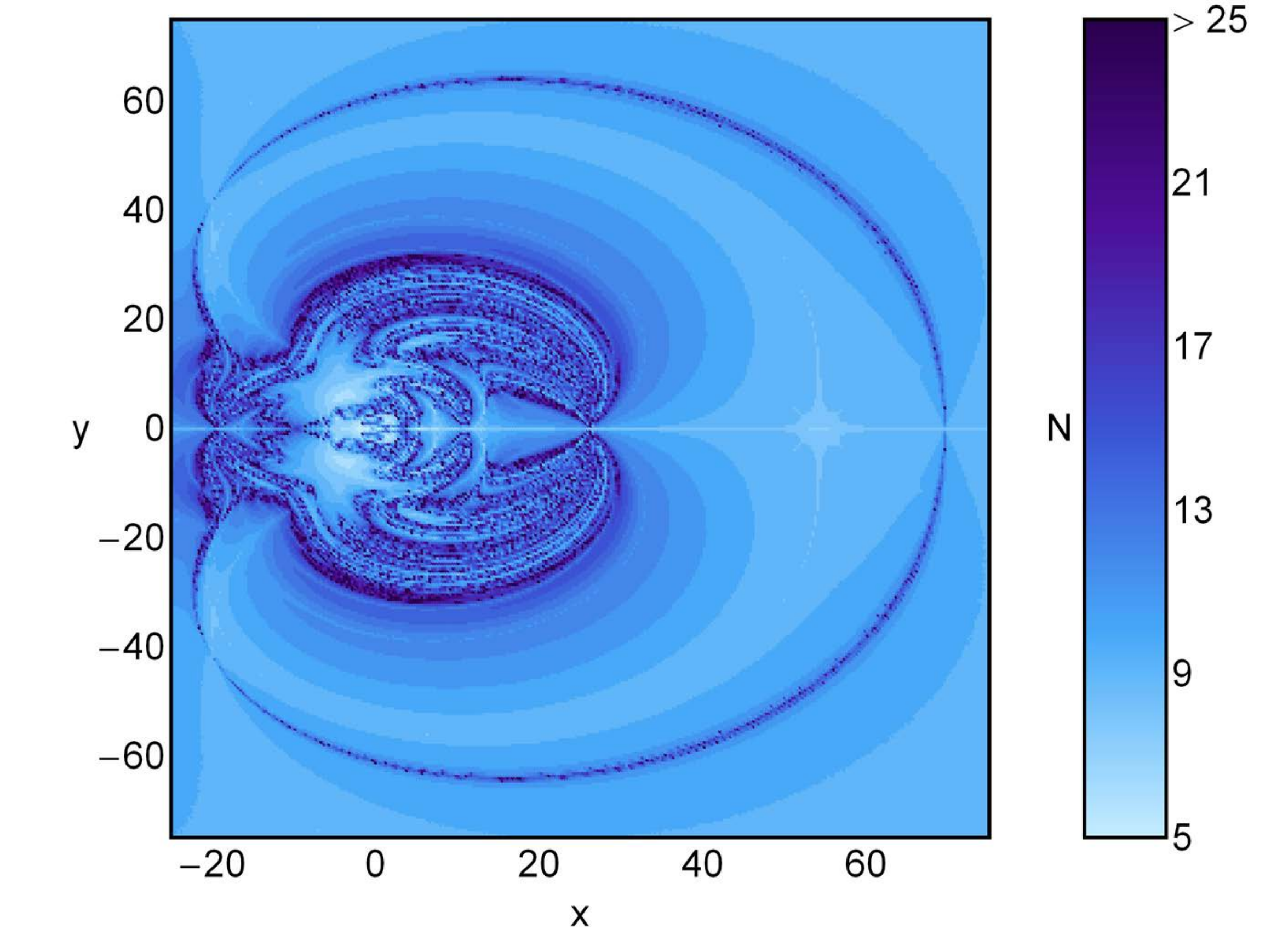}
(i)\includegraphics[scale=.25]{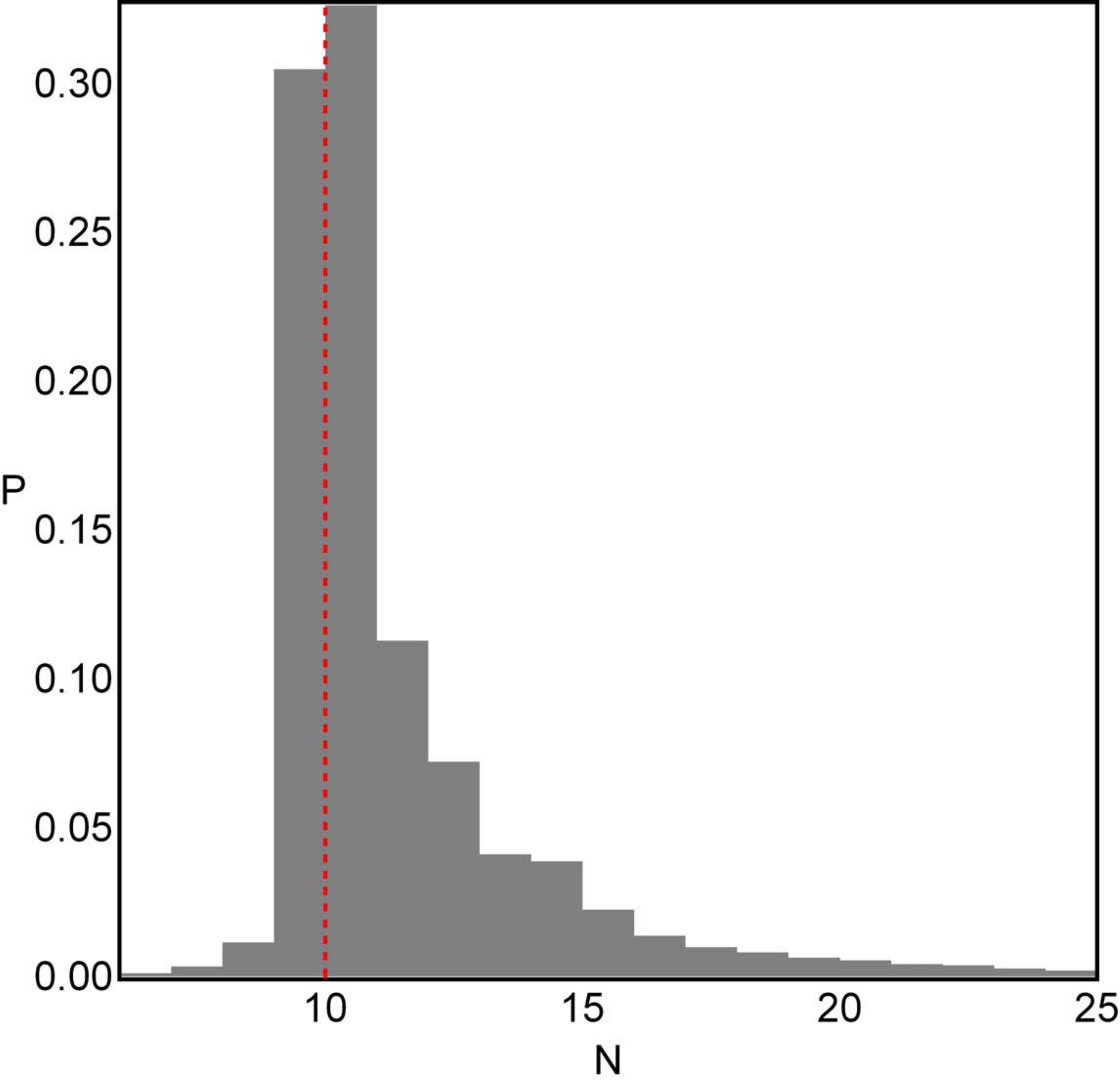}
\caption{The Newton-Raphson basins of attraction on the $xy$-plane for the
case when nine libration points exist for  fixed value of
$\alpha=57 \degree$  and for:
(a) $\beta=1 \degree$; (d) $\beta=1.5\degree$; (g) $\beta=2.5\degree$. The color code for the libration points $L_1$,...,$L_9$ is same as in Fig \ref{NR_Fig_1}; non-converging points (white); (b, e,  h) and (c, f, i) are the distribution of the corresponding number $(N)$ and the  probability distributions of required iterations for obtaining the Newton-Raphson basins of attraction shown in (a, d, g), respectively.
 (Color figure online).}
\label{NR_Fig_2}
\end{figure*}

\begin{figure*}[!t]
\centering
(a)\includegraphics[scale=.27]{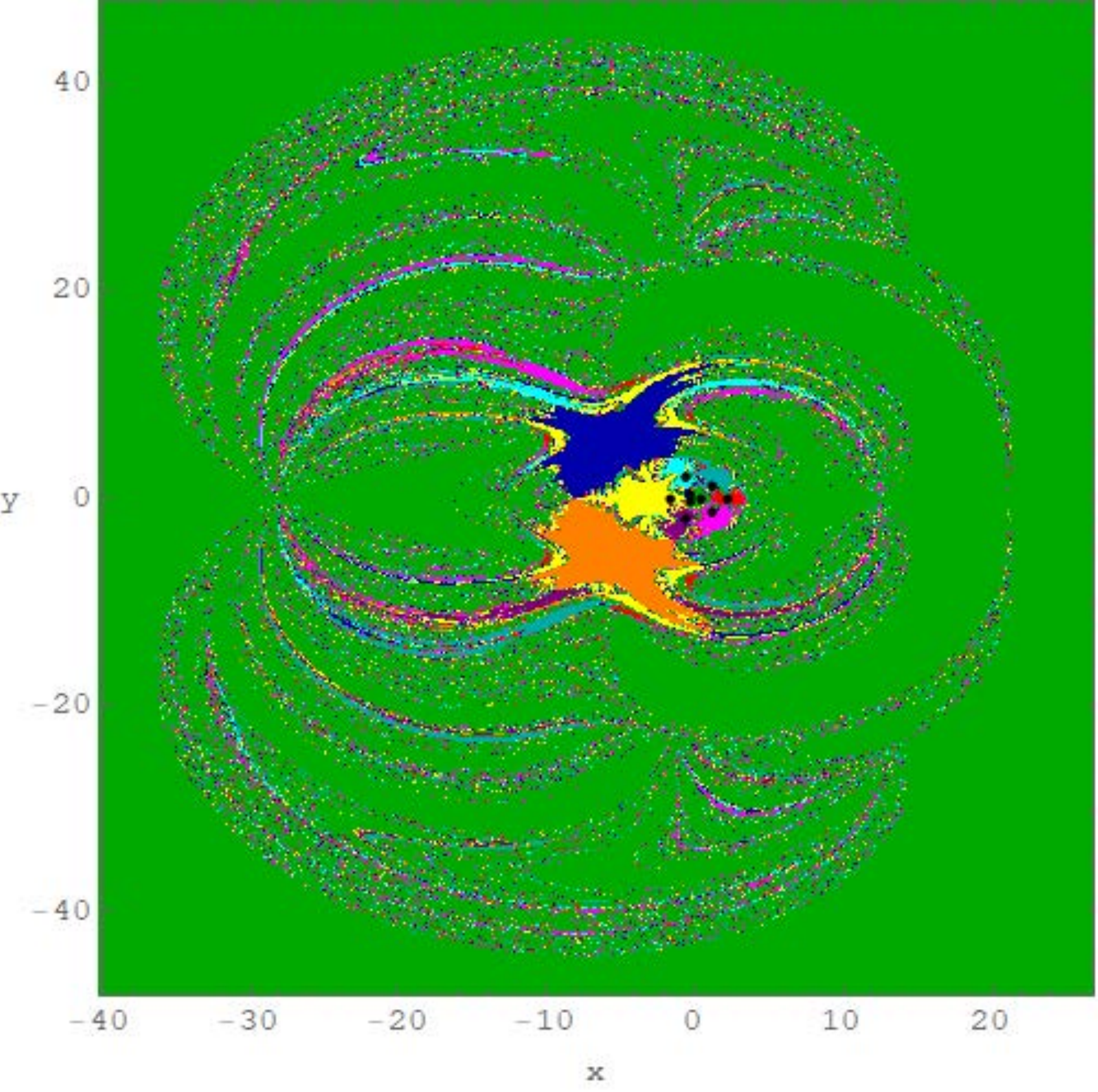}
(b)\includegraphics[scale=.27]{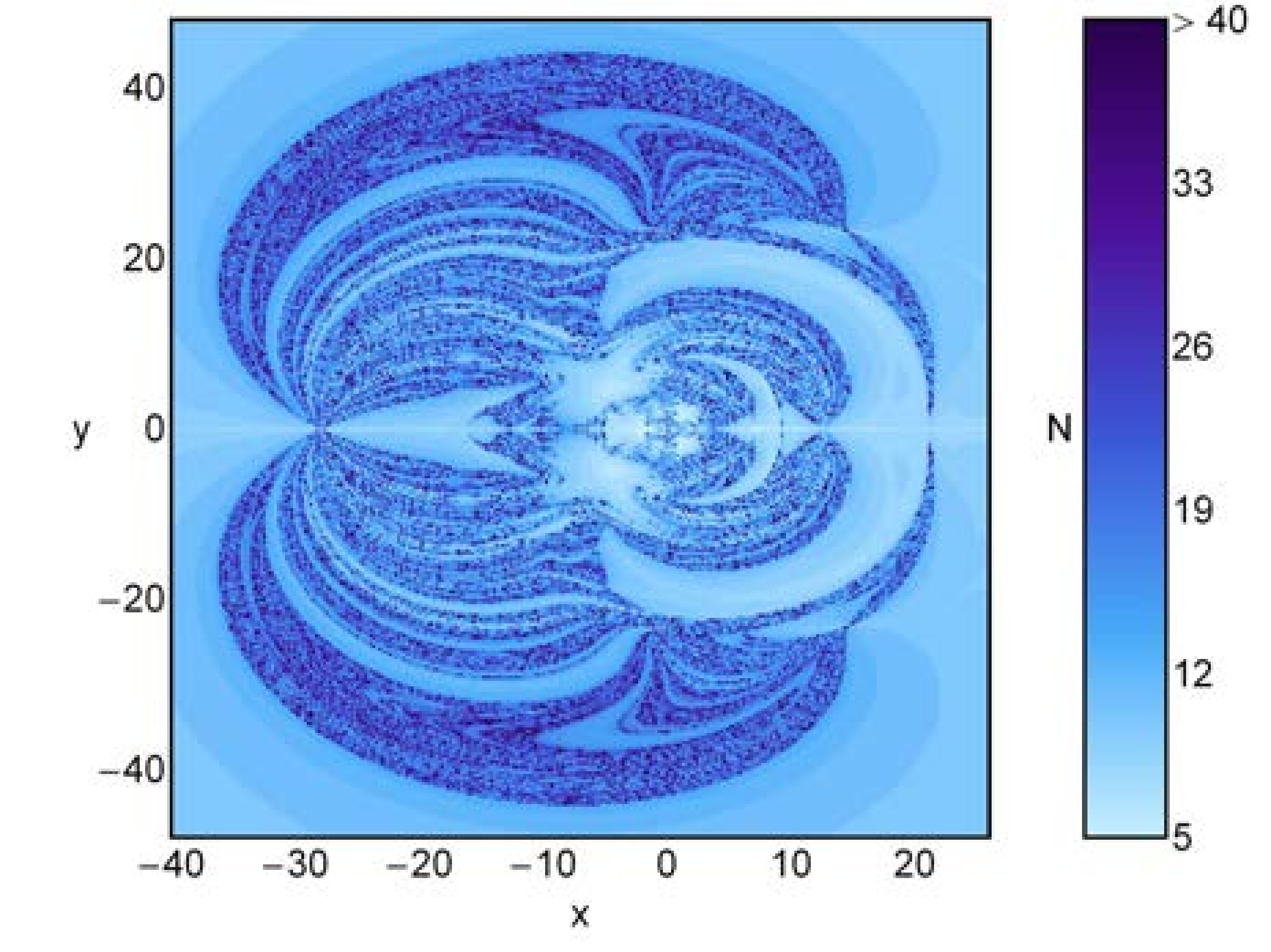}
(c)\includegraphics[scale=.25]{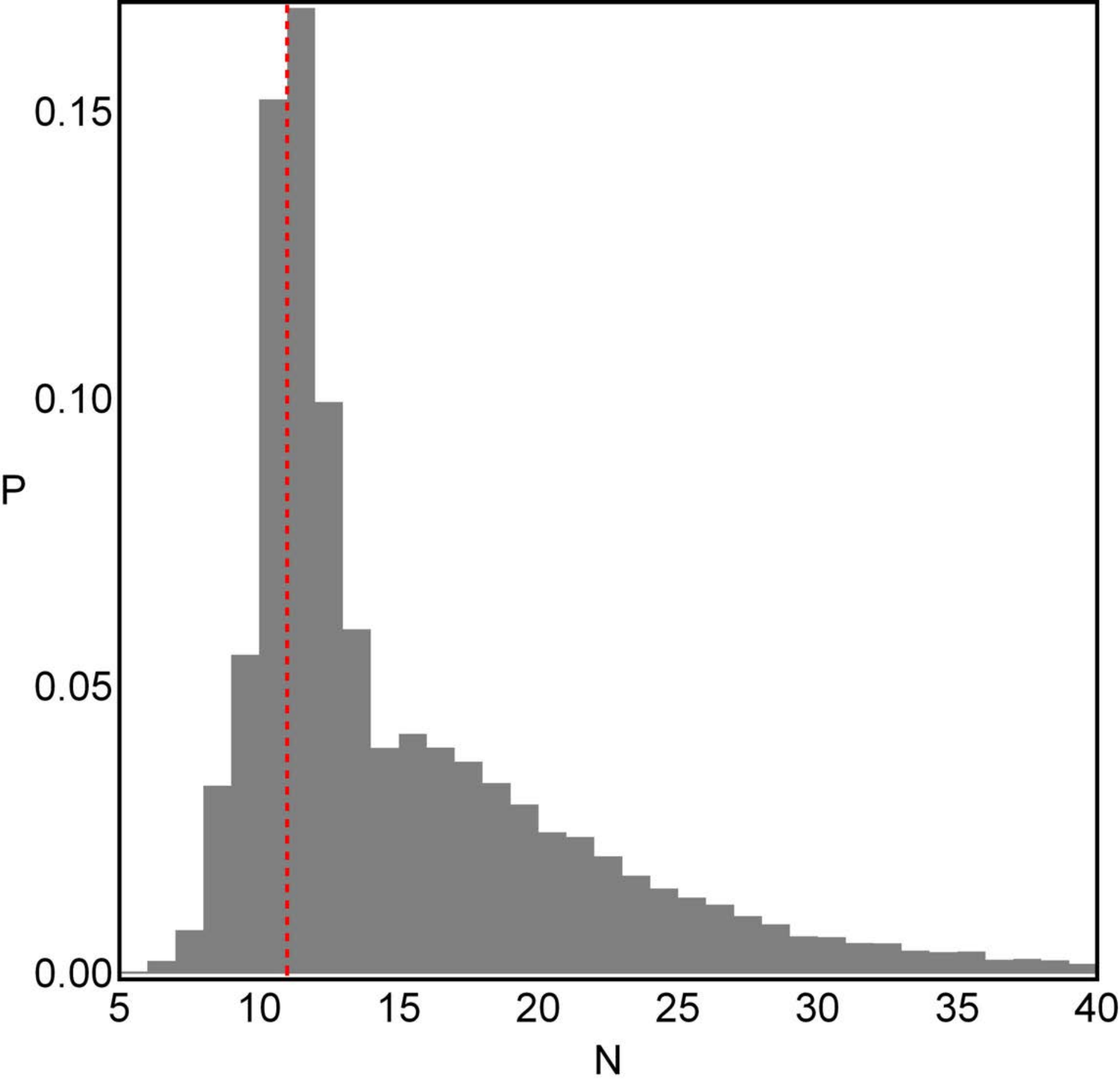}\\
(d)\includegraphics[scale=.27]{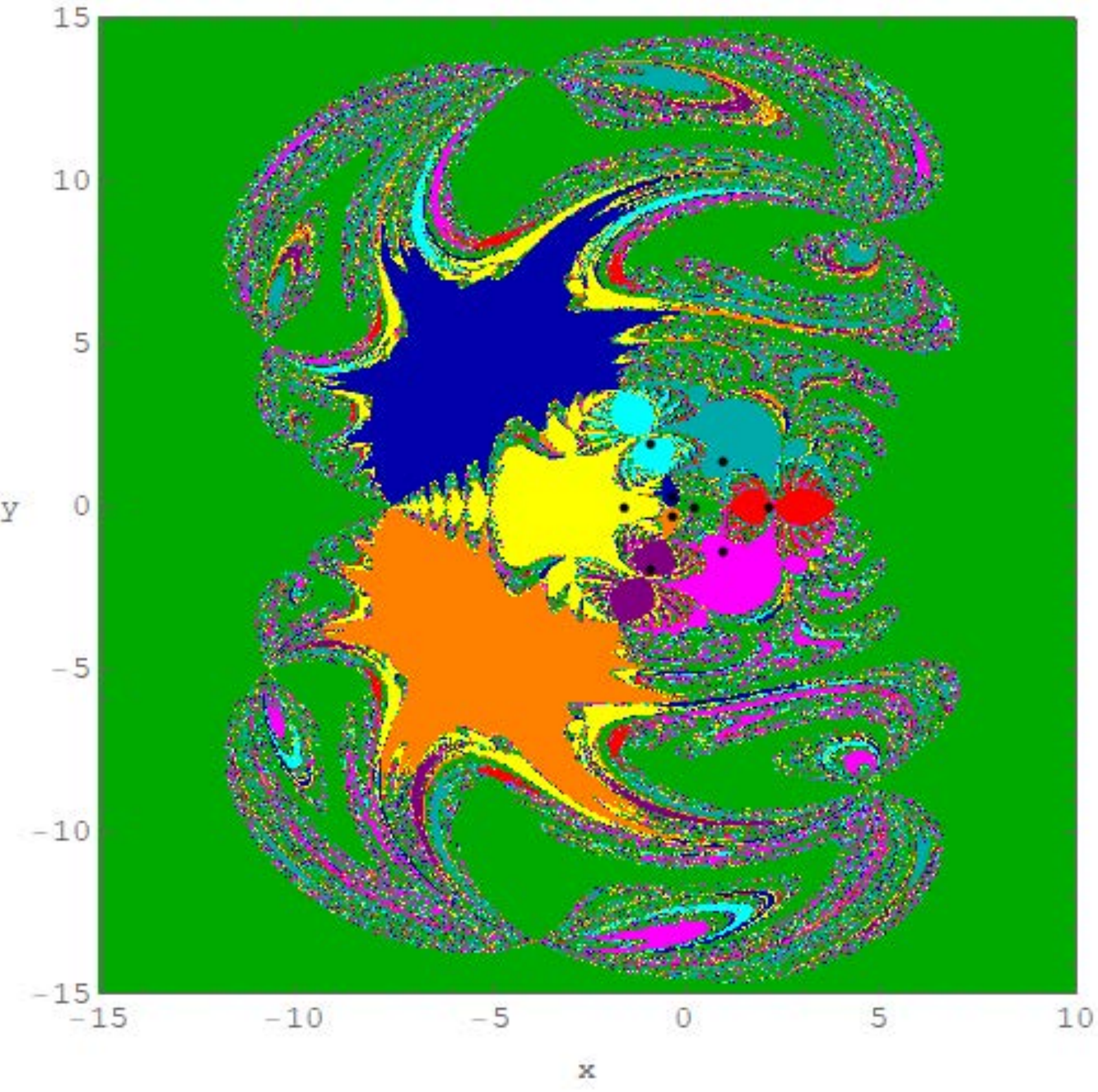}
(e)\includegraphics[scale=.27]{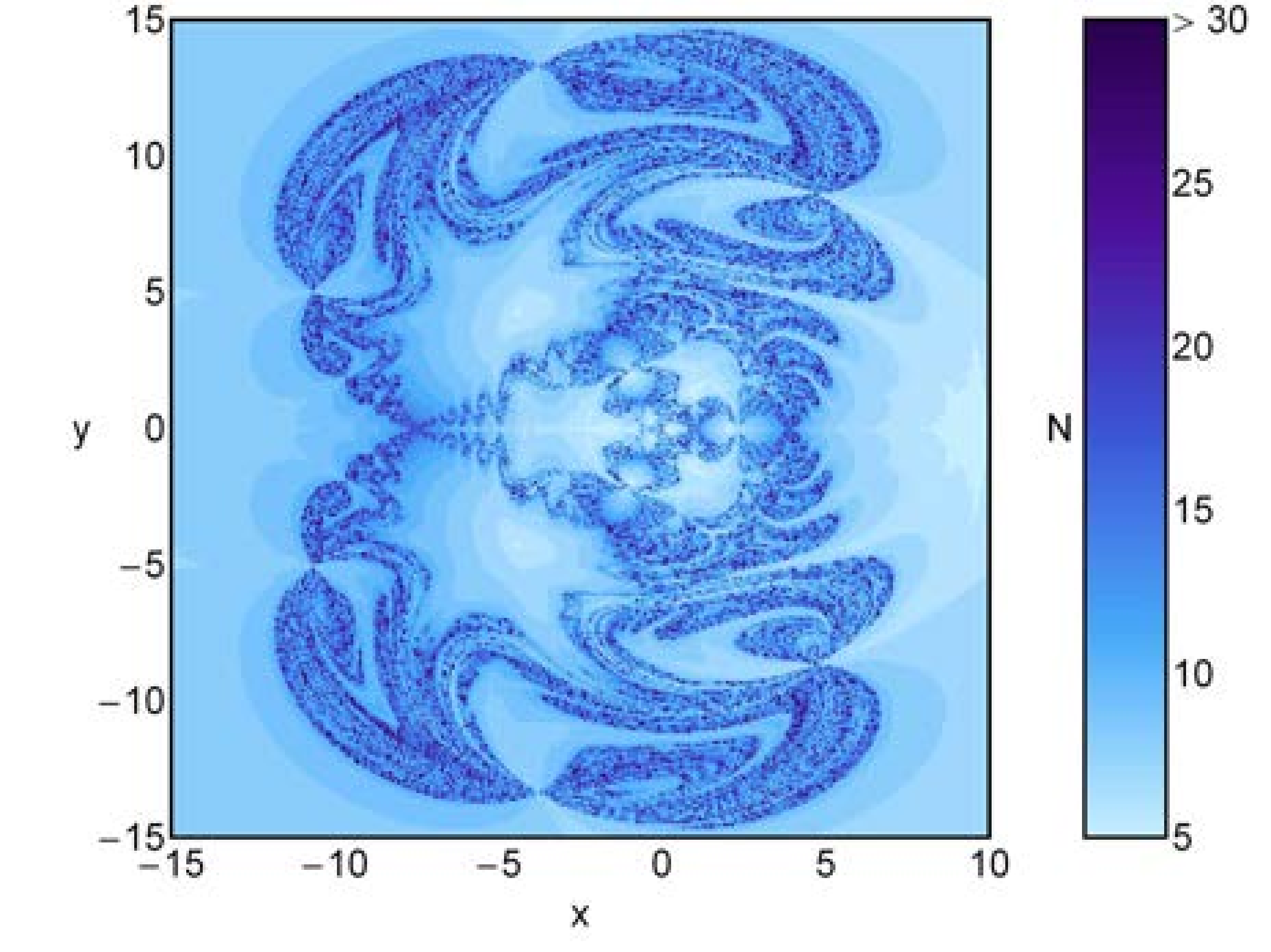}
(f)\includegraphics[scale=.25]{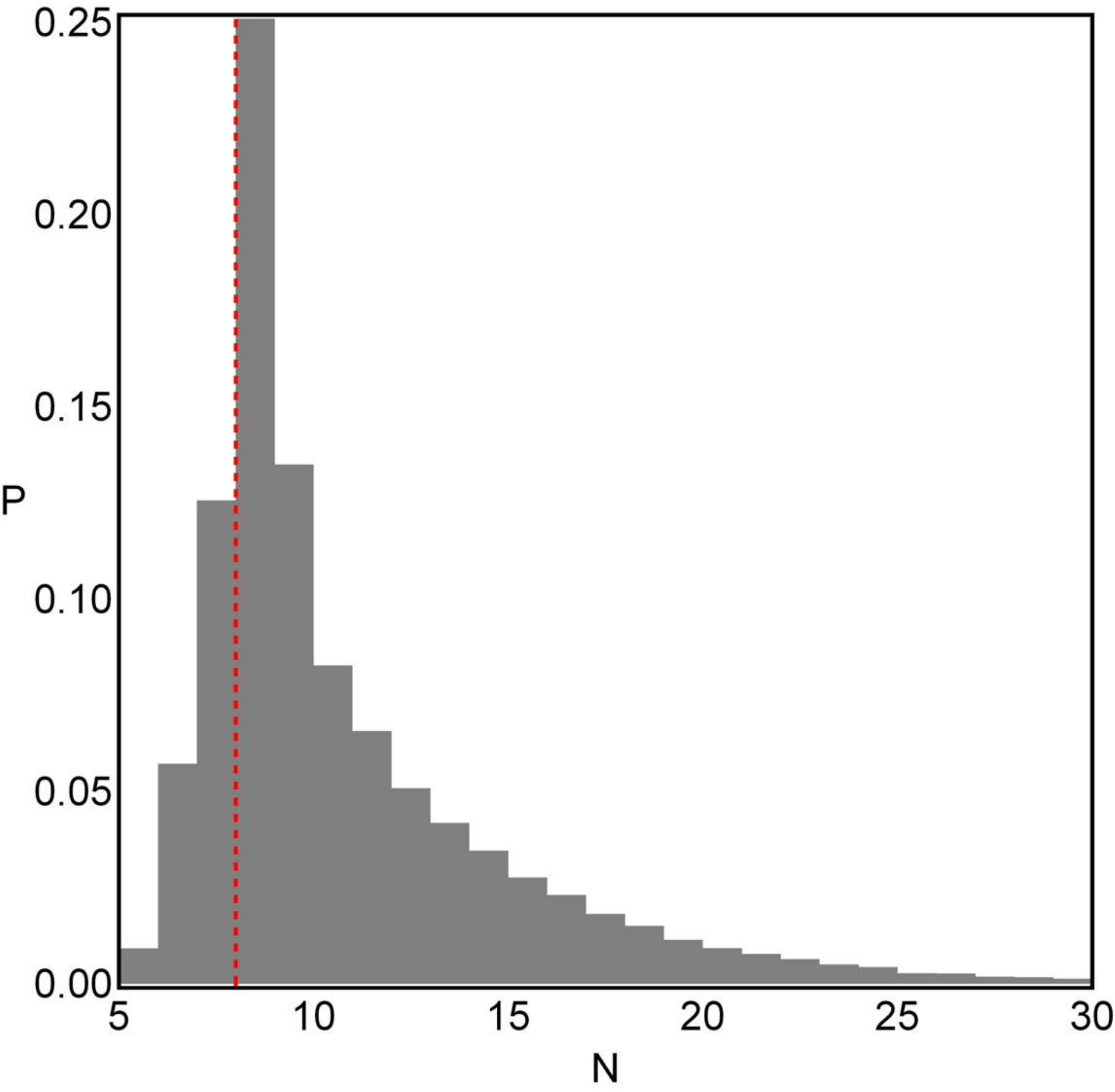}\\
(g)\includegraphics[scale=.27]{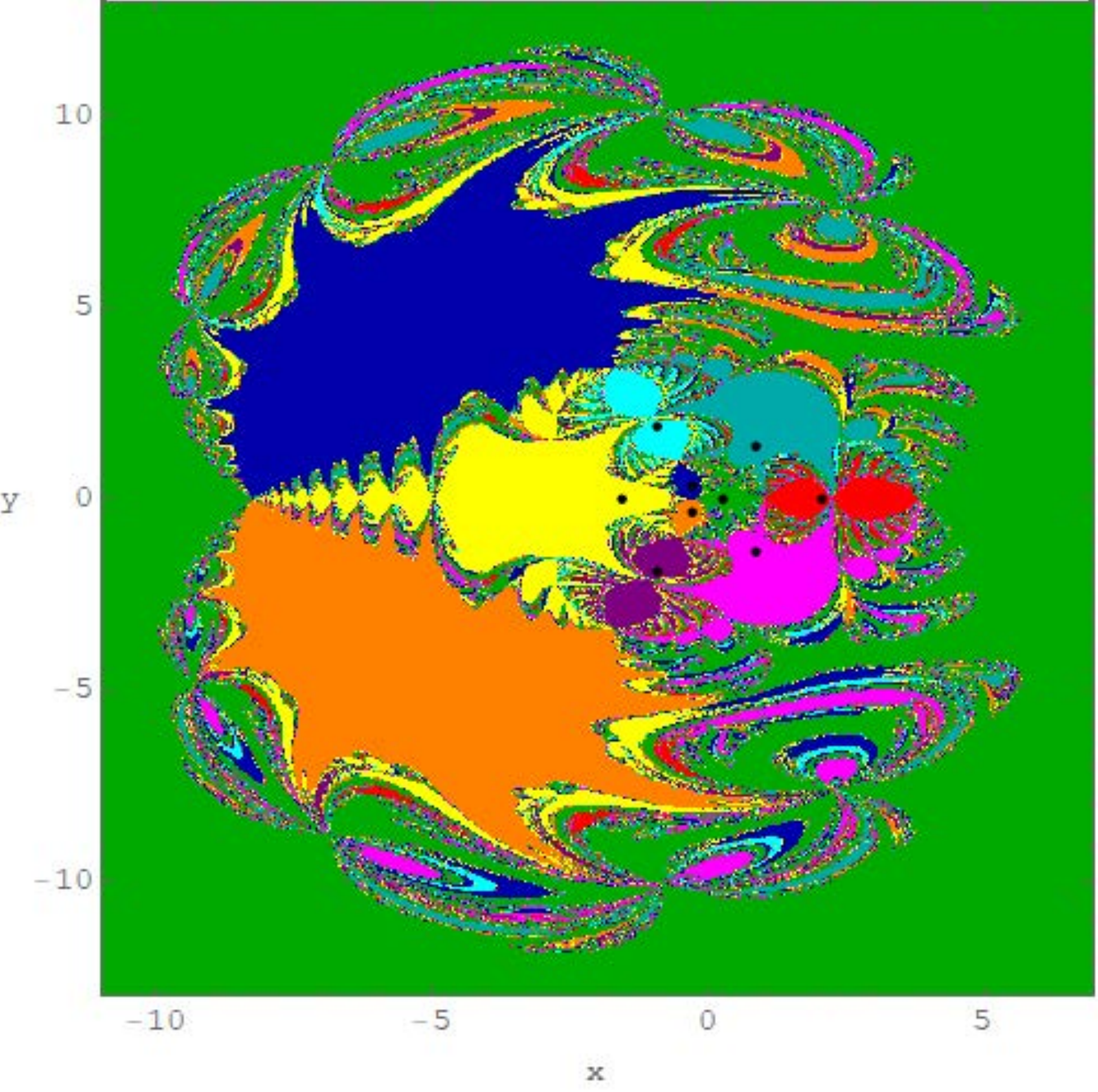}
(h)\includegraphics[scale=.27]{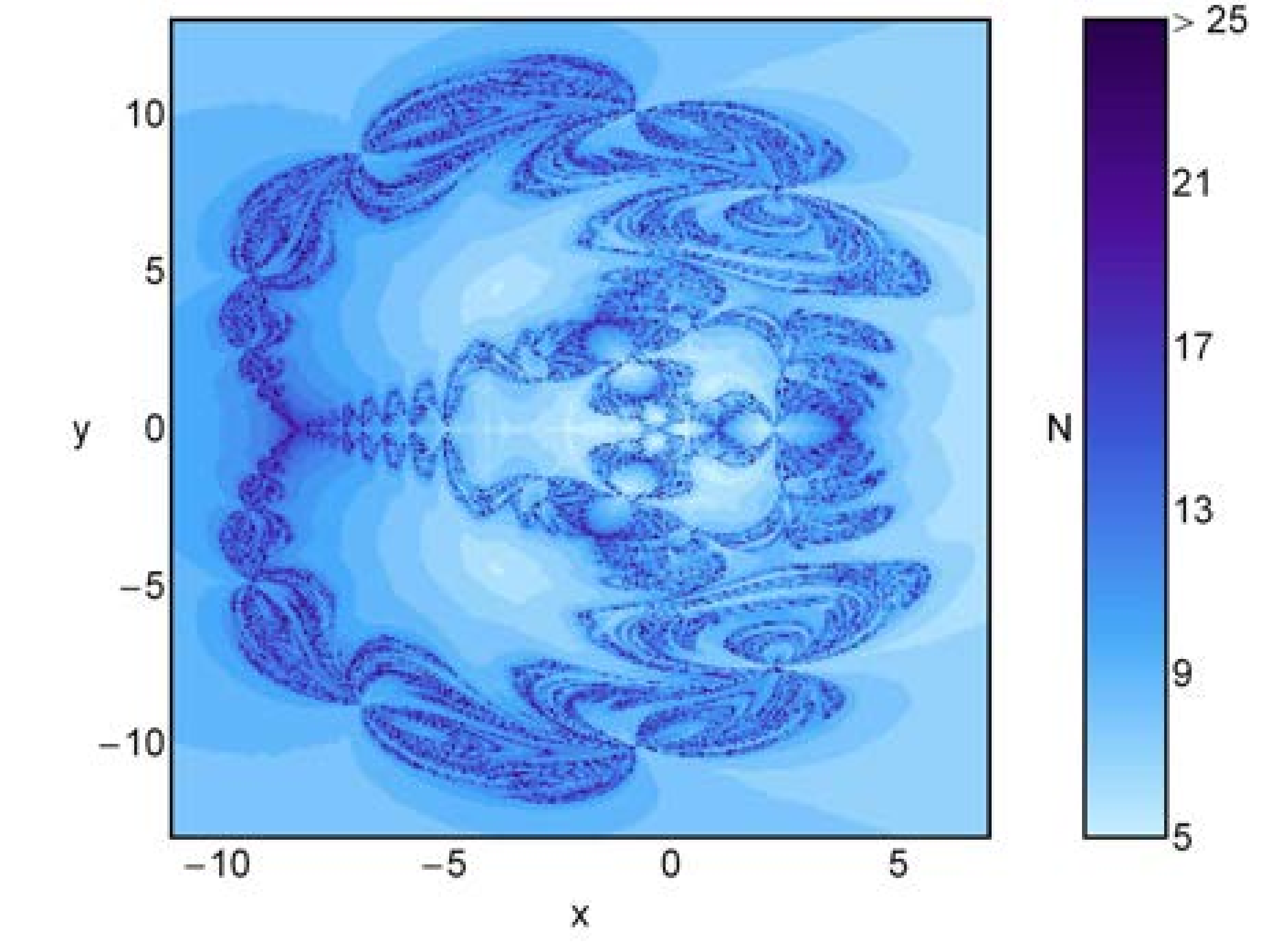}
(i)\includegraphics[scale=.25]{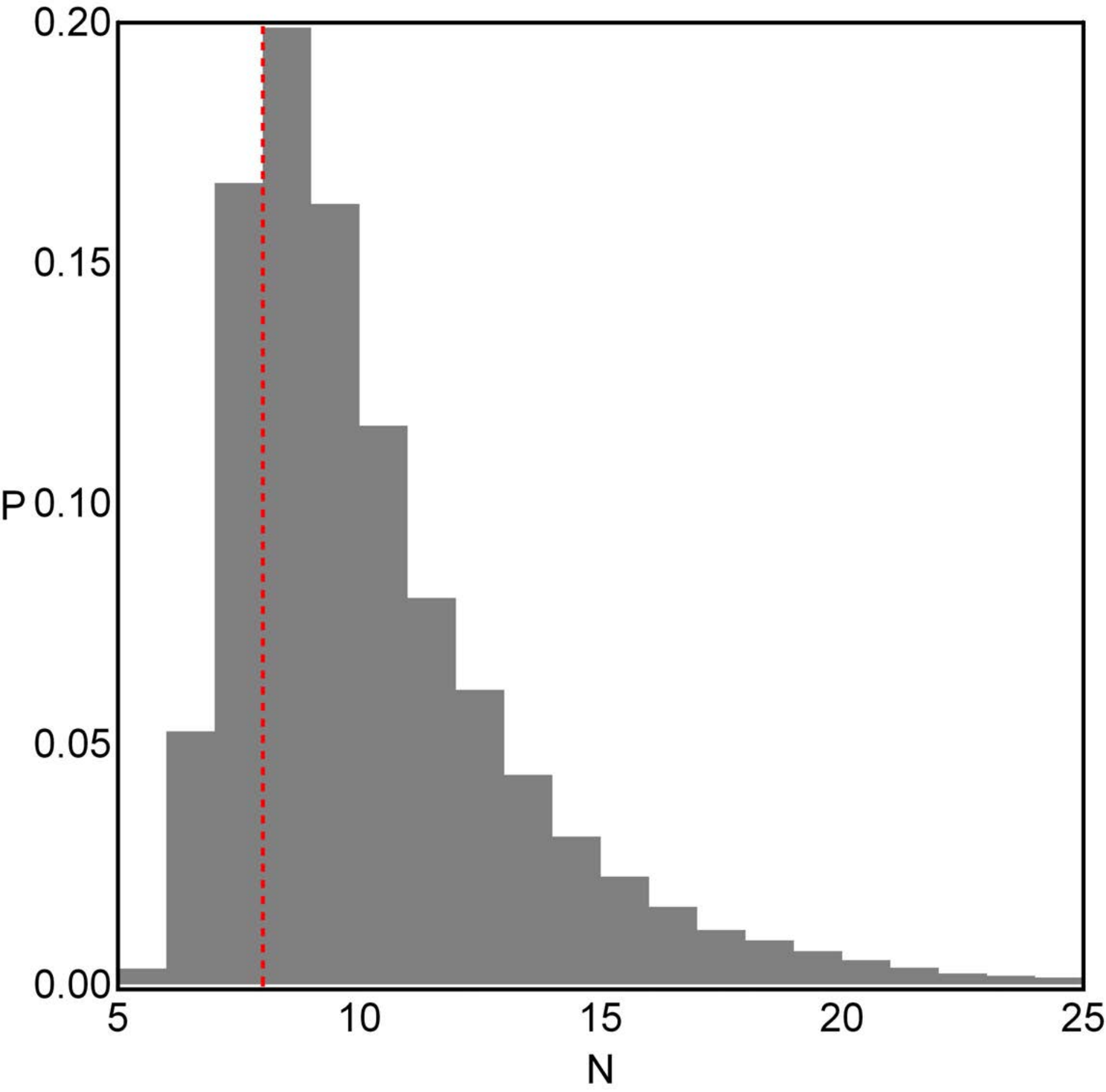}\\
(j)\includegraphics[scale=.27]{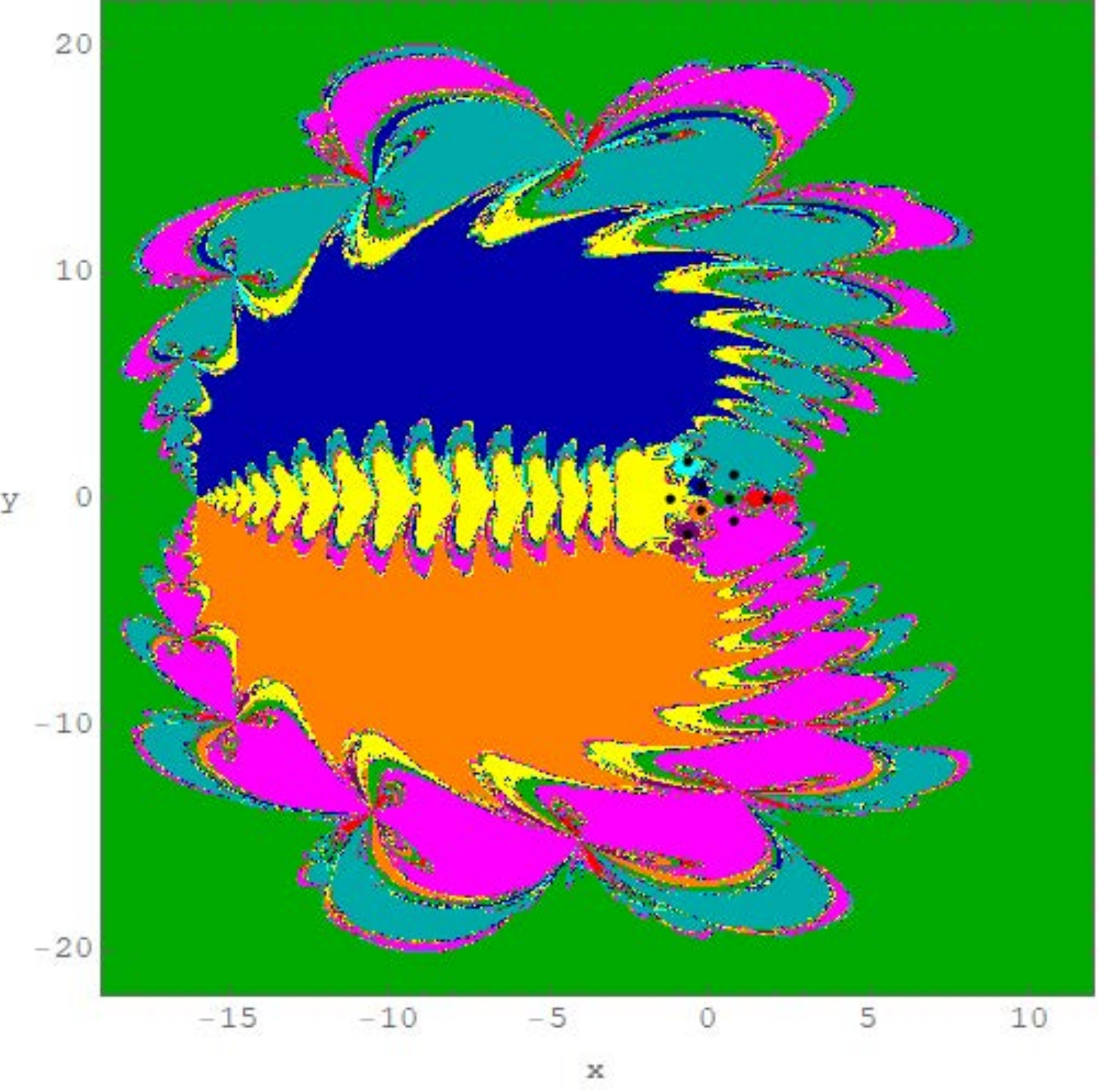}
(k)\includegraphics[scale=.27]{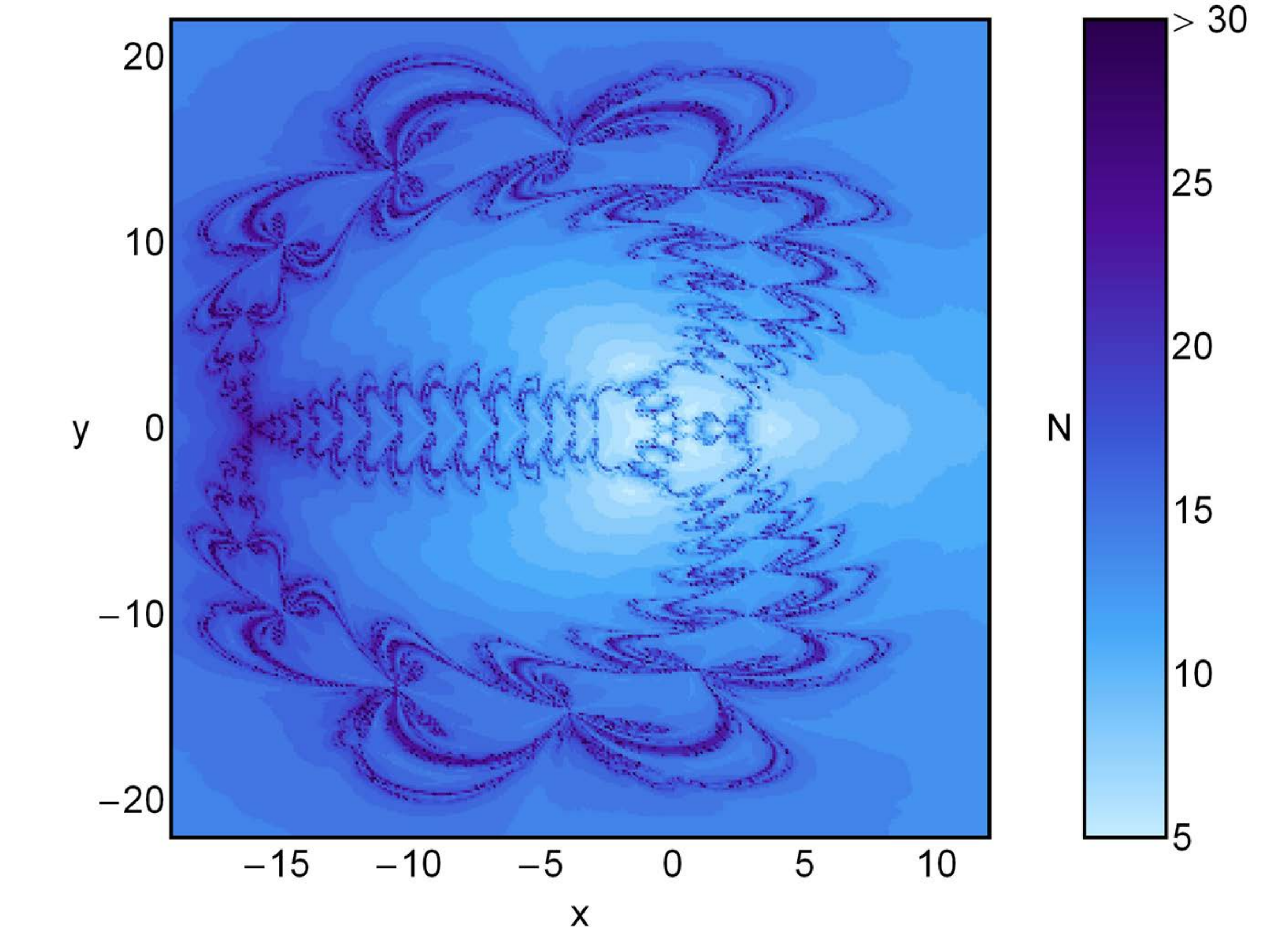}
(l)\includegraphics[scale=.25]{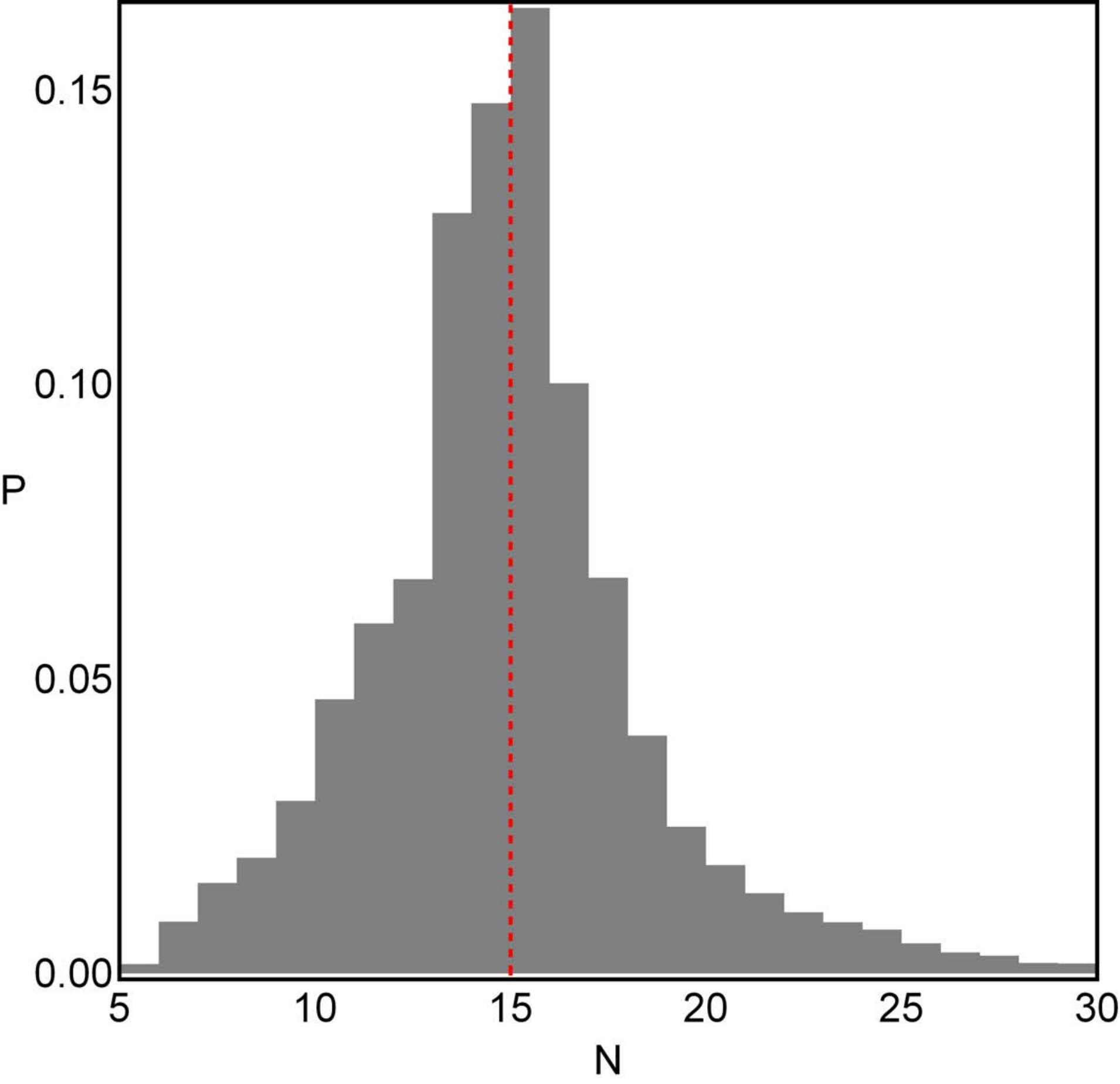}
\caption{The Newton-Raphson basins of attraction on the $xy$-plane for the
case when nine libration points exist for  fixed value of
$\alpha=57 \degree$  and for:
(a) $\beta=5 \degree$; (d) $\beta=10\degree$; (g) $\beta=15\degree$;
(j) $\beta=23 \degree$. The color code for the libration points $L_1$,...,$L_9$ is same as in Fig \ref{NR_Fig_1}; and non-converging points (white);  (b, e,  h, k) and (c, f, i, l) are the distribution of the corresponding number $(N)$ and the  probability distributions of required iterations for obtaining the Newton-Raphson basins of attraction shown in (a, d, g, j), respectively. (Color figure online).}
\label{NR_Fig_4}
\end{figure*}

In Fig. \ref{NR_Fig_2}(a, d, g), the basins of convergence associated with the libration points for the case $\alpha=57\degree$ and $\beta =1\degree, 1.5\degree, 2.5\degree$ respectively are depicted. In addition, Fig. \ref{NR_Fig_2}(b, e, h) and Fig. \ref{NR_Fig_2}(c, f, i) show the distribution of the corresponding number $N$ of required iterations and their corresponding probability distribution to obtain the Newton-Raphson basins of attraction respectively. The observations show that the most of the configuration $(x, y)$ plane is covered by well formed Newton Raphson basins of attraction with highly chaotic basins boundaries. The most notable changes can be summarized as follows:
\begin{itemize}
  \item [*]In all the panels, the extent in the basins of convergence linked to the libration point $L_2$ is infinite whereas, for the rest, the same is finite
  \item [*]The shapes of the basins of convergence, associated with the libration points $L_{4, 6, 8}$ and $L_{5, 7, 9}$ look like  nearly symmetrical (w.r.t. $x$-axis) magnetic fields. Owing to the symmetry, the shape of the basins of convergence linked with the non-collinear libration points are identical and exist in pairs.
  \item[*]Major basins areas for the first panel lie on the left of the origin whereas, the same lie on the right for the panel d. A slight variation in the value of $\beta$ brought a large change in the geometry of the corresponding basins of convergence. Therefore, difficulty to predict the change in the geometry is going to be a herculean task.
  \item[*]It is observed that the boundaries separating the major basins geometries are chaotic in nature. Further, the prediction of convergence of the respective boundaries for the initial conditions of the basins of geometry is very difficult. It is observed that the boundaries, separating the major basins are not only chaotic but also their convergence corresponding to the initial conditions associated with the libration points can't be predicted aptly.
  \item[*]A further increase in the value of $\beta$ ($\beta=2.5\degree$) ultimately resulted in shrinking of the domain of convergence corresponding to the finite extent. Further, a tendril, originating from the antennas of the bug-like geometry of the basins, appears.
  \item[*]The distributions of the corresponding number $N$ for the iterations to obtain the Newton-Raphson basins of attraction are shown using the blue shades (see Fig. \ref{NR_Fig_2}(b, e, h)). An analysis of these figures reveals that the initial conditions inside the attracting regions converge relatively faster than those of the initial conditions which lie in the vicinity of the basins boundaries.
  \item[*]The probability distribution of the iterations illustrated in Fig. \ref{NR_Fig_2}(c, f, i) unveils that the most probable number $N^*$ of the iterations is not constant. Its value decreases with the increase in the value of $\beta$, i.e., $N^*=14, 13, 10$ for the corresponding values of $\beta = 1\degree, 1.5\degree, 2.5\degree$ respectively.
\end{itemize}
%%%%%%%%%%%%
In Fig. \ref{NR_Fig_4}, we have plotted the Newton Raphson basins of attraction associated with the libration points in which the case under investigation takes into consideration of a scenario wherein 9 libration points exist, i.e., when $\alpha =57\degree$. The evolution of the geometry of the basins of attraction for four values of the angle parameter $\beta$ has been illustrated in Fig. \ref{NR_Fig_4} (a, d, g, j). It is seen that in all the cases, the configuration $(x, y)$ plane is covered by many well defined basins of convergence associated with the libration points. The extent of the basins of convergence associated with all the libration points except $L_2$  is finite. Further, in the vicinity of the basins boundaries a highly chaotic mixture of initial conditions is observed. The basins boundaries are highly chaotic and hence the final state (attractor) of the initial conditions inside the area is super sensitive. Particularly, even a slight change of the initial conditions leads to a completely different attractor. This ultimately means that the prediction of the final state (libration point) is almost next to impossible for such above mentioned initial conditions in the basins of boundaries.

An increase in the value of angle parameter $\beta$, for the structure of configuration $(x, y)$ plane, leads to drastic change in the basins of convergence associated with the libration points.

Gradually changes in the configuration $(x, y)$ plane for the increasing value of the angle parameter $\beta$ can be summarized as follows:
\begin{itemize}
  \item In Fig. \ref{NR_Fig_4}a, when the value of $\beta =5\degree$, most of the area is found to be covered by a chaotic sea composed of several initial conditions.
  \item Further, when $\beta = 23\degree$, the chaotic sea of the initial conditions found to be condensing into well defined basins of convergence. This phenomenon continues to condense further gradually with the increase in the value of angle parameter $\beta$ and finally leads to a butterfly wing shaped domain of convergence, corresponding to the libration points $L_4$ and $L_5$ in Fig. \ref{NR_Fig_4}j.
\end{itemize}

In Fig. \ref{NR_Fig_4}(b, e, h, k), number $N$, which corresponds to number of iterations required to converge the particular initial condition to one of the attractor, has been presented through blue colored shades. More dark blue regions mean, more number of iteration points, are involved. Figure \ref{NR_Fig_4}b represents that most of the initial conditions lying in chaotic sea require more number of iterations whereas, the initial conditions inside the domain of basins of convergence require less number of iteration points to converge.

In Fig. \ref{NR_Fig_4}(c, f, i, l), the corresponding probability distribution for the most probable number of required iterations $N^*$ are illustrated. This reveals that the number first decreases from 11 to 8 in first three panels while it increases to 15 in the last one. Hence, it is not possible to predict the value of $N^*$ with the increase in value of angle parameter $\beta$.

\begin{figure*}[!t]
\centering
(a)\includegraphics[scale=.27]{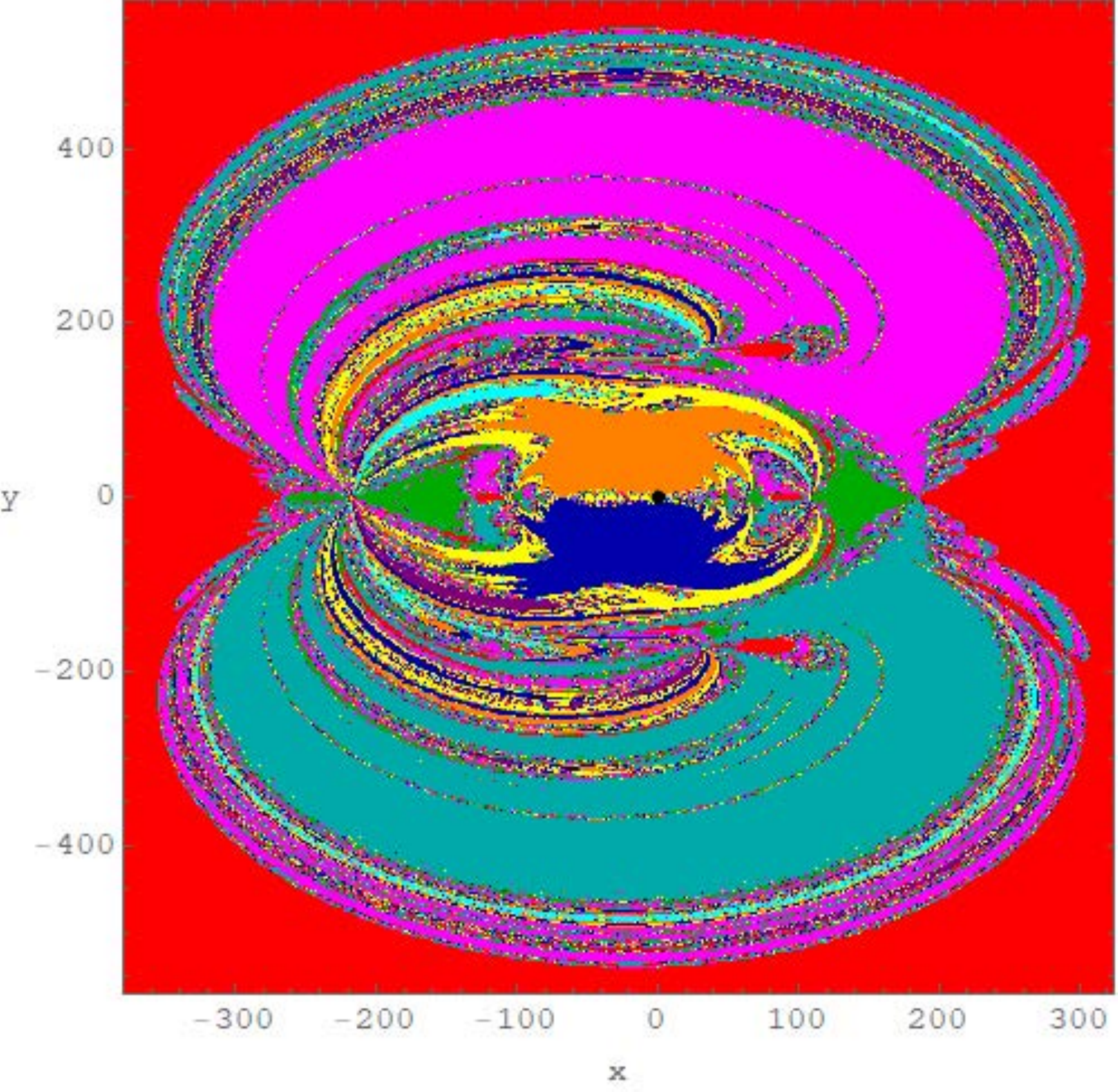}
(b)\includegraphics[scale=.27]{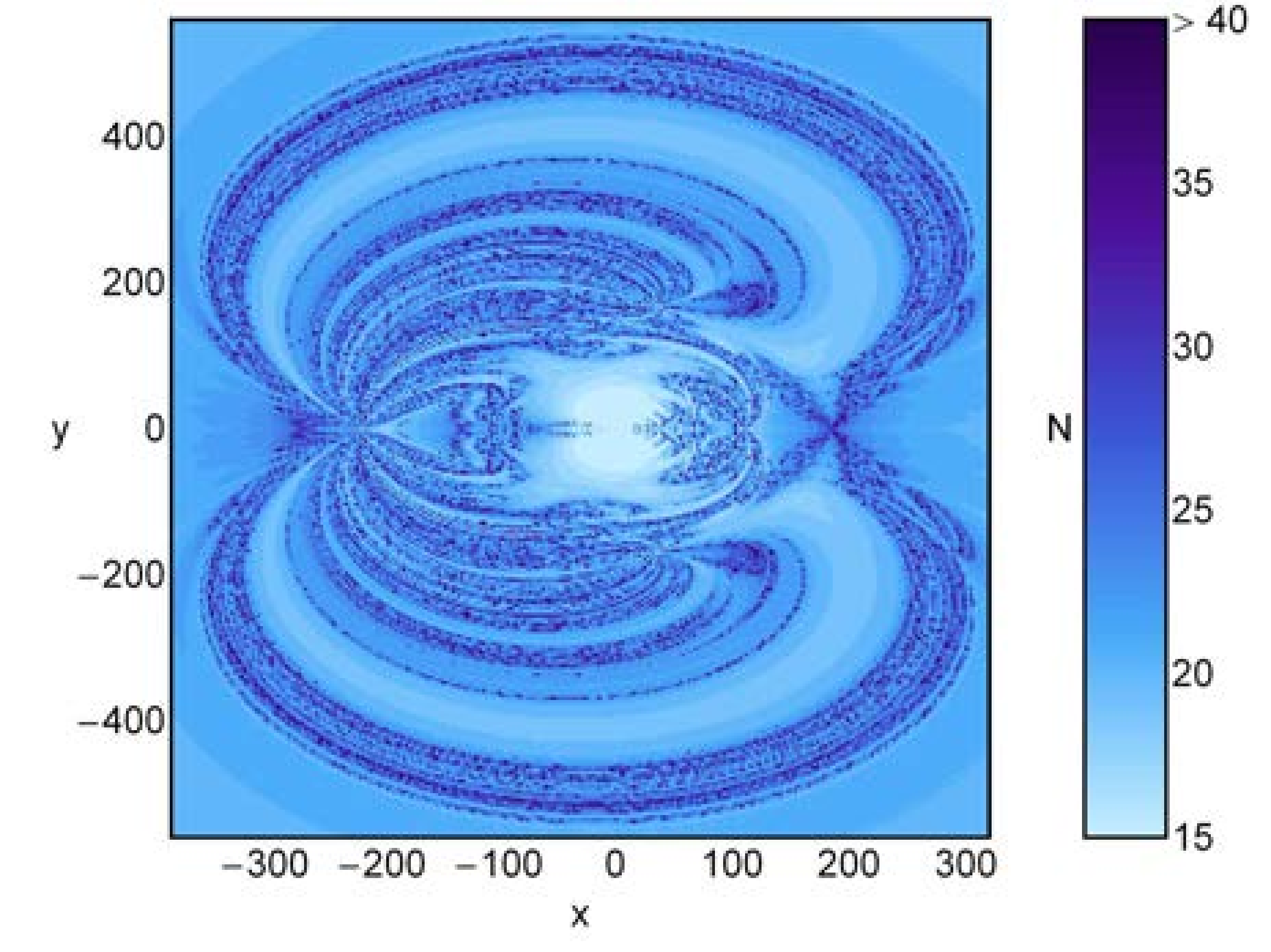}
(c)\includegraphics[scale=.25]{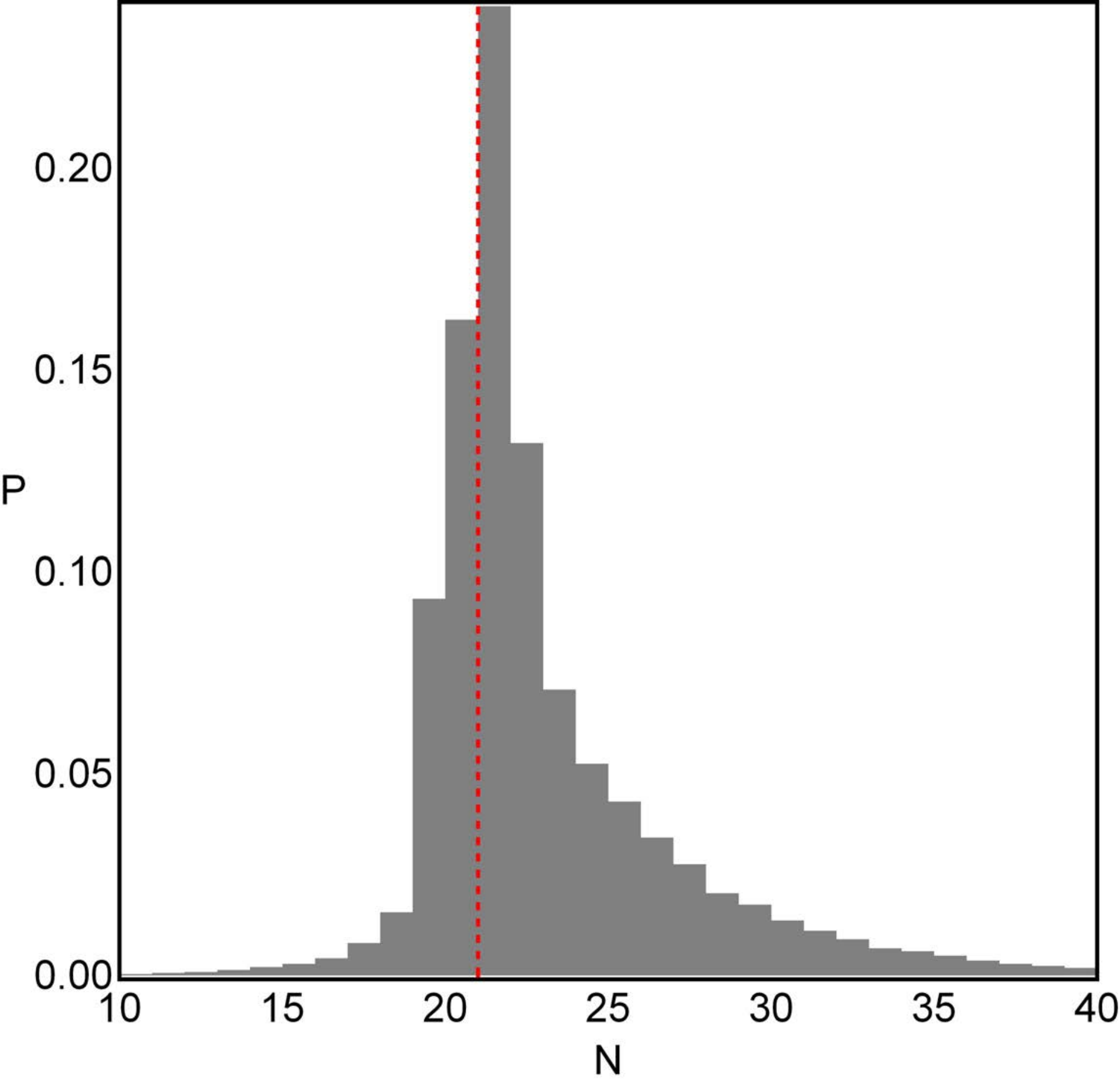}\\
(d)\includegraphics[scale=.27]{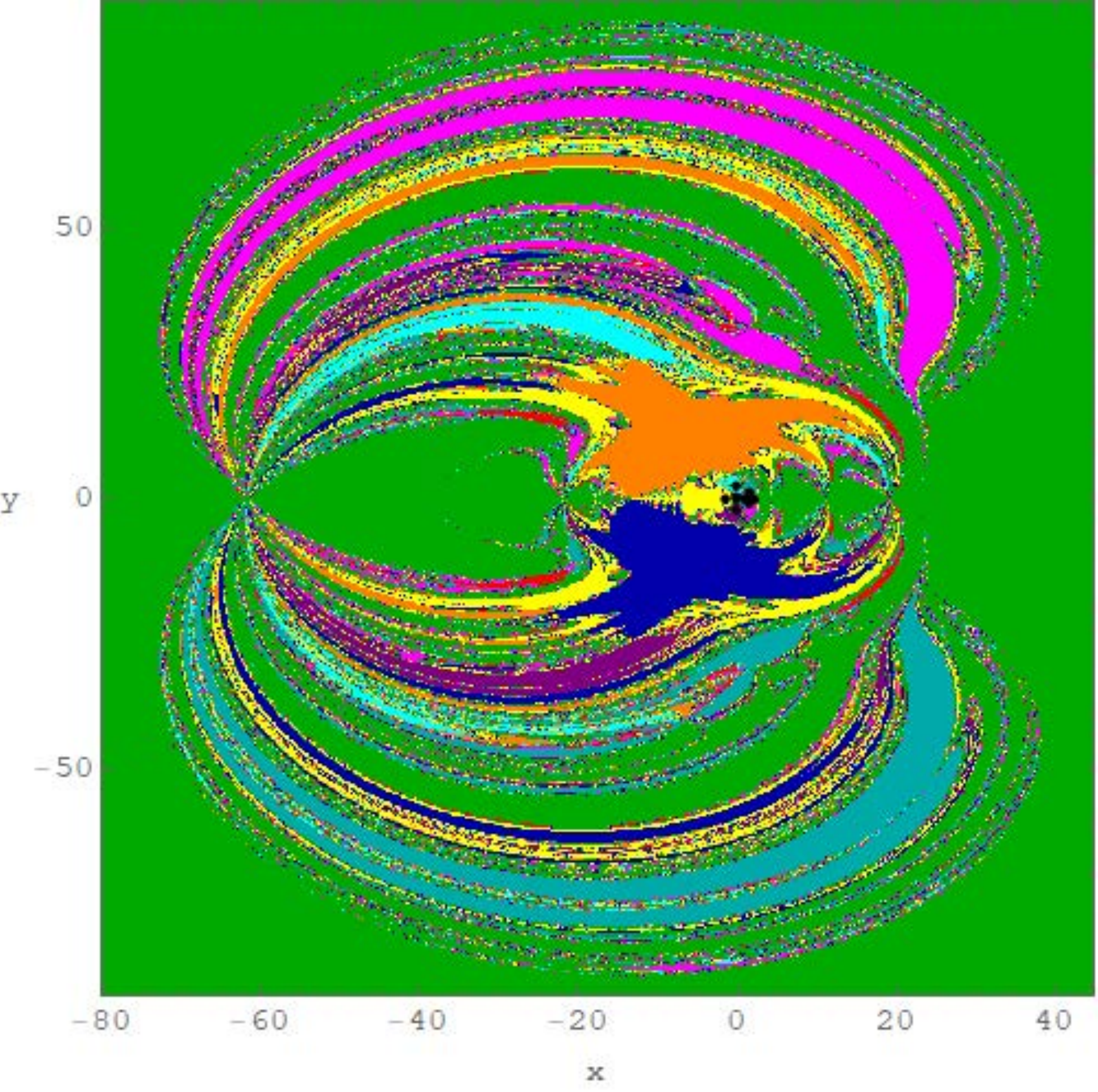}
(e)\includegraphics[scale=.27]{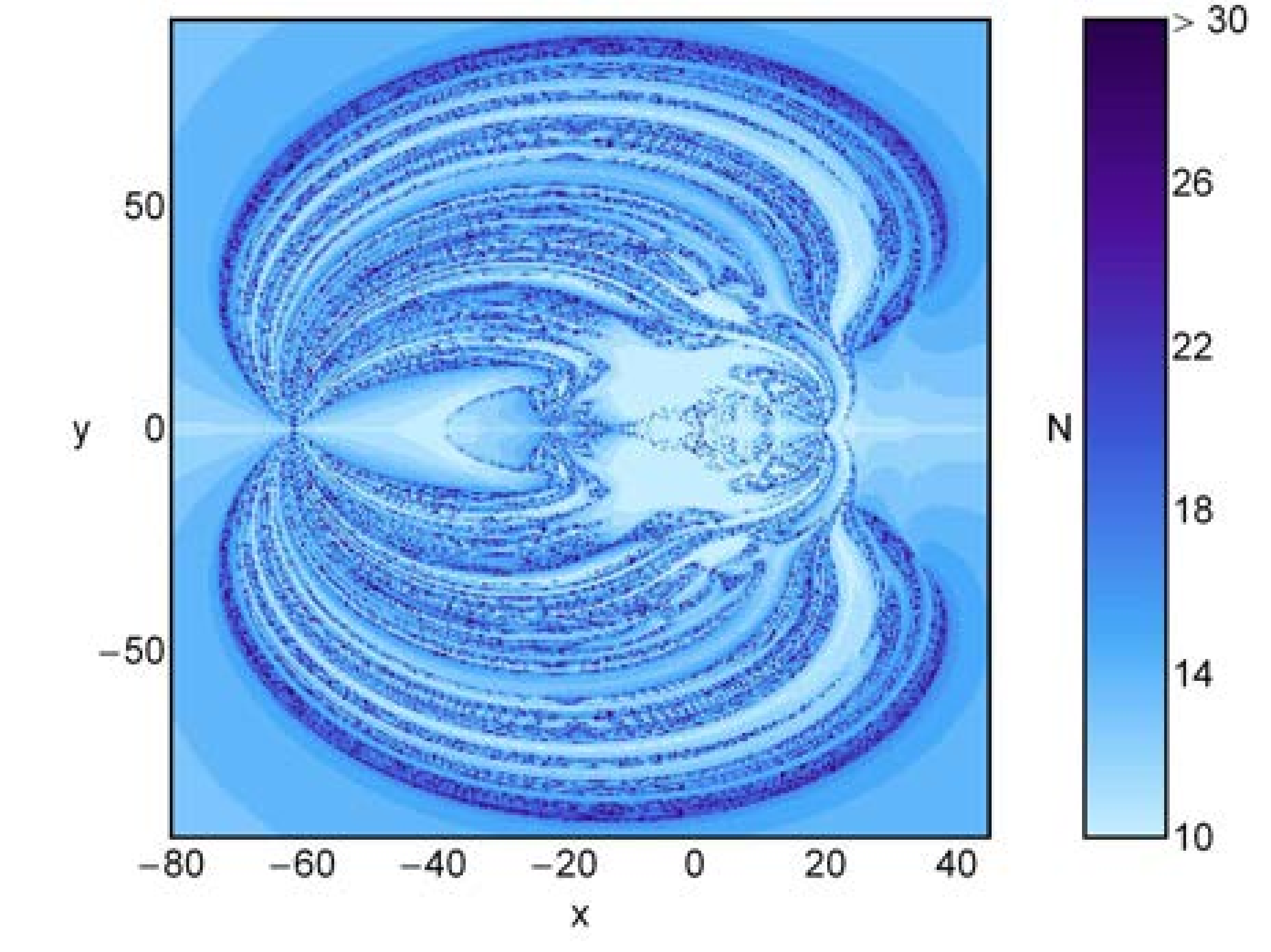}
(f)\includegraphics[scale=.25]{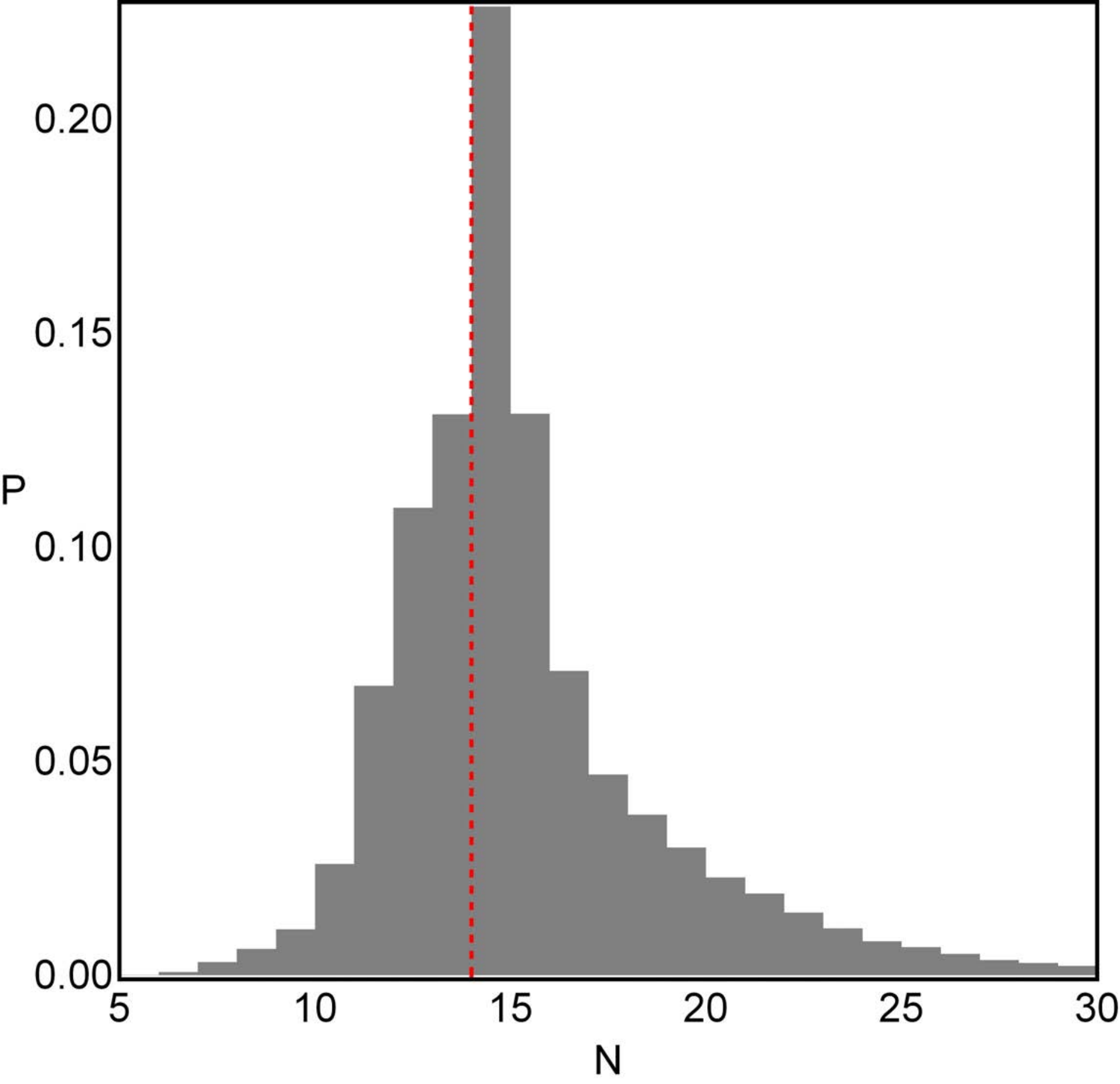}\\
(g)\includegraphics[scale=.27]{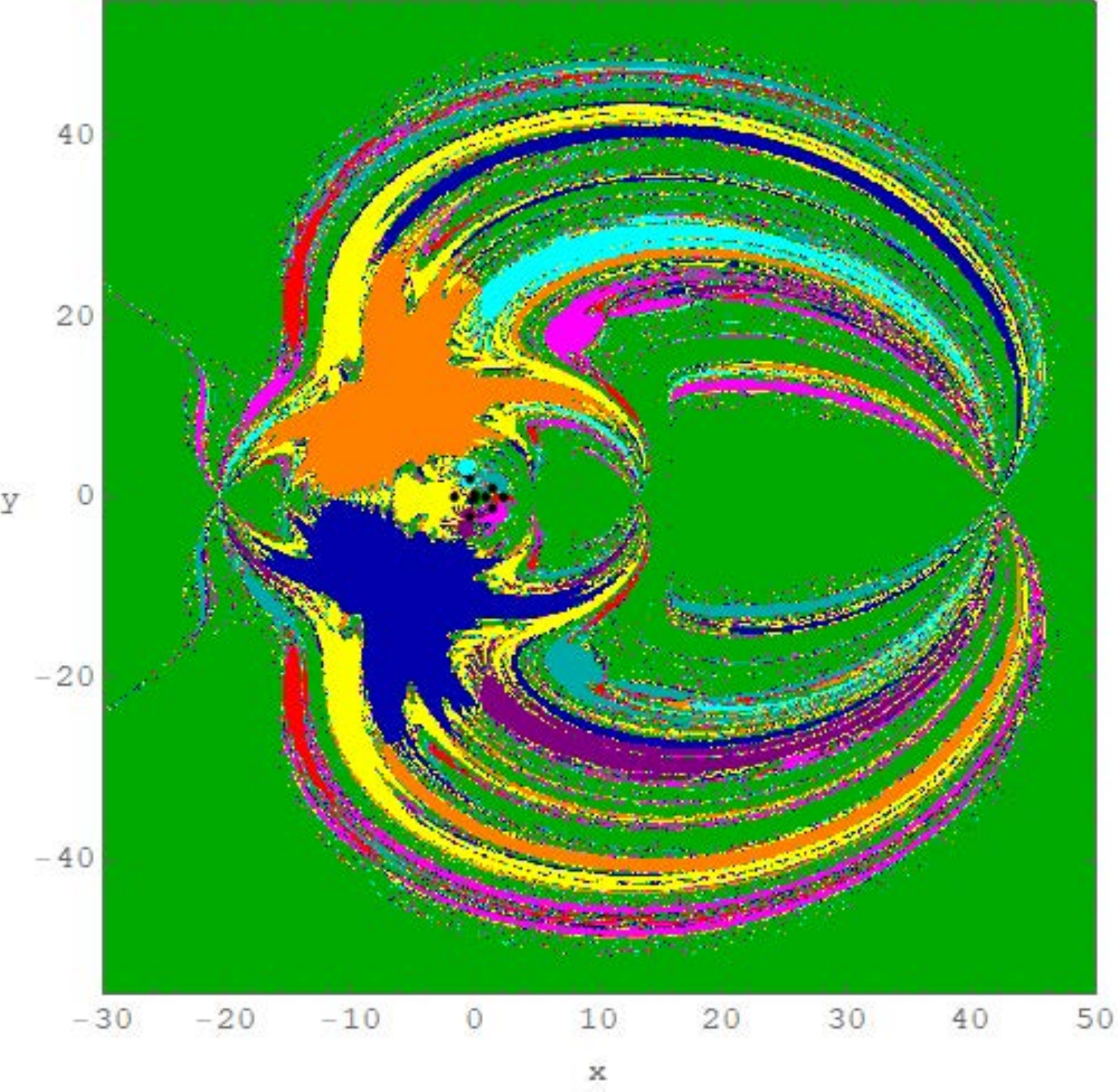}
(h)\includegraphics[scale=.27]{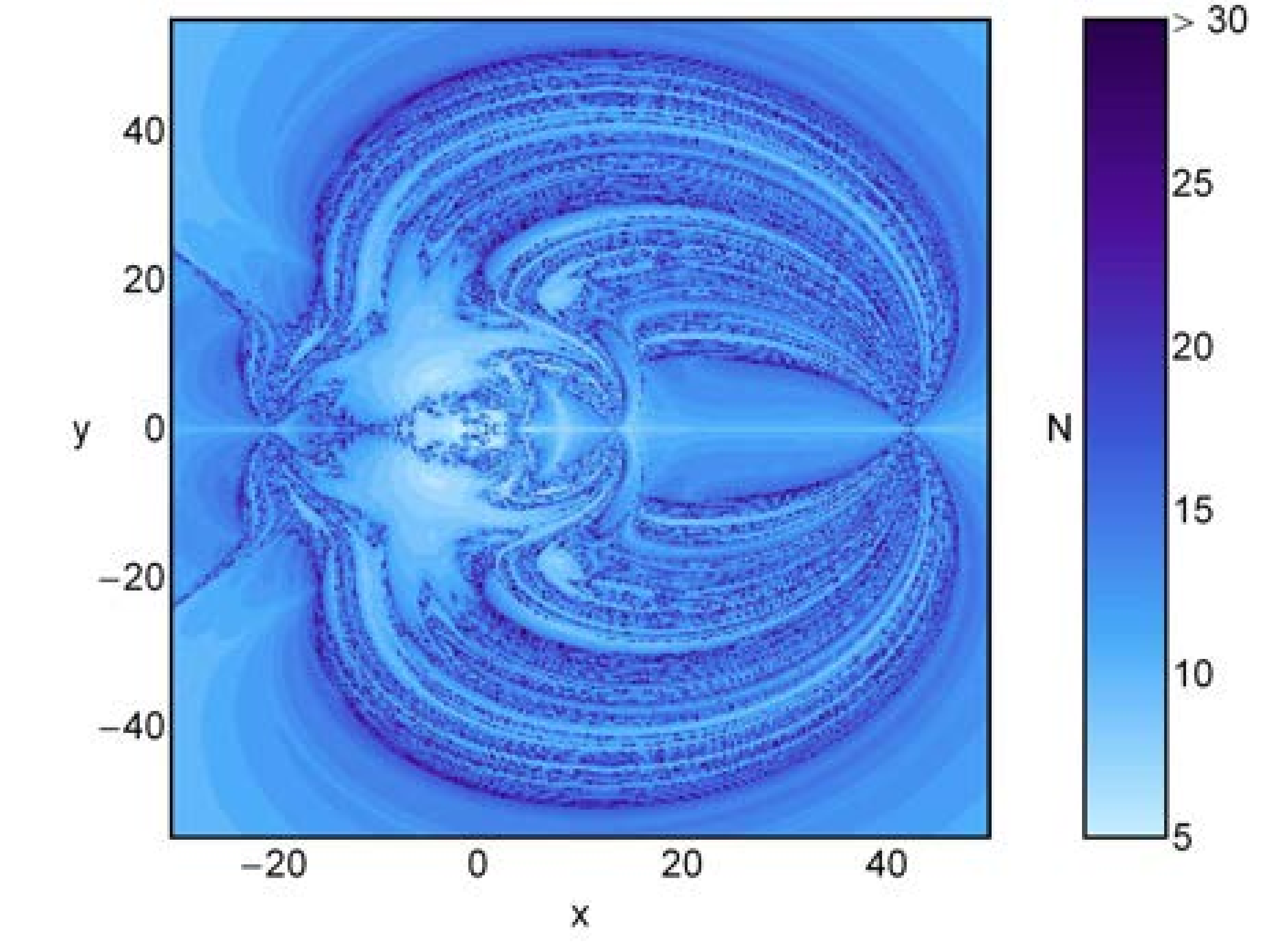}
(i)\includegraphics[scale=.25]{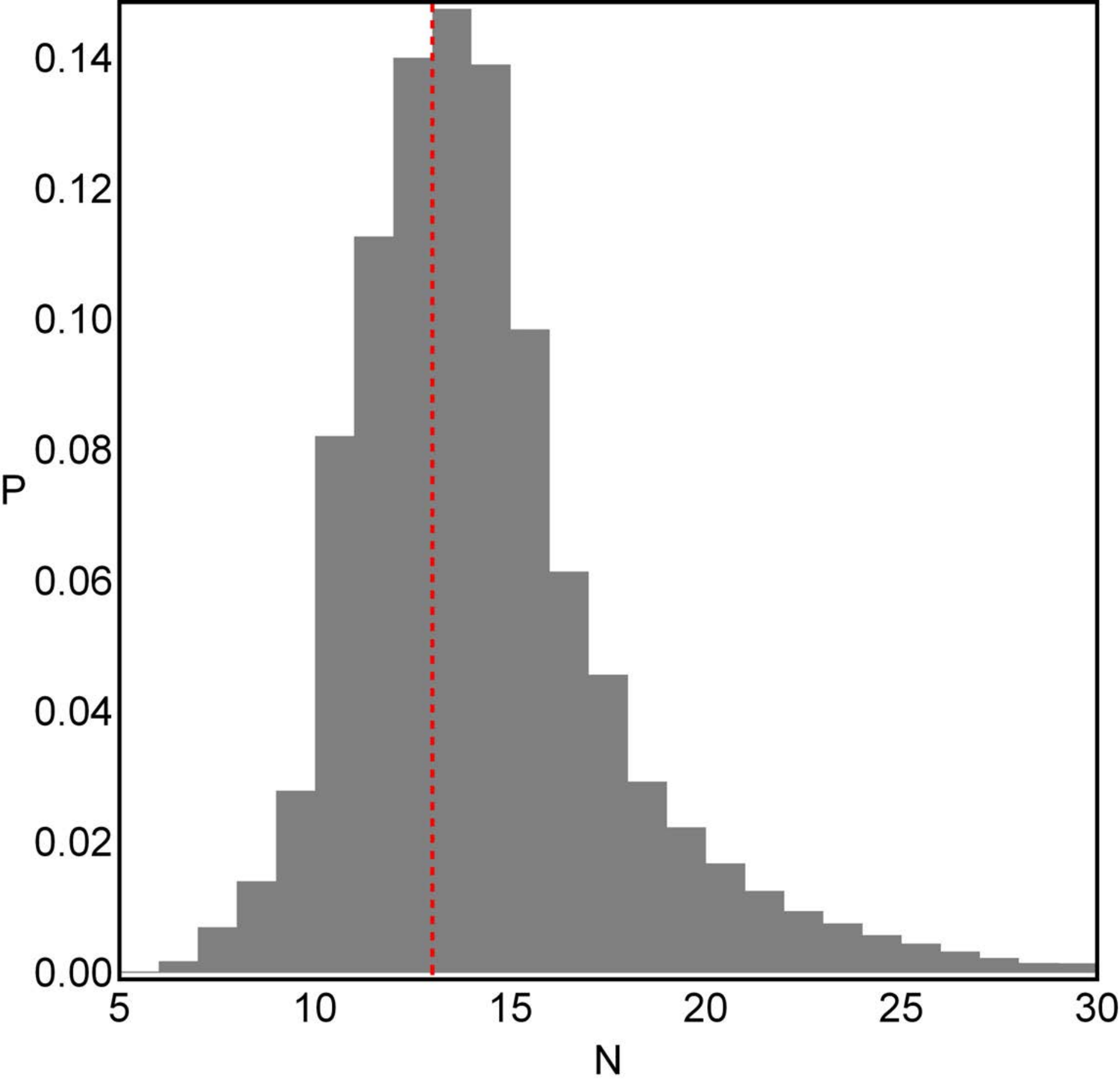}
\caption{The Newton-Raphson basins of attraction on the $xy$-plane for the
case when nine libration points exist for  fixed value of
$\alpha=58 \degree$  and for:
(a) $\beta=1/30 \degree$; (d) $\beta=1\degree$; (g) $\beta=1.5\degree$. The color code for the libration points $L_1$,...,$L_9$ is same as in Fig \ref{NR_Fig_1};  and non-converging points (white);  (b, e,  h) and (c, f, i) are the distribution of the corresponding number $(N)$ and the  probability distributions of required iterations for obtaining the Newton-Raphson basins of attraction shown in (a, d, g), respectively.
 (Color figure online).}
\label{NR_Fig_5}
\end{figure*}

Figure  \ref{NR_Fig_5} (a, d, g) represents the basins of attraction wherein 9 libration points exist. This figure has been drawn for fixed value of $\alpha$ and different values of $\beta$. We may note that extent of basins of convergence associated with libration point $L_1$ is infinite in Fig. \ref{NR_Fig_5}a whereas in other two figures (i.e., Fig. \ref{NR_Fig_5} d, g) infinite extent is for the libration point $L_2$. In Fig. \ref{NR_Fig_5}d, more chaotic area is in the left side of the origin whereas, in Fig. \ref{NR_Fig_5}g, it is in the right hand side of the origin.  In Fig. \ref{NR_Fig_5}(b, e, h), the distribution of the required number of iteration are illustrated. It is revealed that the  most of initial conditions converge to one of the attractors for $N<15$ whereas for the initial conditions falling inside the basin boundaries require more iterations to converge at one of the attractors. As the value of $\beta$ increases the most probable number of required iteration decreases from 22 to 13 (see. Fig. \ref{NR_Fig_5}(c, f, i)).

\begin{figure*}[!t]
\centering
(a)\includegraphics[scale=.27]{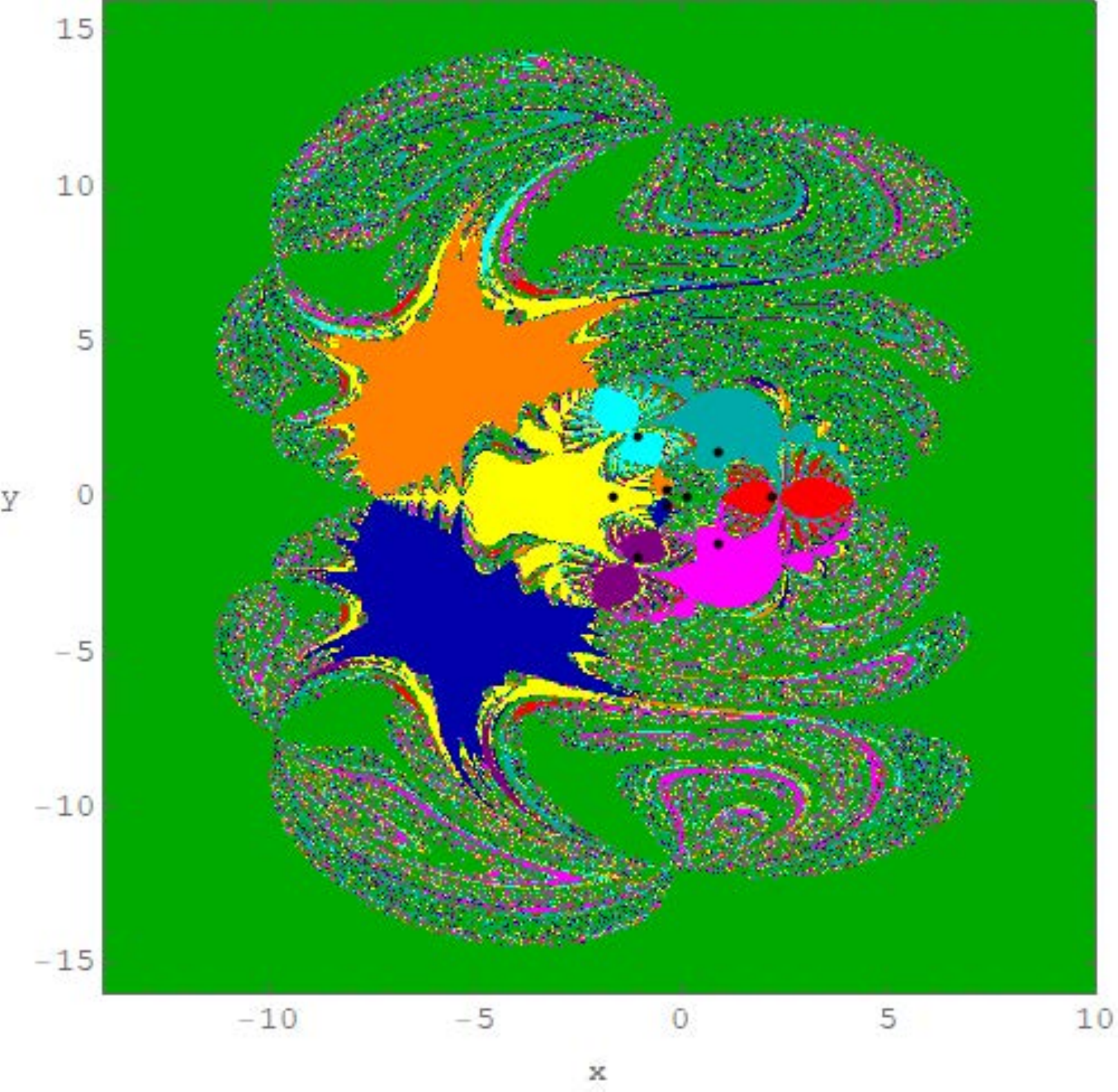}
(b)\includegraphics[scale=.27]{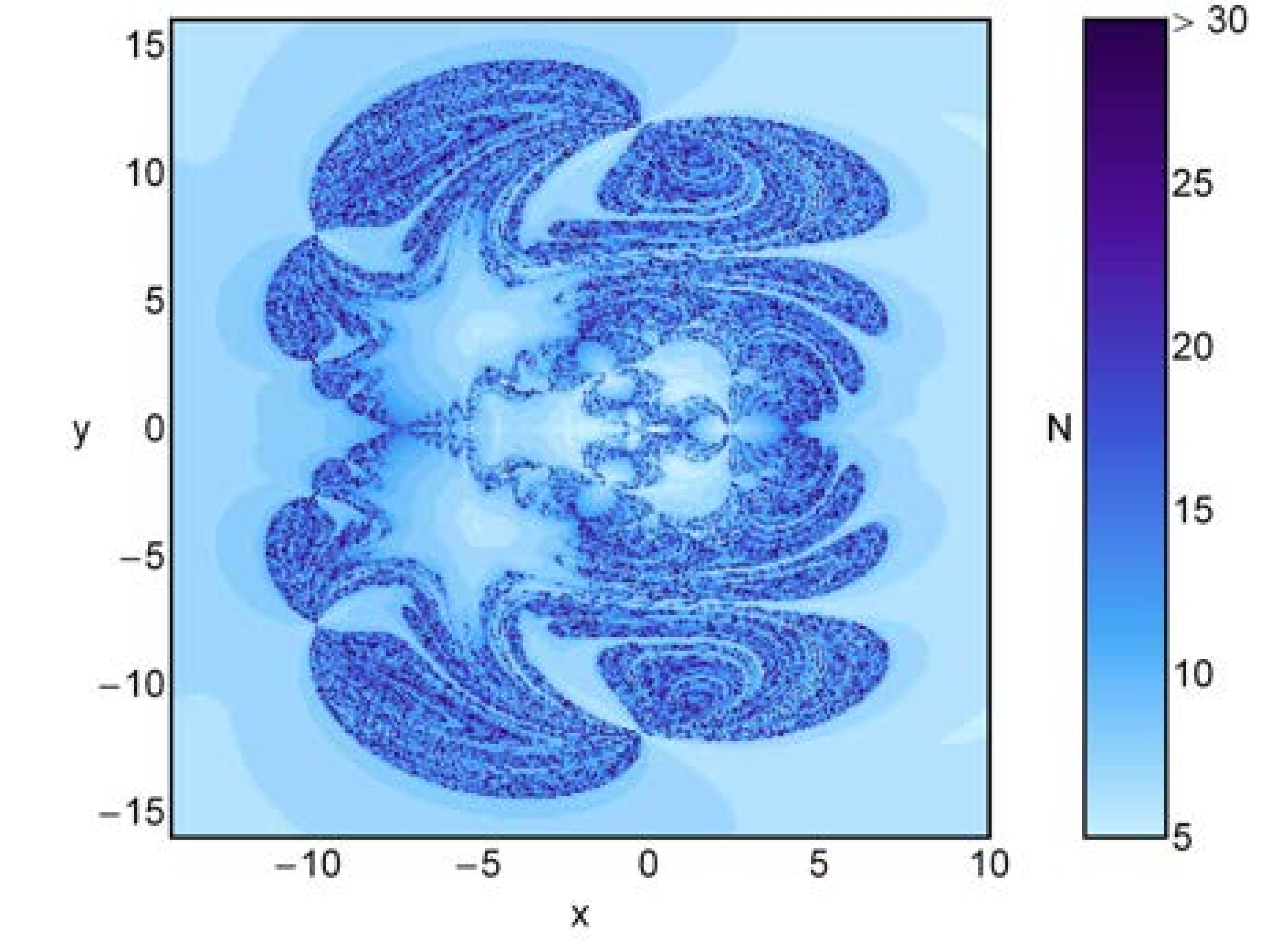}
(c)\includegraphics[scale=.25]{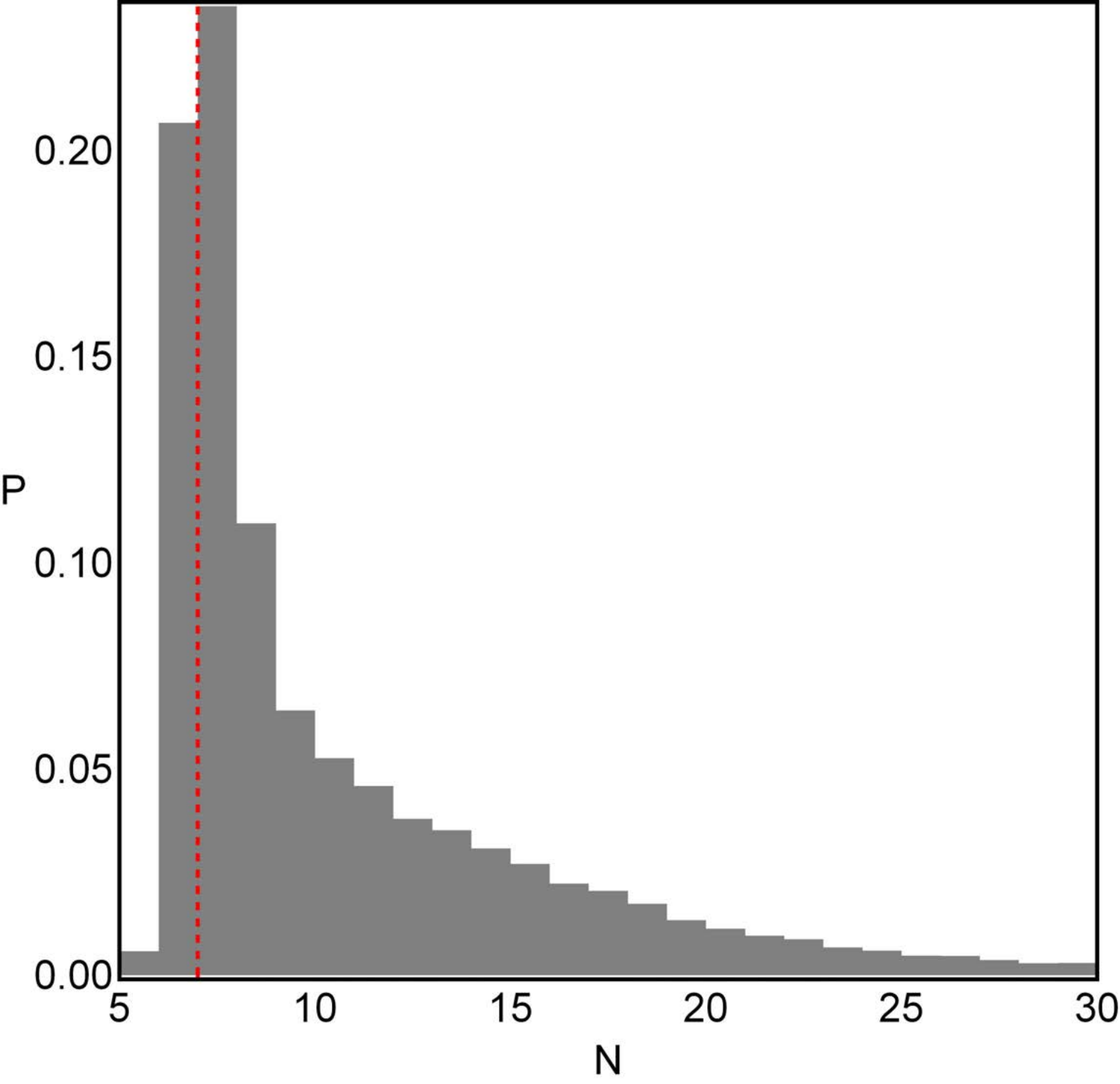}\\
(d)\includegraphics[scale=.27]{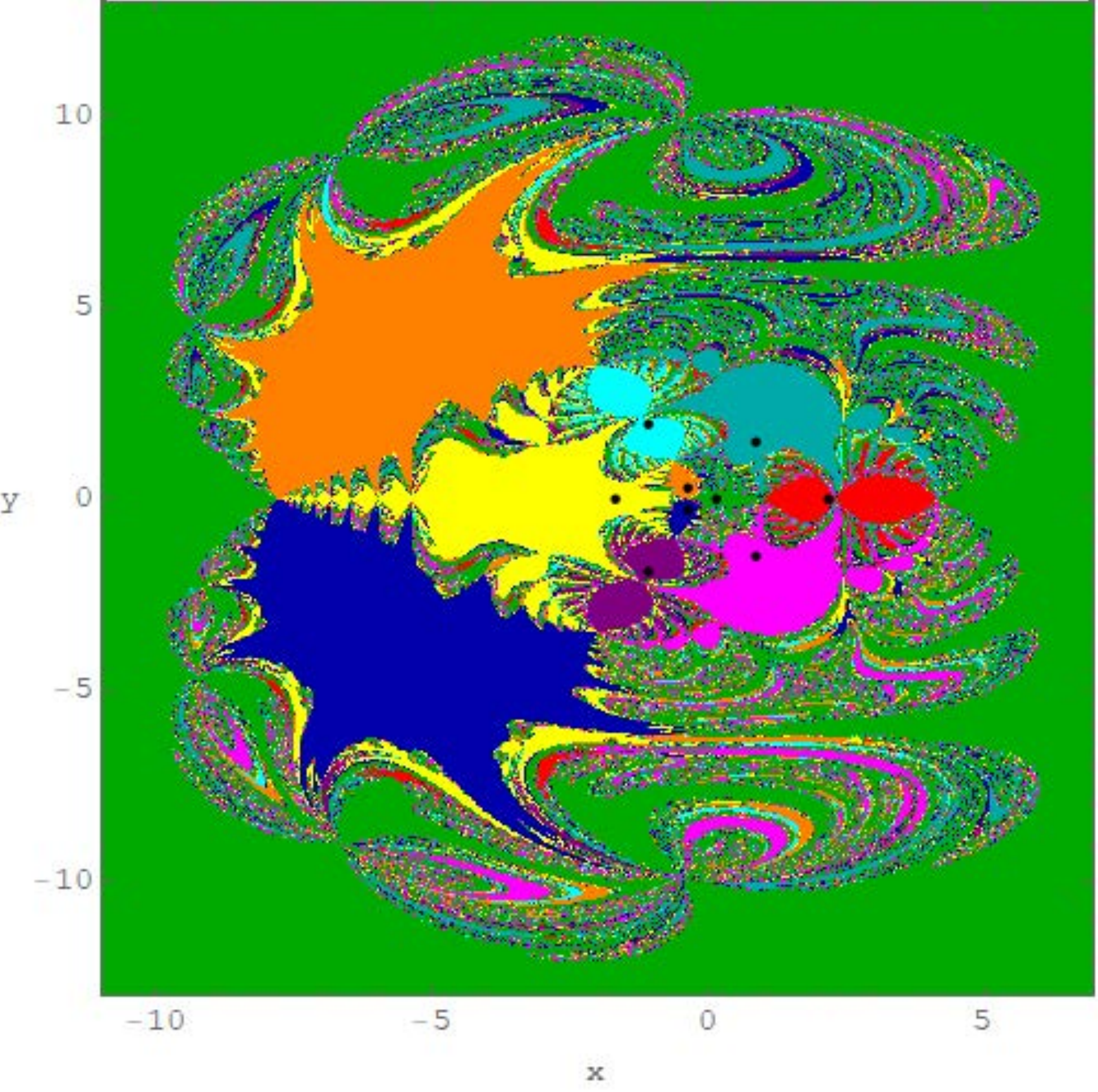}
(e)\includegraphics[scale=.27]{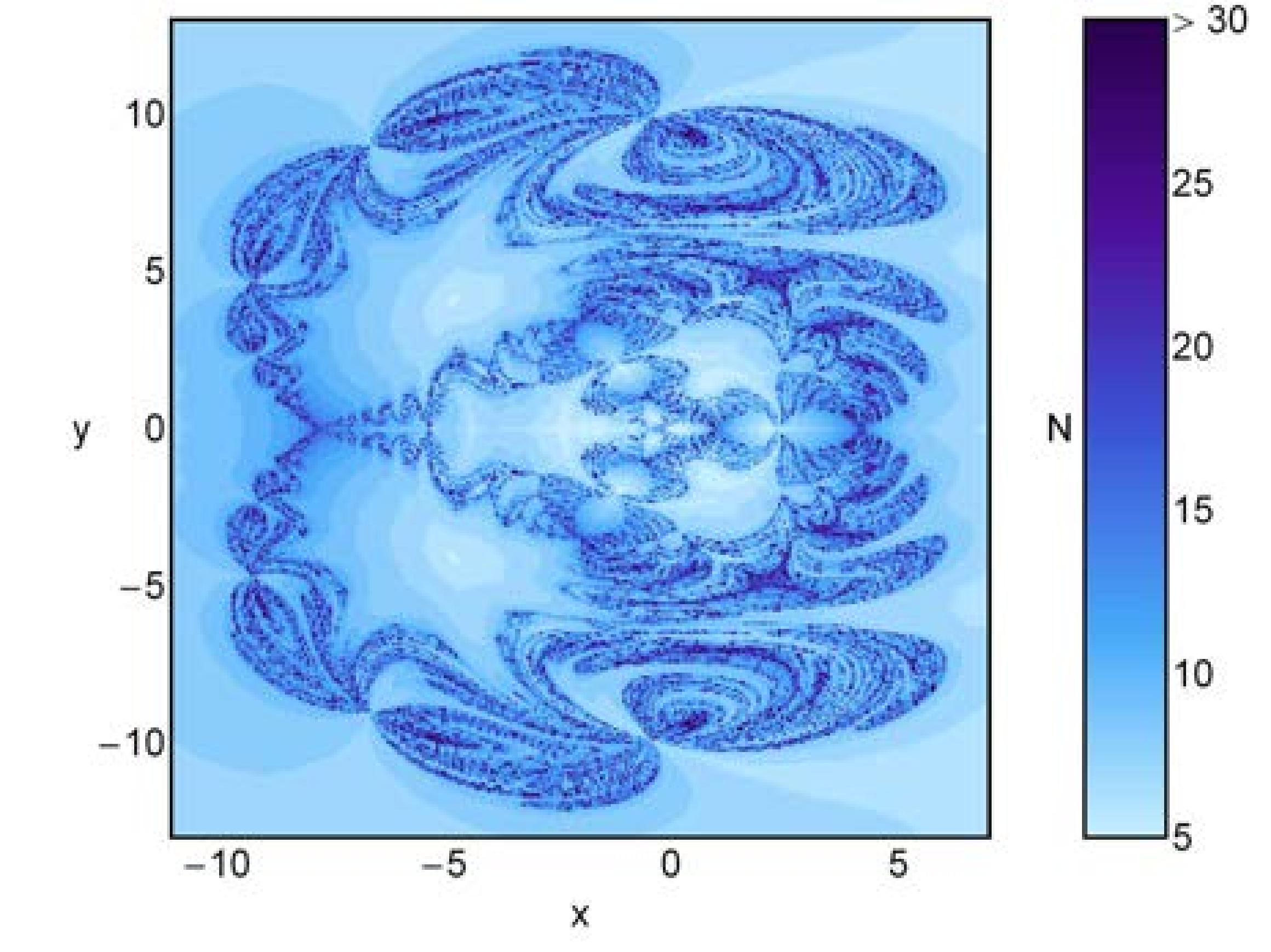}
(f)\includegraphics[scale=.25]{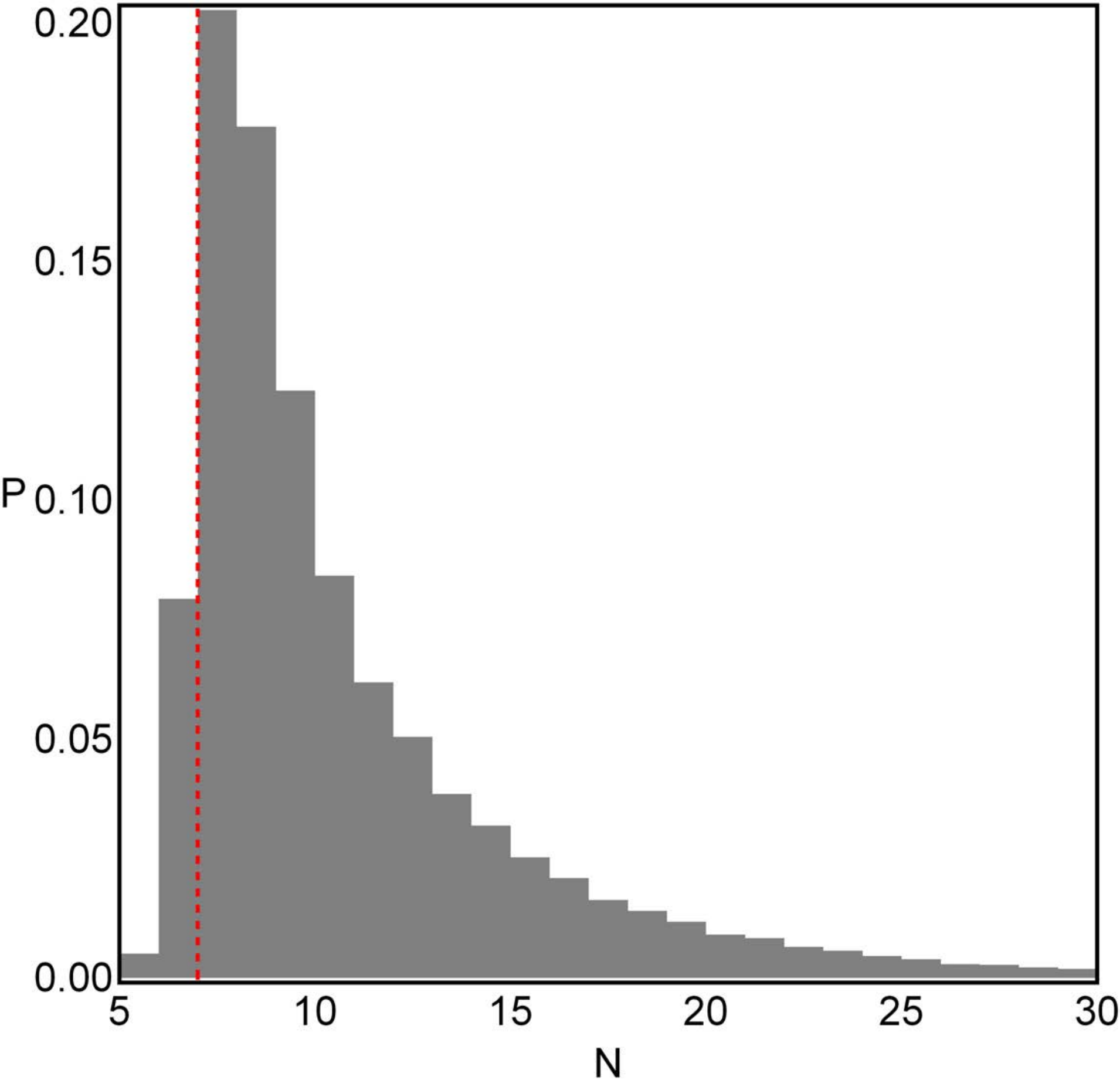}\\
(g)\includegraphics[scale=.27]{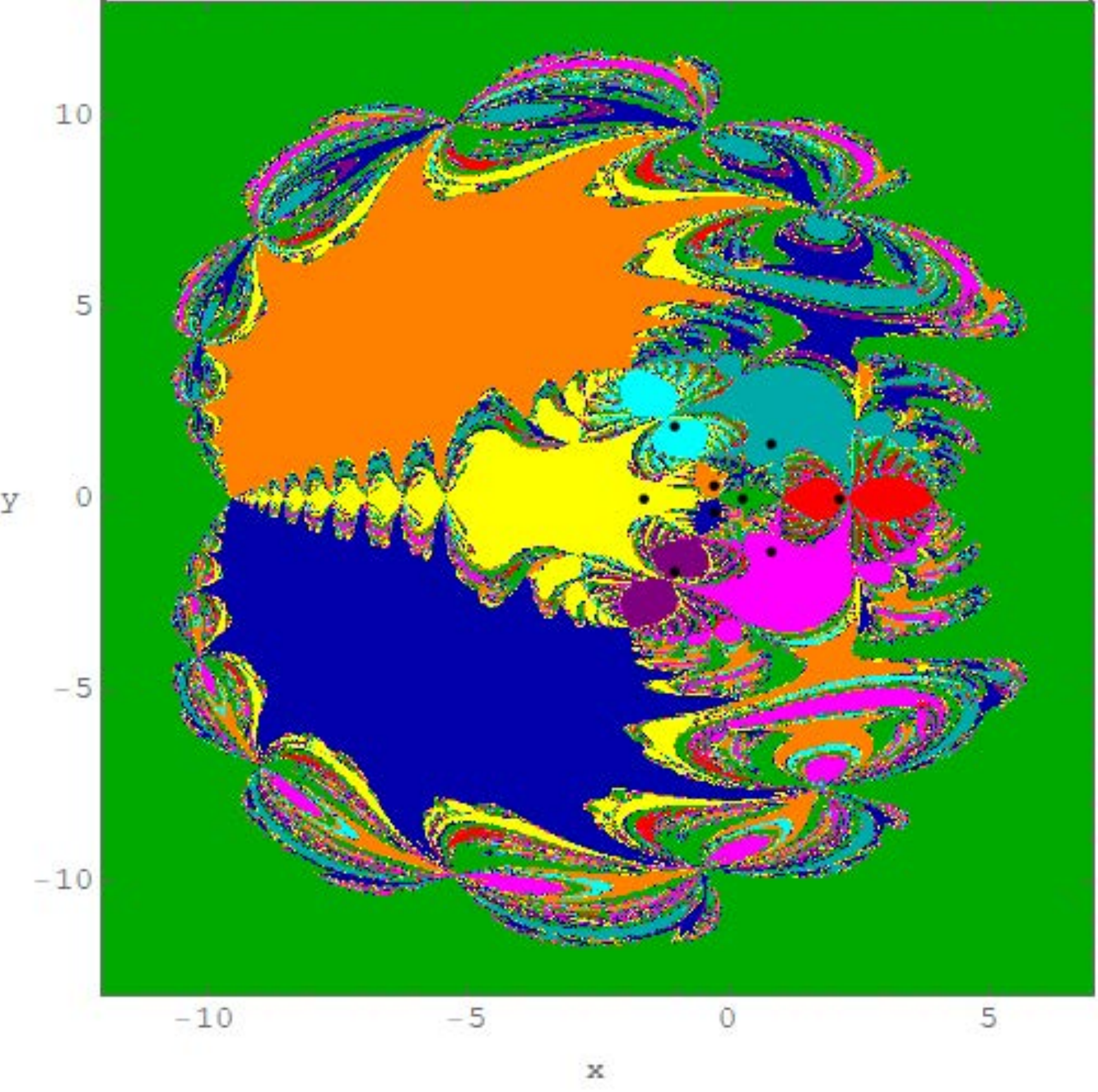}
(h)\includegraphics[scale=.27]{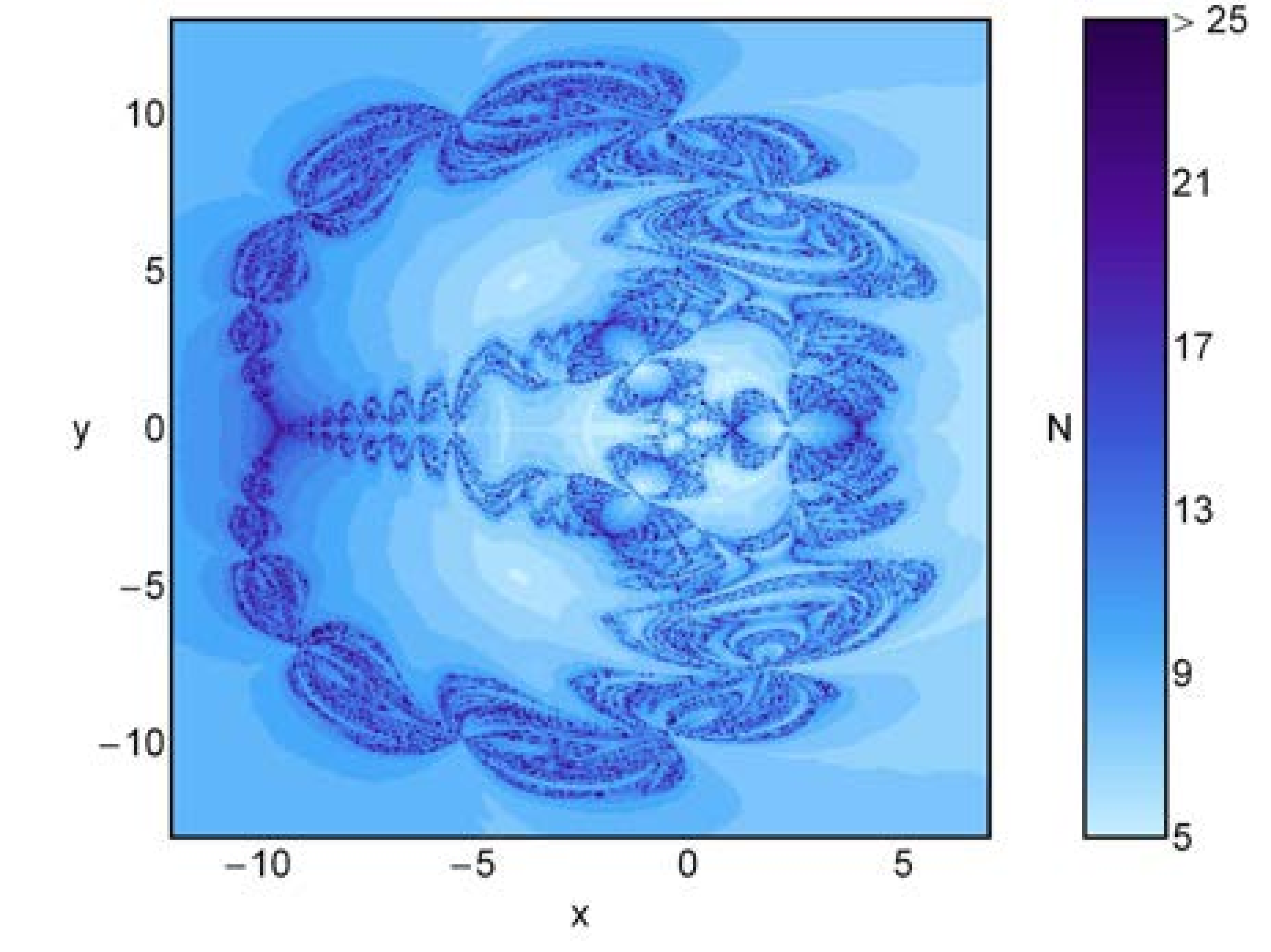}
(i)\includegraphics[scale=.25]{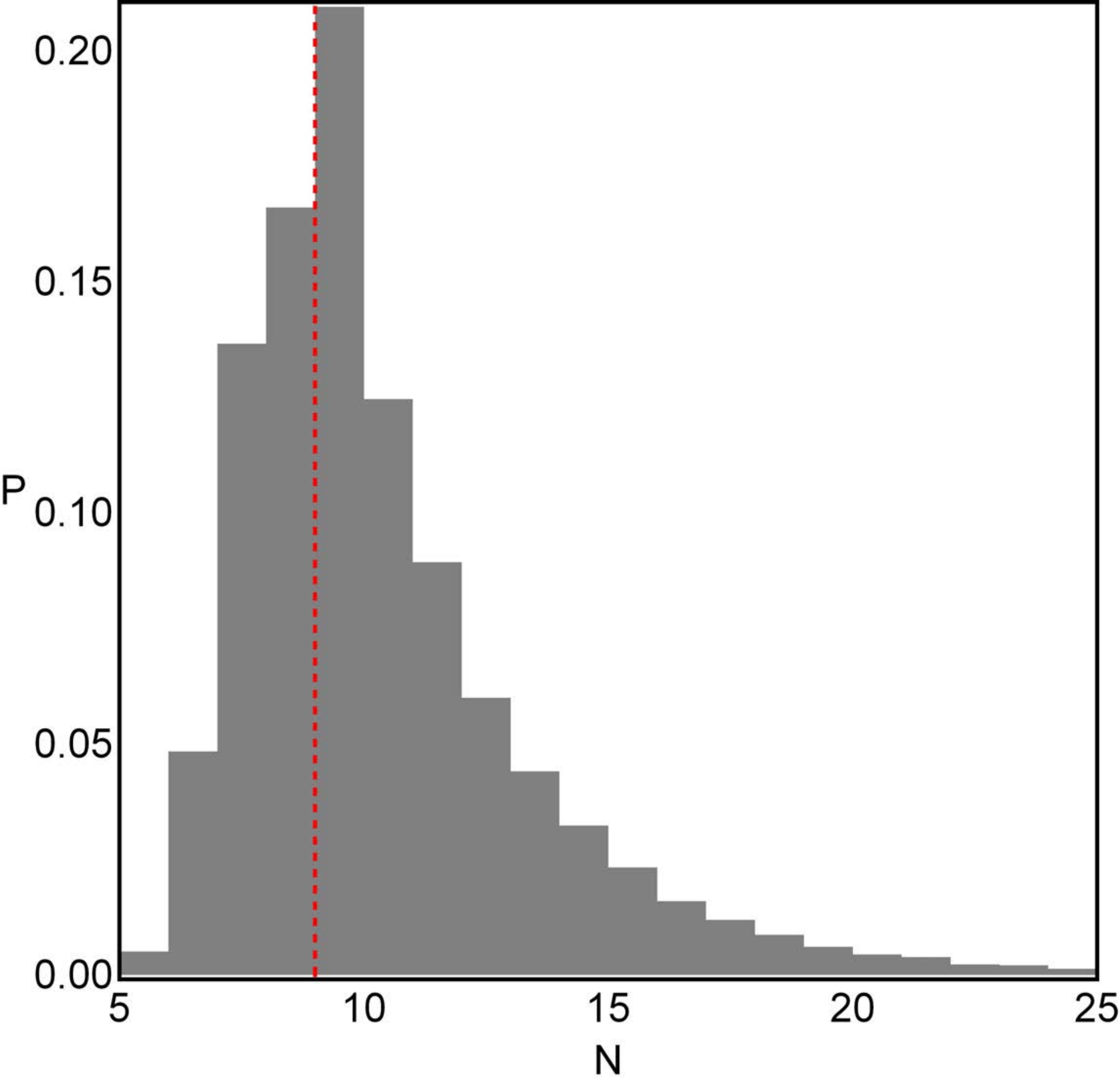}\\
(j)\includegraphics[scale=.27]{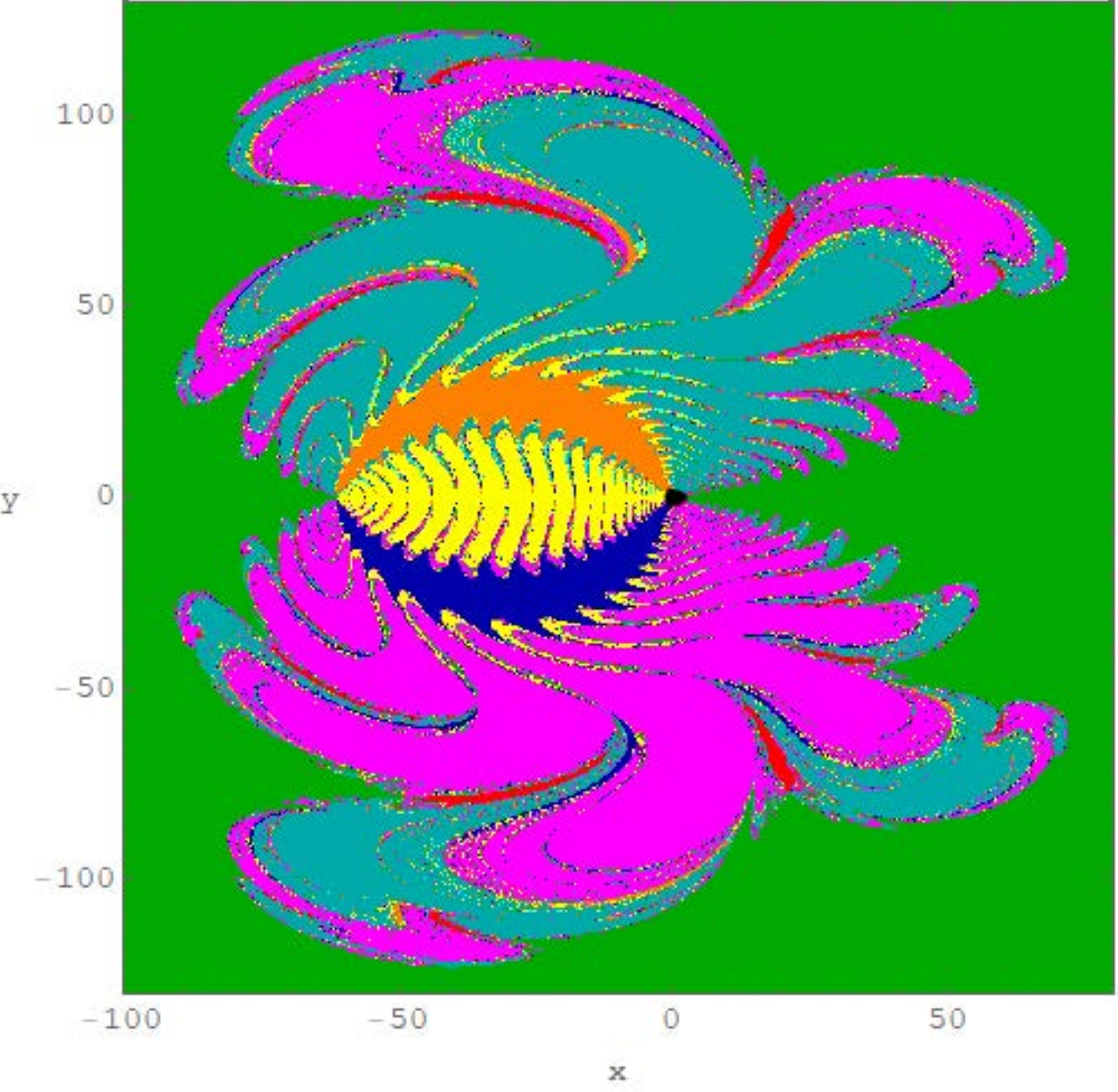}
(k)\includegraphics[scale=.27]{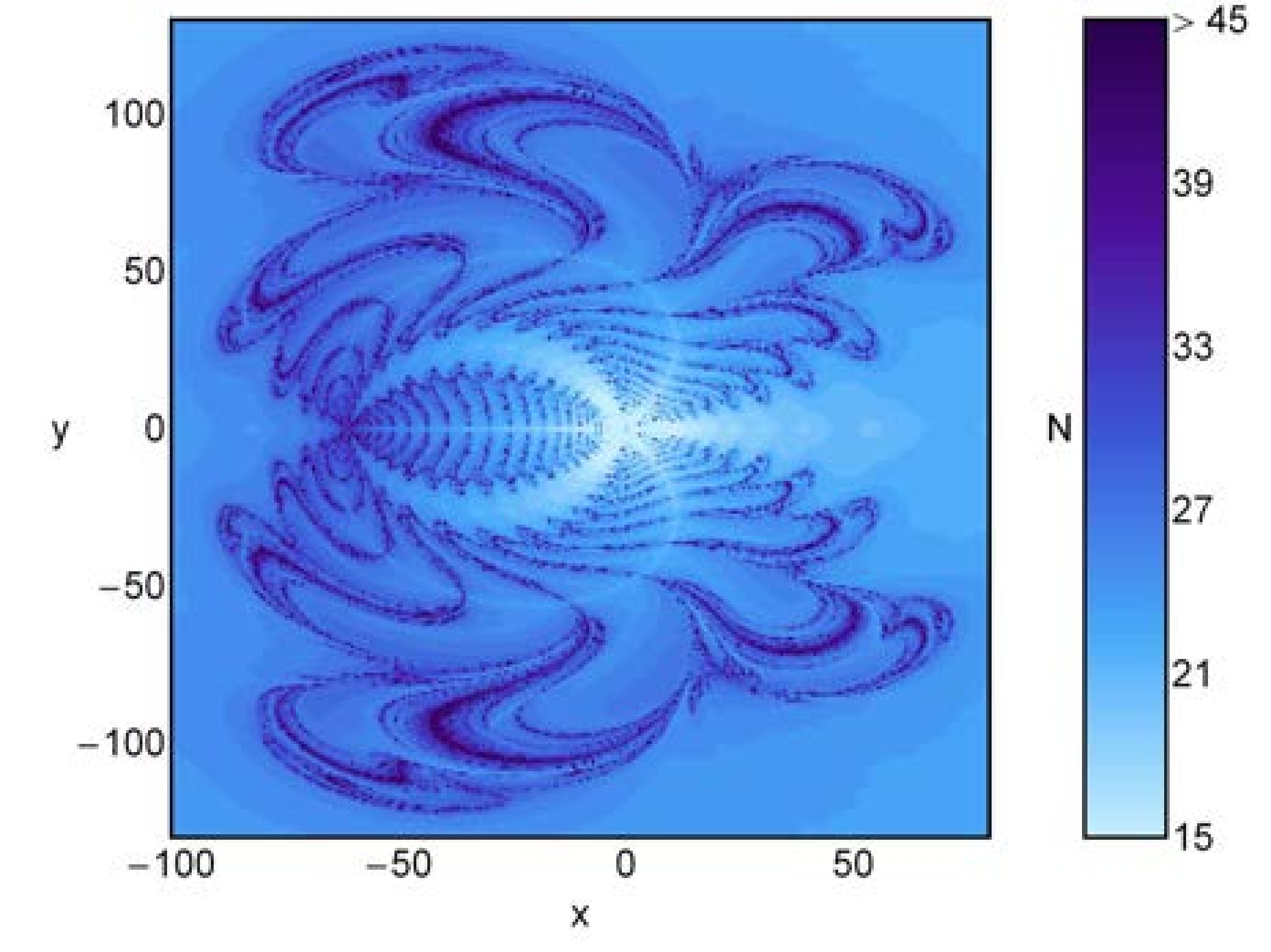}
(l)\includegraphics[scale=.25]{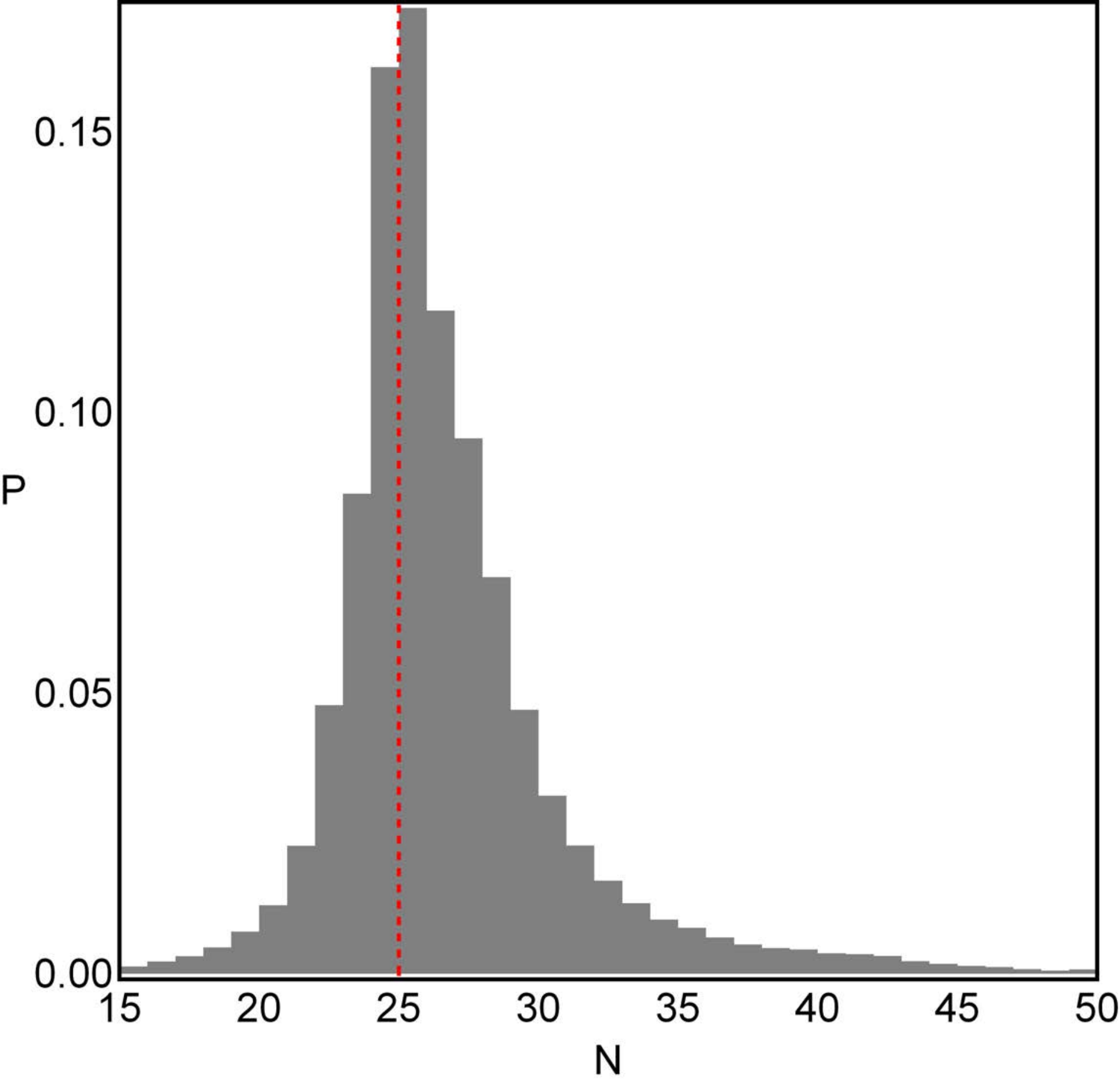}
\caption{The Newton-Raphson basins of attraction on the $xy$-plane for the
case when nine libration points exist for fixed value of $\alpha=58 \degree$  and for:
(a) $\beta=12\degree$; (d) $\beta=15\degree$; (g) $\beta=20\degree$; (j) $\beta=(26-\frac{1}{30})\degree$.
The color code for the libration points $L_1$,...,$L_9$ is same as in Fig \ref{NR_Fig_1}; and non-converging points (white);  (b, e, h, k) and (c, f, i, l) are the distribution of the corresponding number $(N)$ and the  probability distributions of required iterations for obtaining the Newton-Raphson basins of attraction shown in (a, d, g, j), respectively.
(Color figure online).}
\label{NR_Fig_6}
\end{figure*}

In Fig. \ref{NR_Fig_6}, the basins of convergence associated with libration points are presented for fixed value of $\alpha=58\degree$ and four different value of $\beta$ where nine libration points exist.  One can note that there is substantial effect of the angle parameter $\beta$ on the topology of the domain of the basins of convergence. In Fig. \ref{NR_Fig_6}(a, d, g, j), the extent of the basins of convergence is finite corresponding to each libration point except $L_2$. Looking at Fig. \ref{NR_Fig_6}a, we observe that the domain of basins of convergence is well shaped and resemble with island shaped region surrounded by the chaotic sea composed of the initial conditions. There exist three chaotic bugs corresponding to the libration points $L_{1, 6, 7}$ surrounded by the basins boundaries. These basins become regular with the increase in the value of $\beta$.

The distribution of the corresponding number $N$ of iterations, required to obtain the predetermined accuracy, is depicted in Fig. \ref{NR_Fig_6}(b, e, h, k) and majority of initial conditions converge for $N<30$ to one of the attracting domain while in the panel-k, the number of iterations required to converge, for an initial condition to one of the attractors, increases to $N>45$. Moreover, the most probable number $N^*$ of the iterations increases as the parameter $\beta$ increases (see, panels-c, f, i, l).
\subsubsection{when eleven libration points exist}
\label{when eleven libration points exists}
\begin{figure*}[!t]
\centering
(a)\includegraphics[scale=.27]{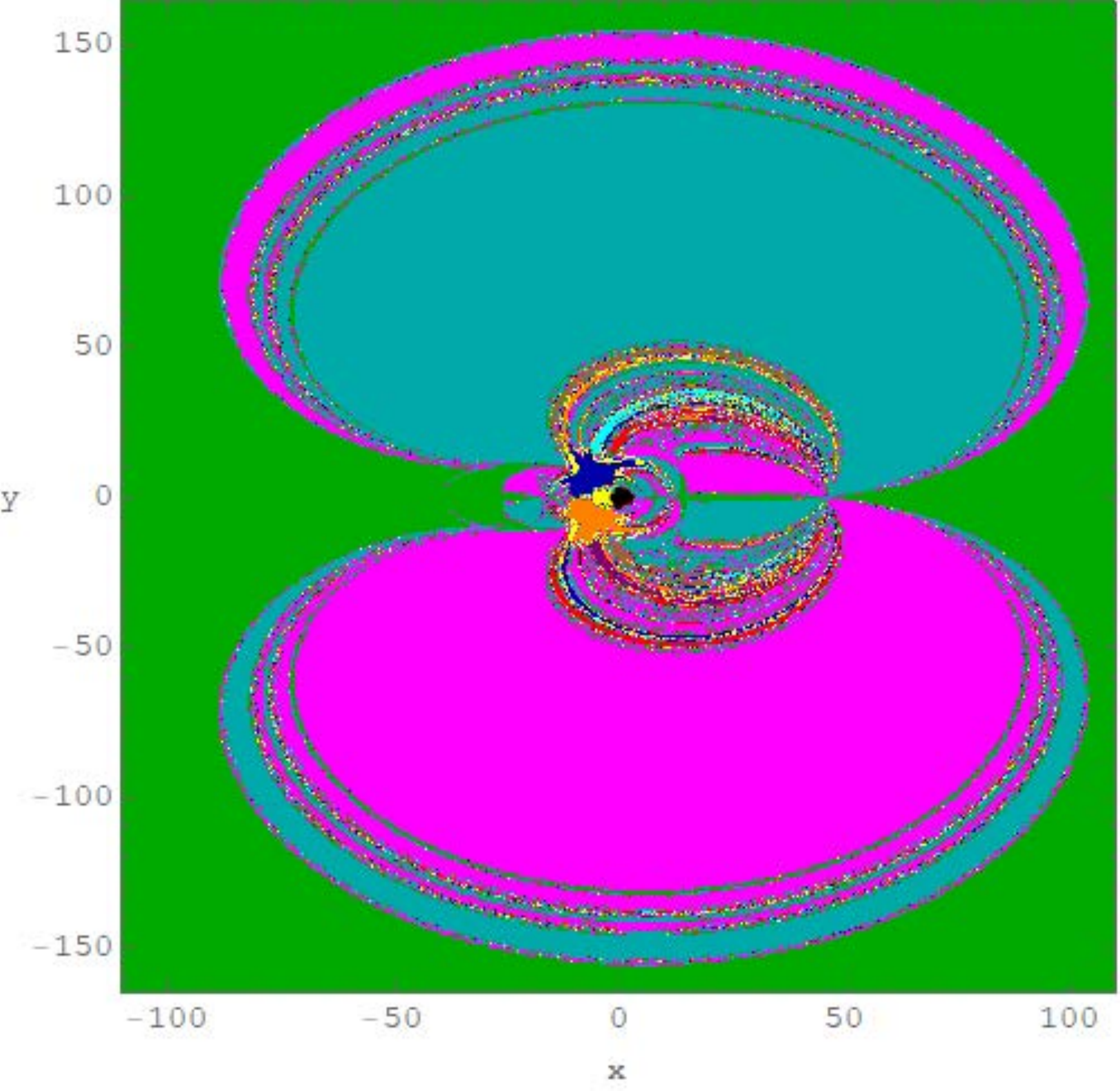}
(b)\includegraphics[scale=.27]{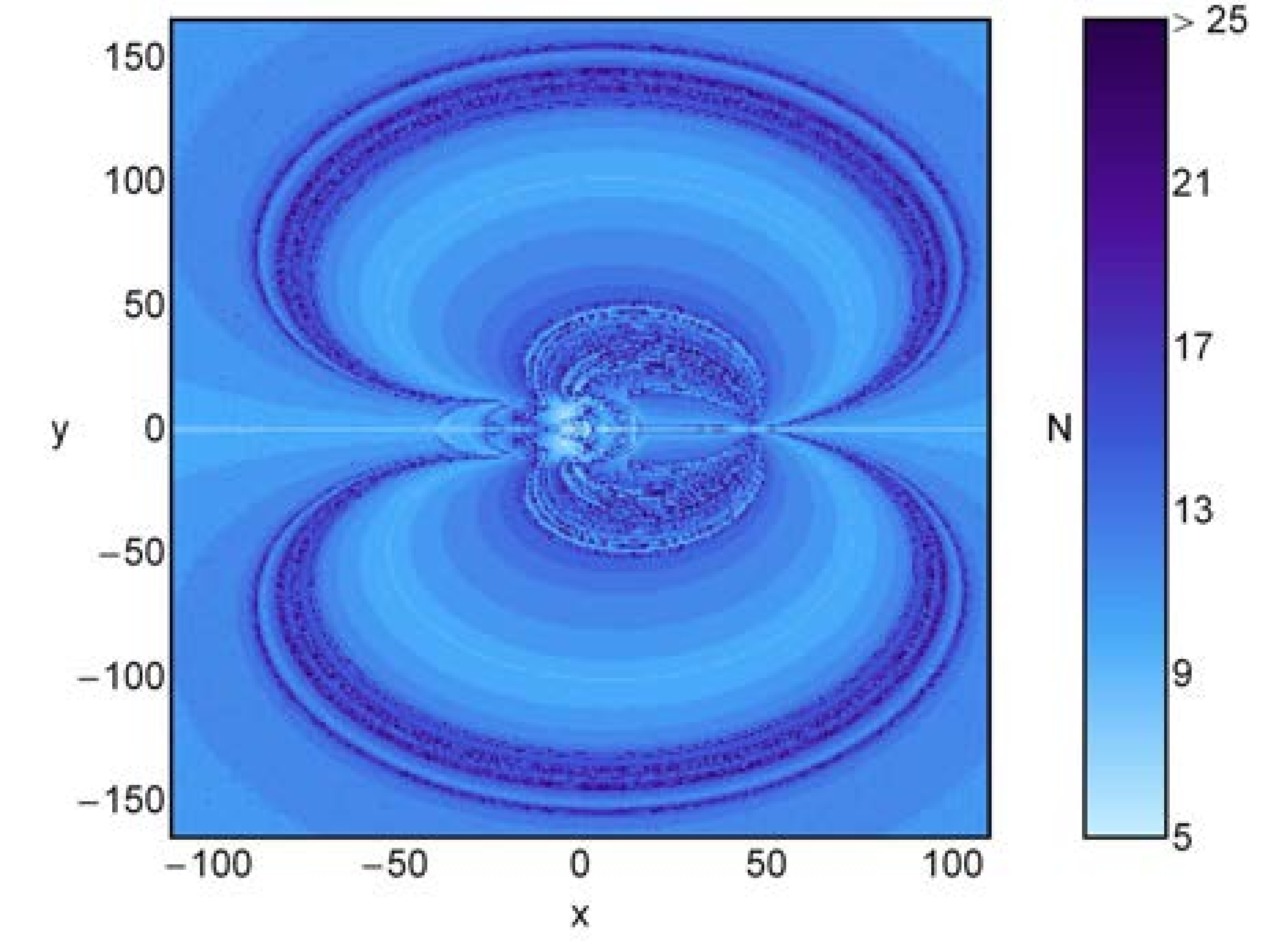}
(c)\includegraphics[scale=.25]{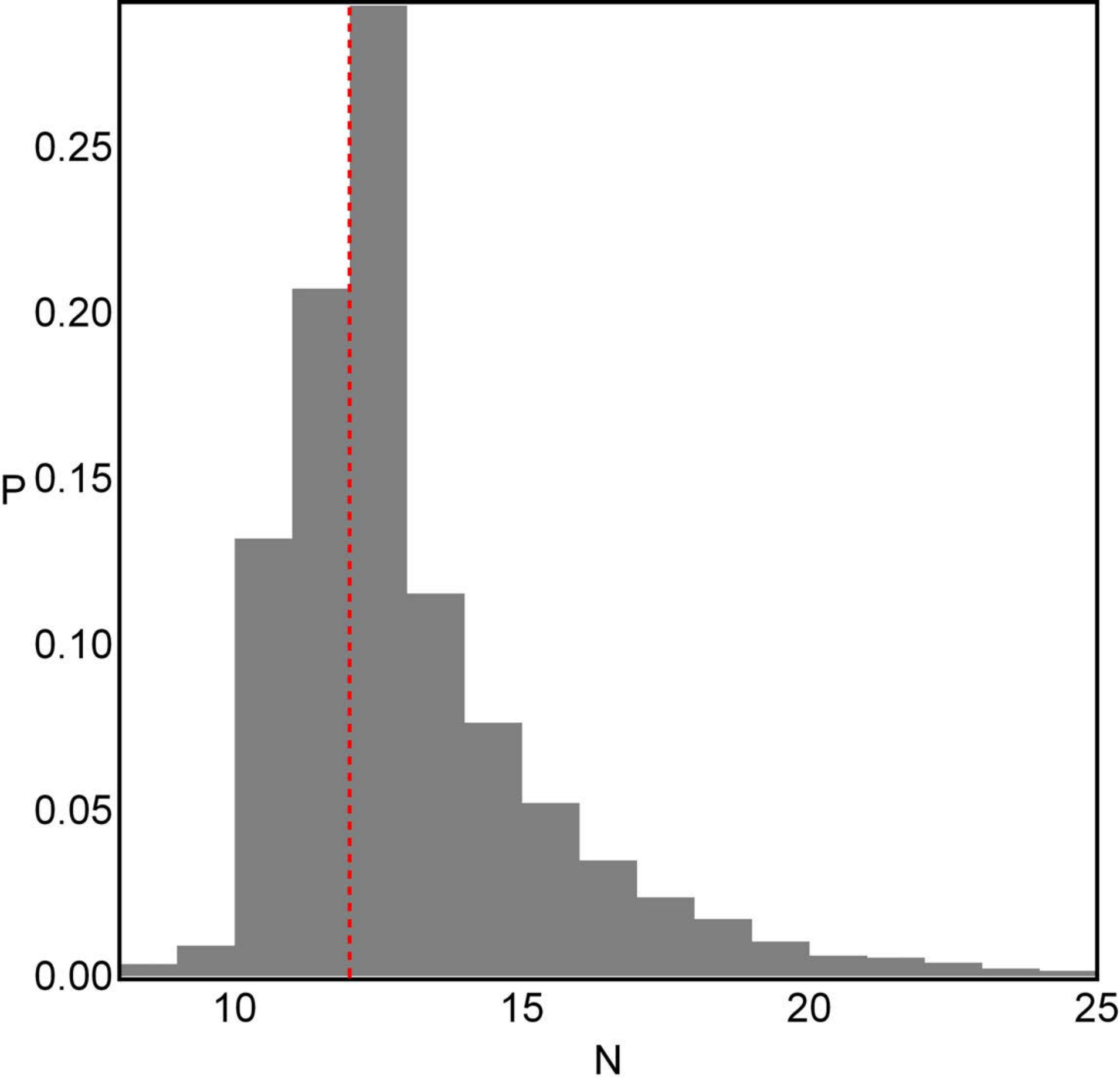}\\
(d)\includegraphics[scale=.27]{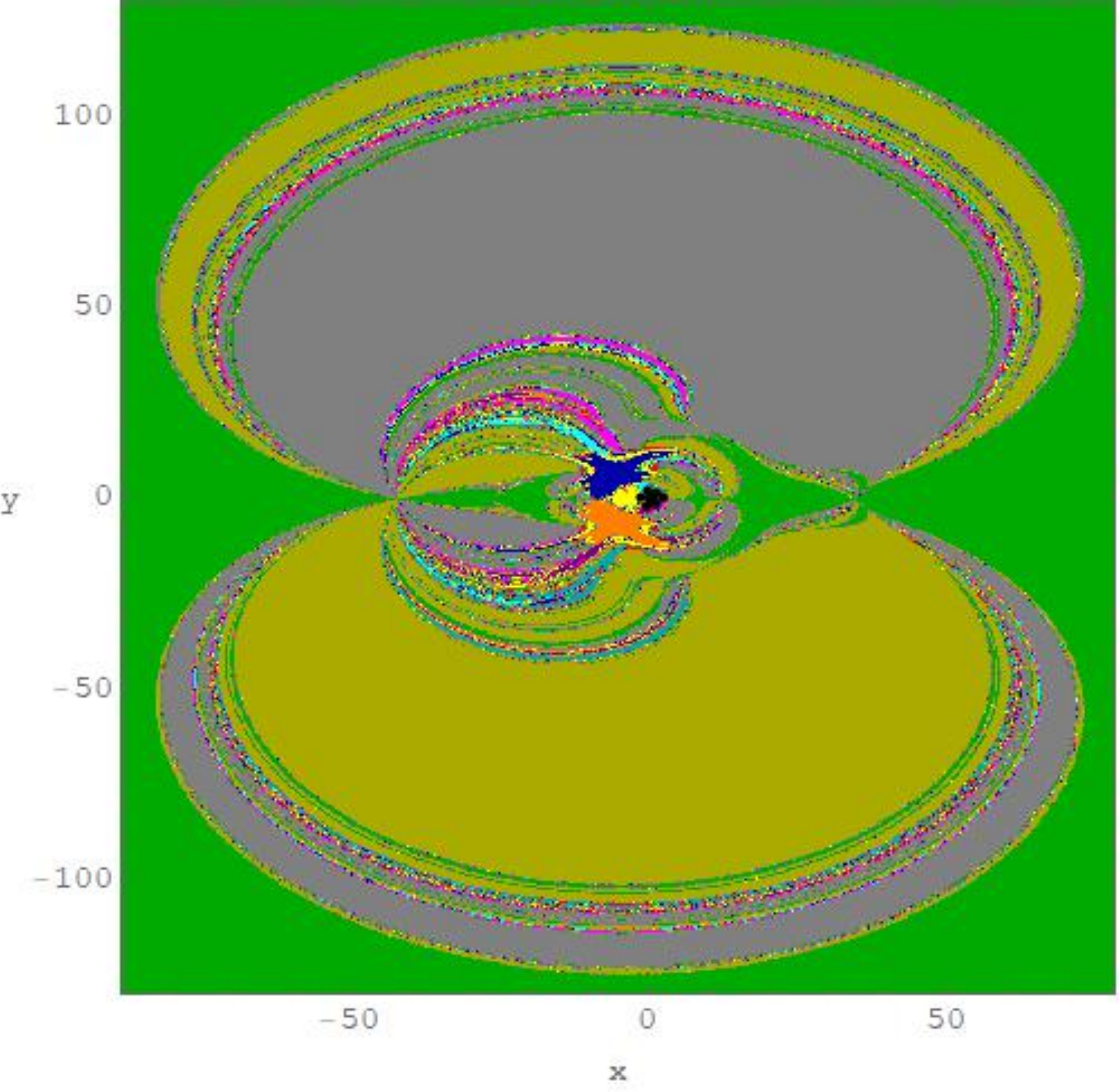}
(e)\includegraphics[scale=.27]{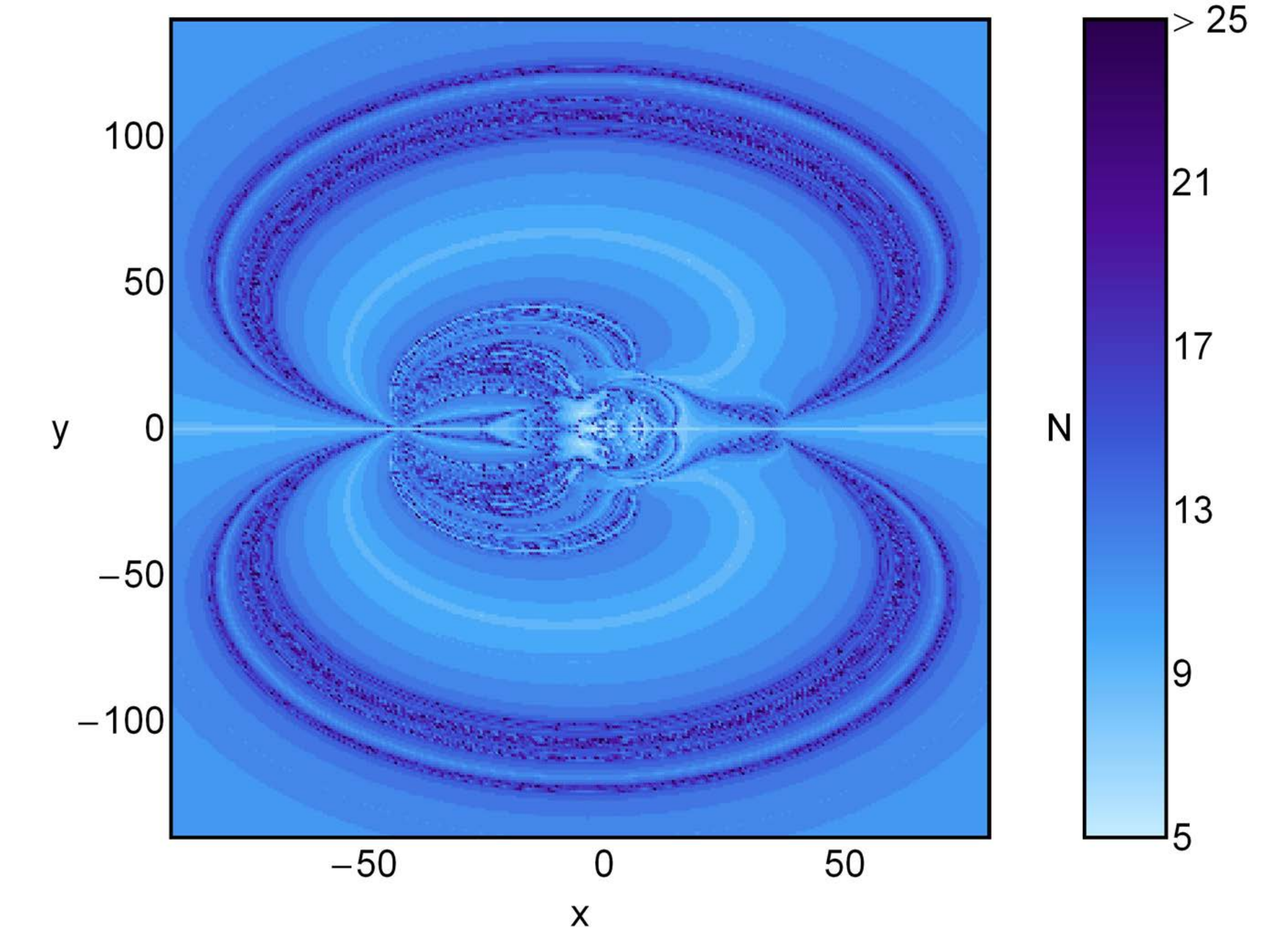}
(f)\includegraphics[scale=.25]{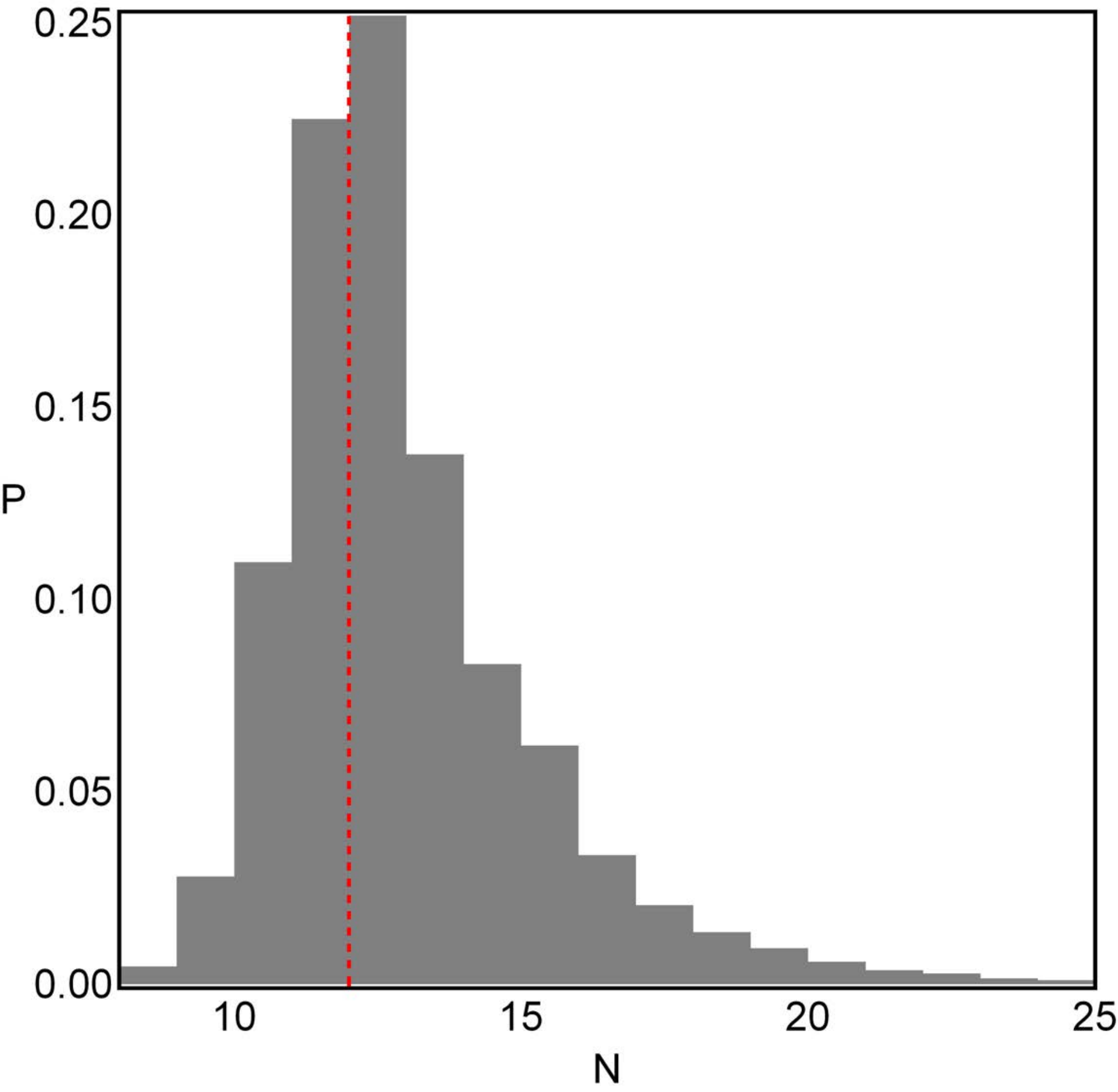}
(g)\includegraphics[scale=.27]{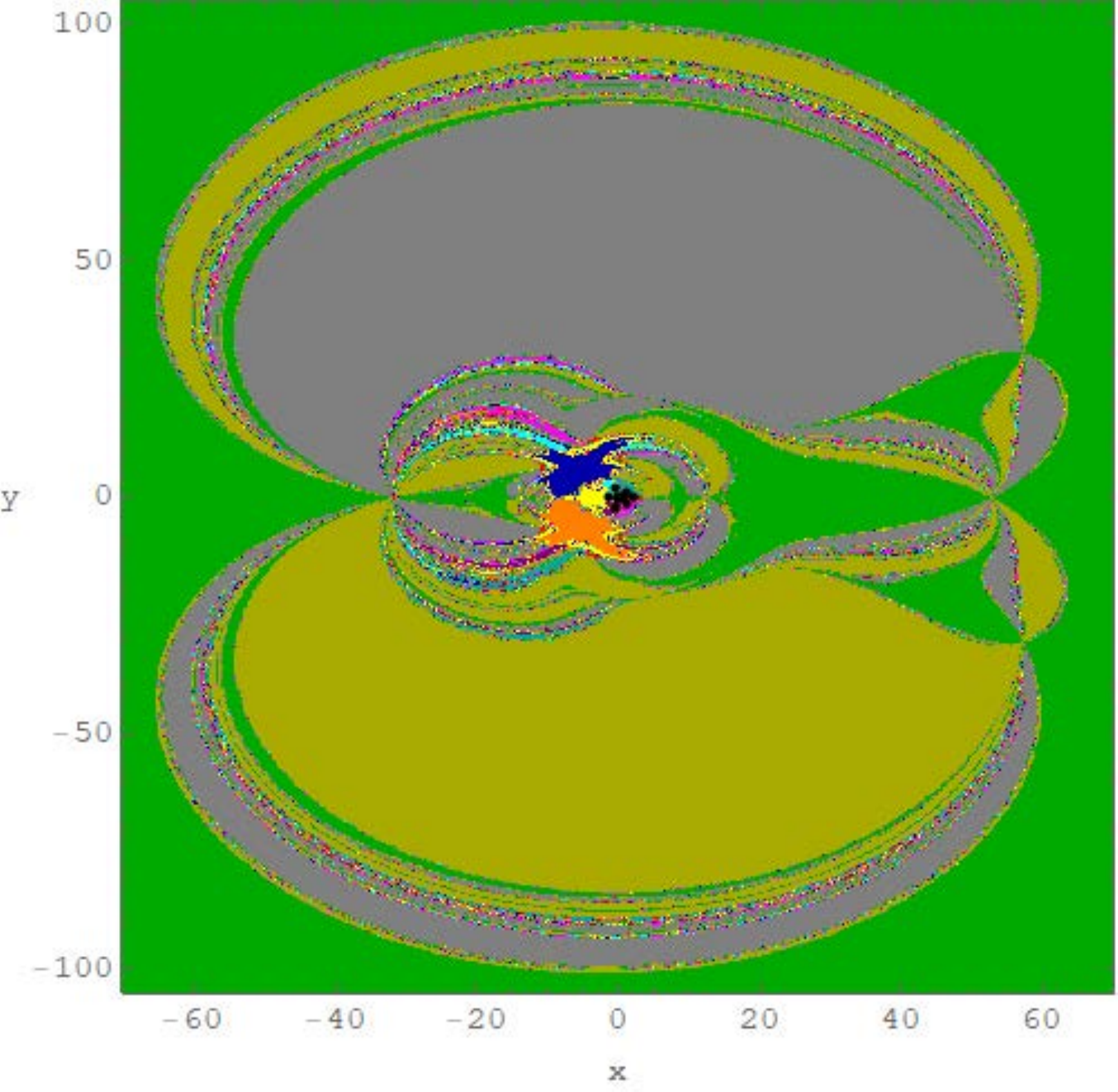}
(h)\includegraphics[scale=.27]{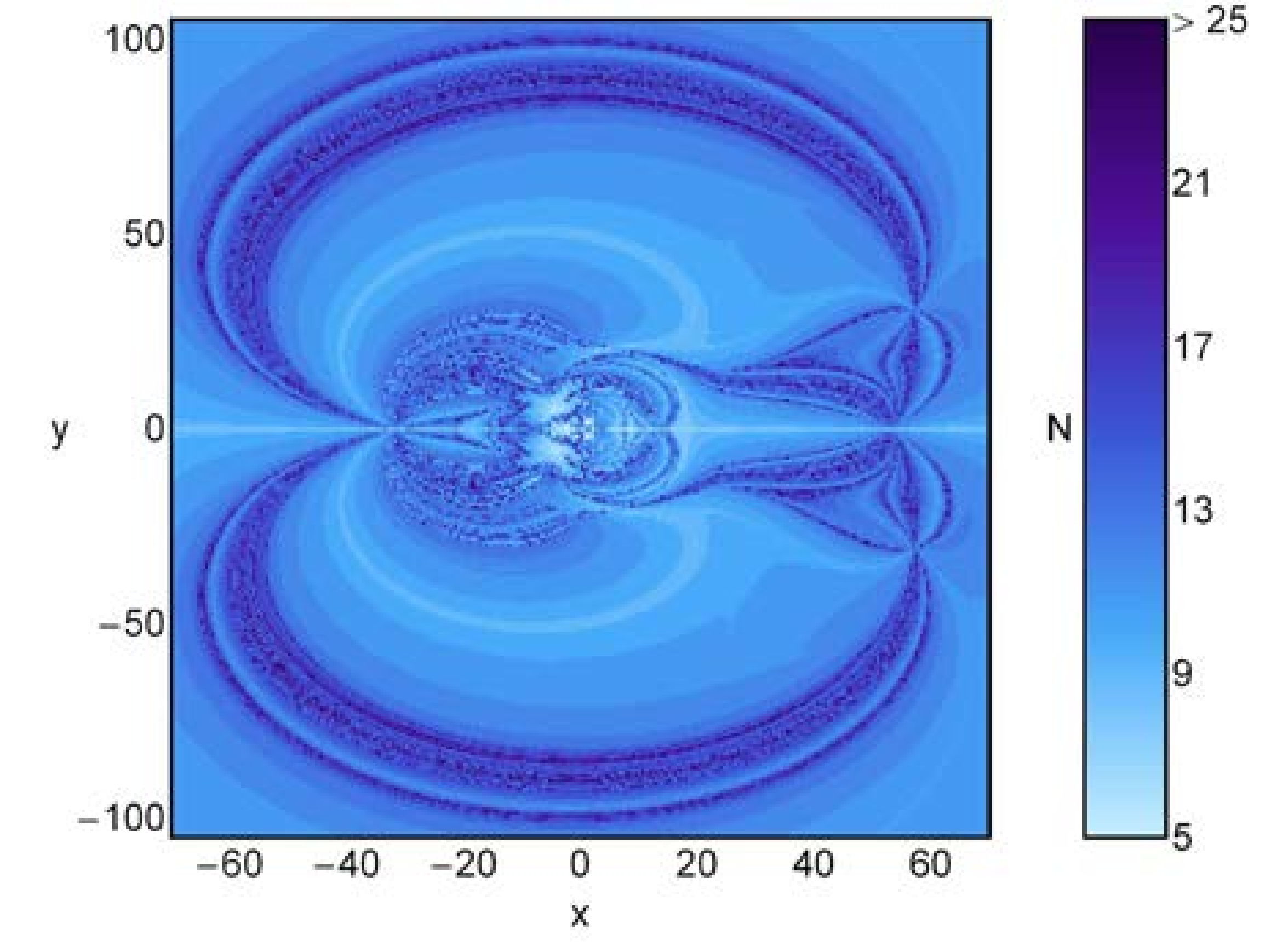}
(i)\includegraphics[scale=.25]{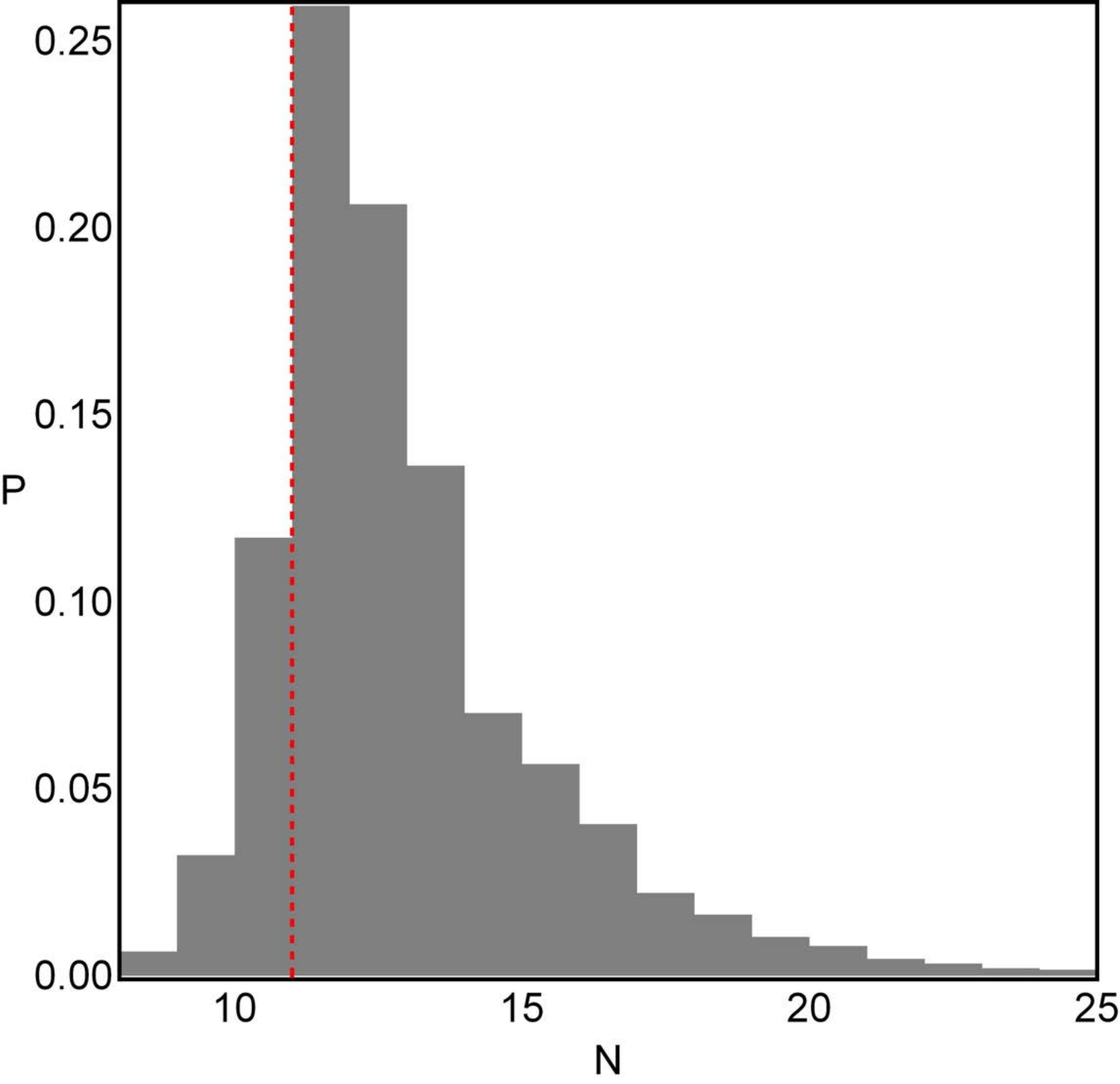}
\caption{The Newton-Raphson basins of attraction on the $xy$-plane for the
case when eleven libration points exist for  fixed value of
$\alpha=57 \degree$  and for:
(a) $\beta=3\degree$; (d) $\beta=3.75\degree$; (g) $\beta=4.25\degree$. The color code for the libration points $L_1$,...,$L_9$ is same as in Fig \ref{NR_Fig_1}, $L_{10}\emph{(olive)}$; $L_{11}\emph{(gray)}$ ; and non-converging points (white);  (b, e,  h) and (c, f, i) are the distribution of the corresponding number $(N)$ and the  probability distributions of required iterations for obtaining the Newton-Raphson basins of attraction shown in (a, d, g), respectively.
 (Color figure online).}
\label{NR_Fig_7}
\end{figure*}

The present subsection deals with the case where eleven libration points exist for different interval of the angle parameter $\beta$. In Fig. \ref{NR_Fig_7}, the basins of convergence are presented for fixed value of $\alpha=57\degree$ and three different values of the angle parameter $\beta$. It can be observed that the domain of the basin of convergence associated with libration point $L_2$ is infinite while for remaining other libration points, the domain of the basins of convergence are finite and well formed in shape. In addition, it is noticed that the majority of the area on the configuration $(x, y)$ plane is covered by the finite domain of the basins of convergence associated with the libration points $L_{10}$ and $L_{11}$ which looks like a butterfly wing and symmetrical with respect to $x-$axis. As we increase the value of $\beta$, the butterfly wing shape region shrinks and its shape changes, so that two antenna shaped area originate and the whole shape of the basins of convergence turn into exotic bugs shaped region with huge wings and antennas (see Fig. \ref{NR_Fig_7}(a, d, g)).

The distribution of the corresponding number $N$ of iterations, required to obtain the predetermined accuracy, is depicted in Fig. \ref{NR_Fig_7}. In panels (b, e, h), majority of initial conditions converge for $N<25$ to one of the attracting domain. Moreover, the most probable number $N^*$ of the iterations decreases as the parameter $\beta$ increases (see, panels-c, f, i).

\begin{figure*}[!t]
\centering
(a)\includegraphics[scale=.27]{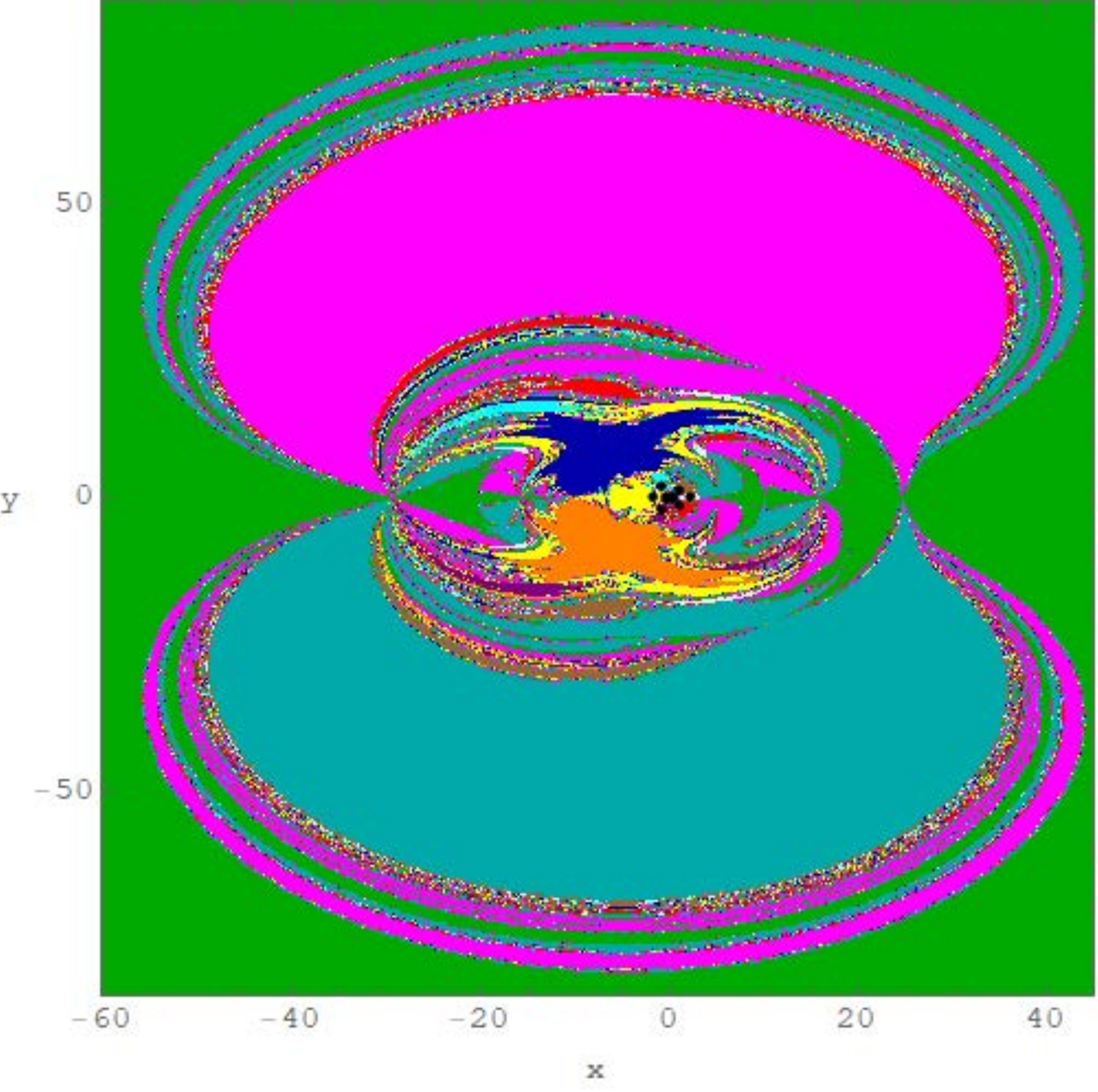}
(b)\includegraphics[scale=.27]{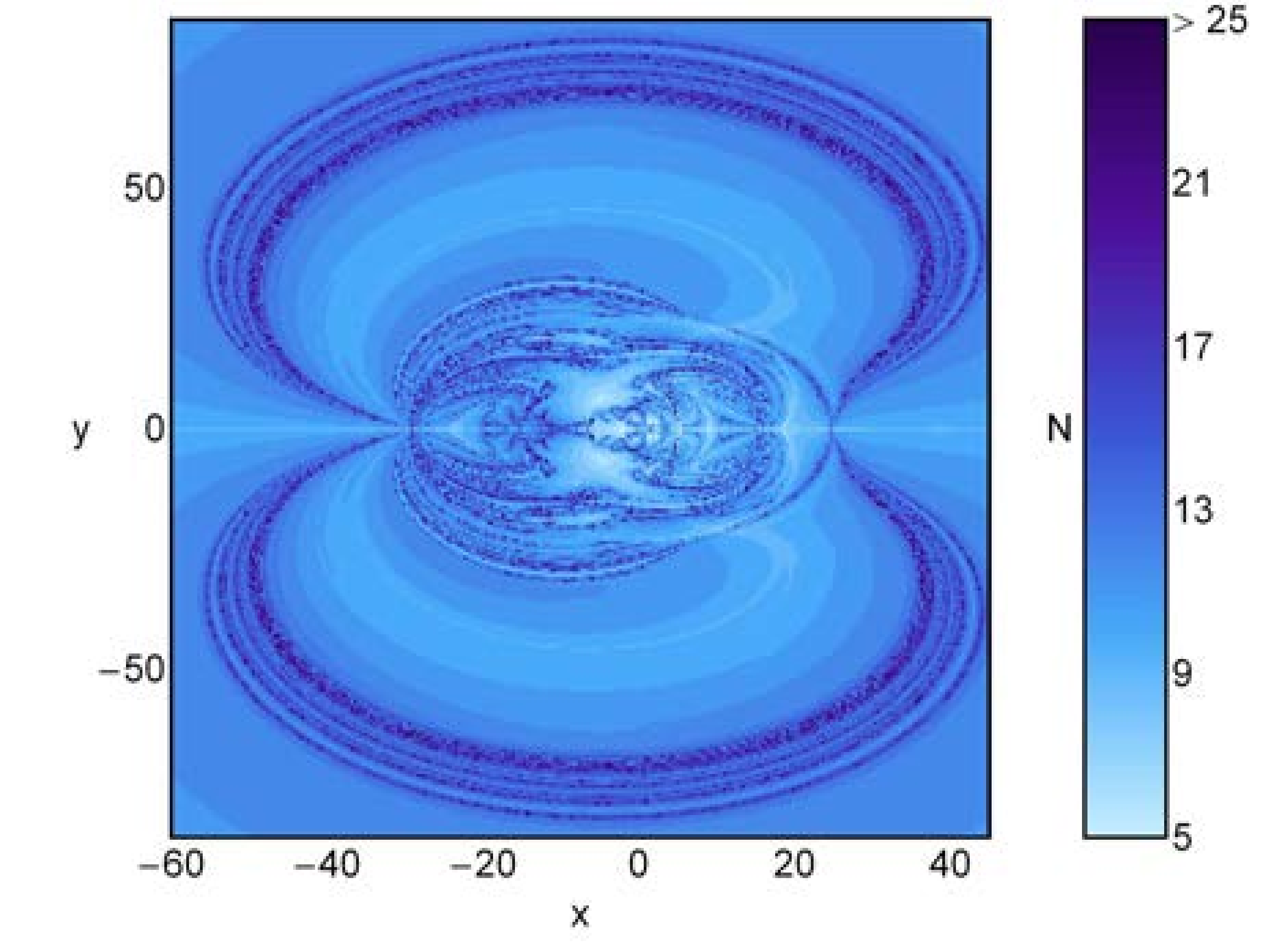}
(c)\includegraphics[scale=.25]{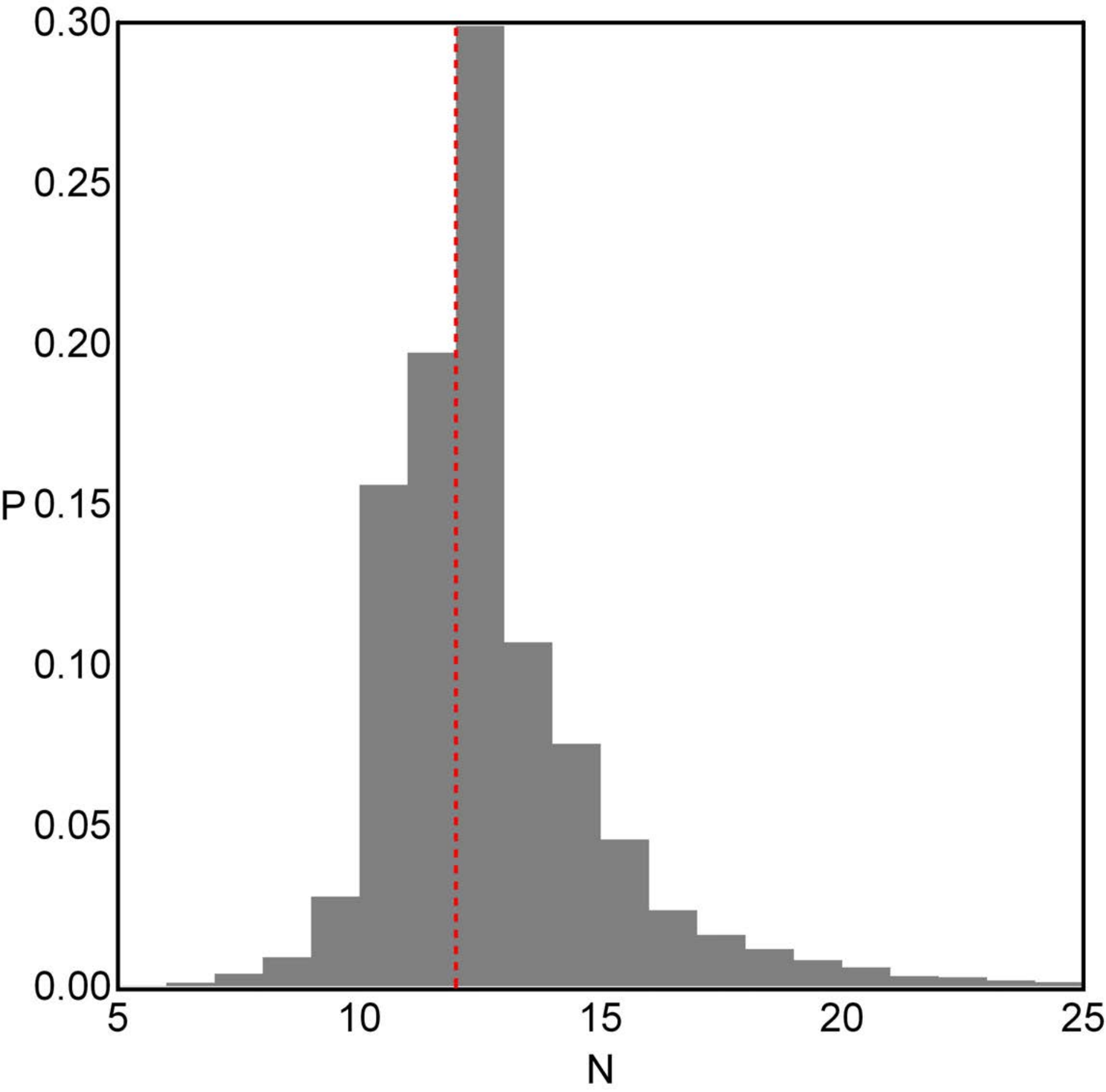}\\
(d)\includegraphics[scale=.27]{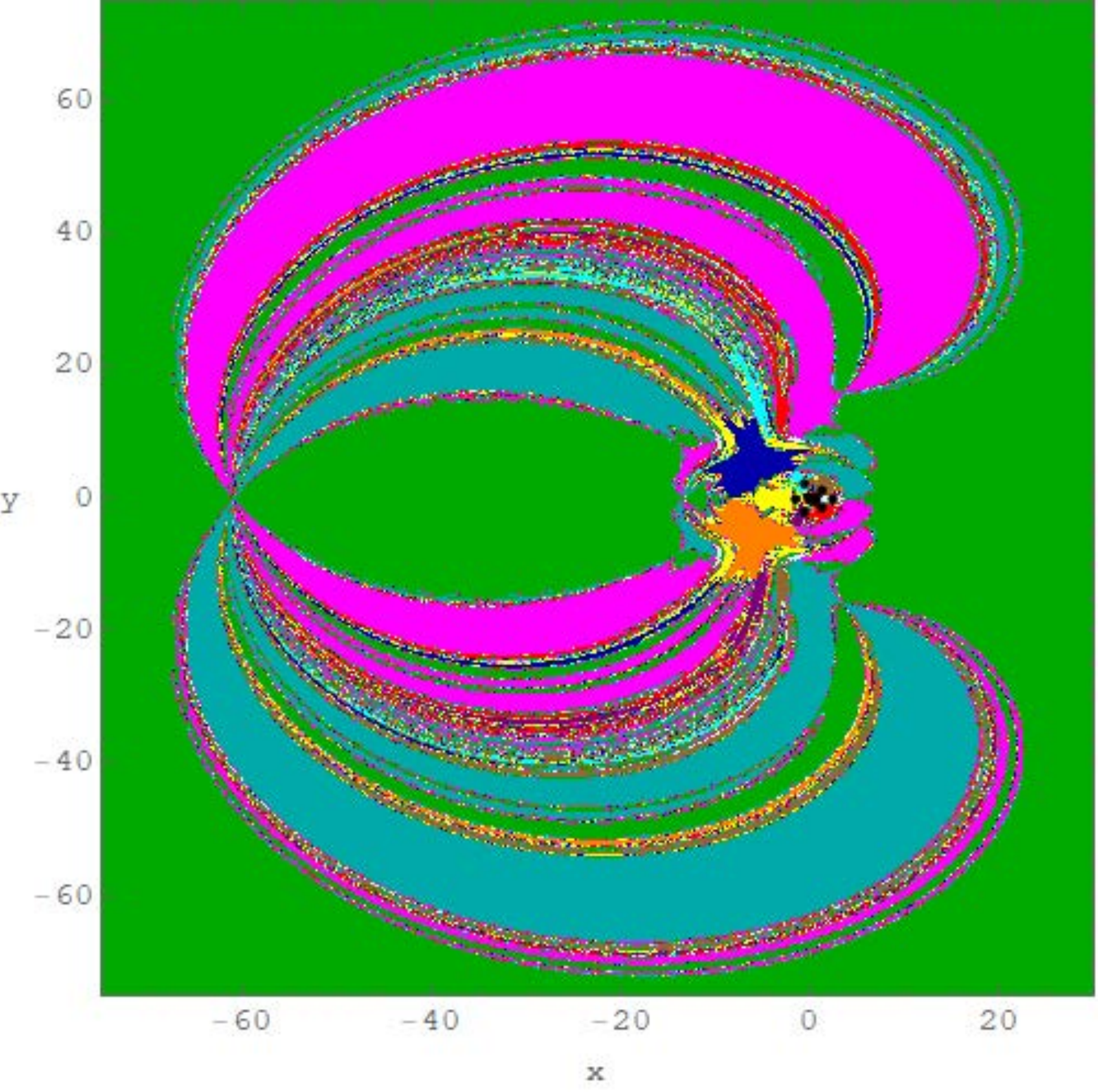}
(e)\includegraphics[scale=.27]{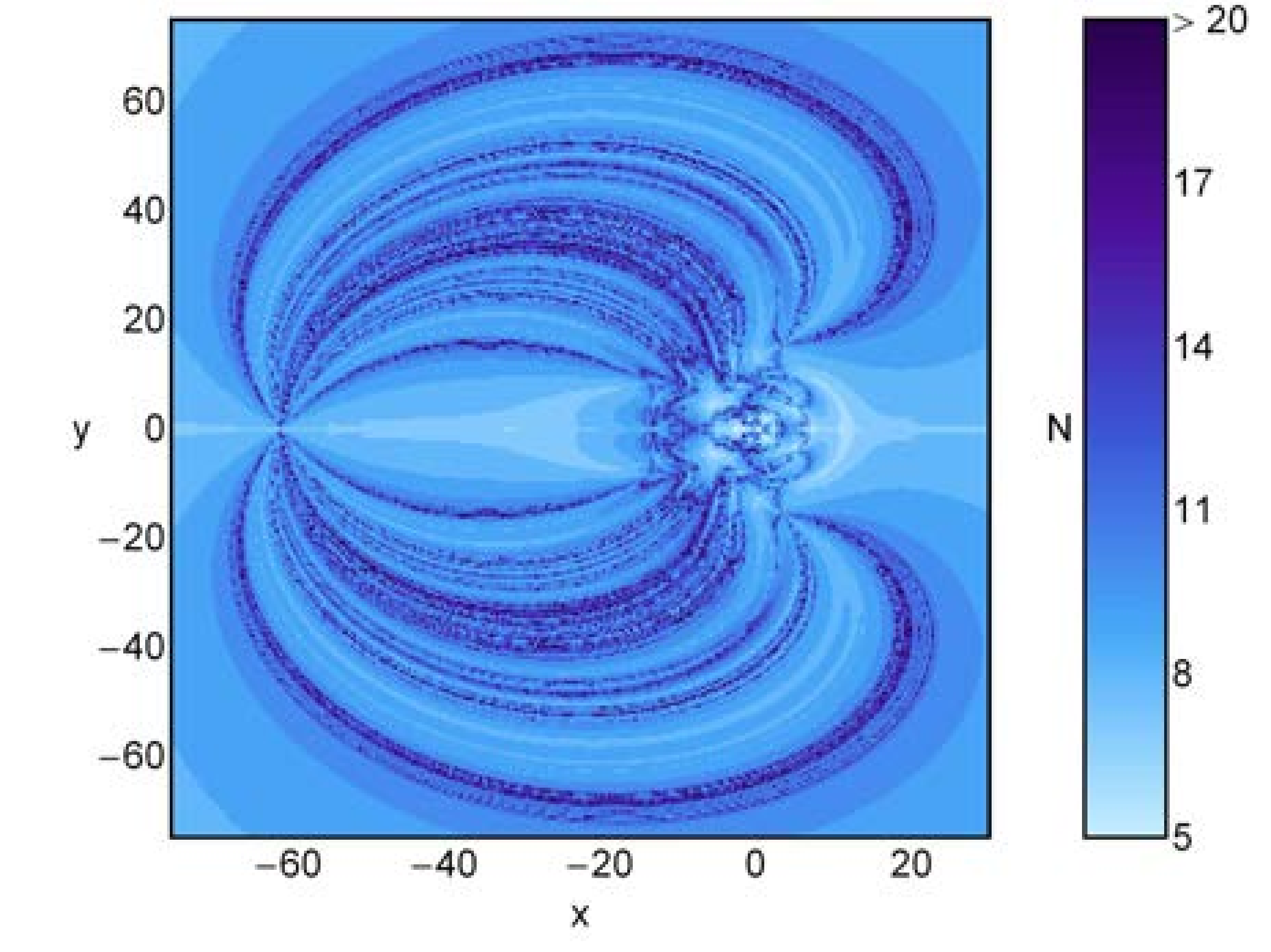}
(f)\includegraphics[scale=.25]{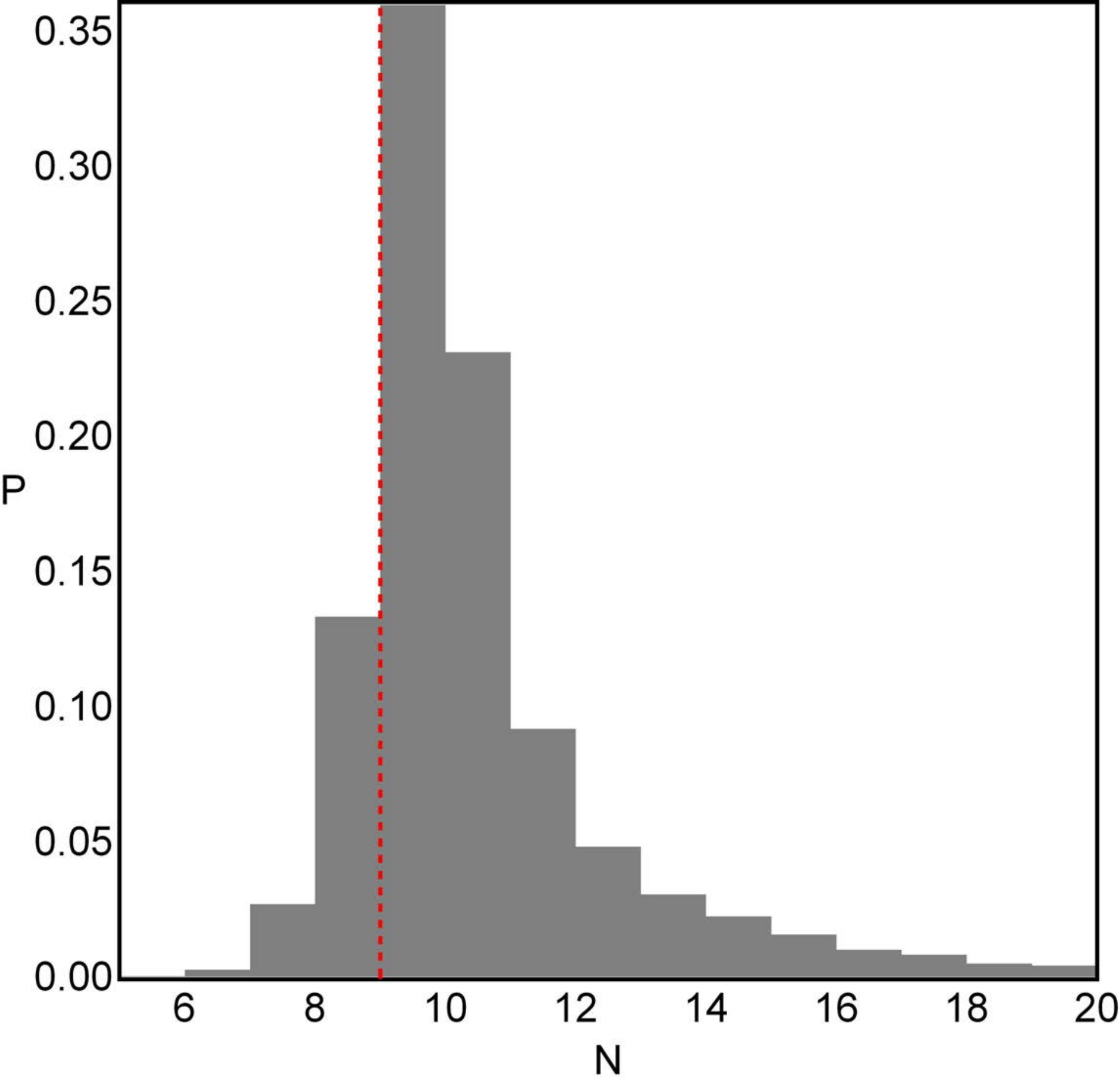}\\
(g)\includegraphics[scale=.27]{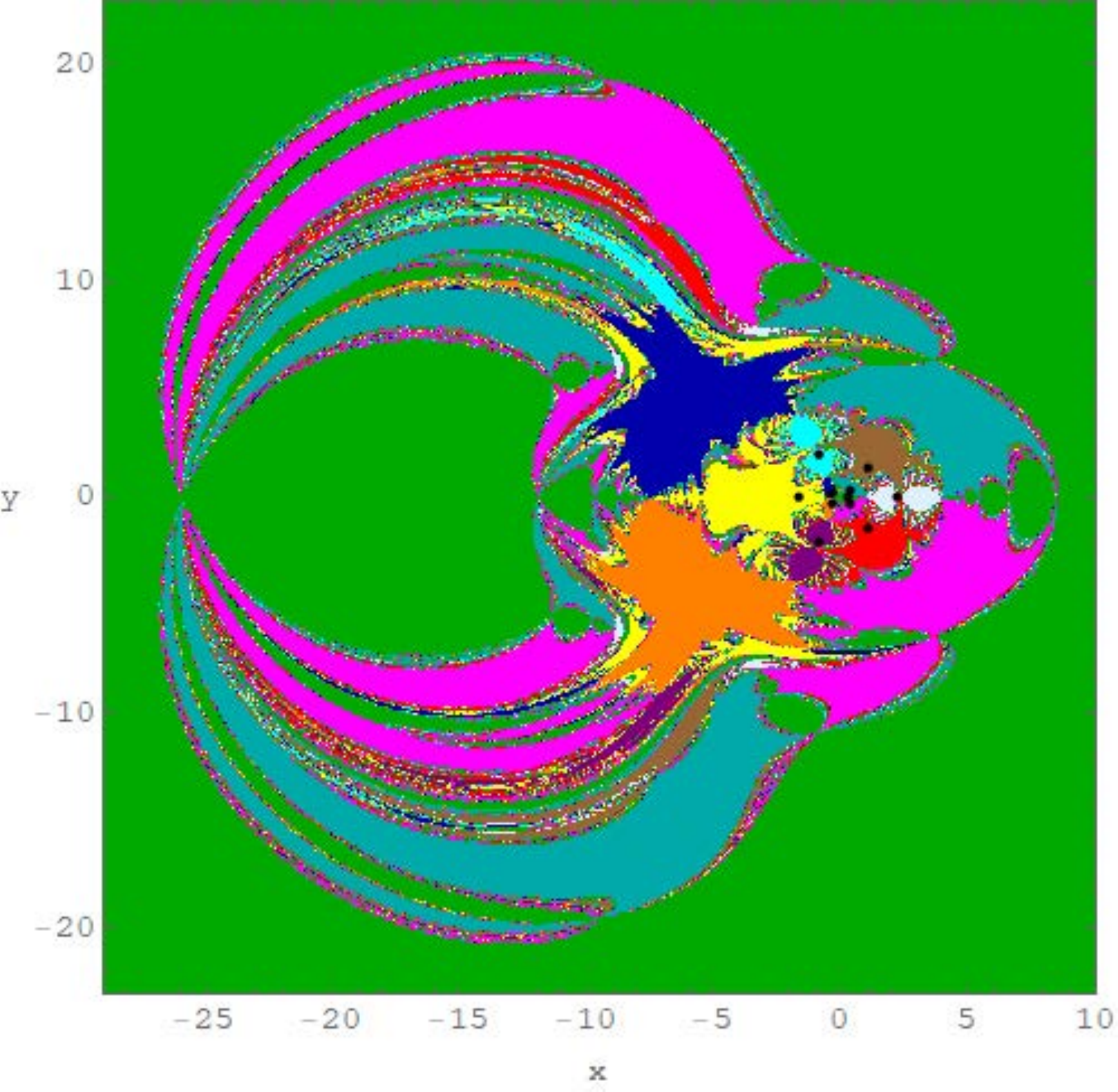}
(h)\includegraphics[scale=.27]{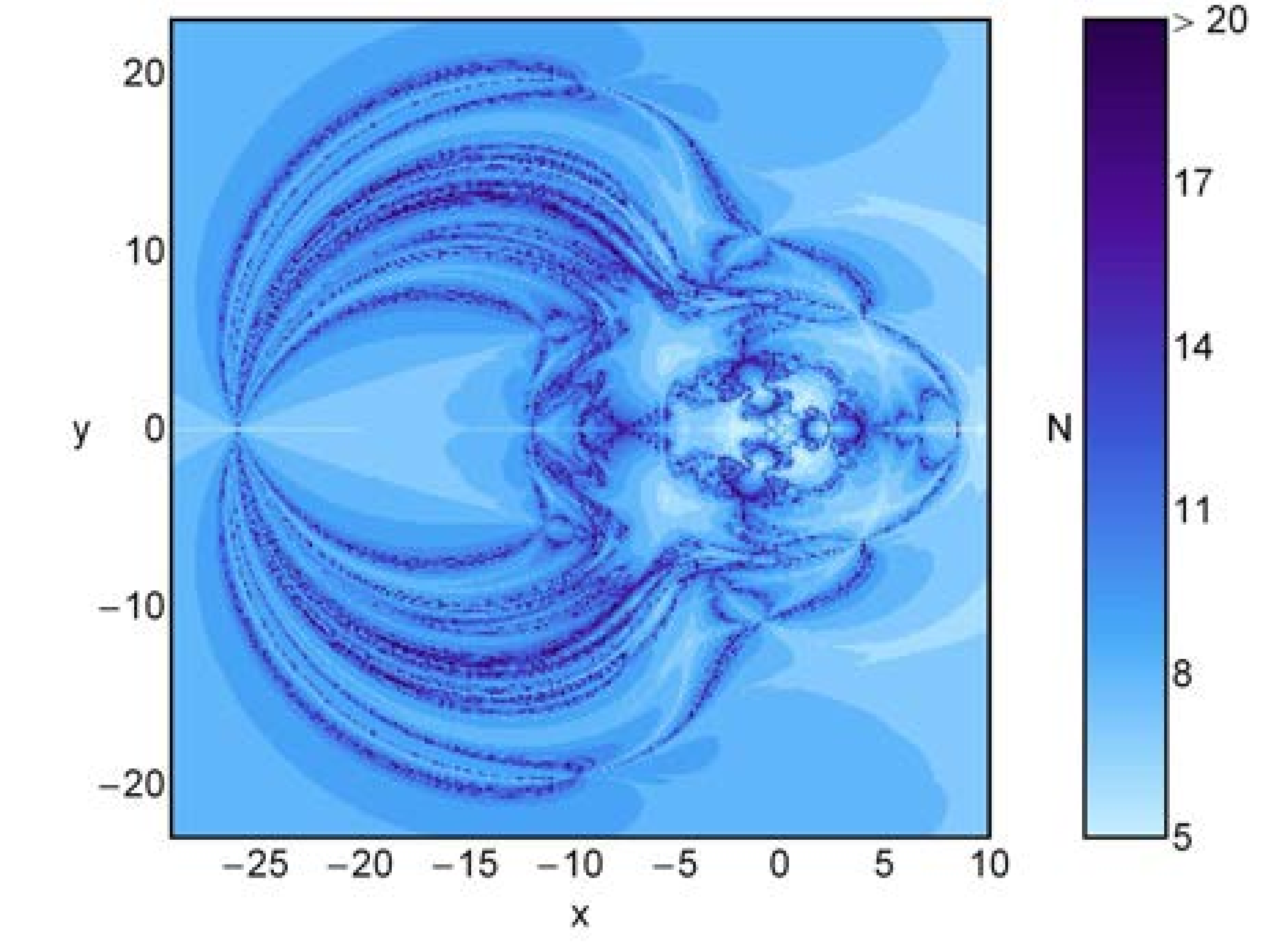}
(i)\includegraphics[scale=.25]{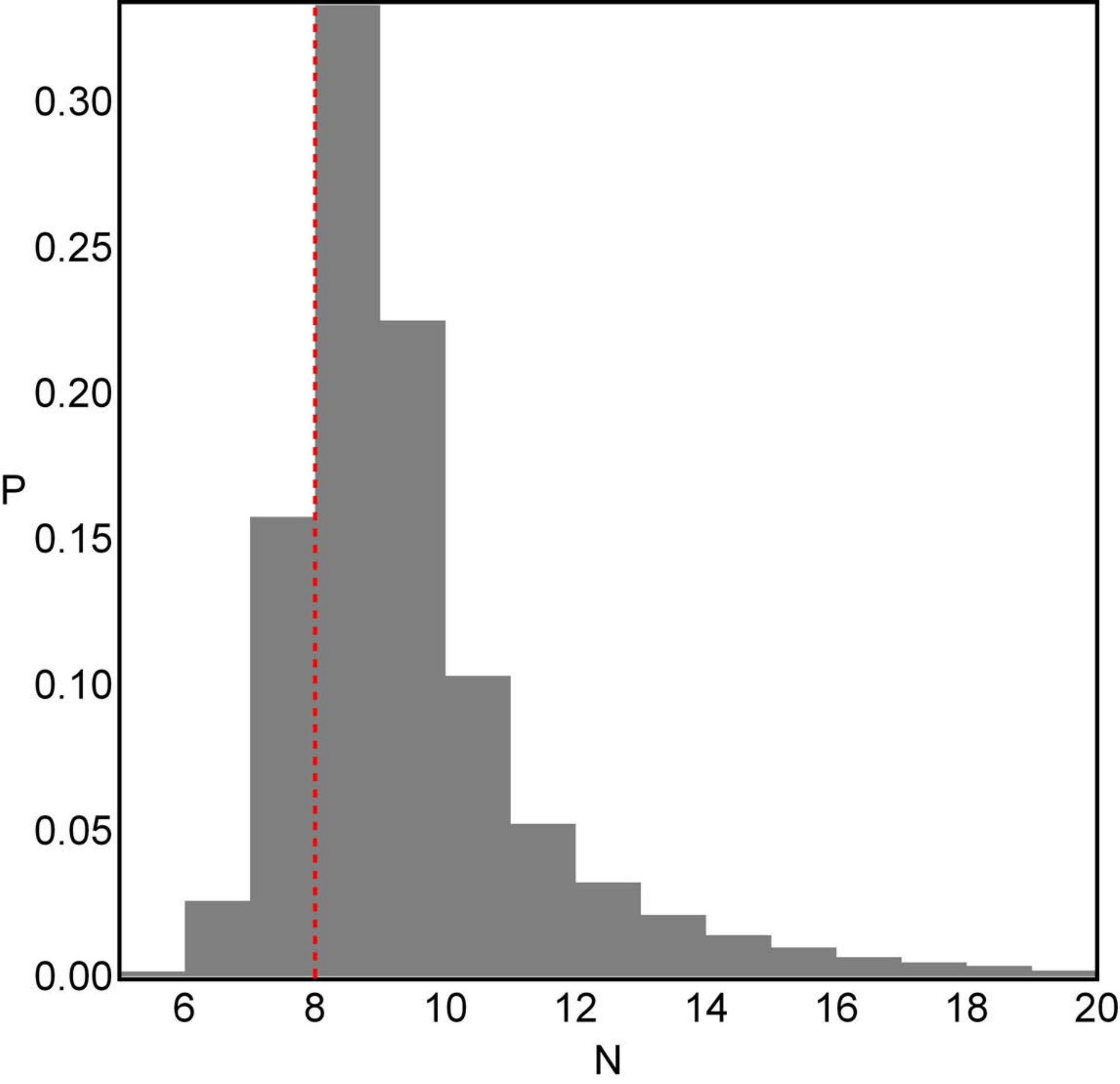}\\
(j)\includegraphics[scale=.27]{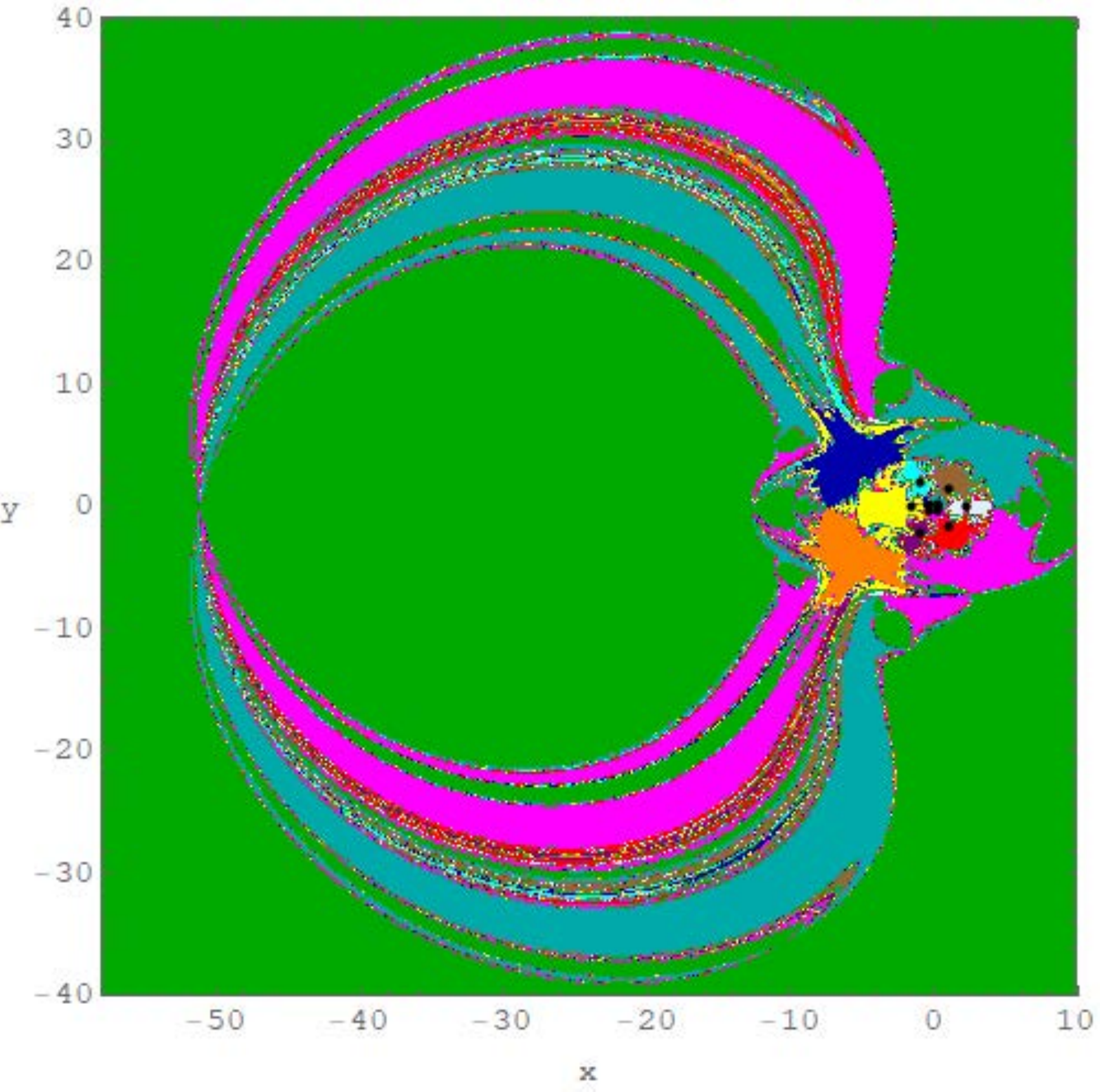}
(k)\includegraphics[scale=.27]{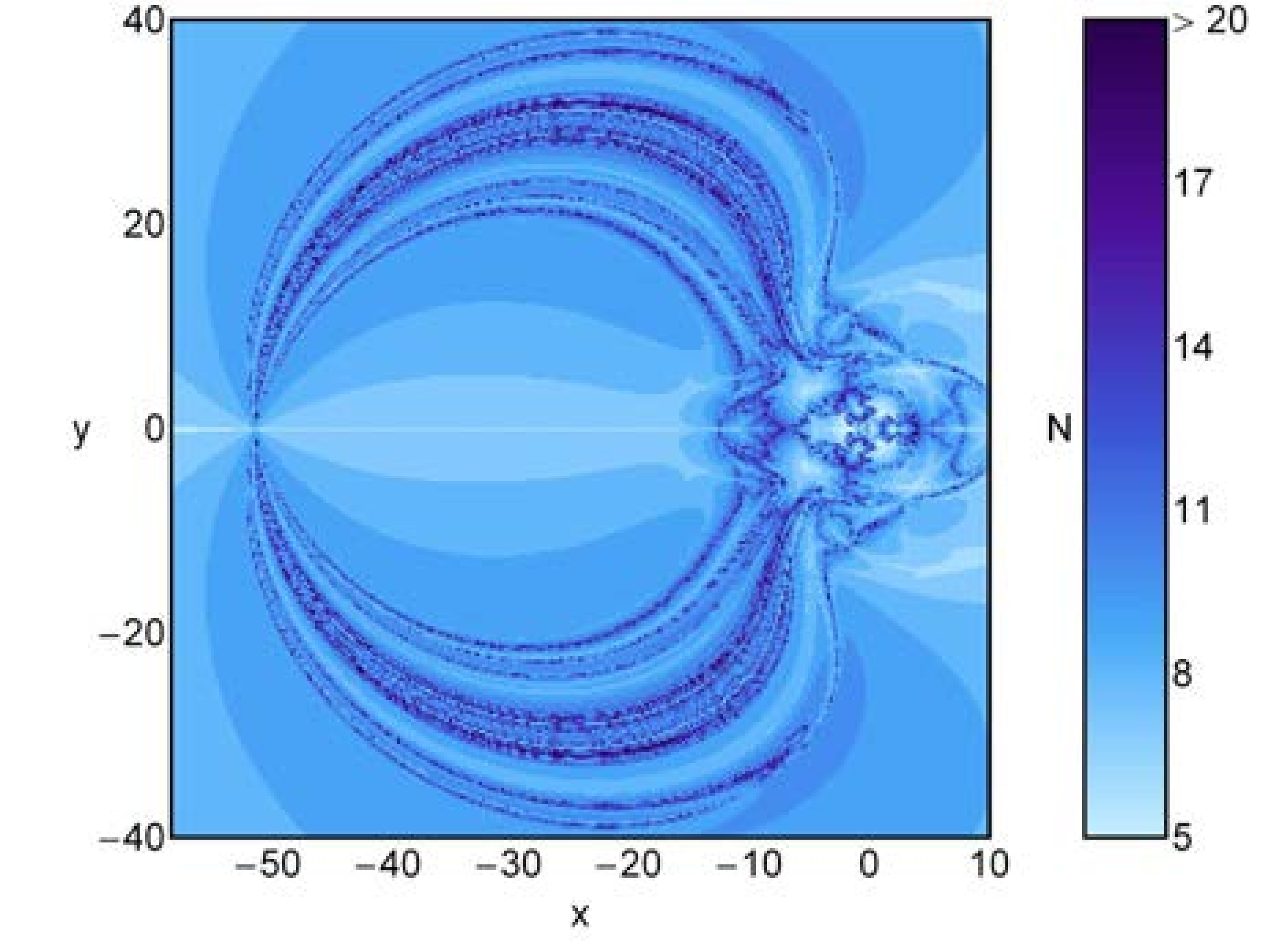}
(l)\includegraphics[scale=.25]{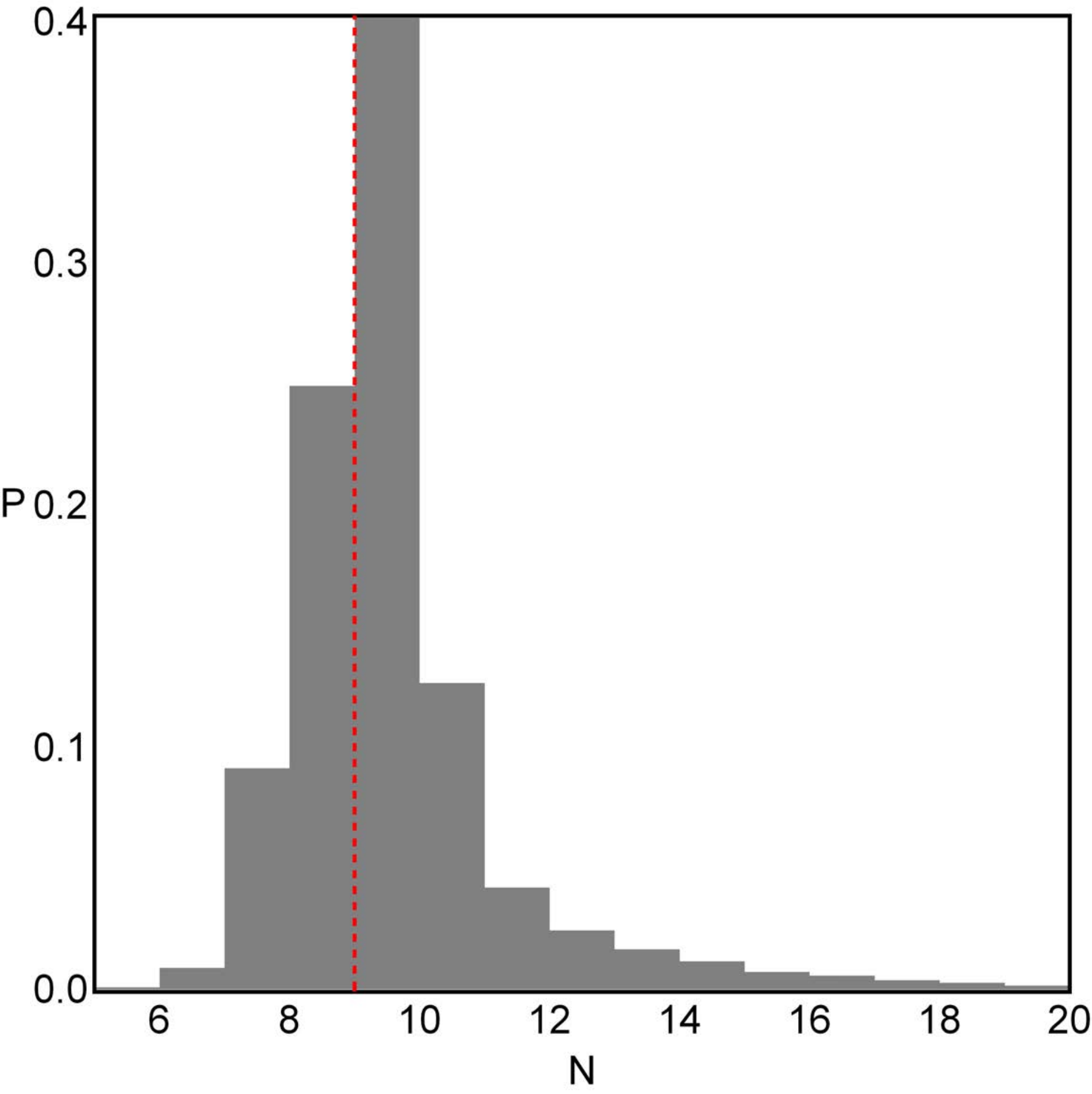}
\caption{The Newton-Raphson basins of attraction on the $xy$-plane for the case when eleven libration points exist for  fixed value of $\alpha=58 \degree$  and for: (a) $\beta=2\degree$; (d) $\beta=4\degree$; (g) $\beta=6\degree$; (j) $\beta=8\degree$.
The color code for the libration points $L_1$,...,$L_{11}$ is same as in Fig \ref{NR_Fig_7}; and non-converging points (white);  (b, e,  h, k) and (c, f, i, l) are the distribution of the corresponding number $(N)$ and the  probability distributions of required iterations for obtaining the Newton-Raphson basins of attraction shown in (a, d, g, j), respectively.
(Color figure online).}
\label{NR_Fig_8}
\end{figure*}

In Fig. \ref{NR_Fig_8}, the basins of convergence are depicted for the fixed value of $\alpha=58\degree$ and various value of the angle parameter $\beta$. We observed that there is significant change in the overall structure of the basins of convergence. In the first three panels of Fig. \ref{NR_Fig_8}(a, d, g, j), the overall structure decreases while in the panel-j, it increases. Therefore, the overall structure composed of the finite domain of convergence will either shrinks or expands with the increase in angle parameter $\beta$. In panels-(c, f, i, l), the corresponding probability distribution is shown while the most probable number of required iterations $N^*$ are illustrated in panels-(b, e, h, k). This reveals that the number $N^*$ first decreases from $13$ to $8$ in first three panels while increases to $9$ in the last panel. Therefore, it is not possible to predict the value of $N^*$ with the increase in value of angle parameter $\beta$.
\subsubsection{when thirteen libration points exist}
\label{when thirteen libration points exists}
\begin{figure*}[!t]
\centering
(a)\includegraphics[scale=.27]{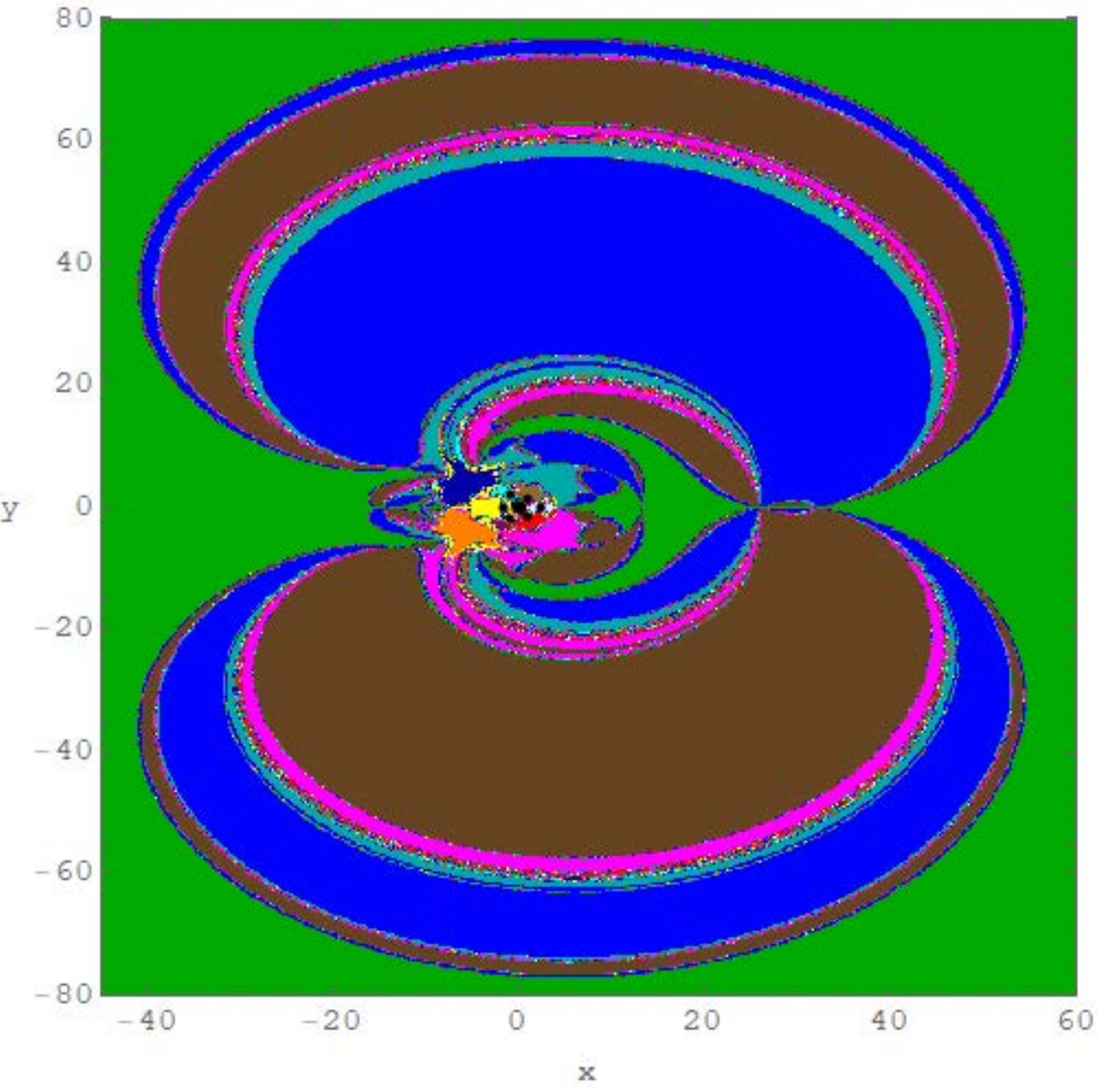}
(b)\includegraphics[scale=.27]{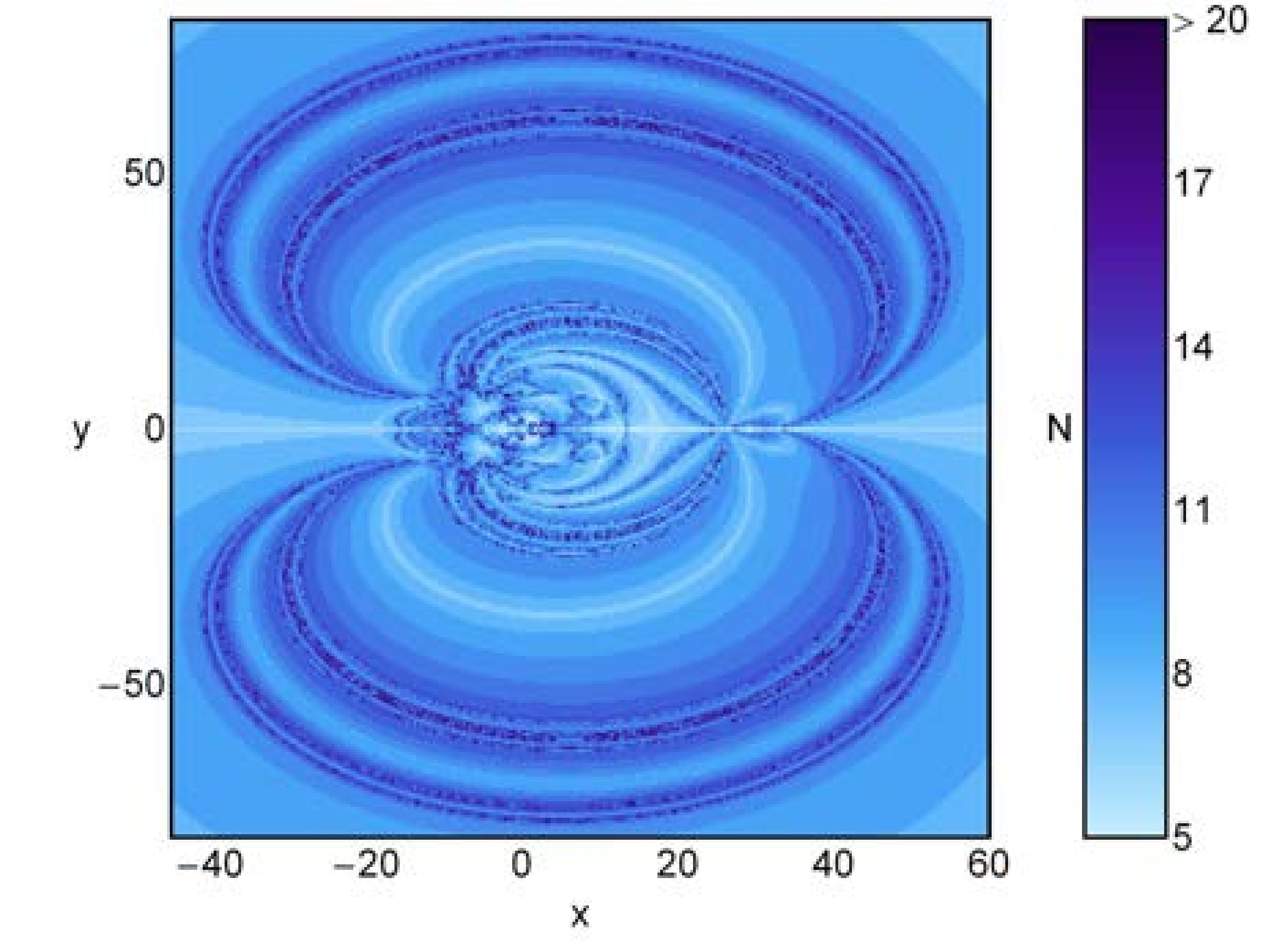}
(c)\includegraphics[scale=.25]{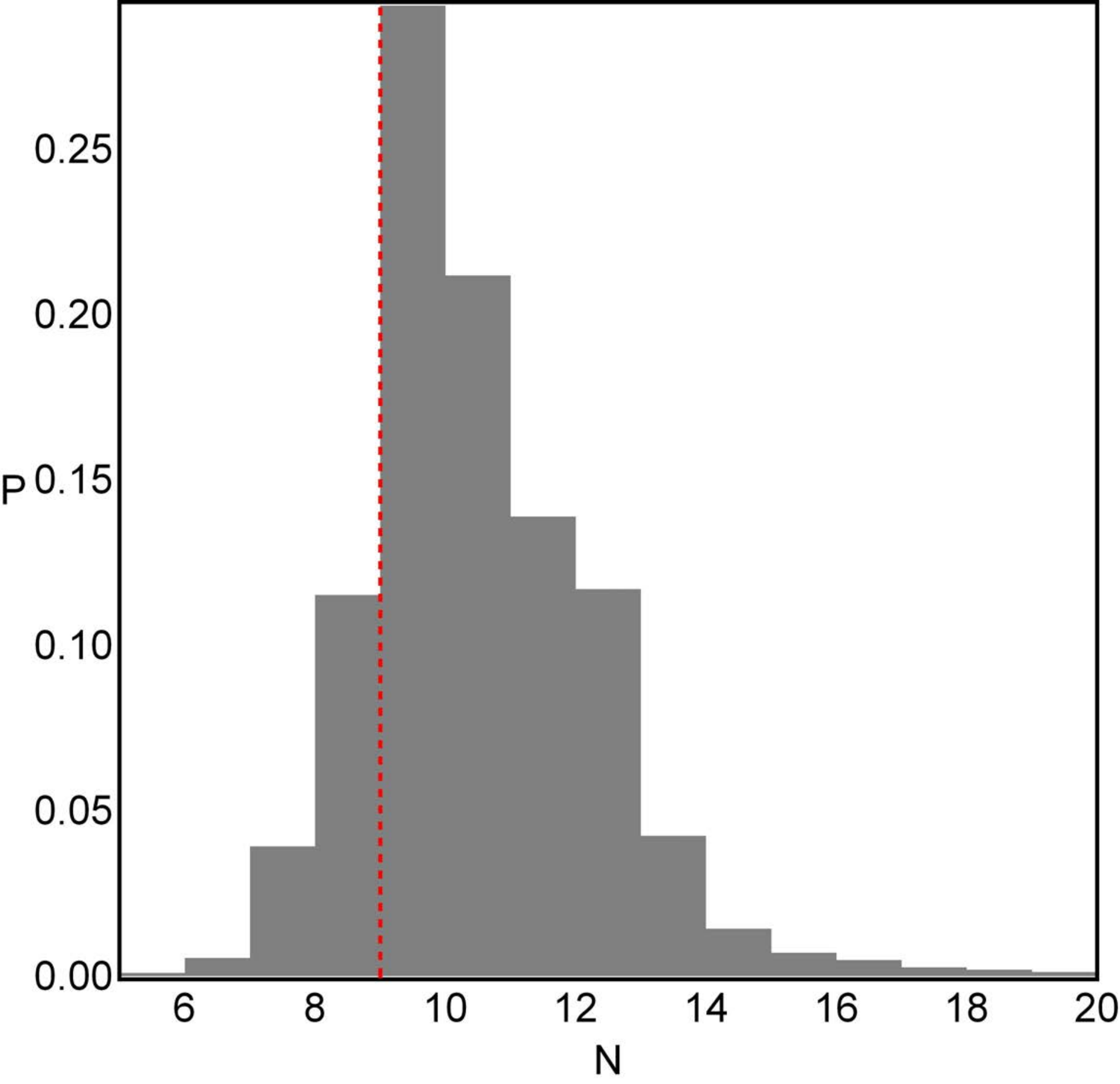}\\
(d)\includegraphics[scale=.27]{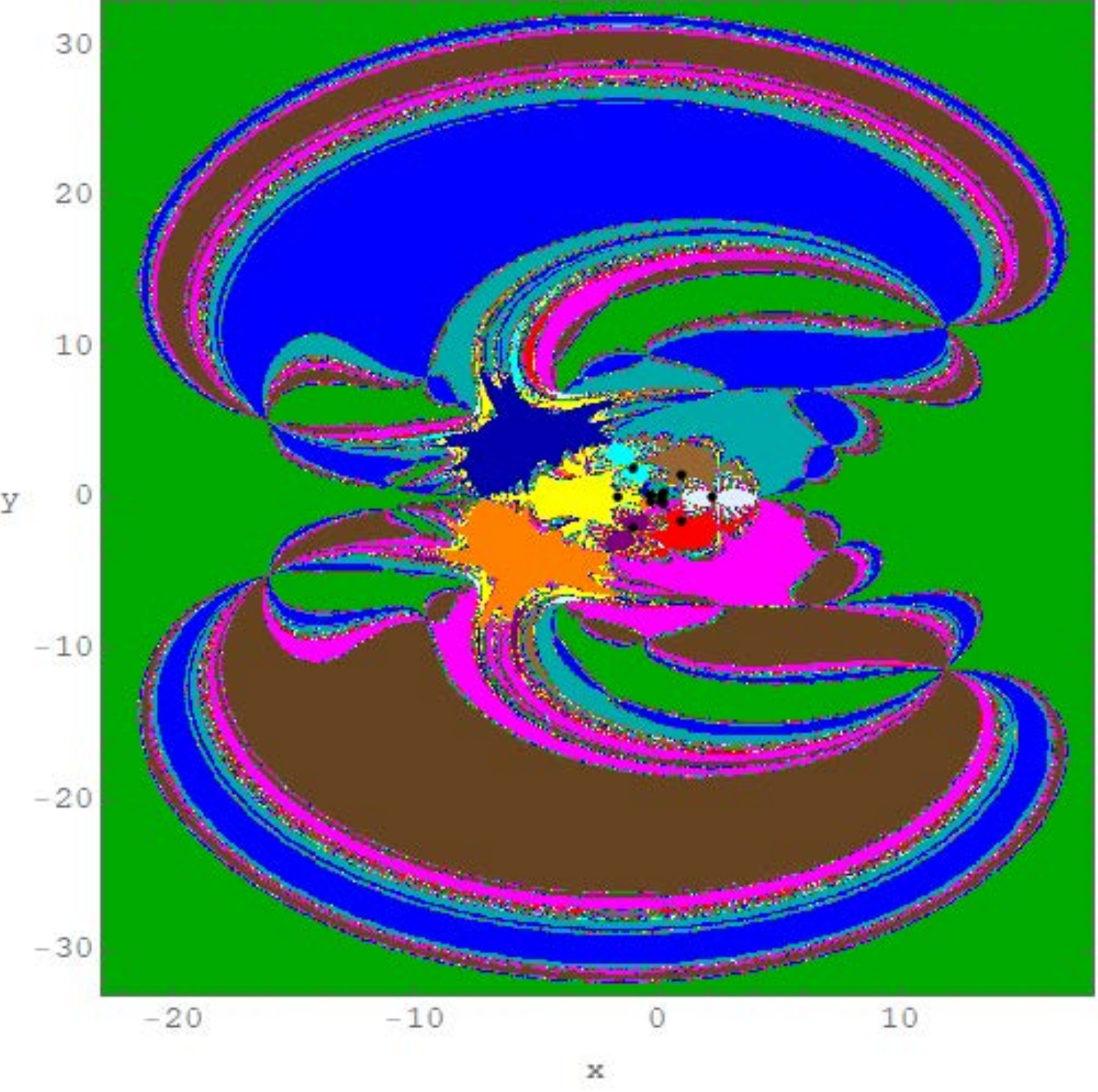}
(e)\includegraphics[scale=.27]{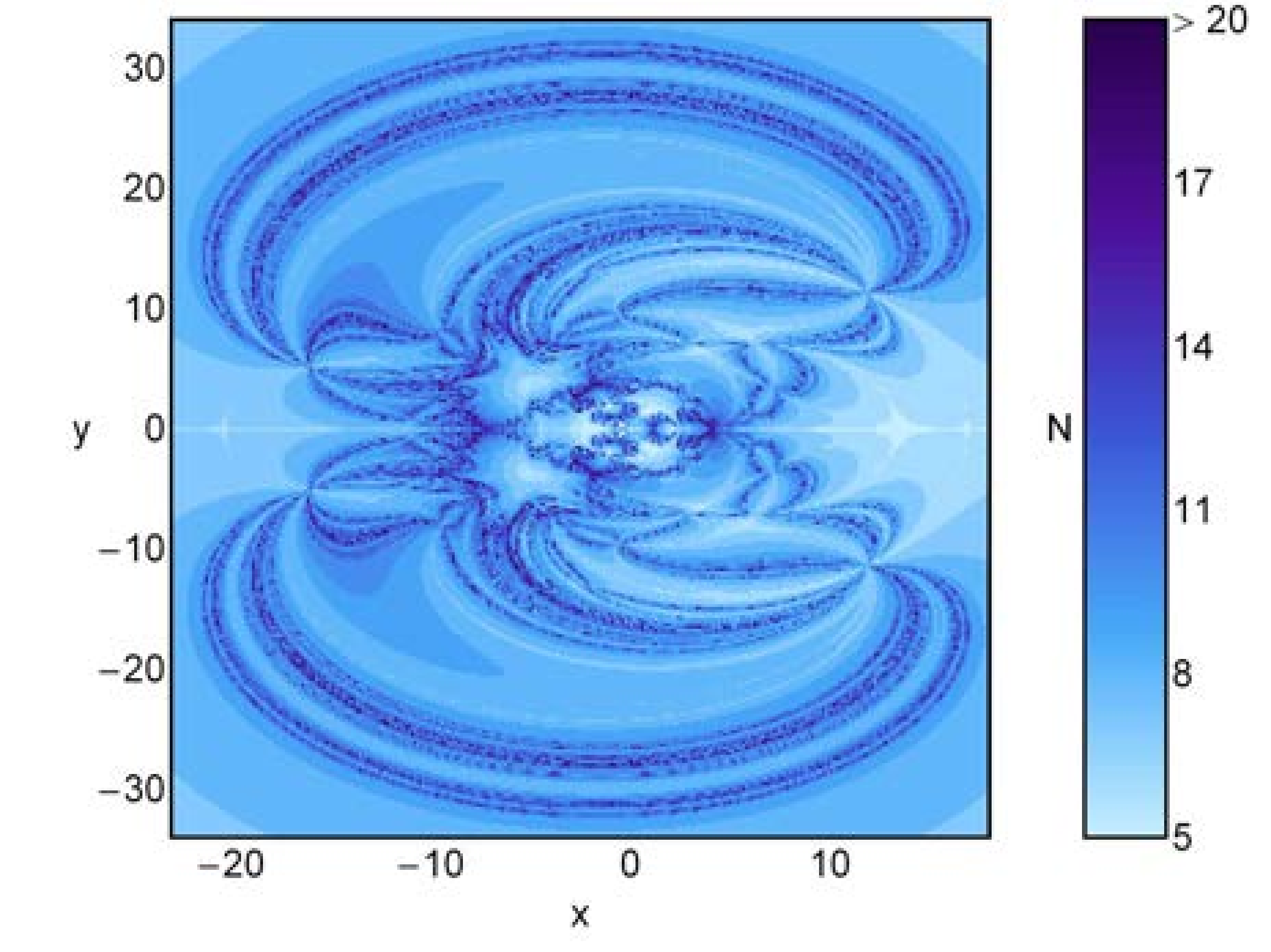}
(f)\includegraphics[scale=.25]{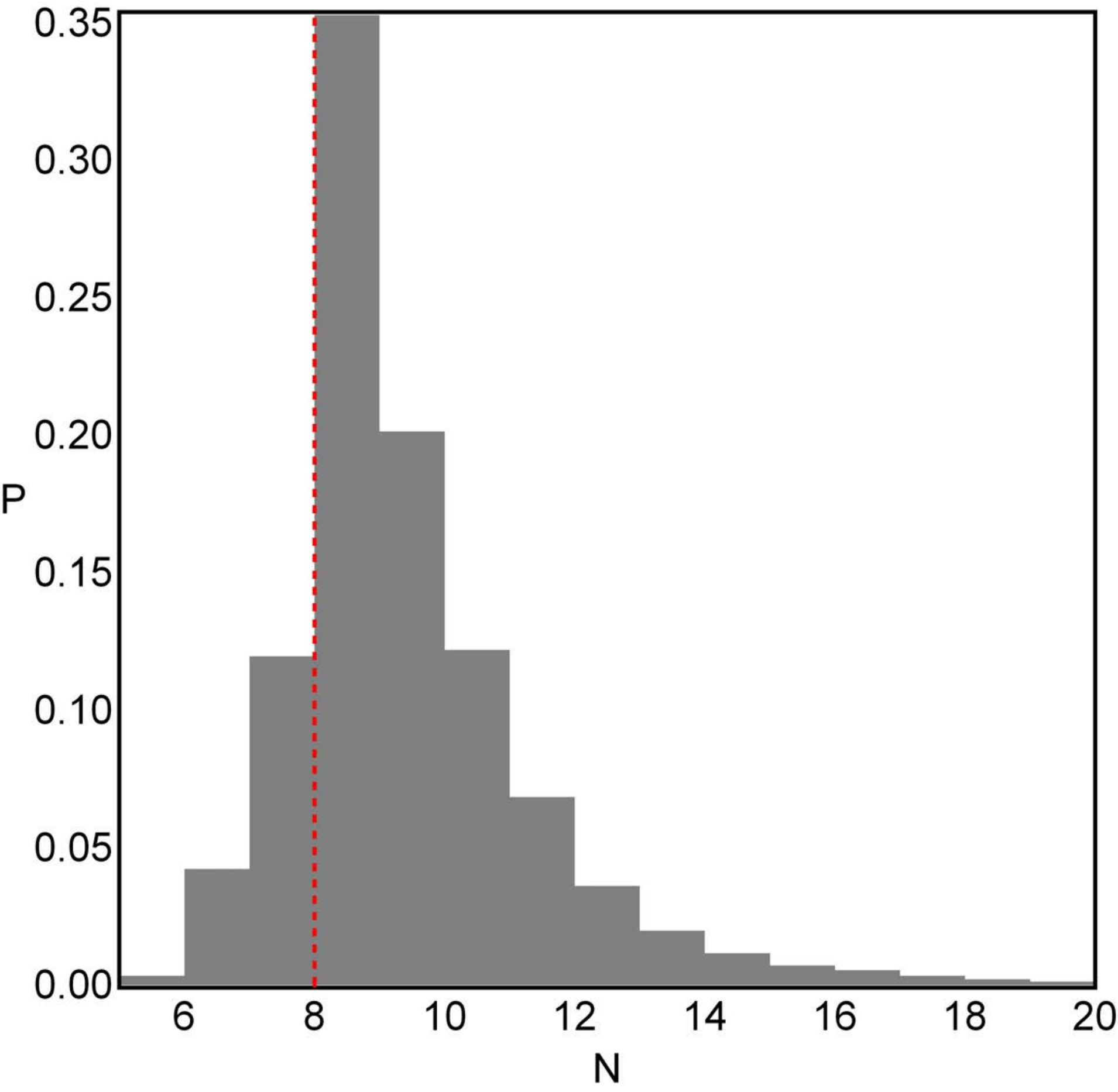}\\
(g)\includegraphics[scale=.27]{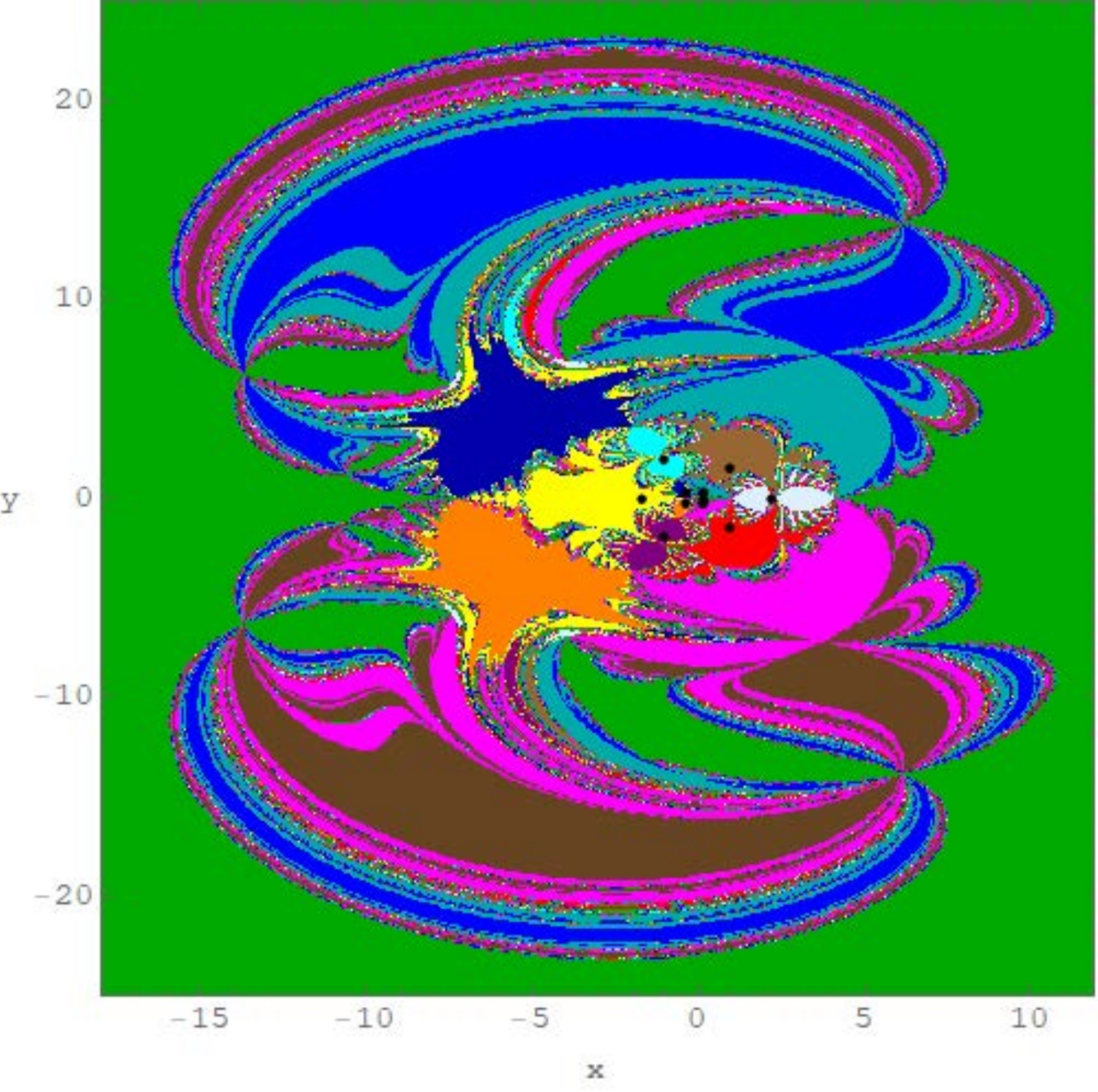}
(h)\includegraphics[scale=.27]{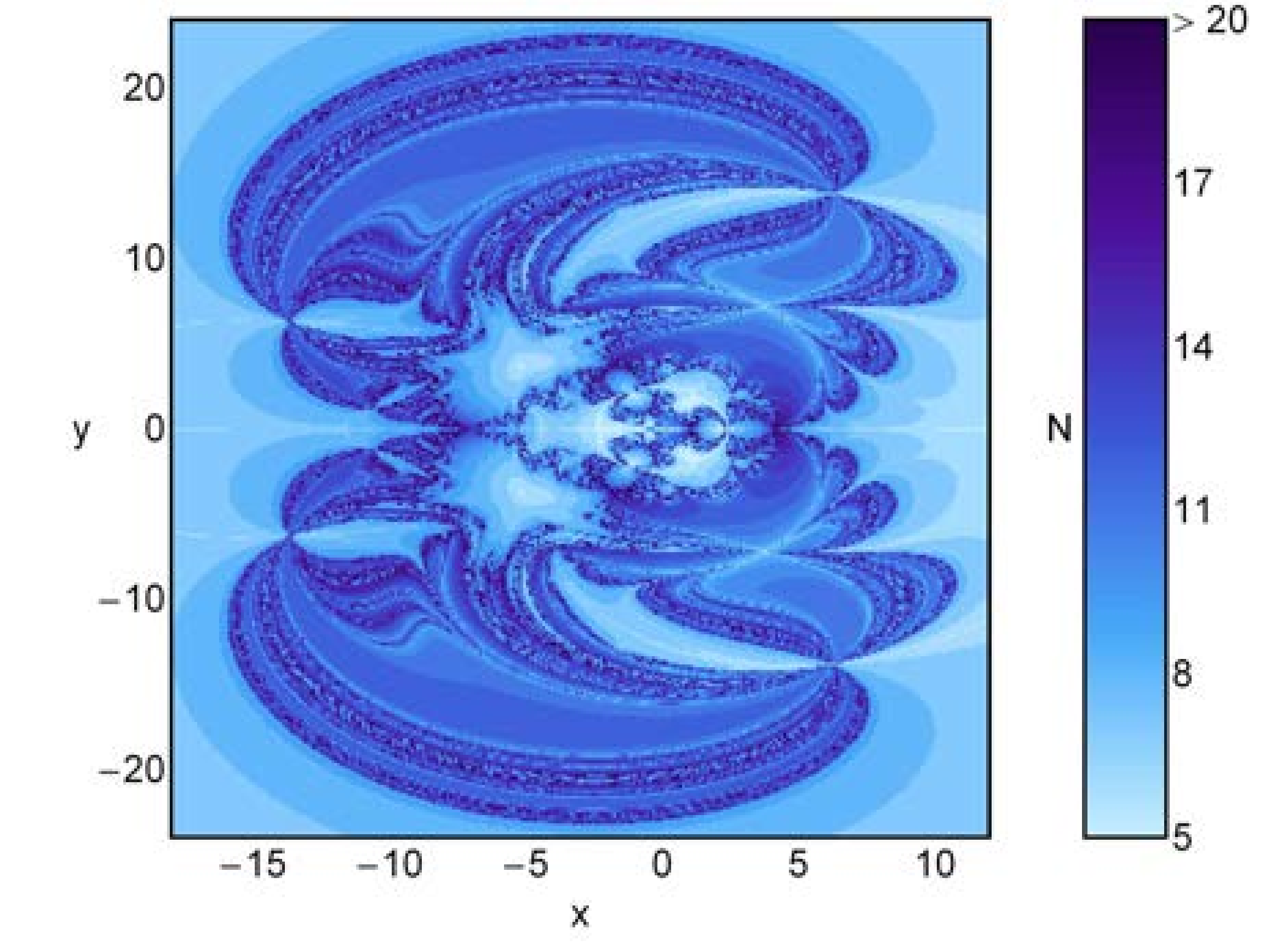}
(i)\includegraphics[scale=.25]{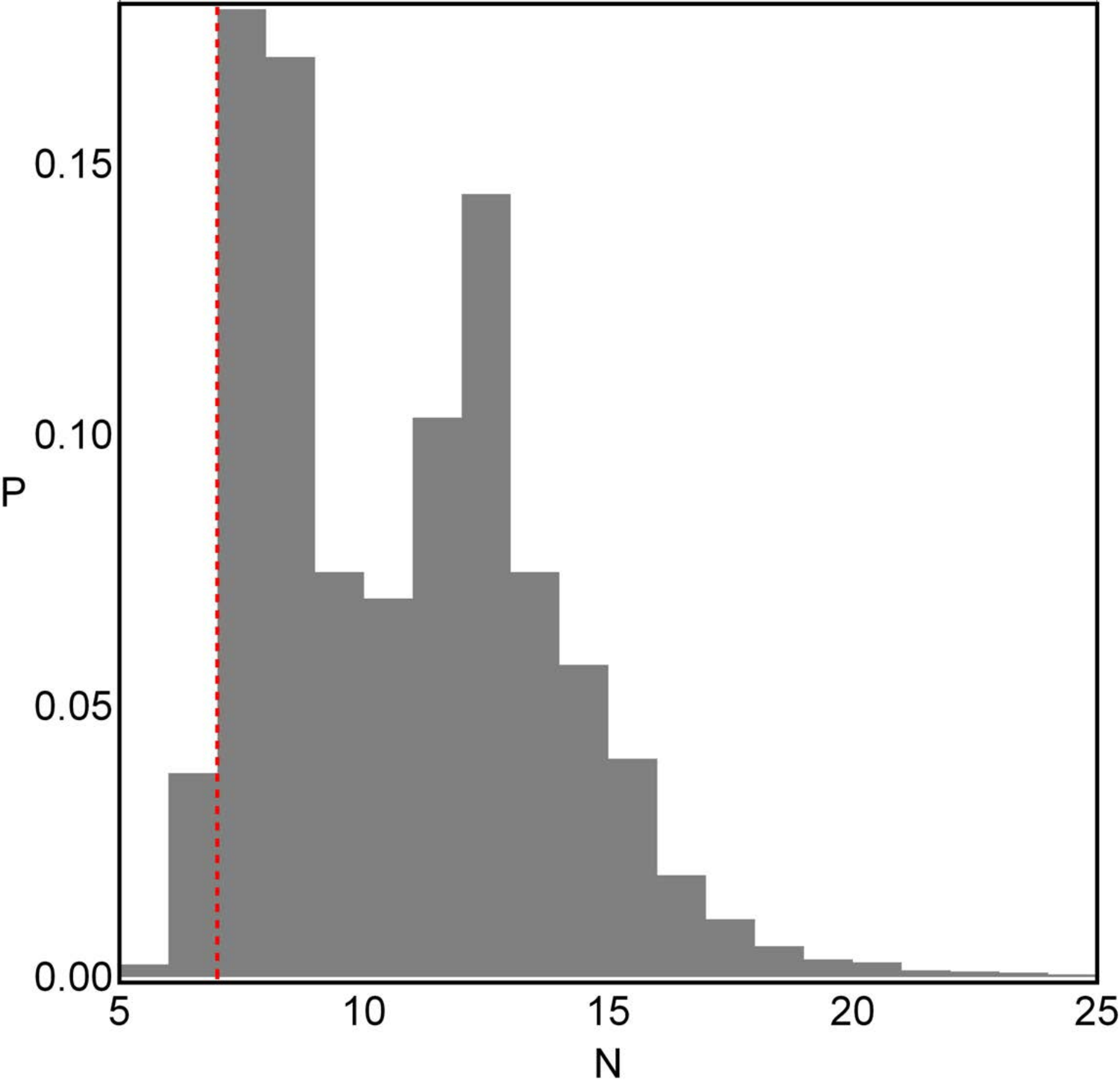}
\caption{The Newton-Raphson basins of attraction on the $xy$-plane for the
case when thirteen libration points exist for fixed value of $\alpha=58 \degree$  and for:
(a) $\beta=8.9\degree$; (d) $\beta=9.5\degree$; (g) $\beta=10\degree$.
The color code for the libration points $L_1$,...,$L_{11}$ is same as in Fig \ref{NR_Fig_7},  $L_{12}\emph{(darker orange)}$;  $L_{13}\emph{(light blue)}$; and non-converging points (white);  (b, e,  h) and (c, f, i) are the distribution of the corresponding number $(N)$ and the  probability distributions of required iterations for obtaining the Newton-Raphson basins of attraction shown in (a, d, g), respectively.
(Color figure online).}
\label{NR_Fig_9}
\end{figure*}

%%%%%%%%%%
We continue our analysis with the case where thirteen libration points exist for fixed value of $\alpha=58\degree$ and three different values of $\beta=8.9\degree, 9.5\degree$ and $10\degree$ are illustrated  in the Figs. \ref{NR_Fig_9}(a, d, g) respectively. One may observe that for this particular value of $\alpha$ and various increasing values of $\beta$, the extent of the basins of convergence associated with the libration point $L_2$ is infinite, whereas for the other libration points $L_{1,3,4,...,13}$ are well formed and finite. We have further observed that the overall structure on the configuration $(x, y)$ plane which is composed of all the various domain of the convergence associated with the libration points shrinks rapidly as the angle parameter $\beta$ increases.  The distribution of the corresponding number $N$ of iterations, required to obtain the predetermined accuracy, is depicted in Fig. \ref{NR_Fig_9}(b, e, h). In the panels (b, e), majority of initial conditions converge for $N<10$ to one of the attracting domain. In panel-h, the number of iterations required to converge, for an initial condition to one of the attractors, increases to $N>15$. Moreover, the most probable number $N^*$ of the iterations is not constant with the increase in parameter $\beta$, on the contrary this number occurs randomly (see, panel-c, f, i).
%%%%
%%%%
%%%%
%%%%
%%%%
%%%%
%%%%
%%%%
%%%%%%%%%%%%%%%%%%%%%%%%%%%%%%%%%
%%%%%%%%%%%%%%%%%%%%%%%%%%%%%%%%%
\subsection{Case:II The second concave case}
\label{Case:II The second concave case}
We further continue our analysis with the second case where we have discussed the second type of concave case, for two different values of the angle parameter $\alpha$ and corresponding permissible range of $\beta$.
\subsubsection{when nine libration points exist}
\begin{figure*}[!t]
\centering
(a)\includegraphics[scale=.27]{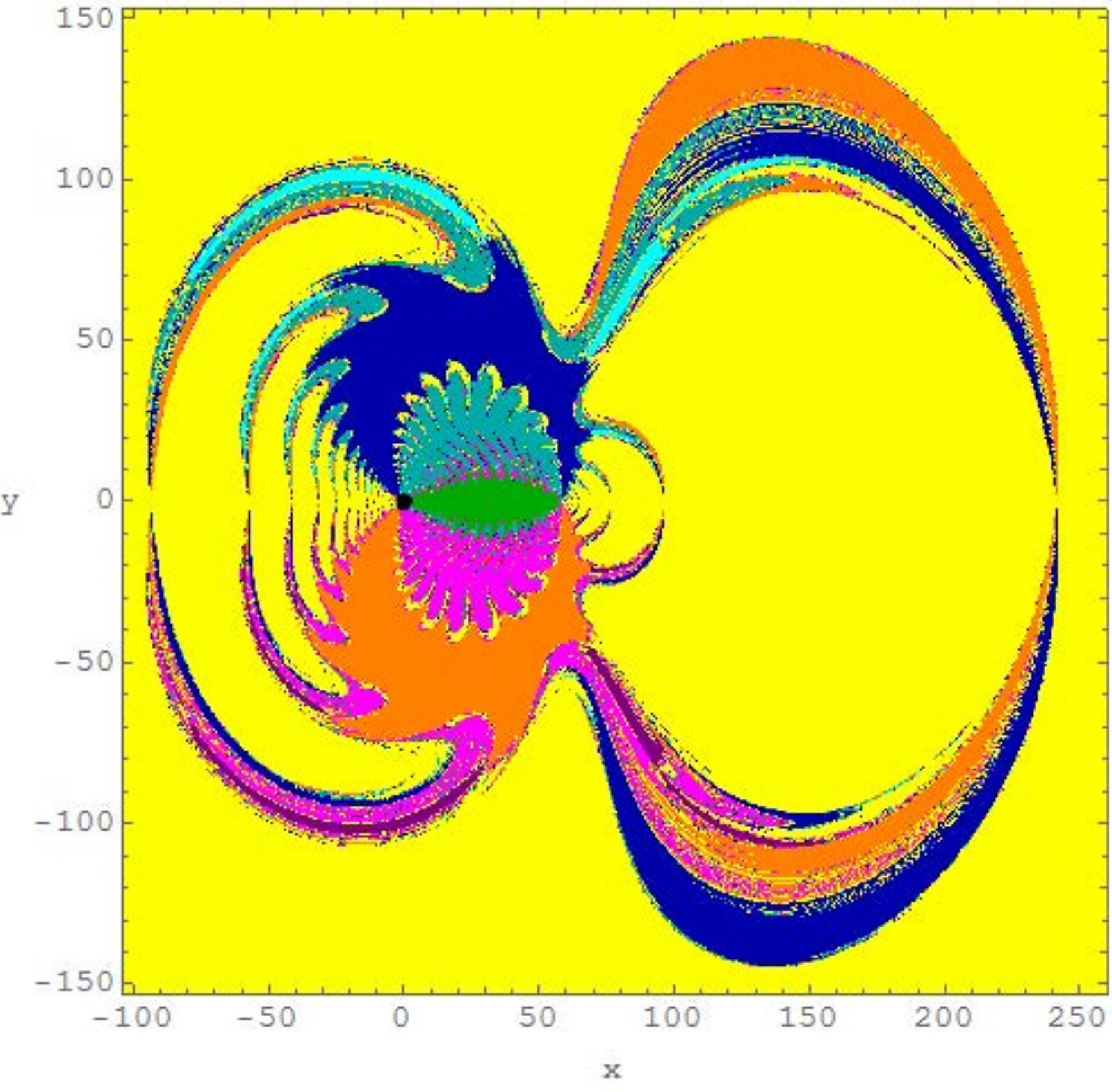}%Fig_61_32+1/30
(b)\includegraphics[scale=.27]{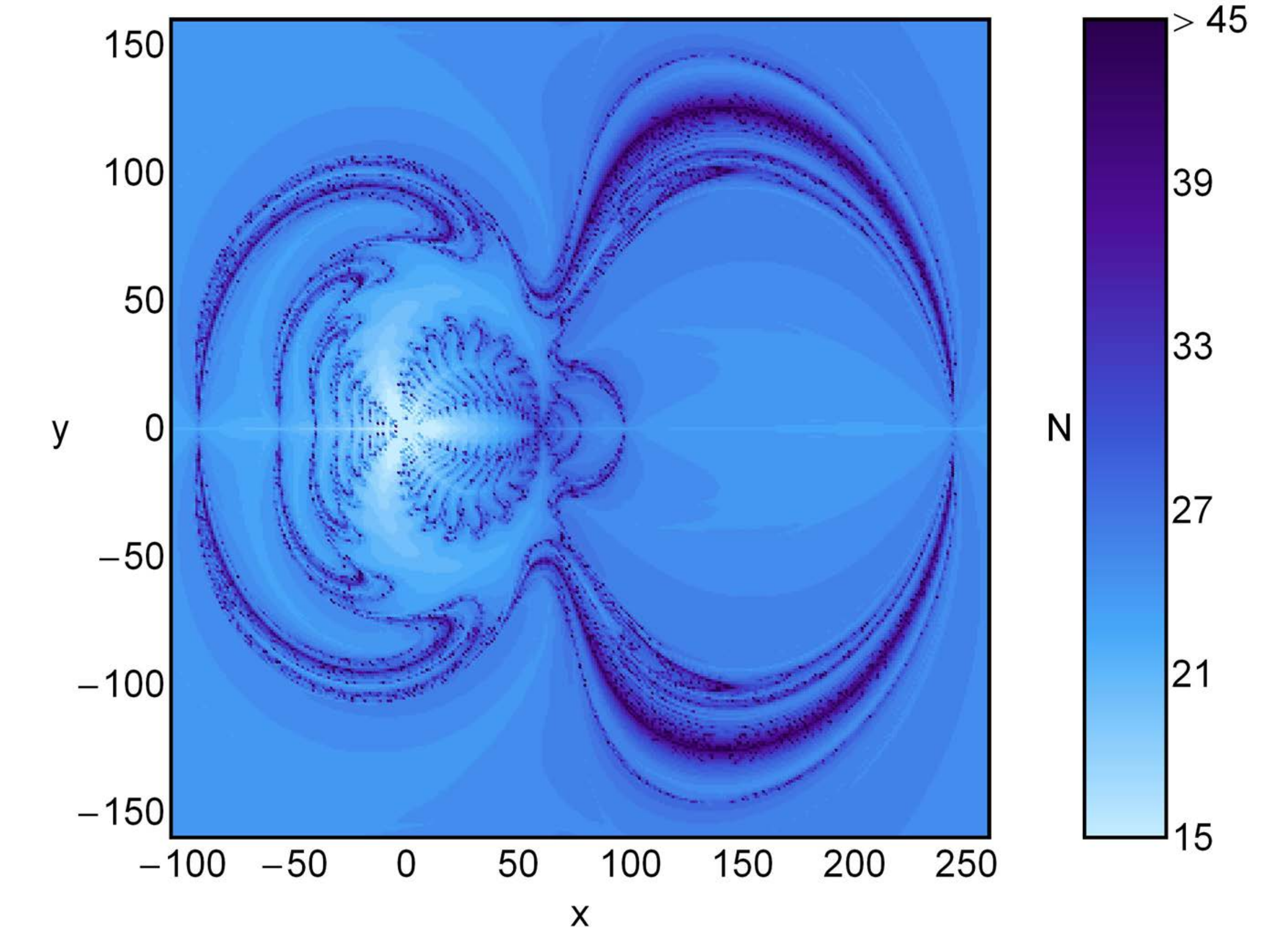}
(c)\includegraphics[scale=.25]{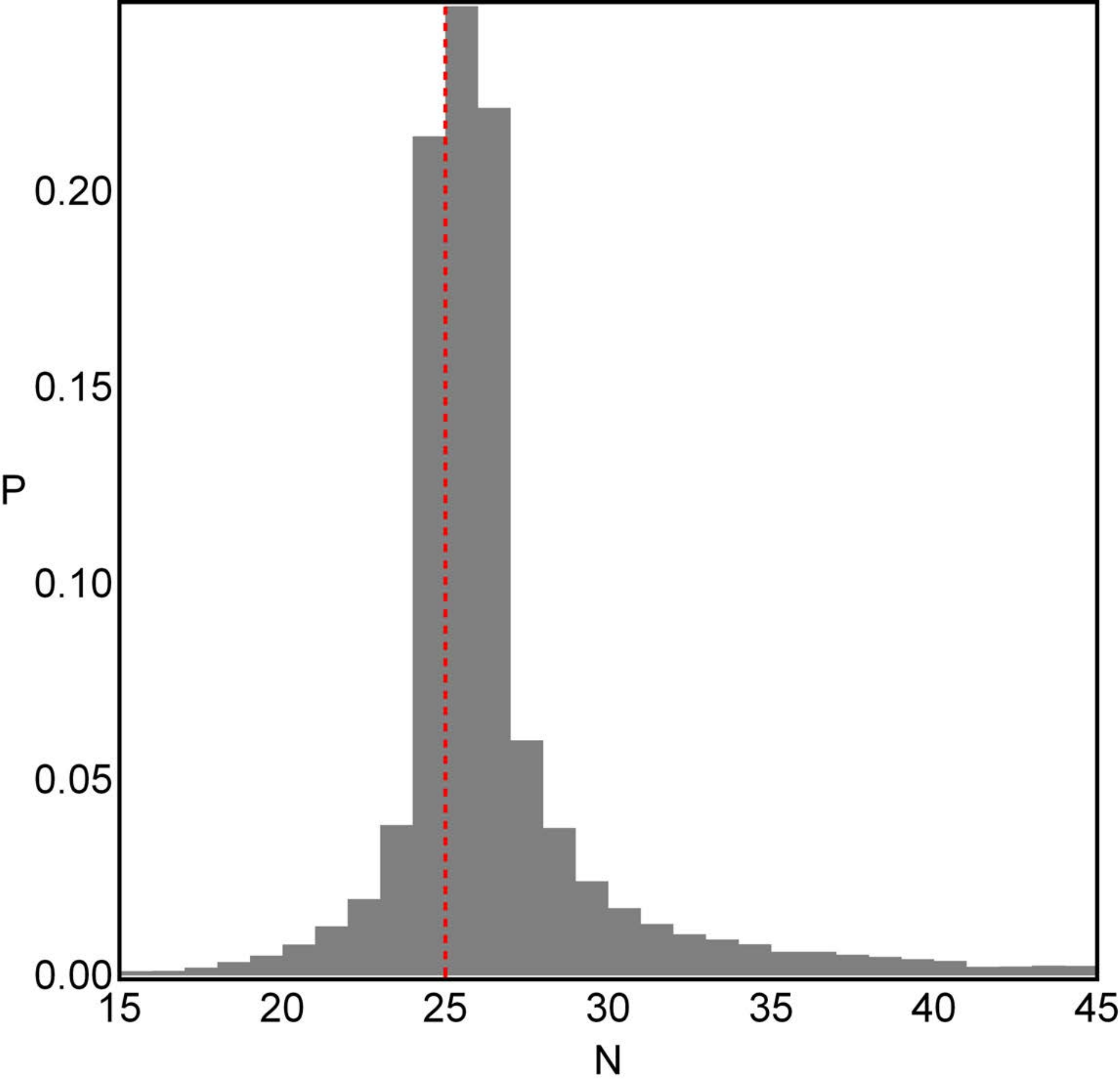}\\
(d)\includegraphics[scale=.27]{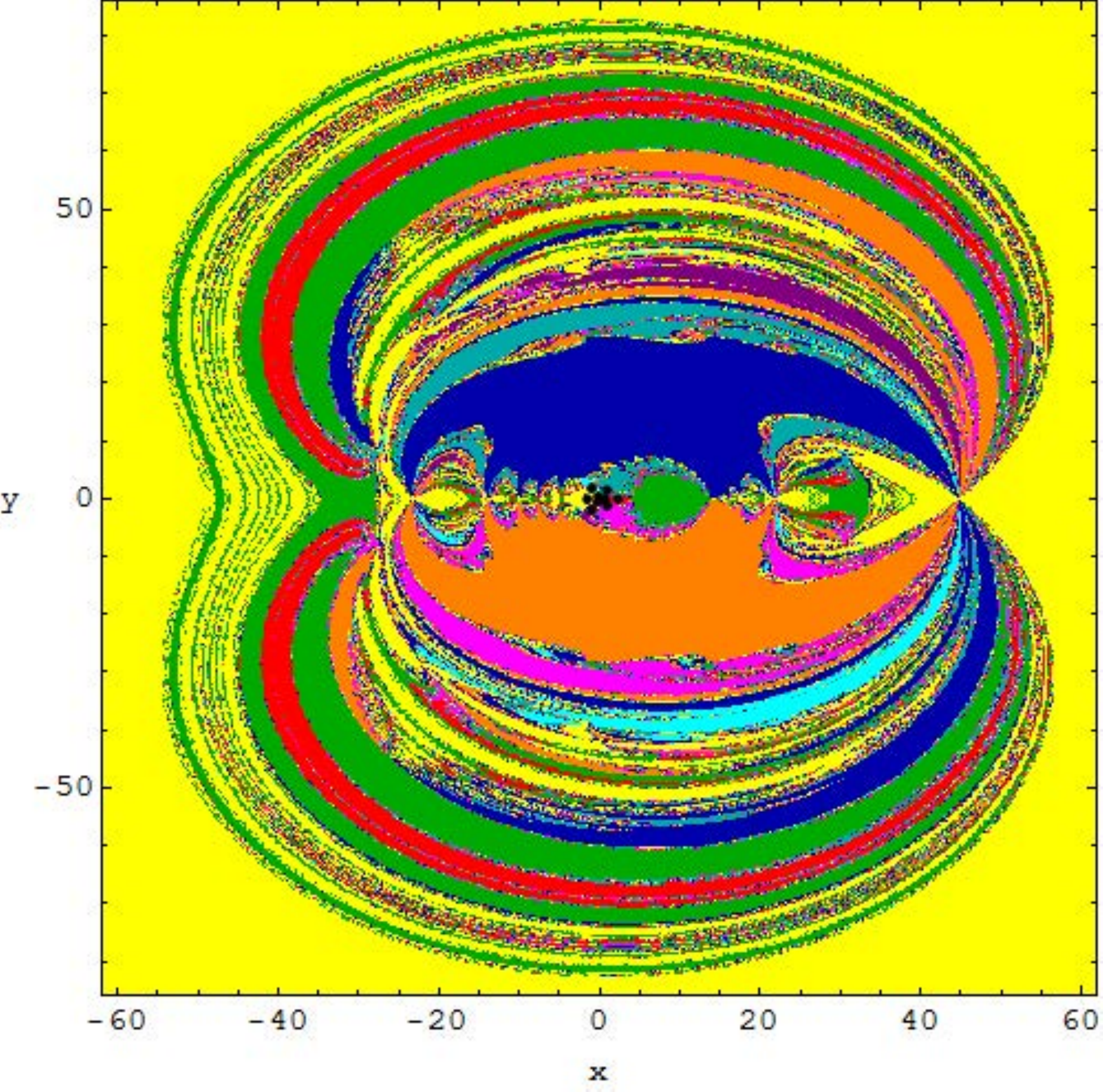}%Fig_61_34
(e)\includegraphics[scale=.27]{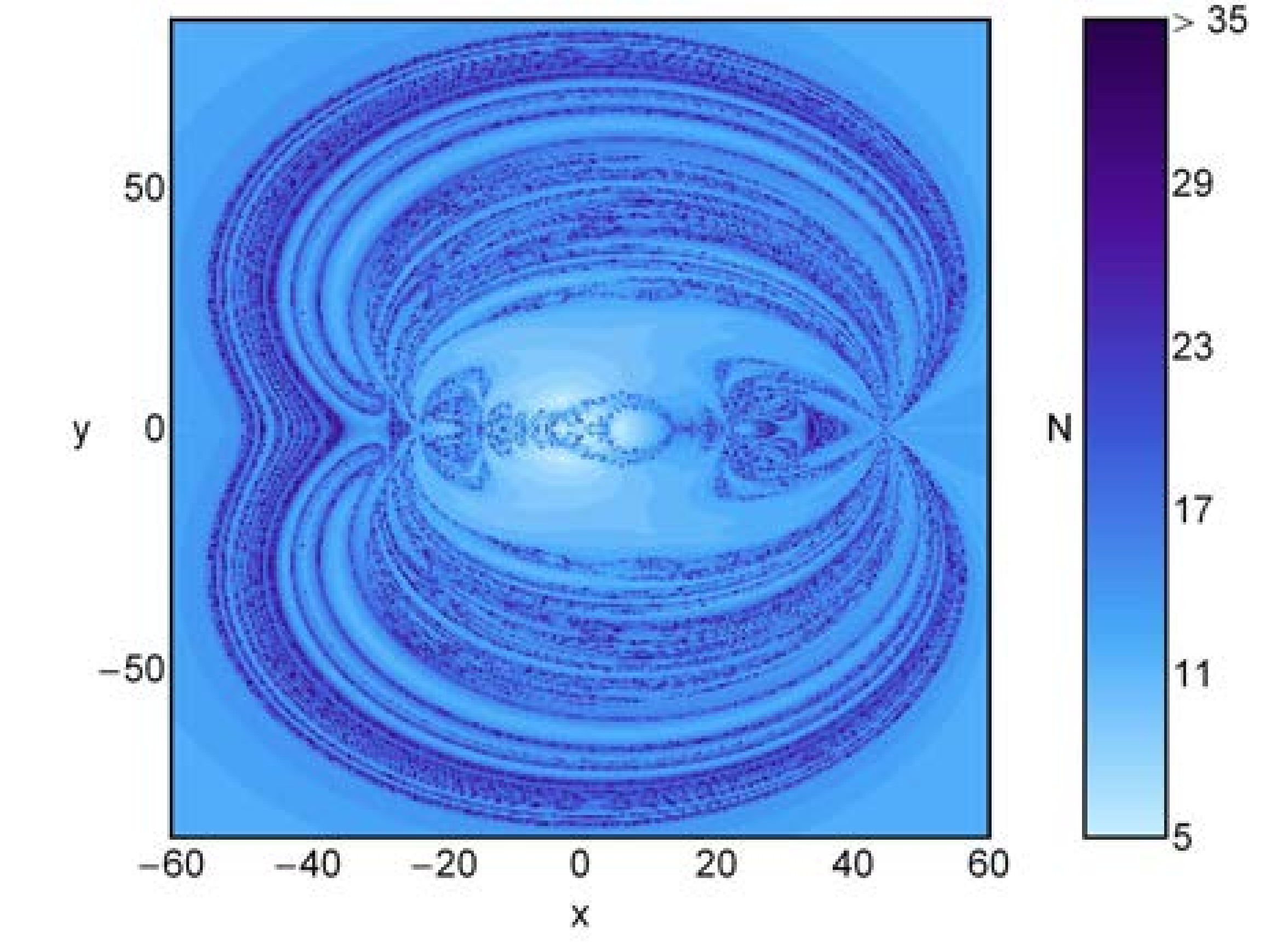}
(f)\includegraphics[scale=.25]{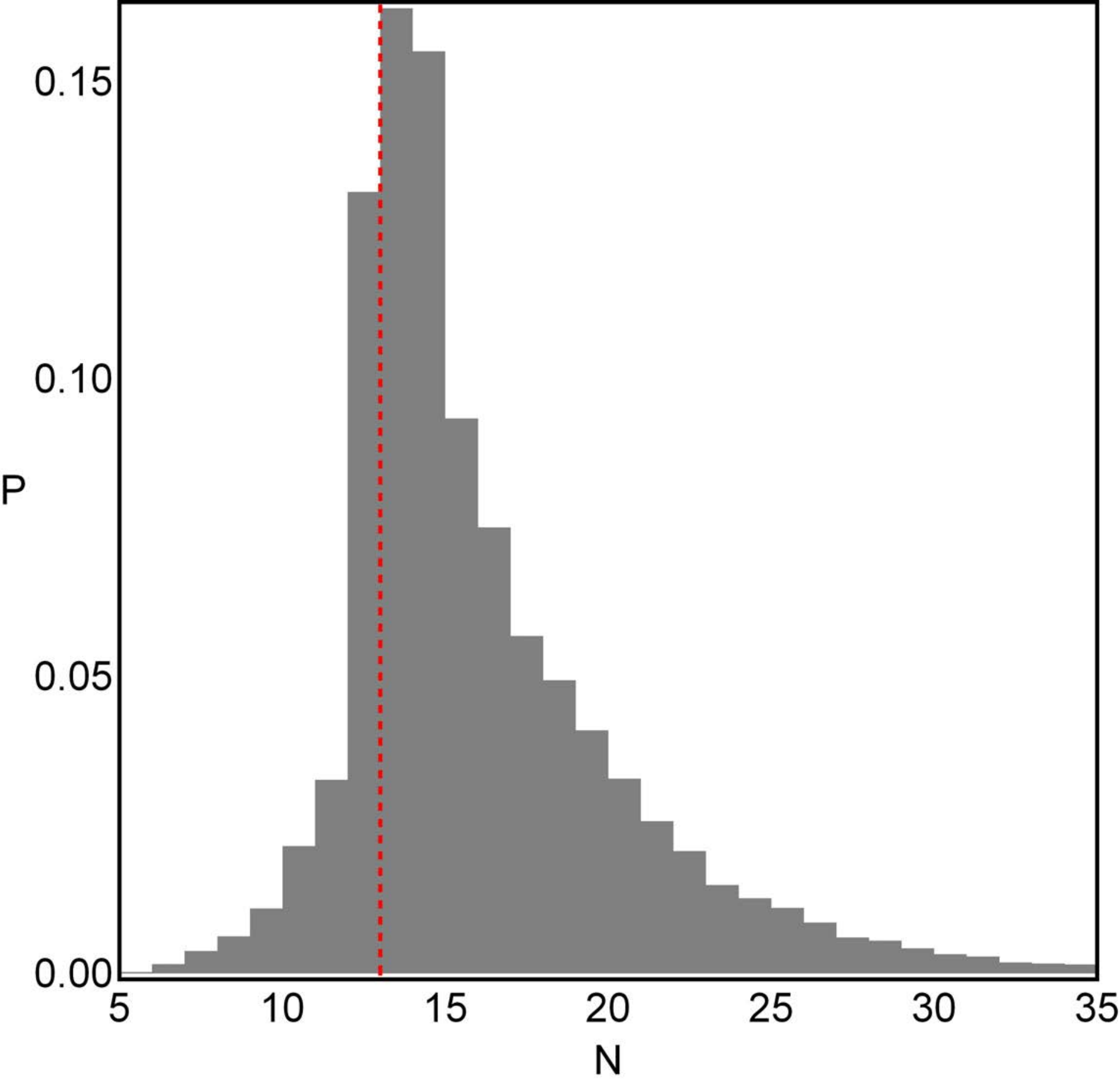}\\
(g)\includegraphics[scale=.27]{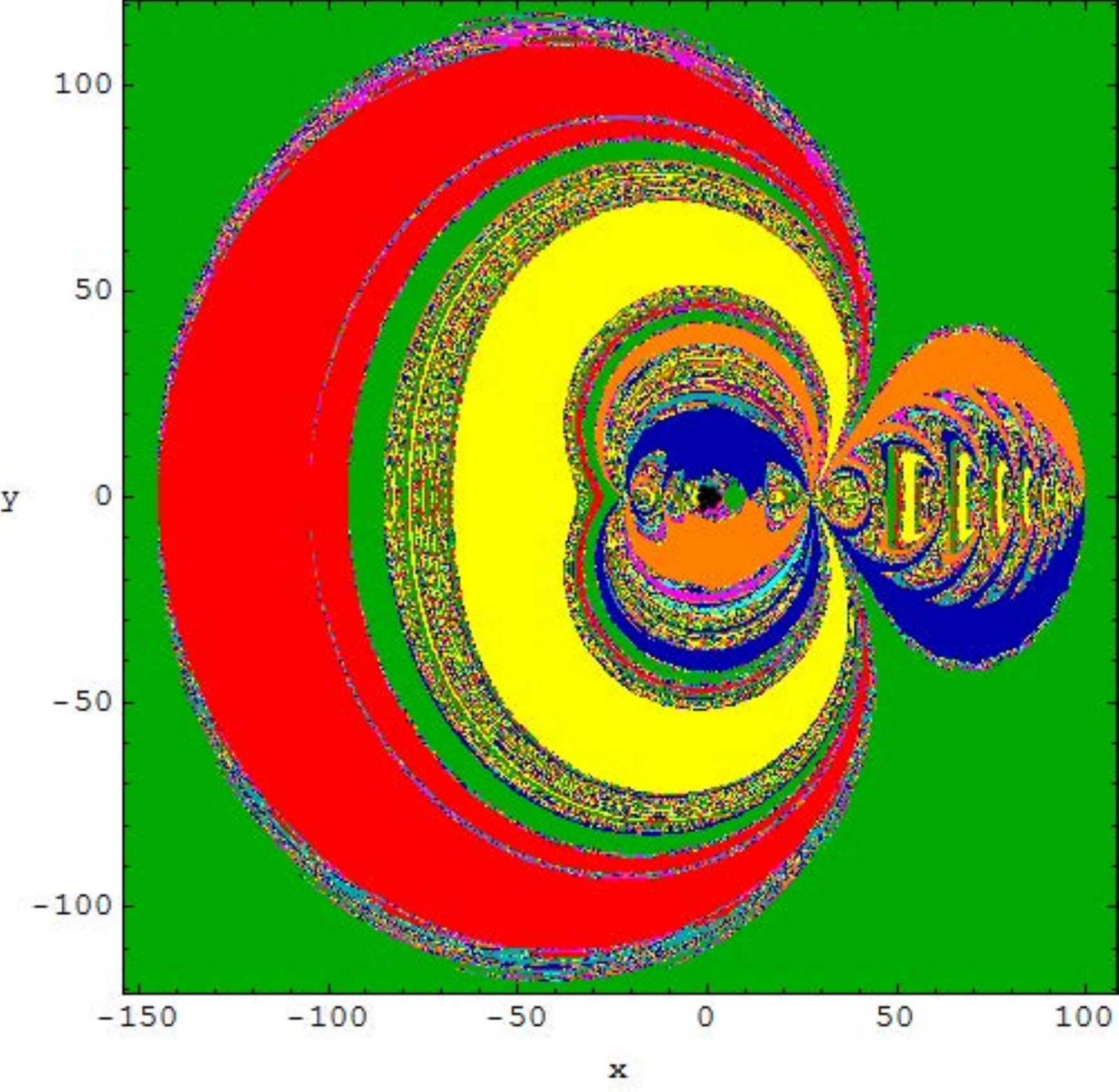}%Fig_61_36
(h)\includegraphics[scale=.27]{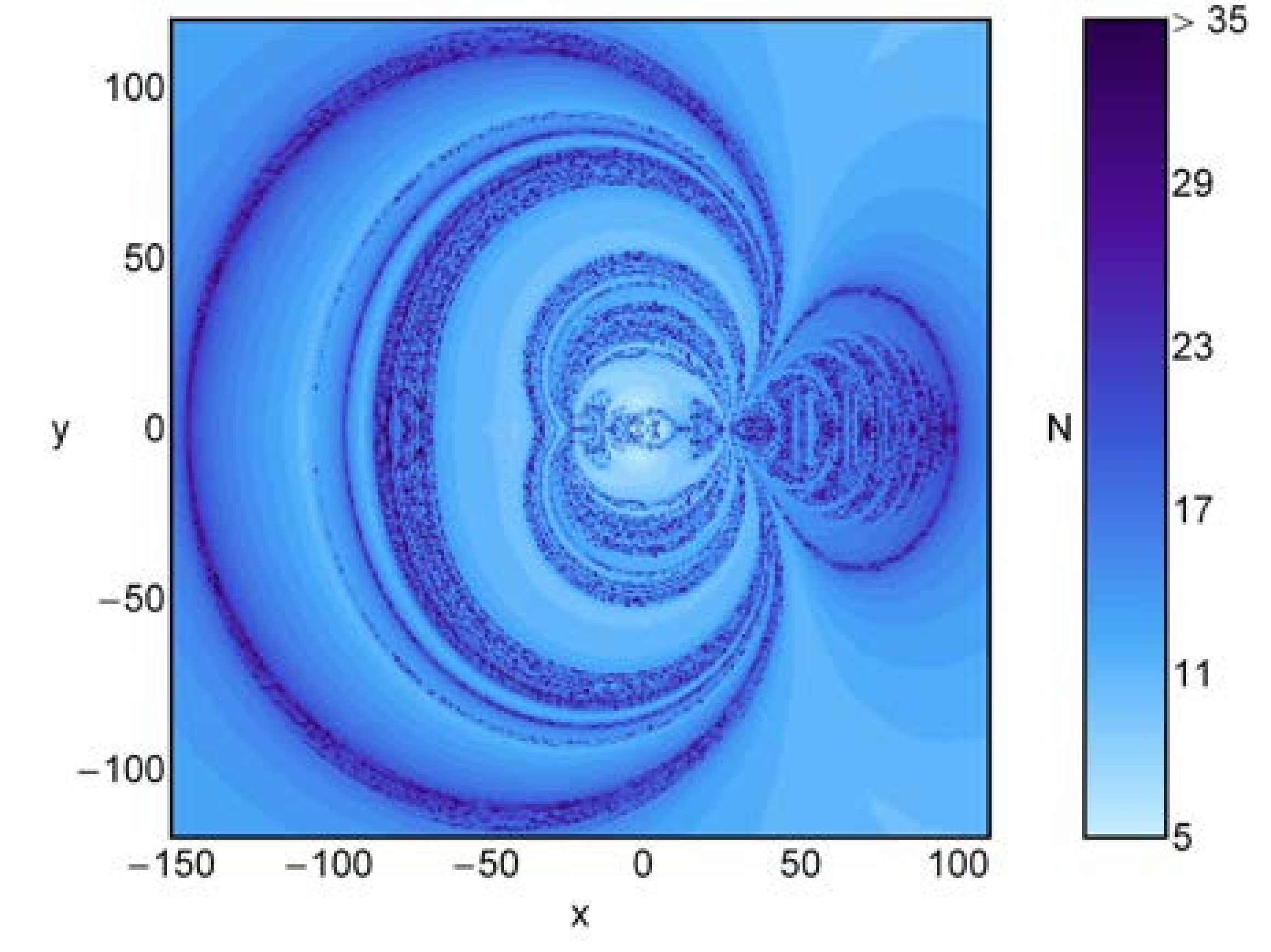}
(i)\includegraphics[scale=.25]{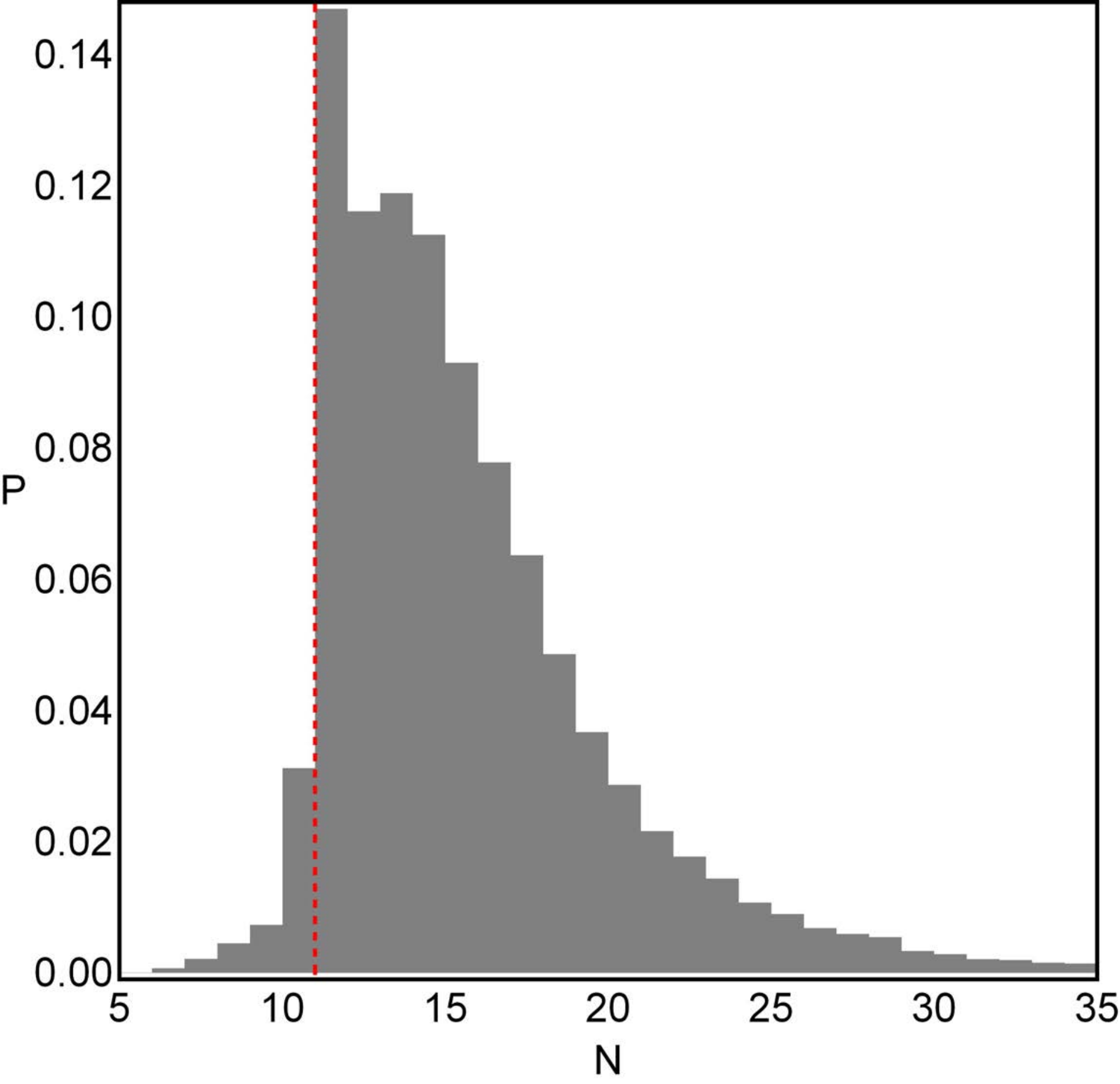}\\
(j)\includegraphics[scale=.27]{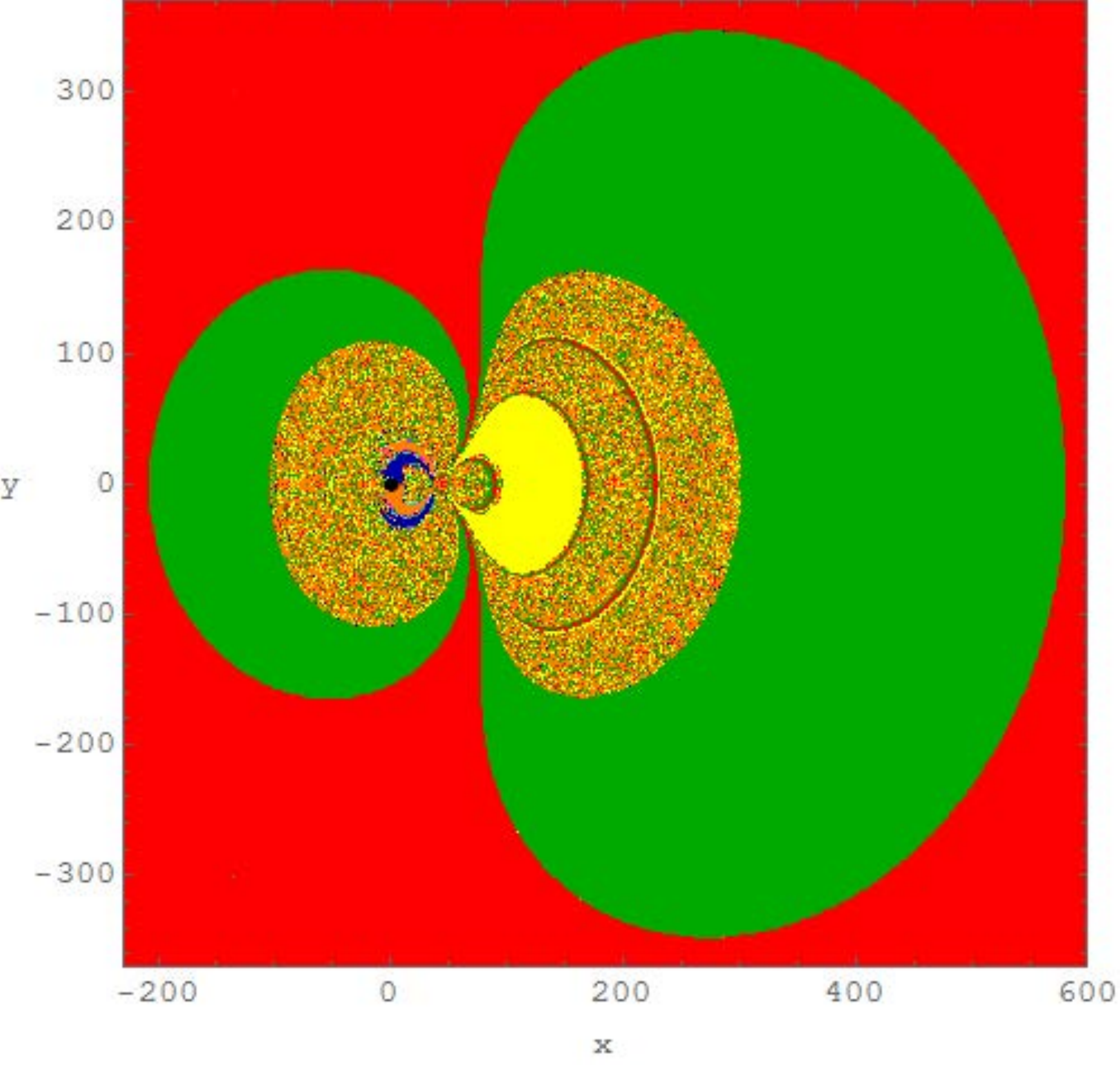}%Fig_61_38.5
(k)\includegraphics[scale=.27]{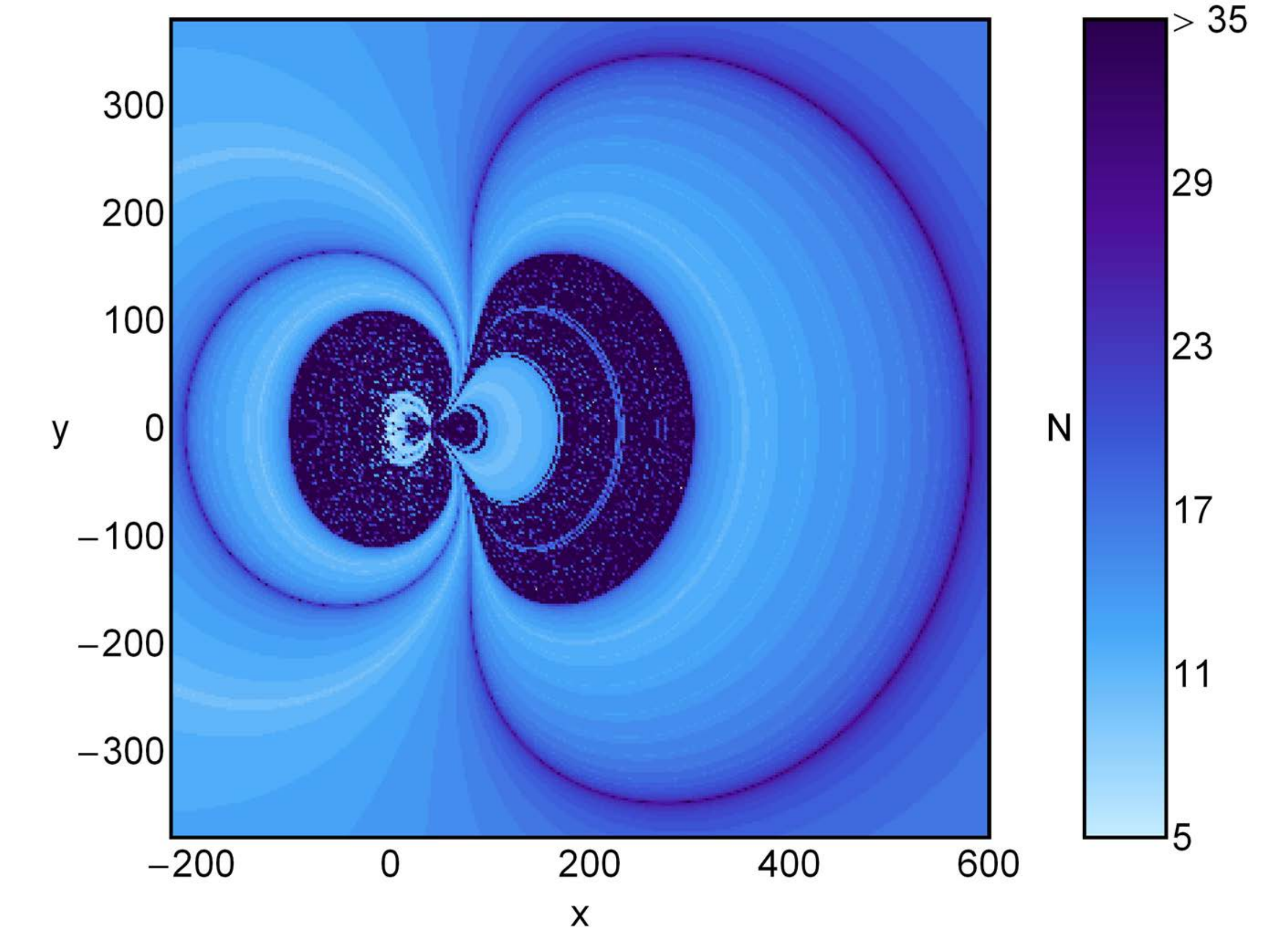}
(l)\includegraphics[scale=.25]{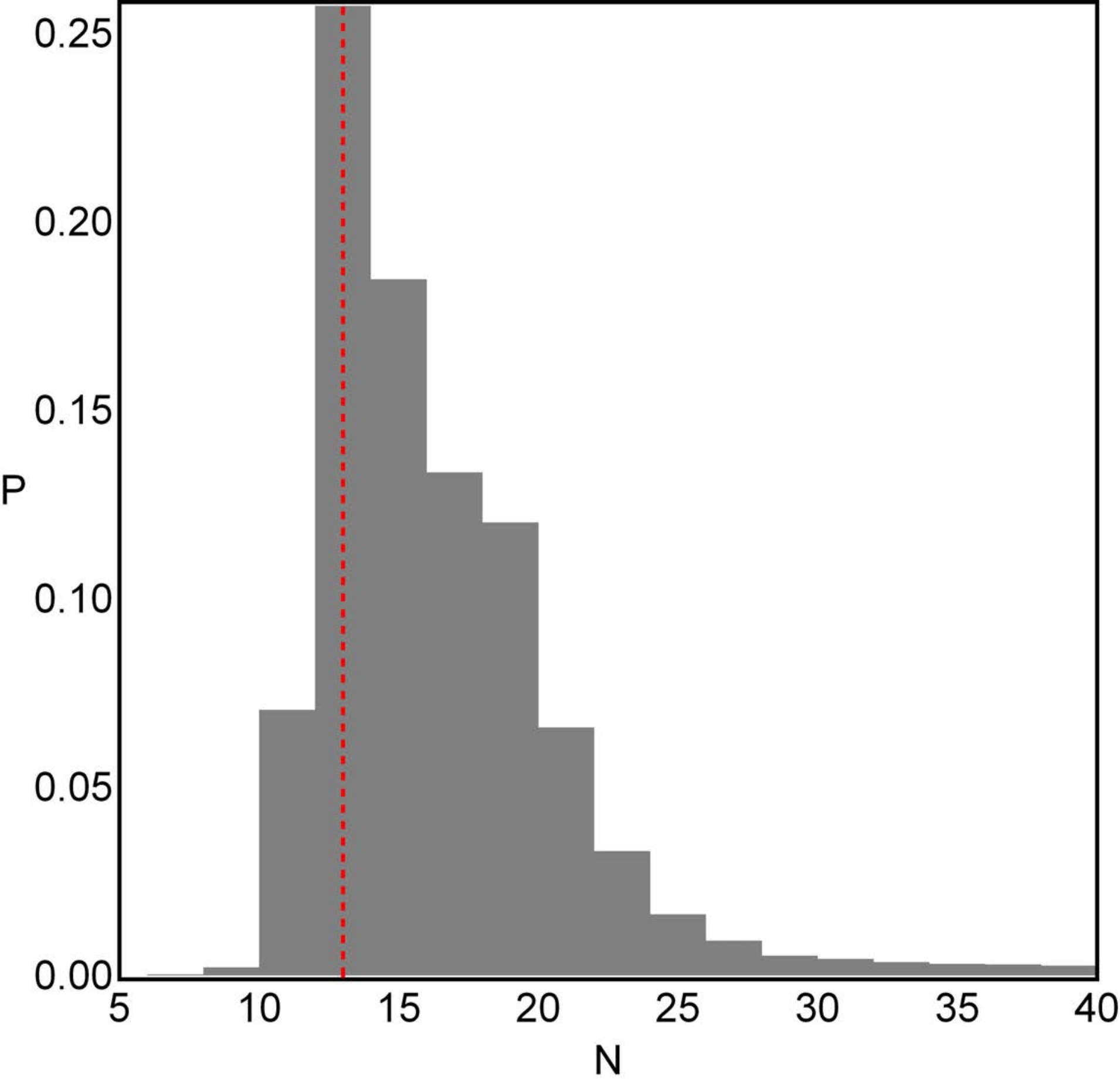}\\
\caption{The Newton-Raphson basins of attraction on the xy-plane for the
case when nine libration points exist for  fixed value of
$\alpha=61 \degree$  and for:
(a) $\beta=(32+\frac{1}{30}) \degree$; (d) $\beta=34\degree$;
(g) $\beta=36\degree$; (j) $\beta=38.5\degree$. The color code for the libration points $L_1$,...,$L_9$ is same as in Fig \ref{NR_Fig_1}; and non-converging points (white);  (b, e,  h, k) and (c, f, i, l) are the distribution of the corresponding number $(N)$ and the  probability distributions of required iterations for obtaining the Newton-Raphson basins of attraction shown in (a, d, g,  j), respectively.
 (Color figure online).}
\label{NR_Fig_C1}
\end{figure*}

In this case, we investigate the Newton-Raphson basins of convergence associated with the libration points for the fixed values of $\alpha=61\degree$ and $\beta$ varies in the interval $[(32+\frac{1}{30})\degree,38.568\degree]$. In Fig. \ref{NR_Fig_C1} ($a, d, g, j$), we have shown the basins of attraction for four different value of $\beta$.  In Fig. \ref{NR_Fig_C1}$(a, d)$,  it is observed that a bug shaped region with many legs occurs which corresponds to the collinear libration point $L_2$. Moreover, the overall shape of the basins of attraction corresponding to the libration points whose domain of convergence look like bugs with multiple butterfly wings. Further, it may be noted that the domain of convergence corresponding to the libration point $L_3$ extend to infinity whereas, the area of all other basins of convergence associated with other libration points is finite. Furthermore, a large portion of the configuration plane is covered by well-formed basins of attraction. The boundaries between the different basins of convergence are highly chaotic which is composed of the different initial conditions. Thus, the choice of a initial point is very sensitive i.e., a slight change in initial conditions leads to entirely different destination and it is extremely difficult to predict in advance.

In Fig. \ref{NR_Fig_C1}$(g, j)$, we have observed a drastic change in the topology of the basins of convergence for the different values of angle parameter $\beta$. In both the panels, it is unveiled that the majority of area of the configuration plane is covered by well formed basins of convergence associated with the libration points. Also, a good amount of the areas of the configuration plane is covered by the region which is highly chaotic in nature. This chaotic region is composed of the various initial conditions which converge to different attractors.

In Fig. \ref{NR_Fig_C1}$(g)$, the basin of convergence associated with the collinear libration point $L_2$ is infinite while in Fig. \ref{NR_Fig_C1}$(j)$, the infinite extent of the basin of convergence is associated with the collinear libration point $L_1$.

Moreover, only the vicinity of boundaries of these basins of convergence exhibit the highly fractal structure which behave alike "chaotic sea". This implies that in these areas it is almost impossible to judge which of the libration point, each initial condition attracted by.

From these two panels, we may observe a very interesting phenomenon related to the basins of convergence. The overall topology of the basins of convergence looks reversed mirror image of each other, however their scale are different. Thus, it can be stated that if we increase the value of the angle parameter $\beta$, the structure  of the configuration plane changes drastically, i.e., it does not follow the specific pattern.

 In Fig. \ref{NR_Fig_C1}$(b, e, h, k)$, using tone of blue, we have shown the distribution of the corresponding number $(N)$ of iterations to obtain the desired accuracy. It is also observed that initial conditions which lie in the domain of basins of convergence have faster rate of convergence when $N<20$ whereas for $N>20$, the rate of convergence is slower. In panel (k), it is observed that required number of iterations is very less i.e., $N\leq 5$, for those initial conditions which lie in the basins of convergence corresponding to the libration points $L_{1, 2, 3, 8, 9}$ whereas the initial conditions falling inside the area which is chaotic mixture os initial conditions, the required number of iterations increases i.e., $N>35$. Similarly, in Fig. \ref{NR_Fig_C1} panels $(c, f, i, l)$, the corresponding probability distribution of iterations is shown. If we increase the value of angle parameter $\beta$, the most probable number $N^*$ of iterations is reduced from $ 25$ to $11$ when $\beta$ increases from $(32+\frac{1}{30})\degree$ to $36\degree$ whereas, for $\beta=38.5\degree$, the number $N^*$ slightly increases.

\begin{figure*}[!t]
\centering
(a)\includegraphics[scale=.27]{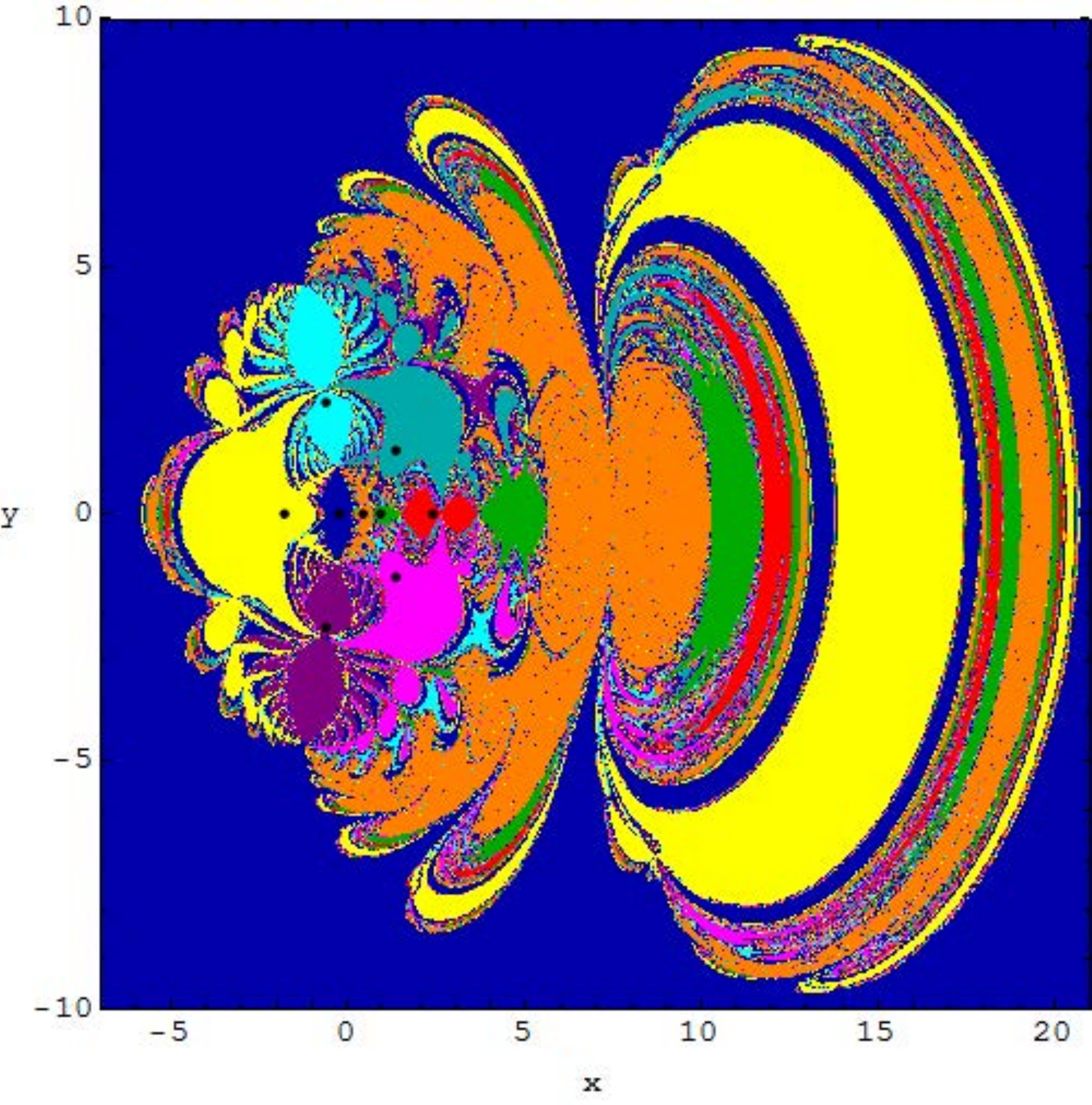}%Basin_ 44_403
(b)\includegraphics[scale=.27]{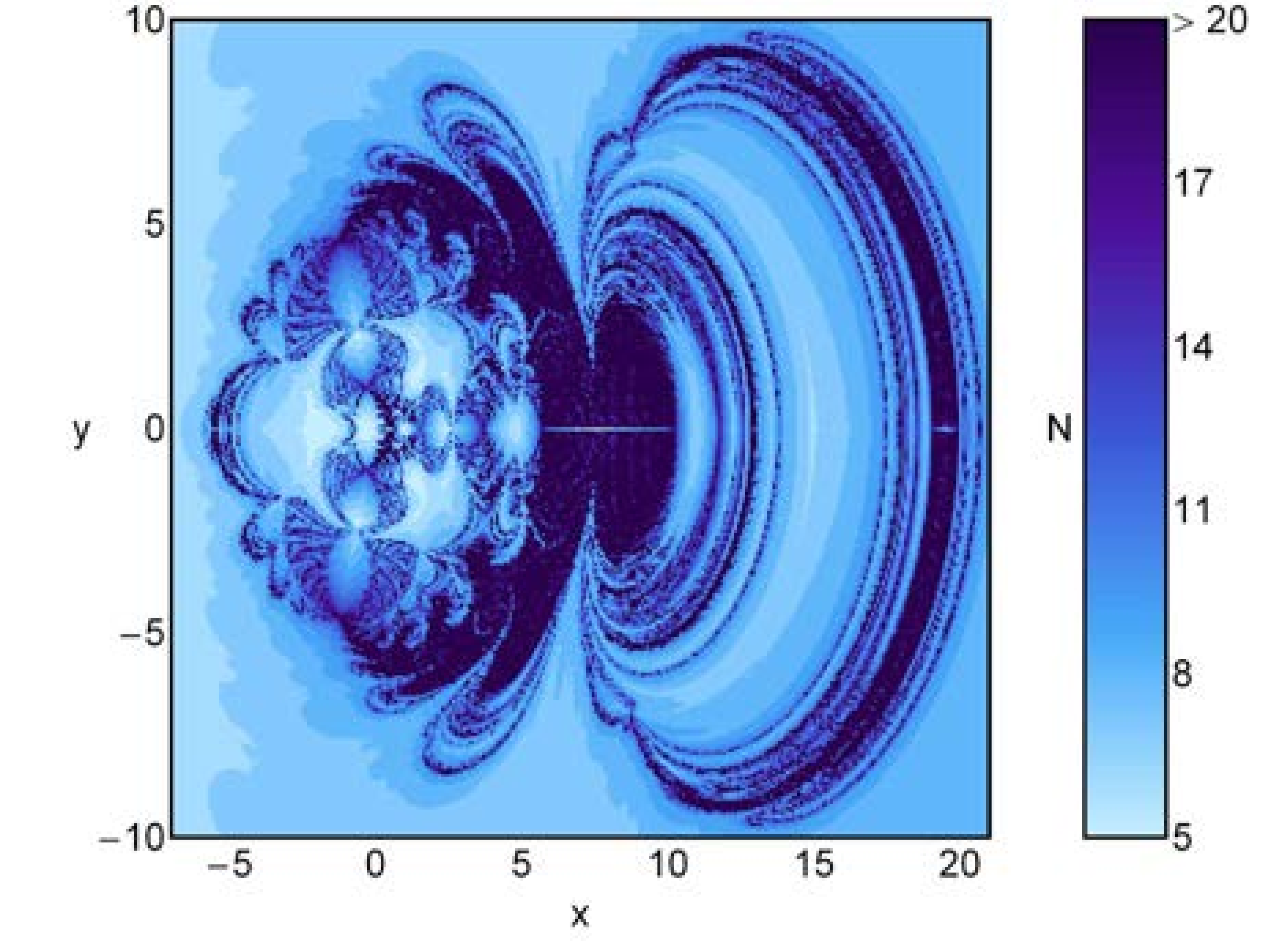}
(c)\includegraphics[scale=.25]{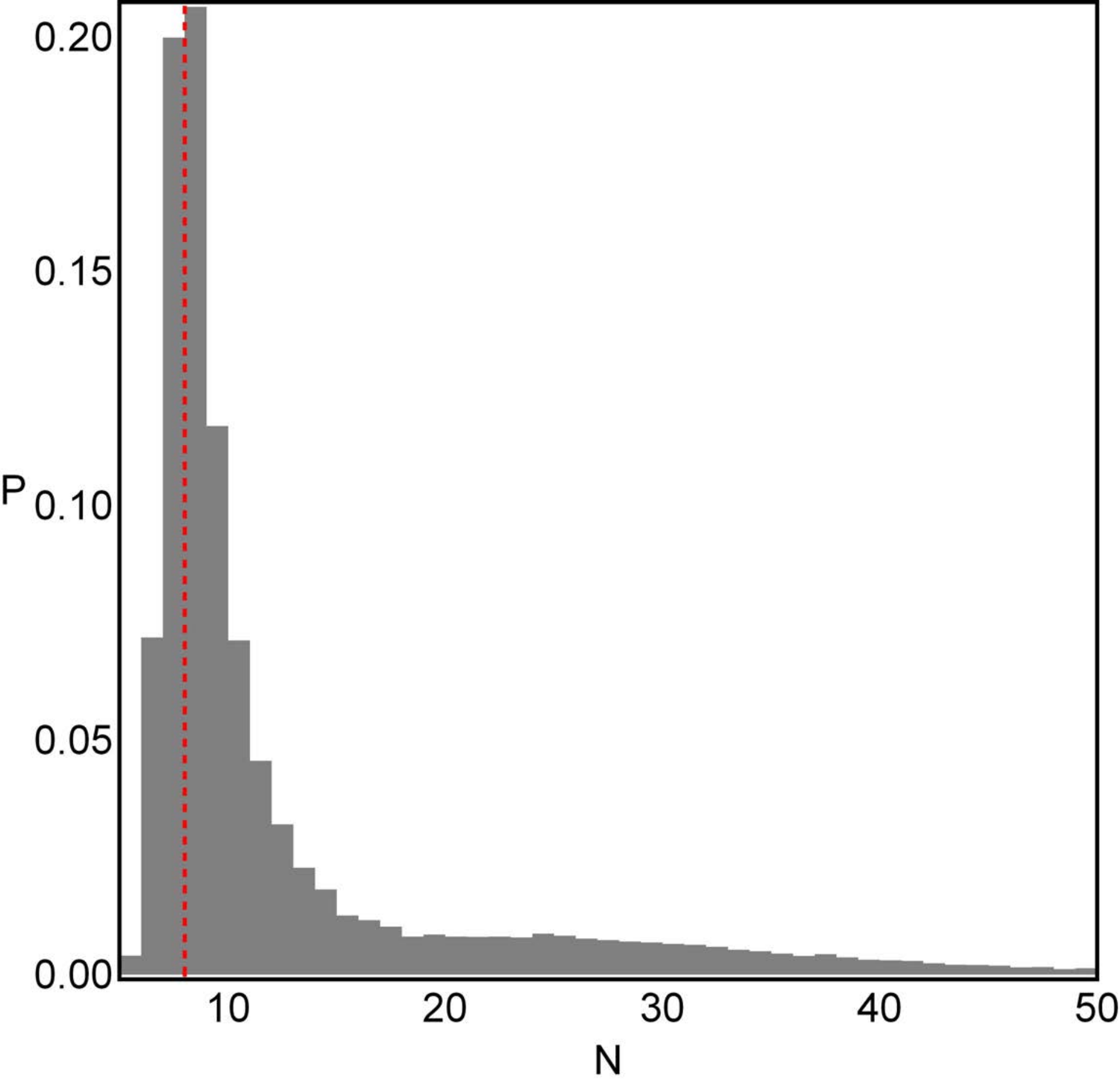}\\
(d)\includegraphics[scale=.27]{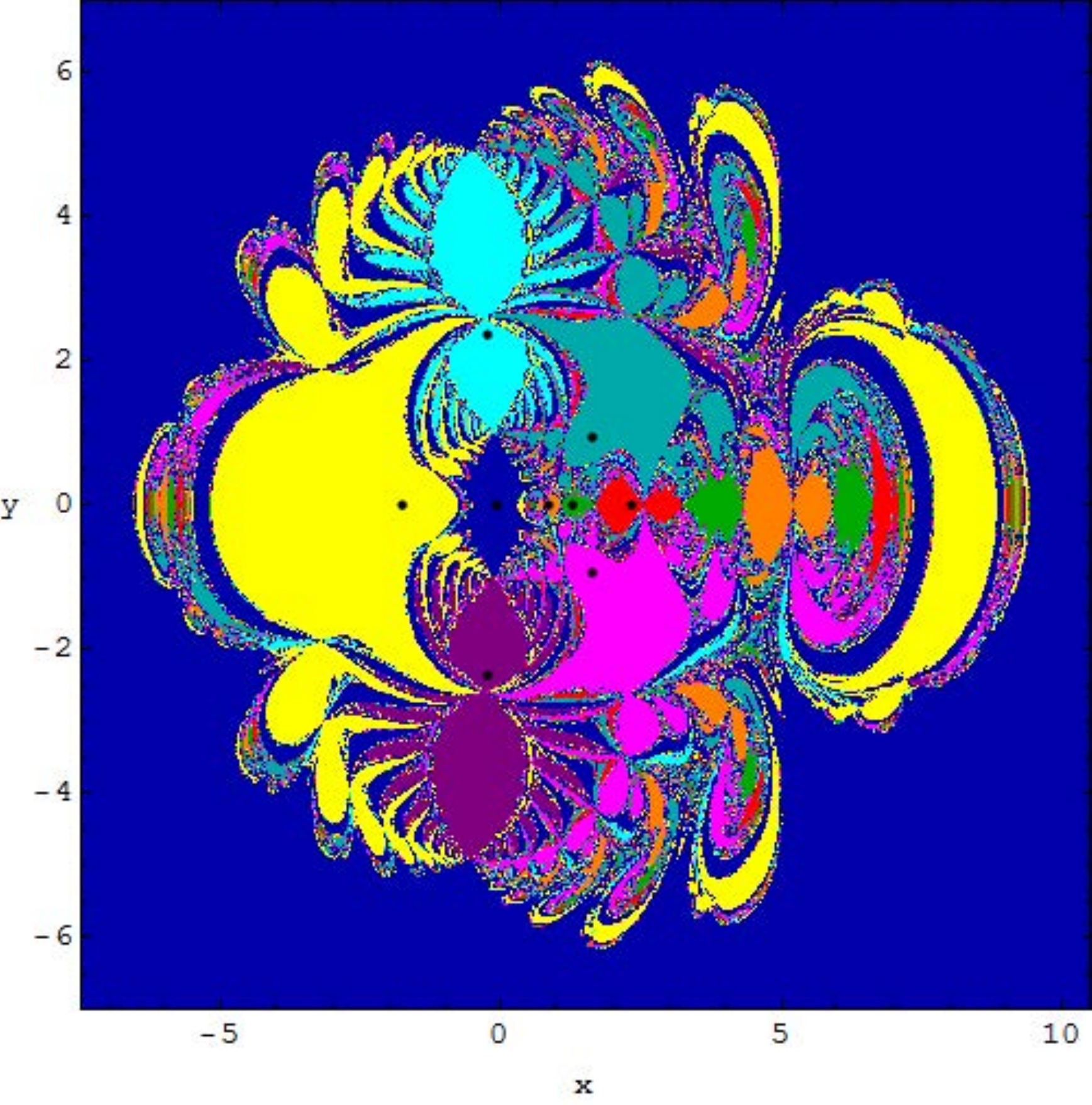}%Fig_61_50
(e)\includegraphics[scale=.27]{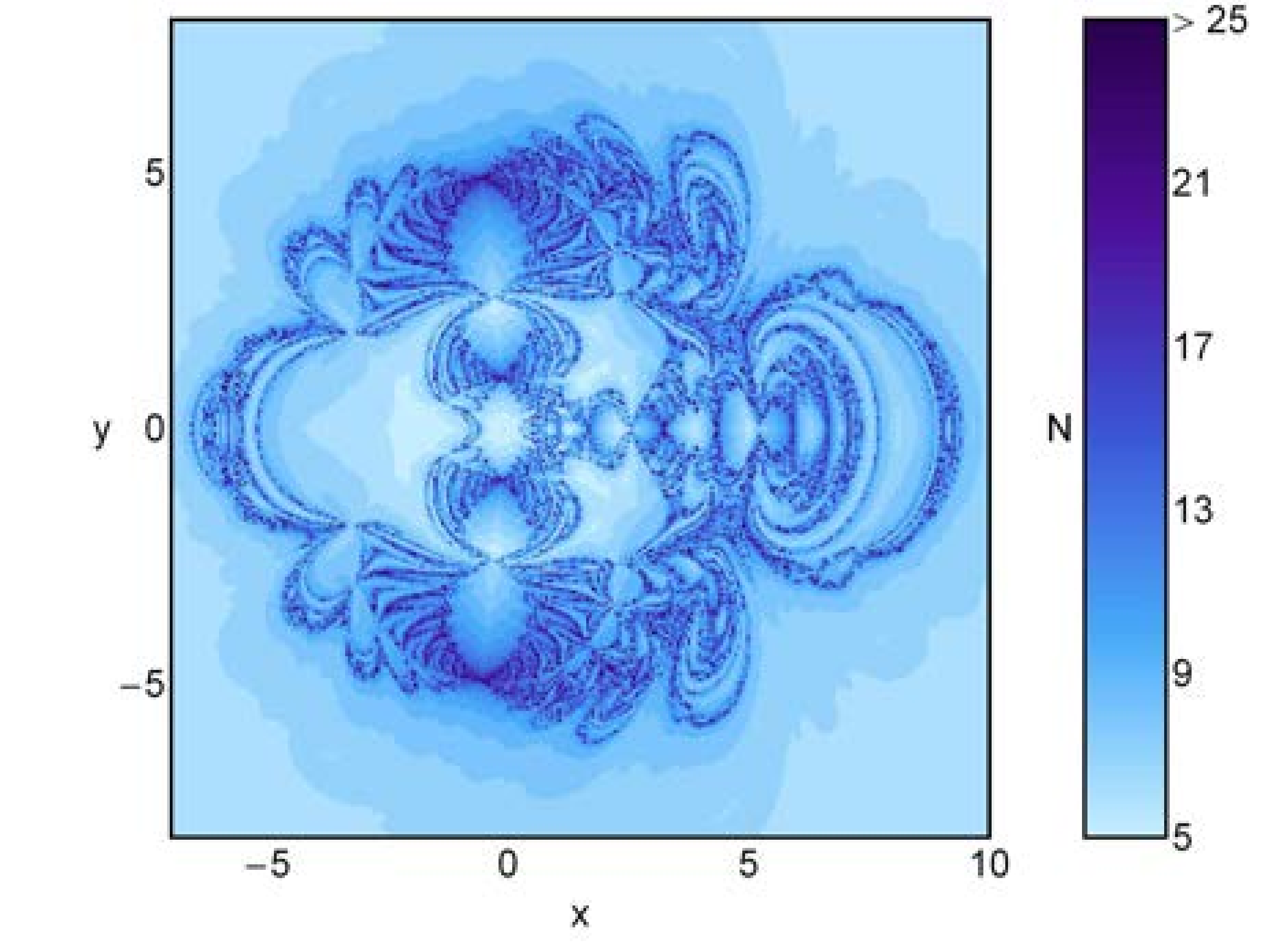}
(f)\includegraphics[scale=.25]{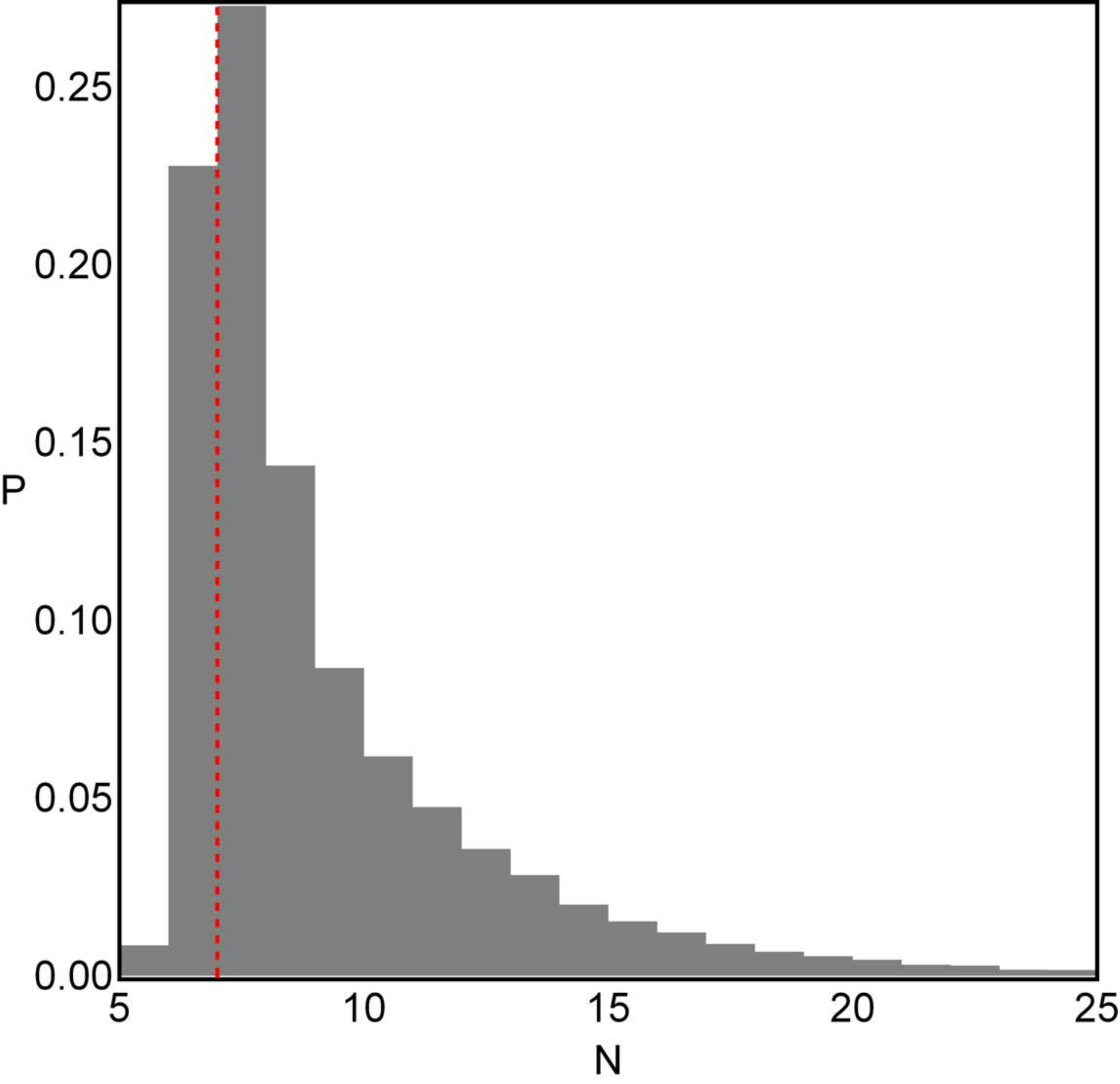}\\
(g)\includegraphics[scale=.27]{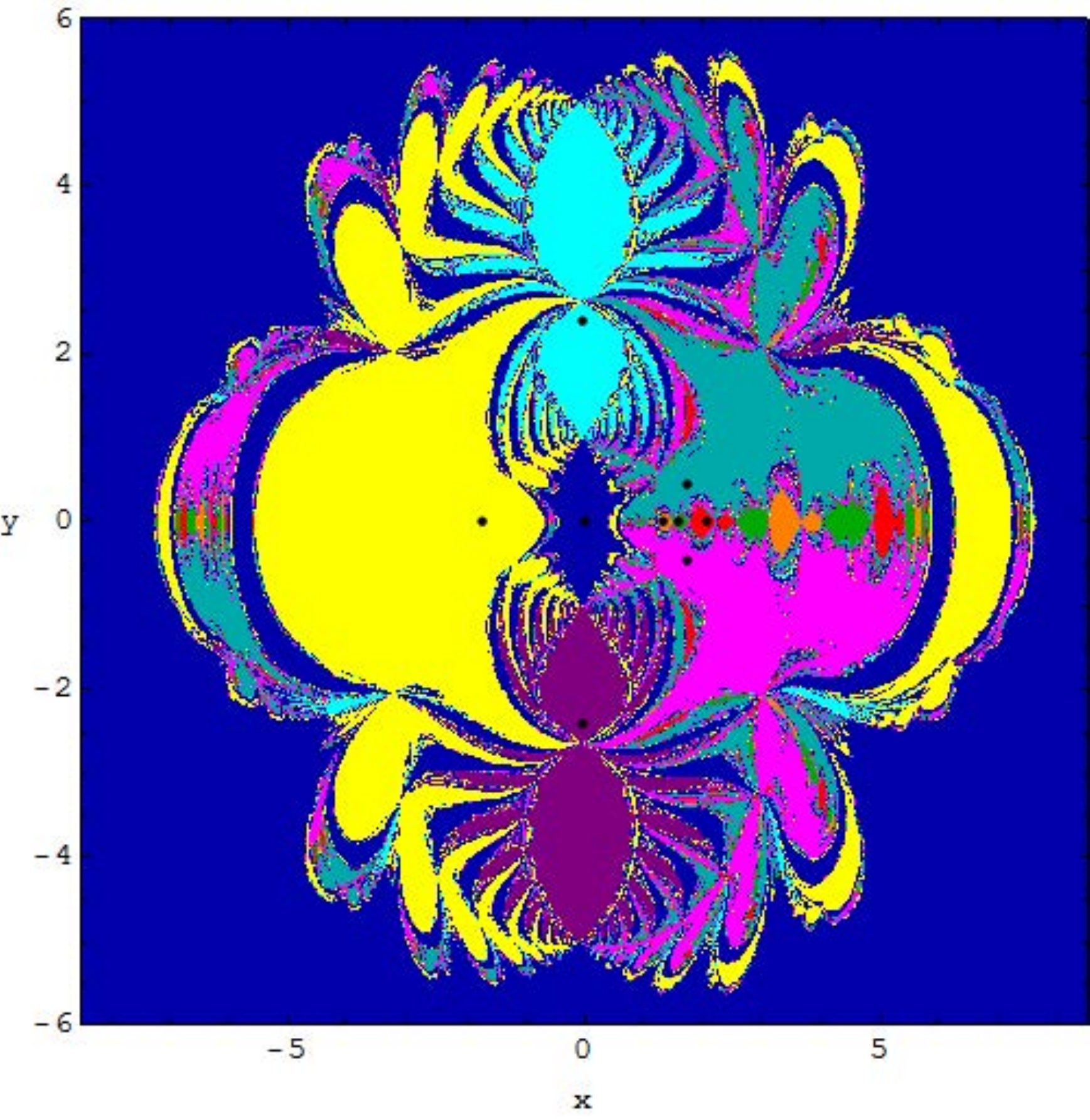}%Fig_61_56
(h)\includegraphics[scale=.27]{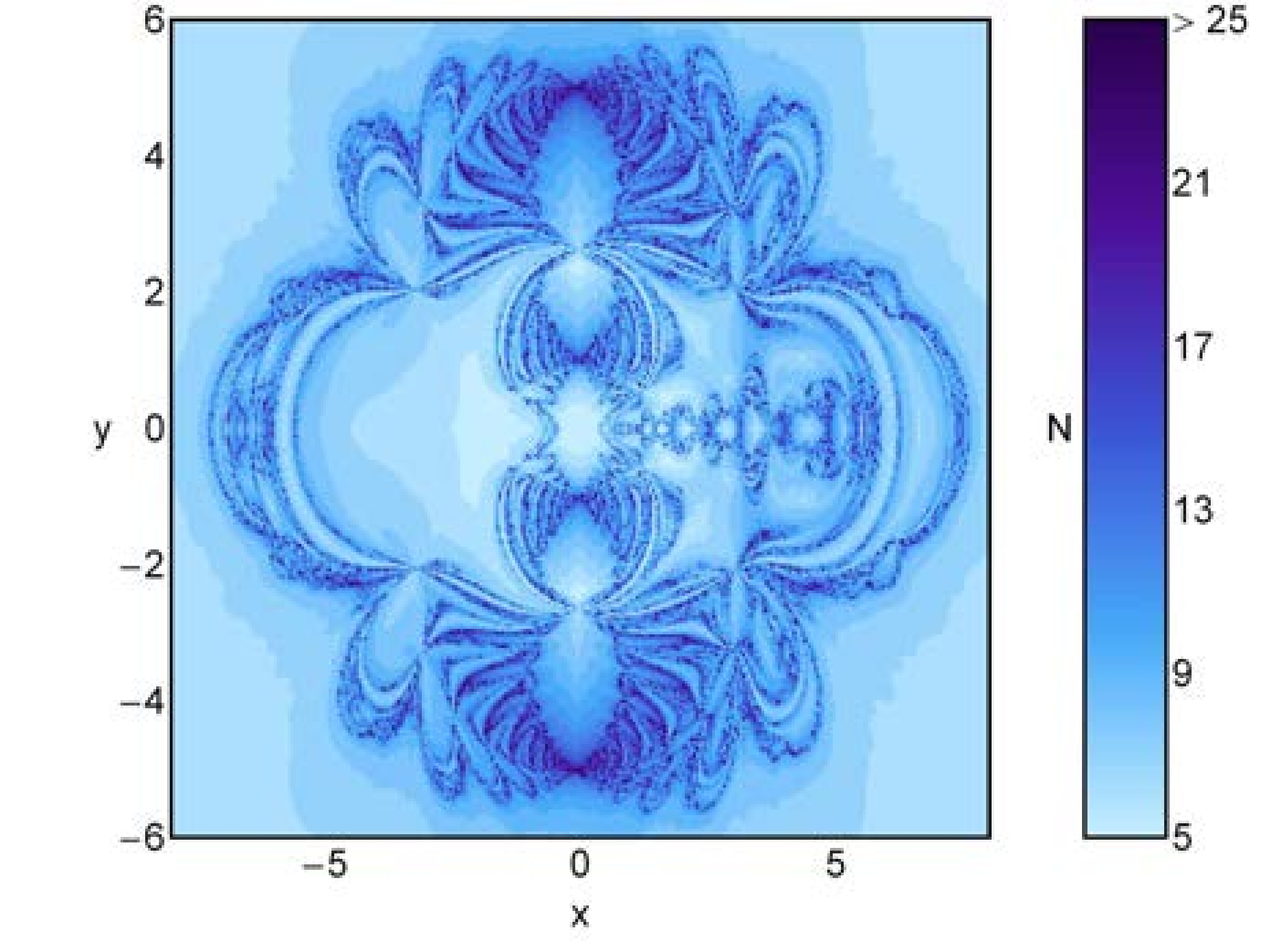}
(i)\includegraphics[scale=.25]{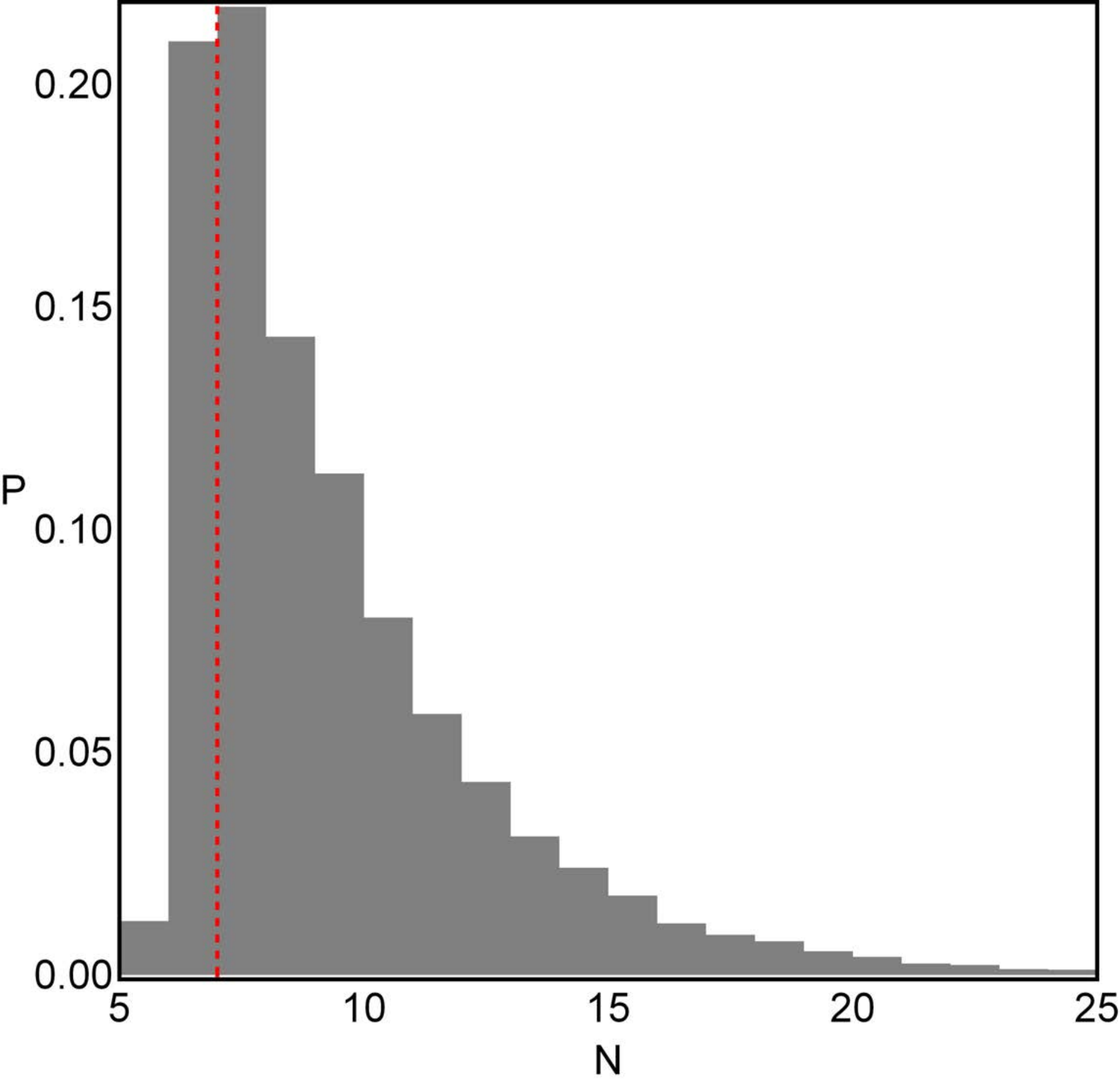}\\
(j)\includegraphics[scale=.27]{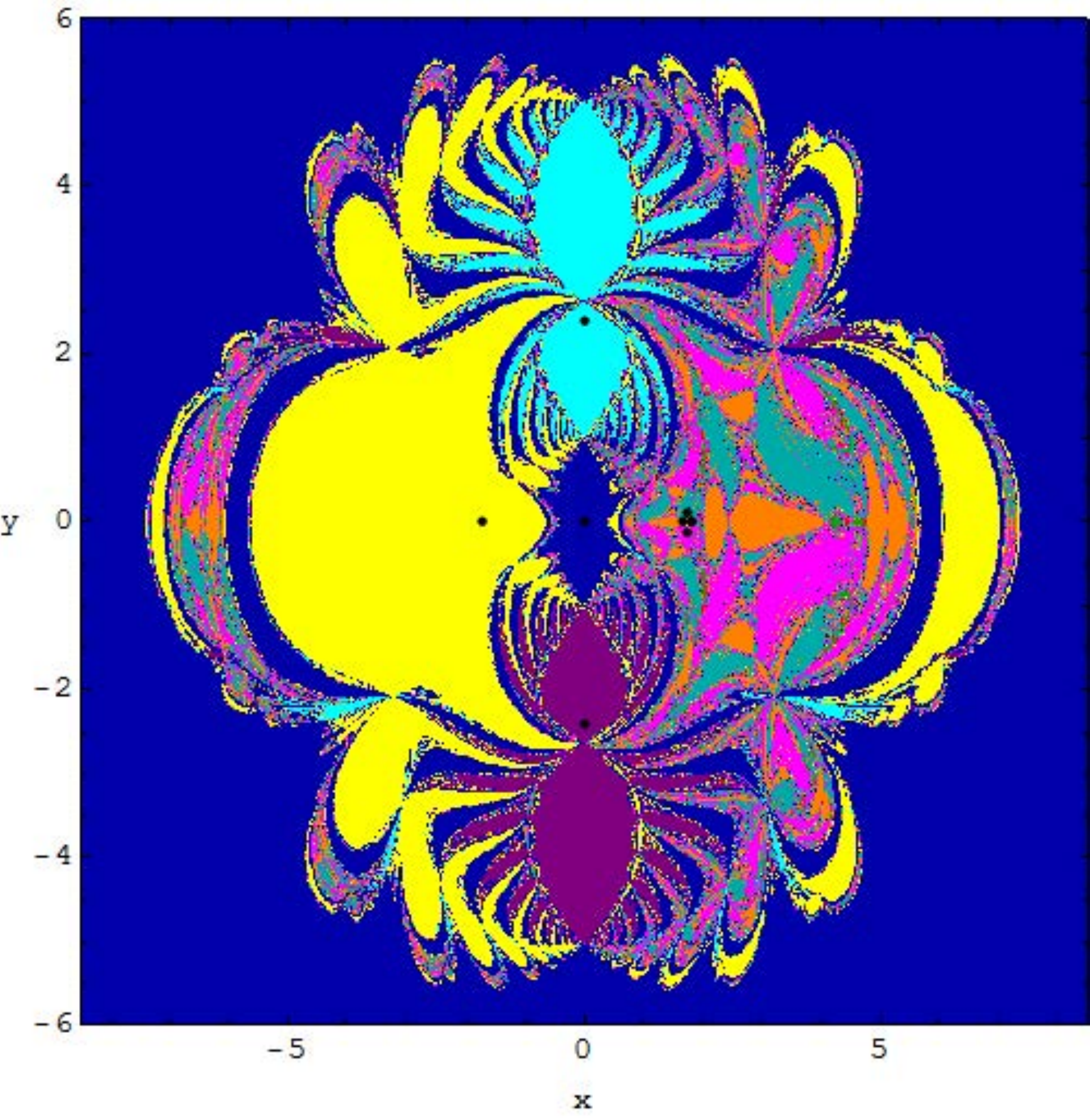}%Fig_61_60-1/30
(k)\includegraphics[scale=.27]{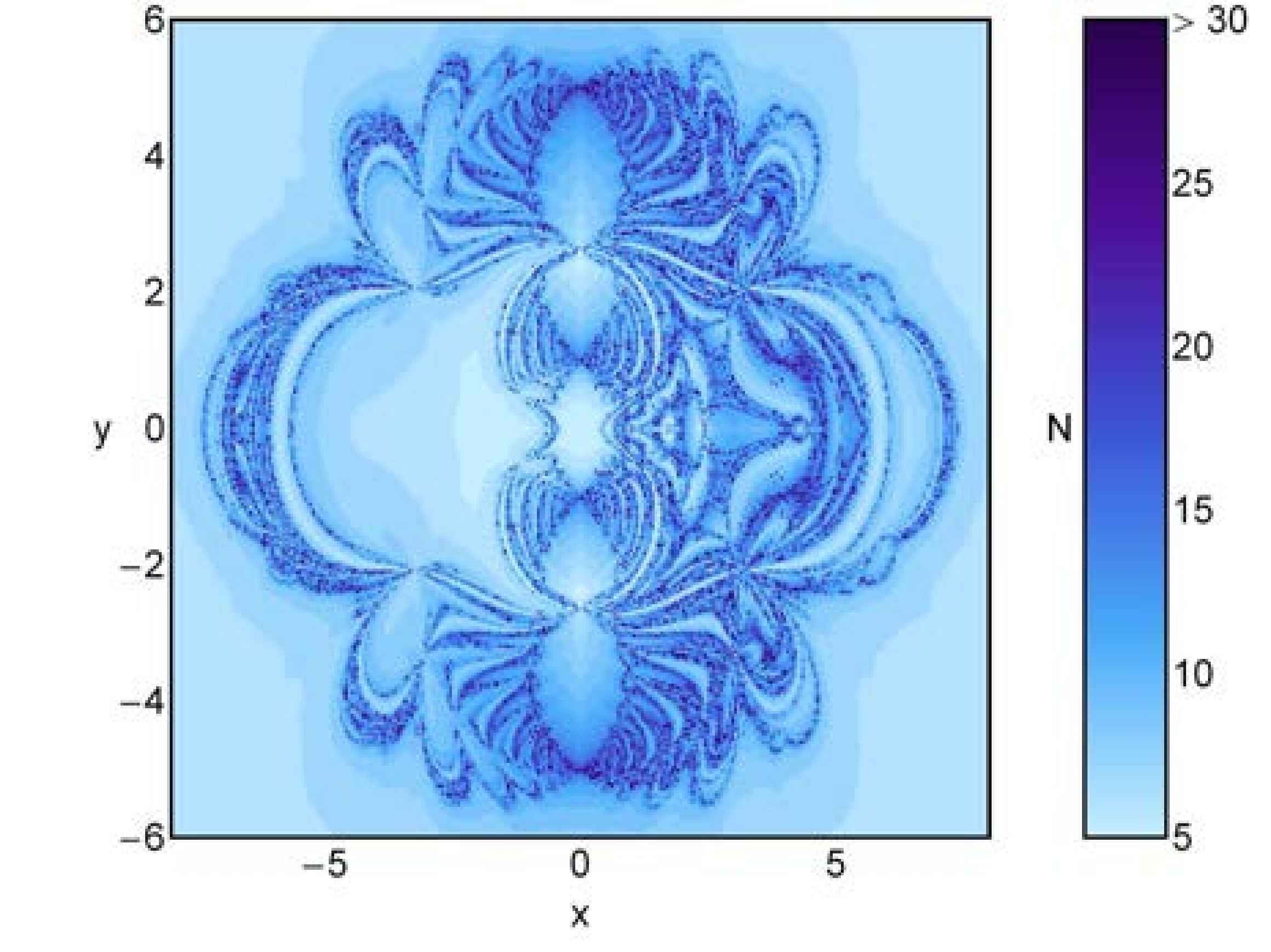}
(l)\includegraphics[scale=.25]{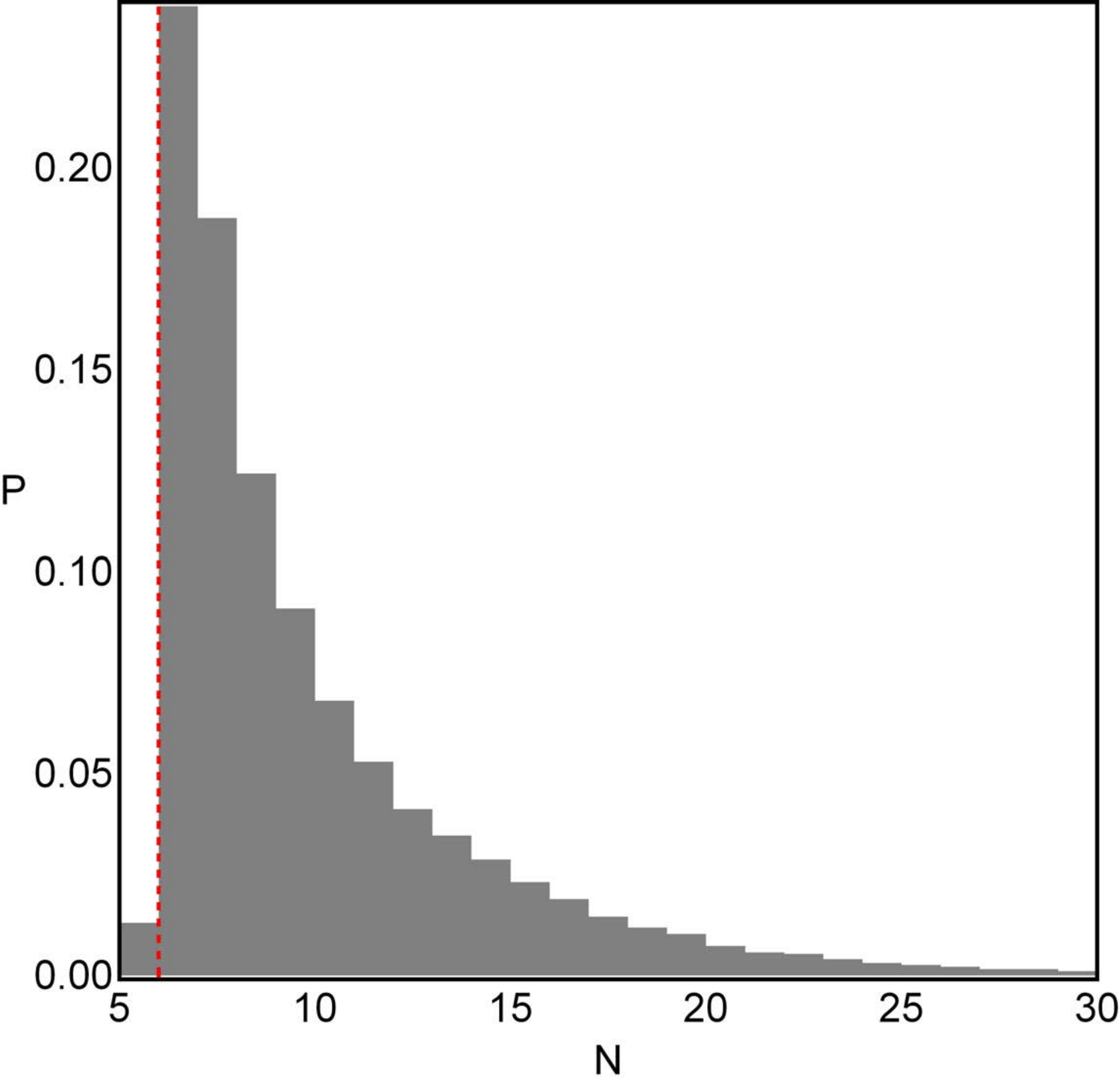}\\
\caption{The Newton-Raphson basins of attraction on the xy-plane for the
case when nine libration points exist for  fixed value of
$\alpha=61 \degree$  and for:
(a) $\beta=44.403\degree$; (d) $\beta=50\degree$;
(g) $\beta=56\degree$; (j) $\beta=(60-\frac{1}{30})\degree$.The color code for the libration points $L_1$,...,$L_9$ is same as in Fig \ref{NR_Fig_1}; and non-converging points (white);  (b, e,  h, k) and (c, f, i, l) are the distribution of the corresponding number $(N)$ and the  probability distributions of required iterations for obtaining the Newton-Raphson basins of attraction shown in (a, d, g, j), respectively; and non-converging points (white).
 (Color figure online).}
\label{NR_Fig_C2}
\end{figure*}

 In Fig. \ref{NR_Fig_C2} $(a, d, g, j)$, we have illustrated the Newton-Raphson basins of convergence for four different values of the angle parameter $\beta$. In this case, the changes in the configuration plane due to the variation of angle parameter in last three panels are not so prominent as it was in the previous case still it can be observed that the domain of  basins of convergence corresponding to the libration points $L_{6,7}$ increases as $\beta$ increases. The domain of convergence corresponding to the libration point $L_9$ is infinite whereas for all other libration points it is finite. The exotic bugs with many legs and antennas like shape of the basin of attraction corresponding to libration points $L_6$ and $L_7$ increase continuously as $\beta$ increases. The domain of the basins of convergence corresponding to the libration point $L_3$  which lies in the plane $x<0$ increases while it decreases for the plane $x>0$ as $\beta$ increases.

 In Fig. \ref{NR_Fig_C2} $(e, h, k)$, the corresponding number $(N)$ of required iterations for desired accuracy is shown. It is noticed that more than $95\%$ of the initial conditions on the configuration plane of this iterative scheme requires not more than $20$ iterations to obtain the desire accuracy whereas for the panel (b), the required number of iterations to converge the libration point $L_8$ is greater than $20$. The probability distribution of iterations is shown in Fig. \ref{NR_Fig_C2} $(c, f, i, l)$. The value of required number $(N)$ of iterations decreases as the value of angle parameter $\beta$ increases. It was measured $N^*=6$ as the least value calculated for $\beta=(60-\frac{1}{30})\degree$.

\begin{figure*}[!t]
\centering
(a)\includegraphics[scale=.27]{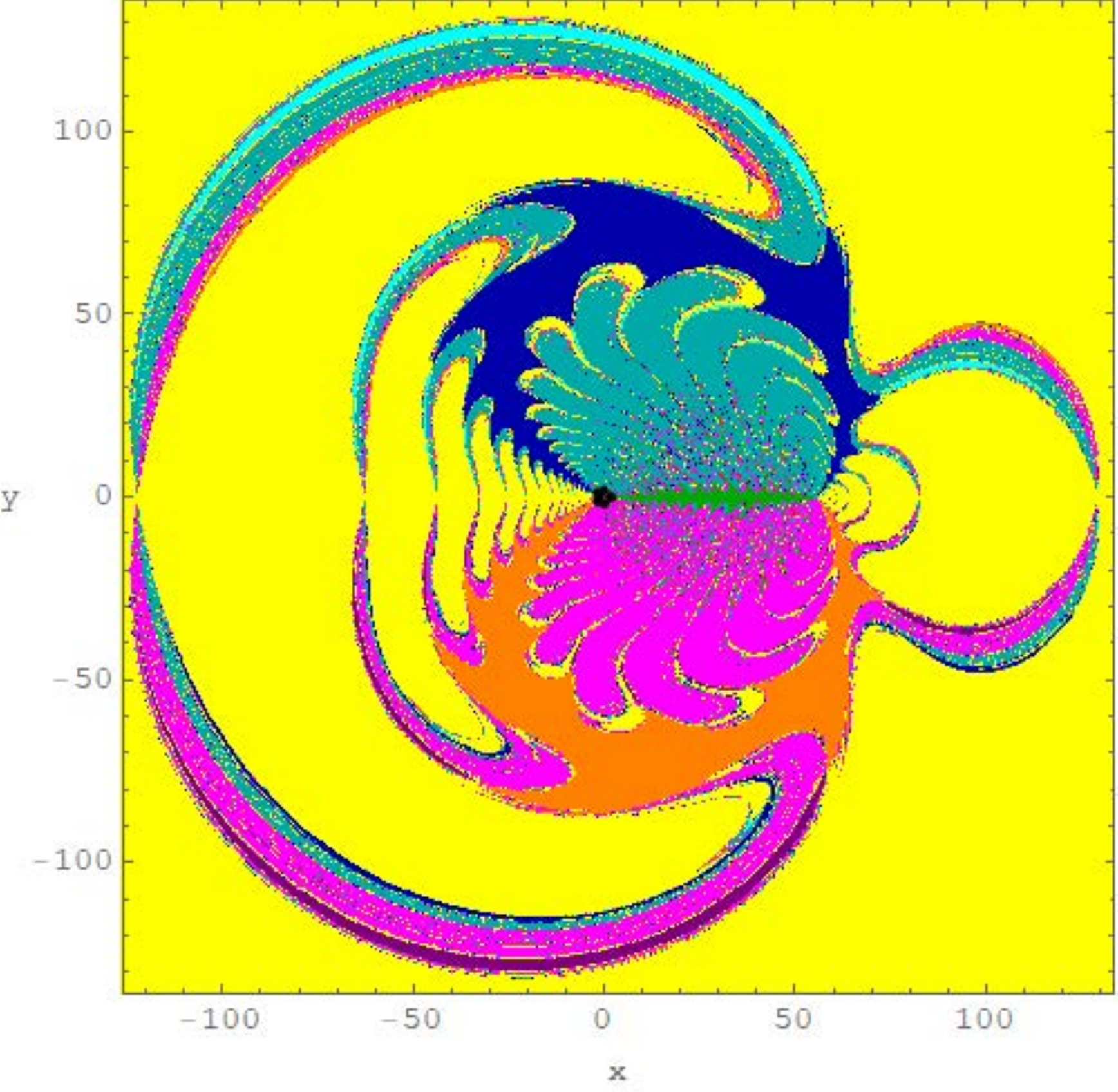}%Fig_73_56+1/30
(b)\includegraphics[scale=.27]{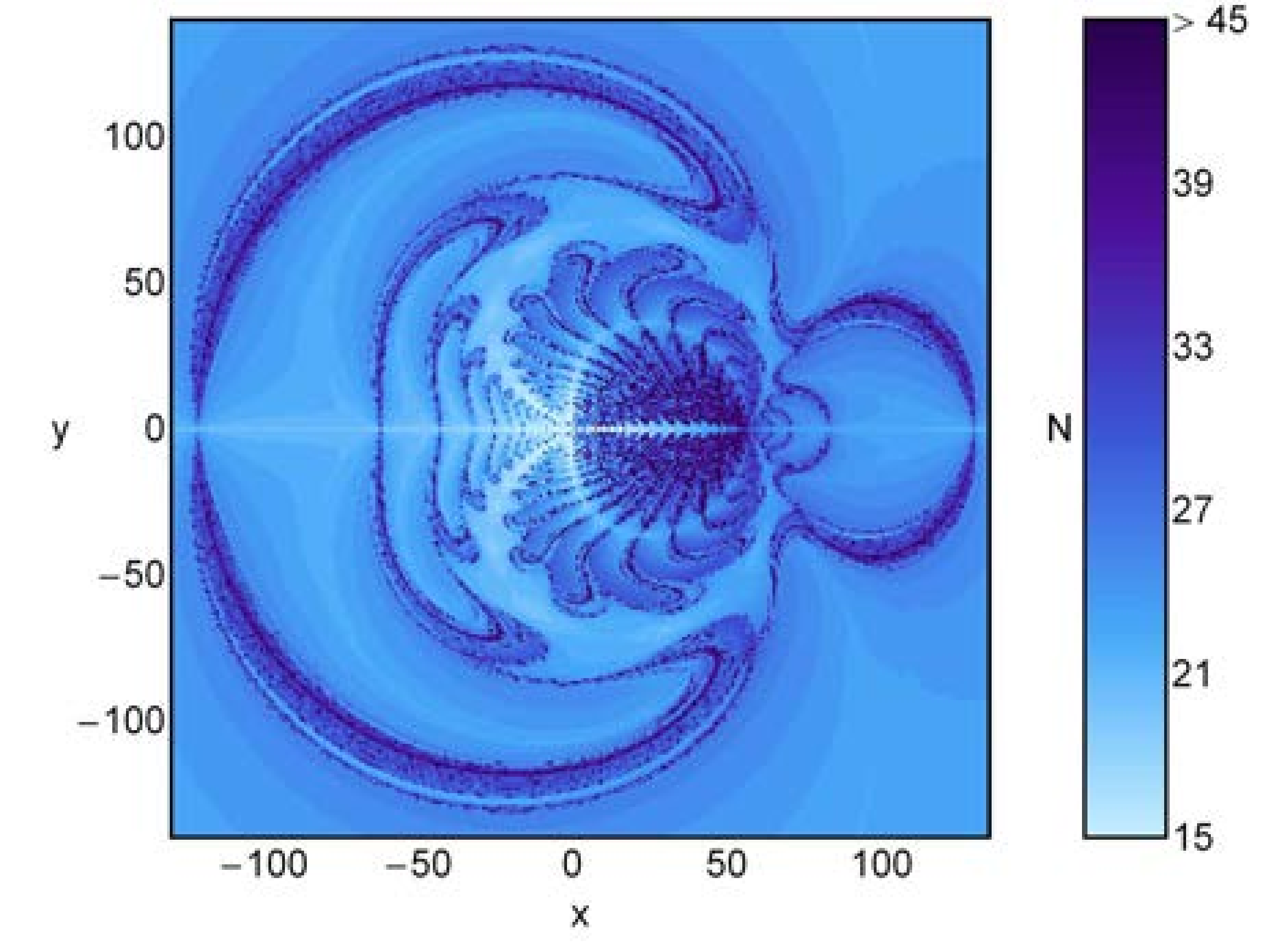}
(c)\includegraphics[scale=.25]{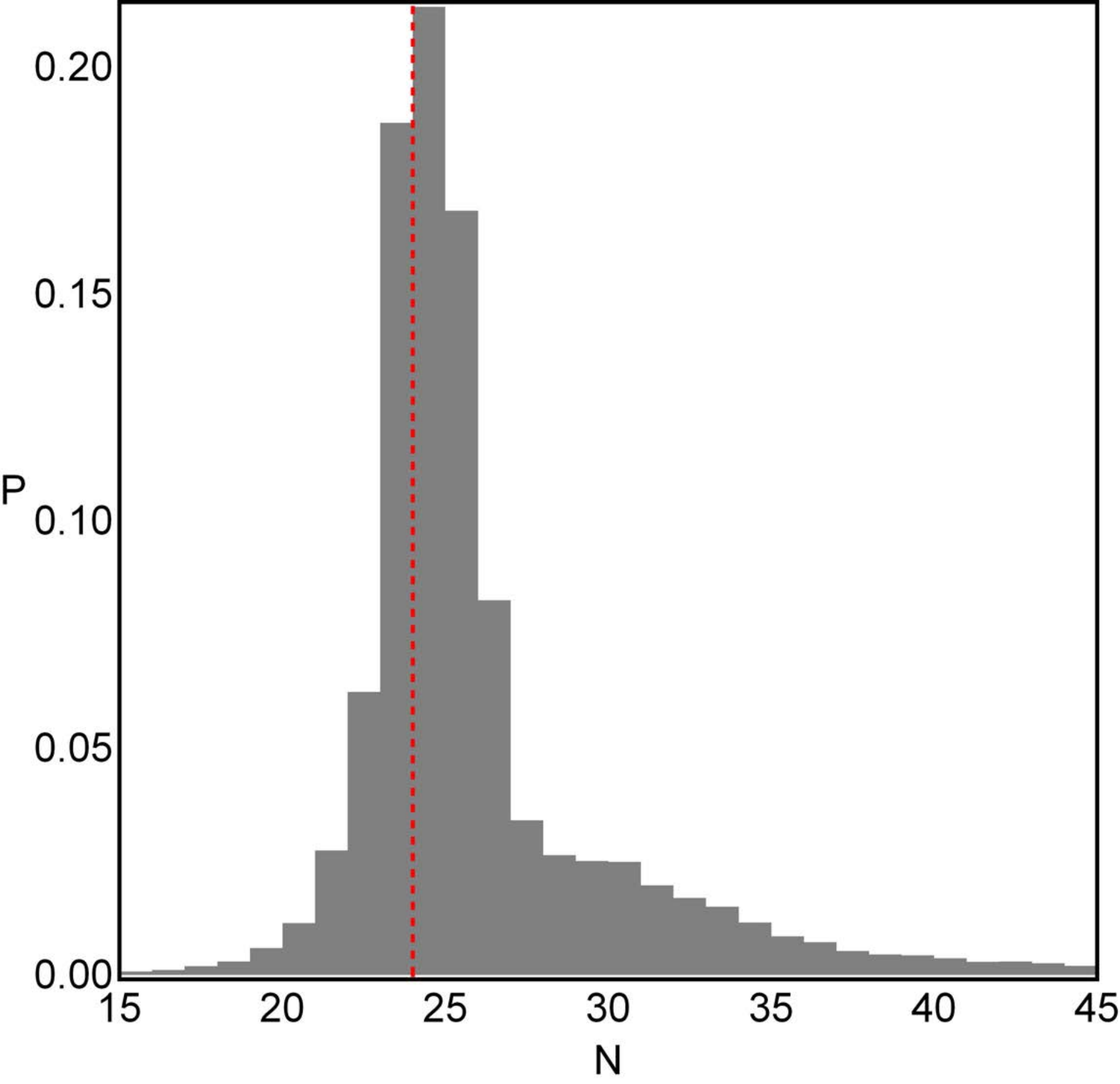}\\
(d)\includegraphics[scale=.27]{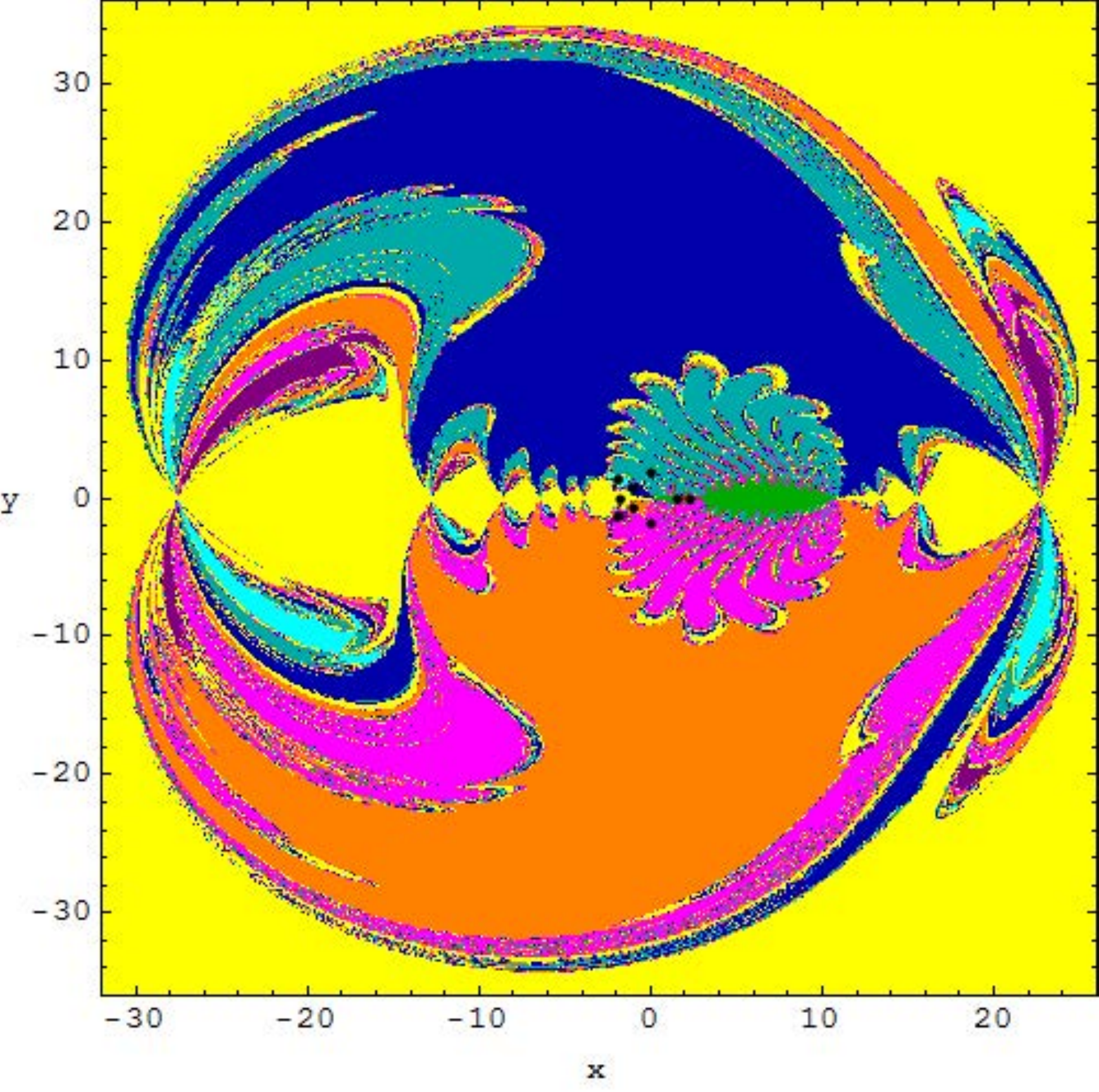}%Fig_73_57
(e)\includegraphics[scale=.27]{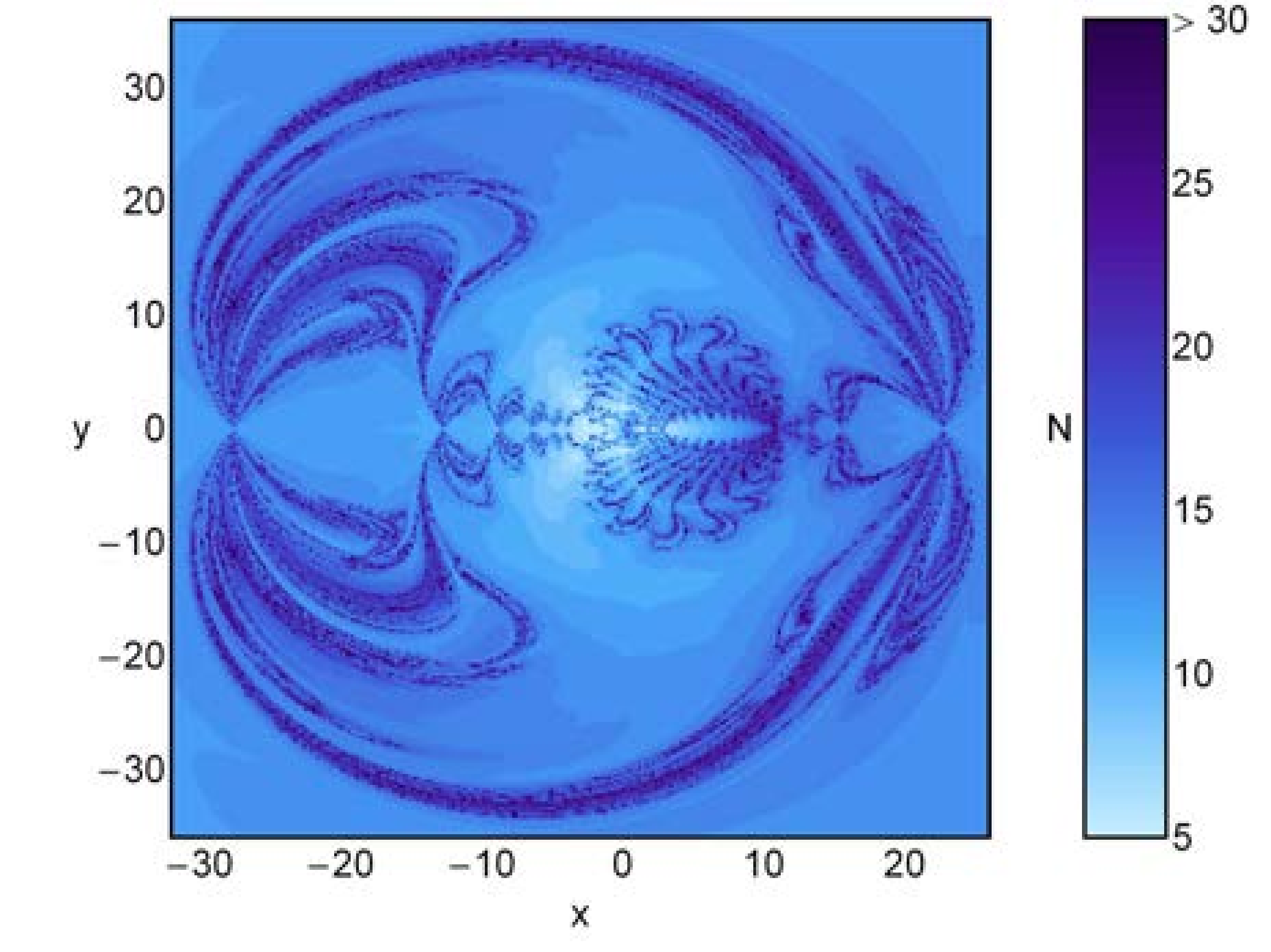}
(f)\includegraphics[scale=.25]{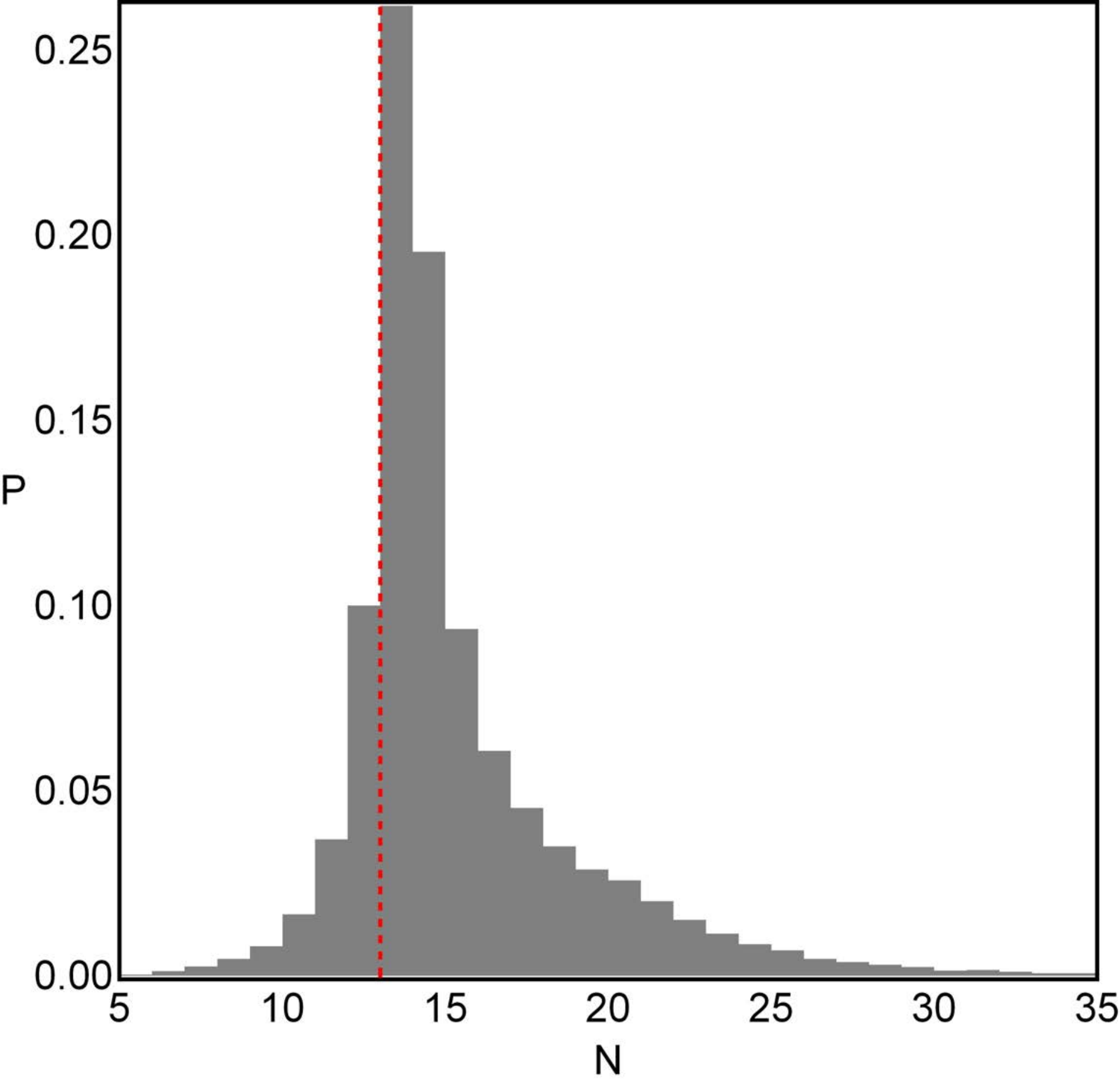}\\
(g)\includegraphics[scale=.27]{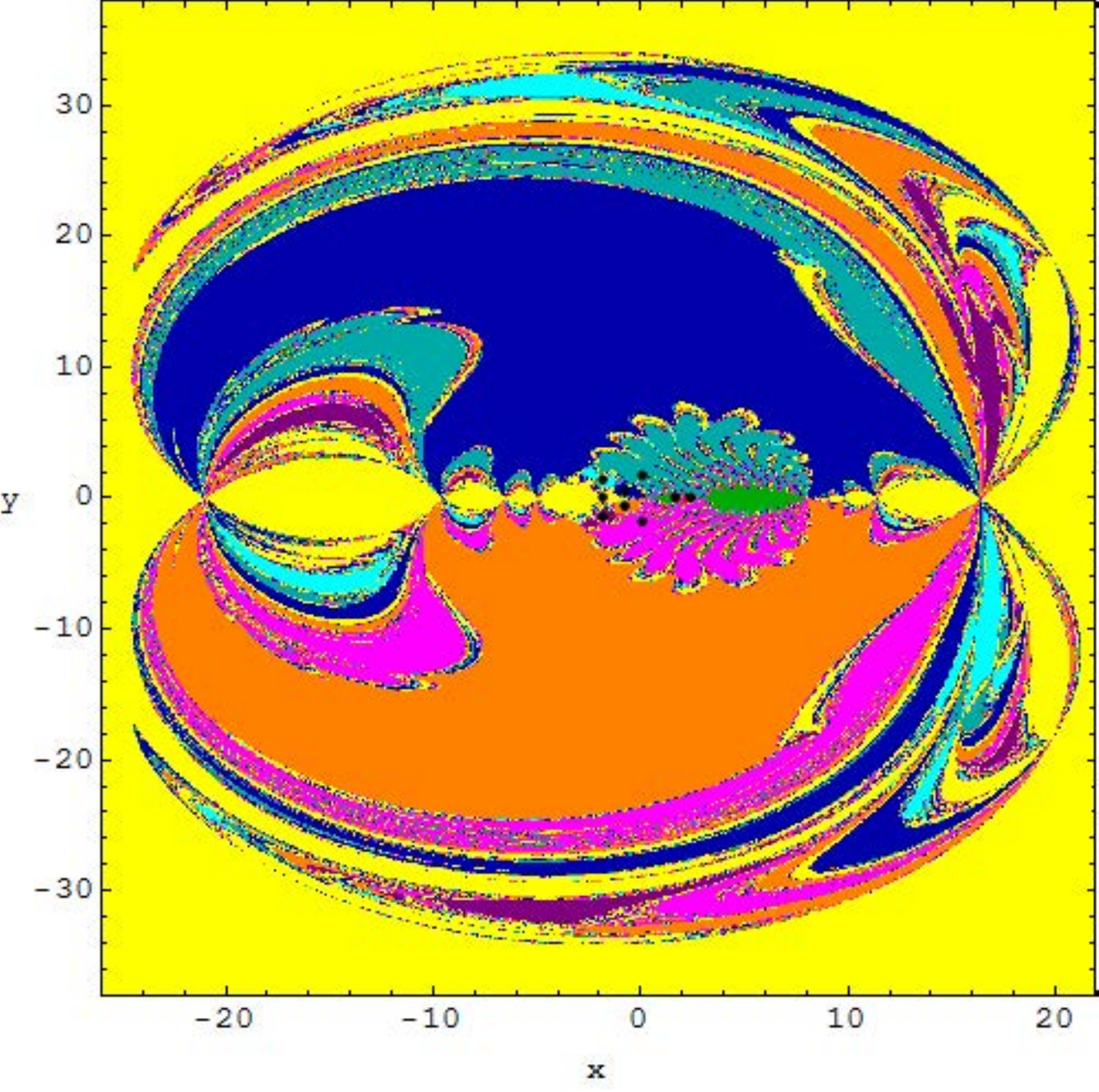}%Fig_73_58
(h)\includegraphics[scale=.27]{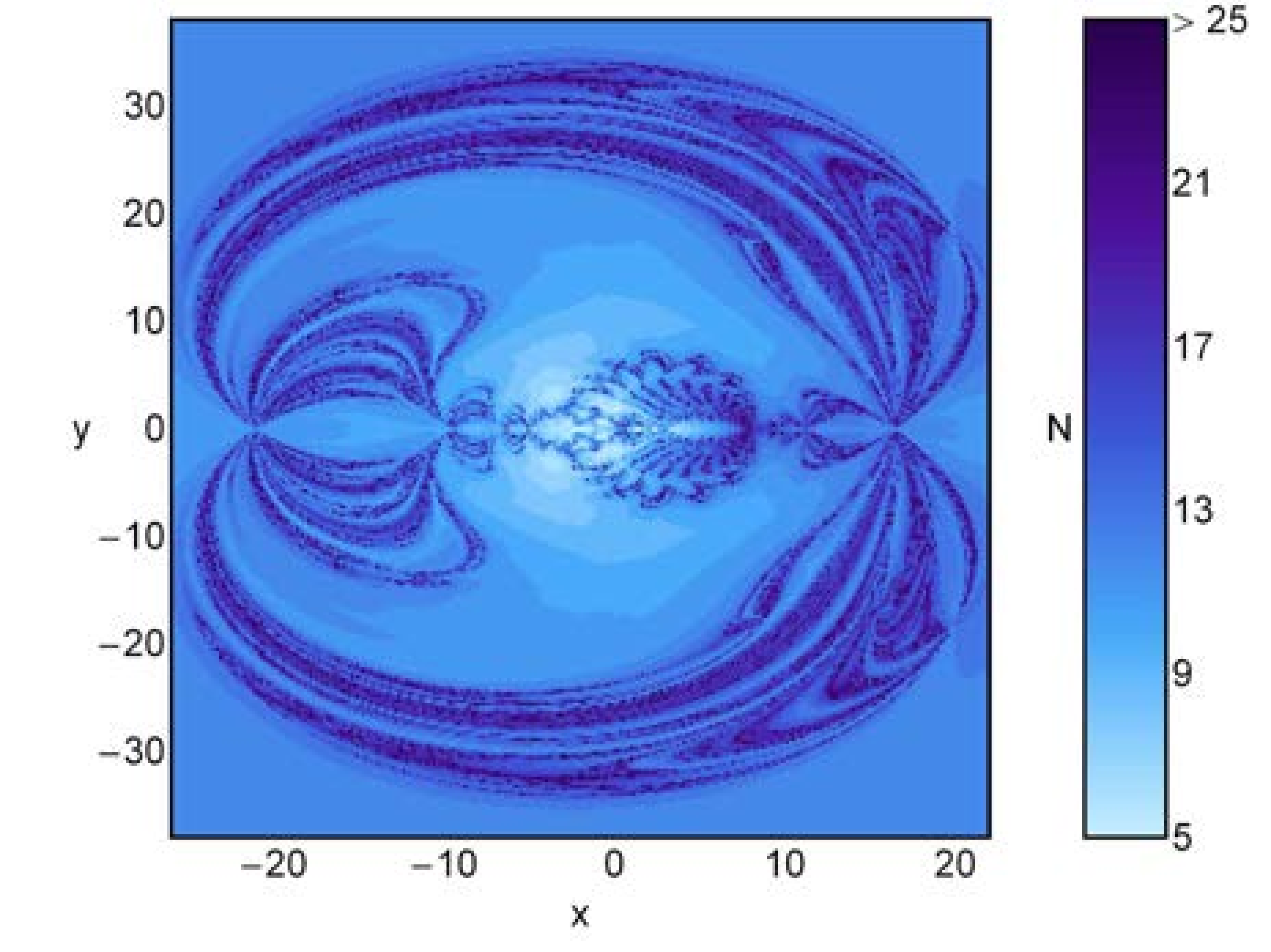}
(i)\includegraphics[scale=.25]{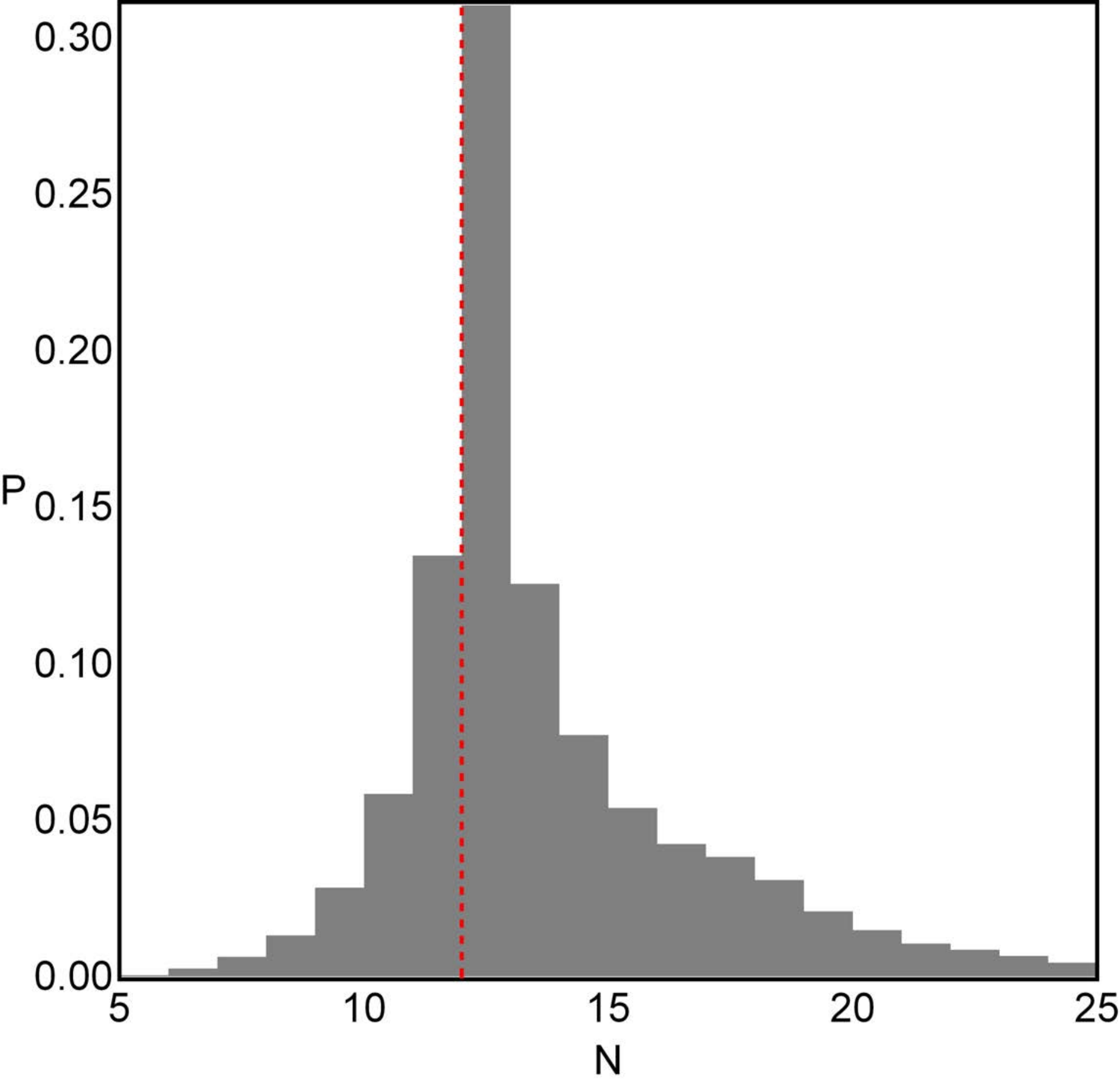}\\
(j)\includegraphics[scale=.27]{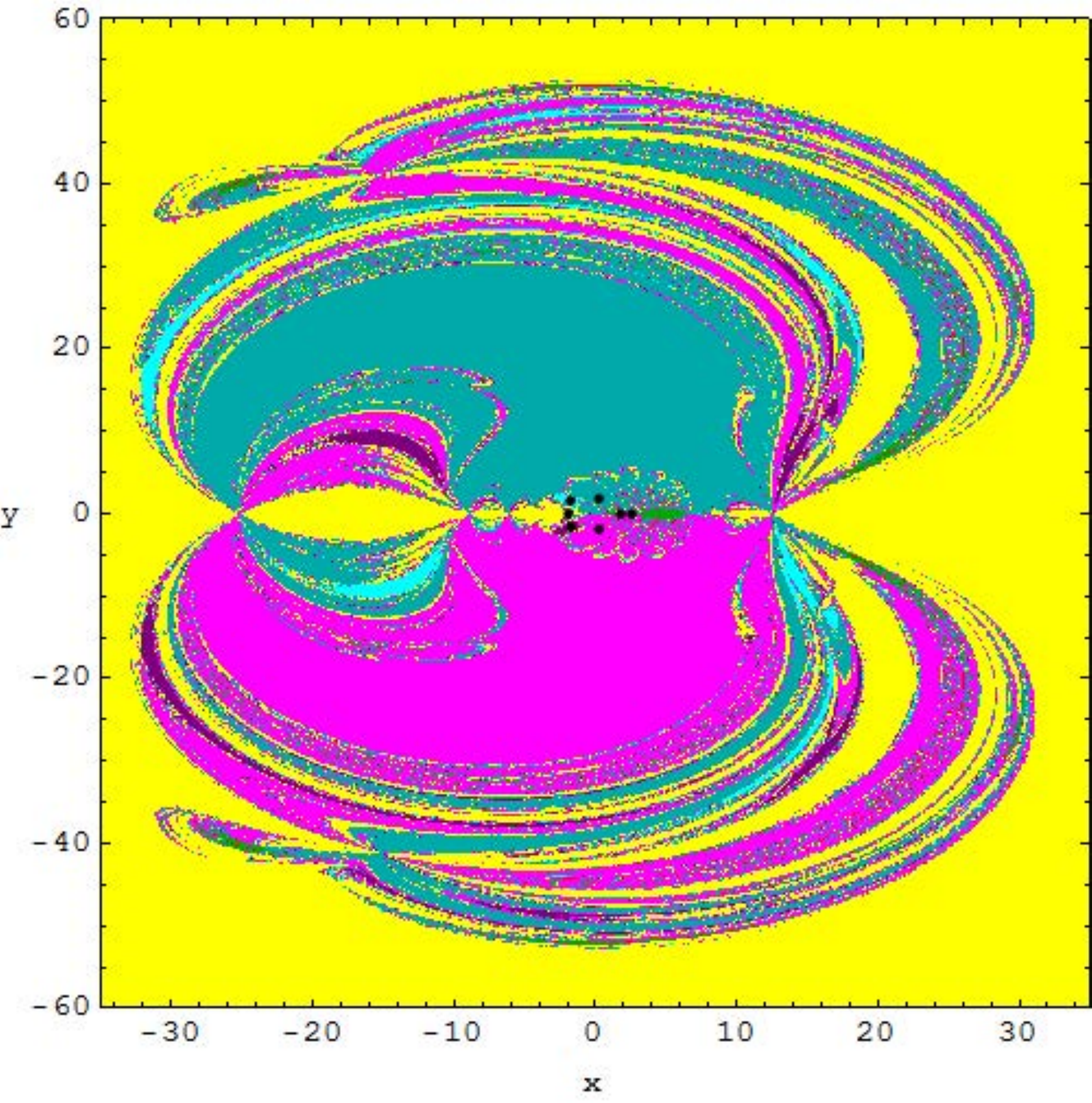}%Fig_73_59
(k)\includegraphics[scale=.27]{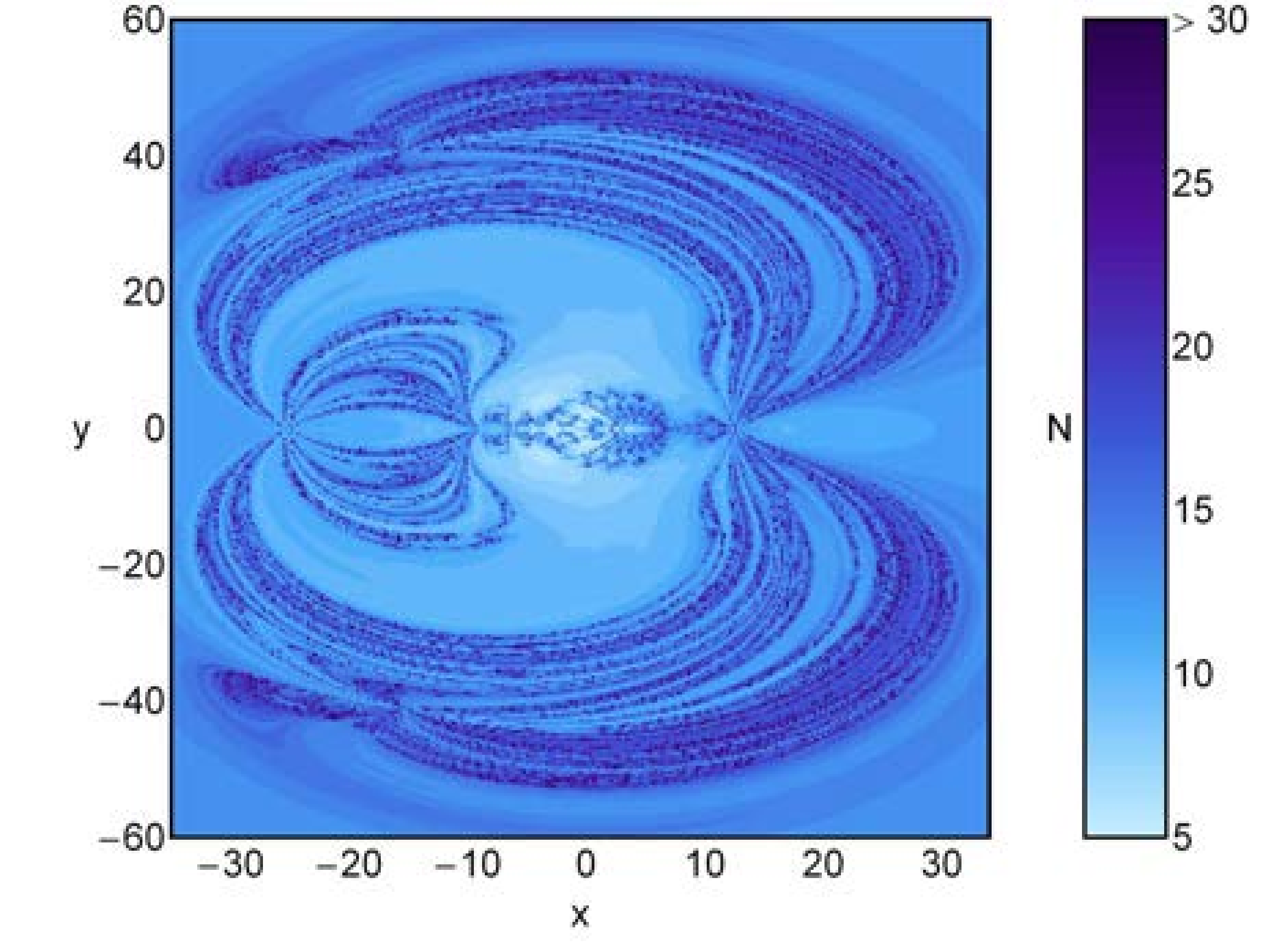}
(l)\includegraphics[scale=.25]{P123.pdf}\\
\caption{The Newton-Raphson basins of attraction on the xy-plane for the
case when nine libration points exist for  fixed value of
$\alpha=73 \degree$  and for:
(a) $\beta=(56+\frac{1}{30})\degree$; (d) $\beta=57\degree$;
(g) $\beta=58\degree$; (j) $\beta=59\degree$.The color code for the libration points $L_1$,...,$L_9$ is same as in Fig \ref{NR_Fig_1}; and non-converging points (white);  (b, e,  h, k) and (c, f, i, l) are the distribution of the corresponding number $(N)$ and the  probability distributions of required iterations for obtaining the Newton-Raphson basins of attraction shown in (a, d, g, j), respectively.
 (Color figure online).}
\label{NR_Fig_C3}
\end{figure*}

In the following figure, we have drawn the basins of convergence associated with the nine libration points in the second concave case. In Fig. \ref{NR_Fig_C3}, we have shown the basins for the fixed value of angle parameter $\alpha=73\degree$ and for four different values of $\beta$. We have observed that there is a drastic change in the topology of the finite domain of basins of convergence associated with the libration points as $\beta$ increases. For $\beta=(56+\frac{1}{30})\degree$, Fig. \ref{NR_Fig_C3}a, it has been observed that all the non-collinear libration points are symmetrical with respect to $x$-axis. We have also observed that the extent of basins of convergence corresponding to the collinear libration point $L_{3}$ is infinite while, for remaining libration points, the domain of basins of convergence are finite. Further, it is also observed that most of the area in the configuration plane $(x, y)$ is covered by the finite domain of the basins of convergence associated with the libration points $L_{4, 5}$ and both are symmetrical with respect to the $x-$axis. The region corresponds to $L_{1}$ is chaotic with very low density. The basins corresponds to $L_{2}$ form a caterpillar shape along the $x$-axis and some chaotic region as well. The libration points $L_{6, 7}$ look like butterfly with wings and many antennas having low density. The basins correspond to $L_{8, 9}$ form a bat shaped region.\\
Fig. \ref{NR_Fig_C3} panel-$(d, g, j)$ has been plotted for $\beta=57\degree, 58\degree, 59\degree$ respectively. It has been observed that in panel-d, the basins of convergence take a shape of babugosha whereas in panel-g it looks like a watermelon. Is is also observed that the domain of basins of convergence corresponding to the libration points $L_{8, 9}$ increases continuously as $\beta$ increases.\\
In Fig. \ref{NR_Fig_C3} panels-(b, e, h, k), we have shown the corresponding number $N$ of iterations to obtain the predefined accuracy. It is also observed that most of the initial conditions which lie in the domain of the basins of attraction have relatively faster rate of convergence for $N<20$
whereas, for $N>20$ have slower rate. Fig. \ref{NR_Fig_C3} panels-(c, f, i, l) show the corresponding probability distribution of the iterations and most probable number $N^*$ of the iterations is found. We have observed that the value of $N^*$ does not follow any pattern corresponding to the increasing values of $\beta$.
%%%%%%%%%%%%%%%%%%%%%%%%%%%%%%%%%%%%%
\subsubsection{when eleven libration points exist}
\begin{figure*}[!t]
\centering
(a)\includegraphics[scale=.27]{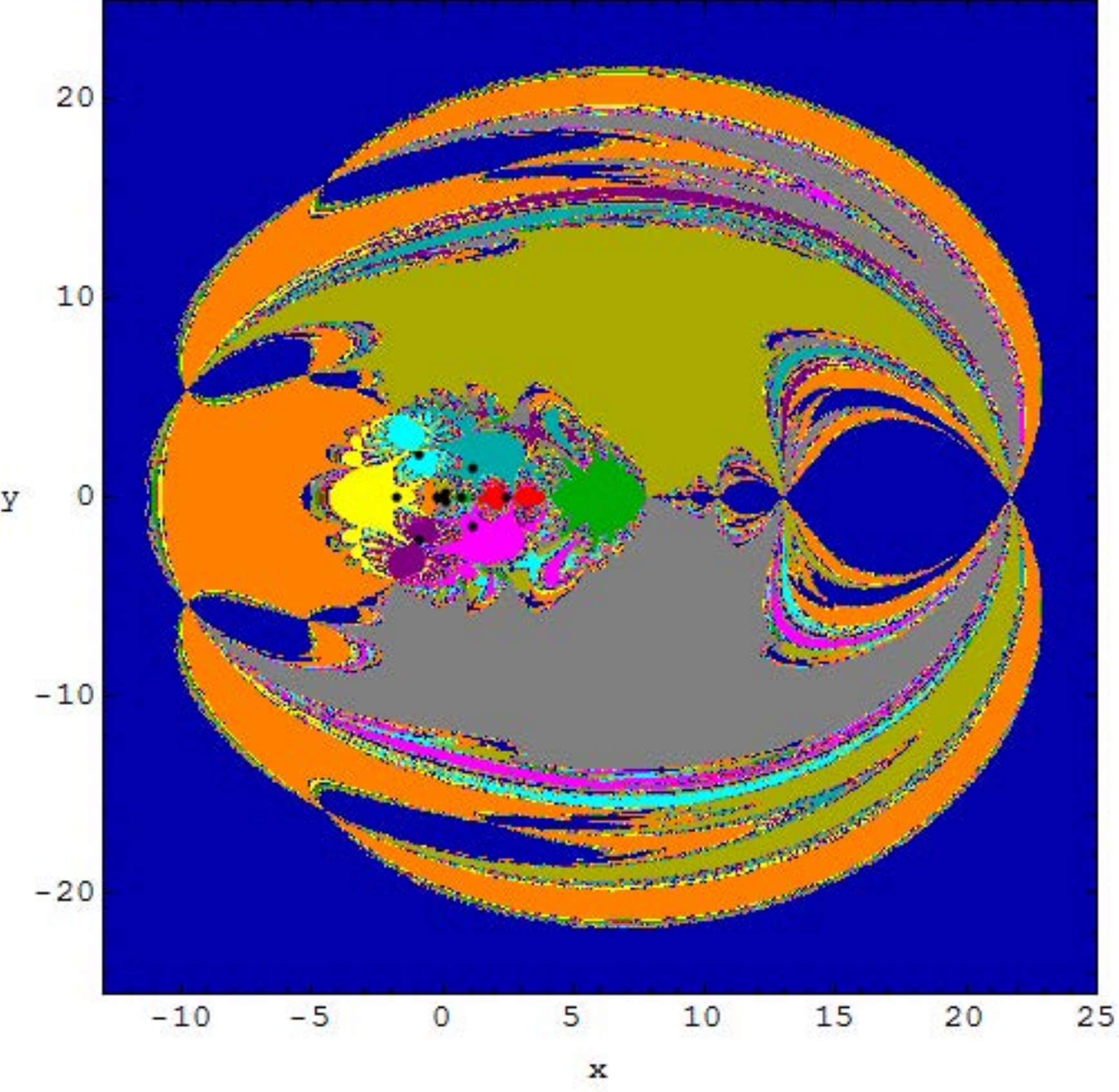}%Basin_39
(b)\includegraphics[scale=.27]{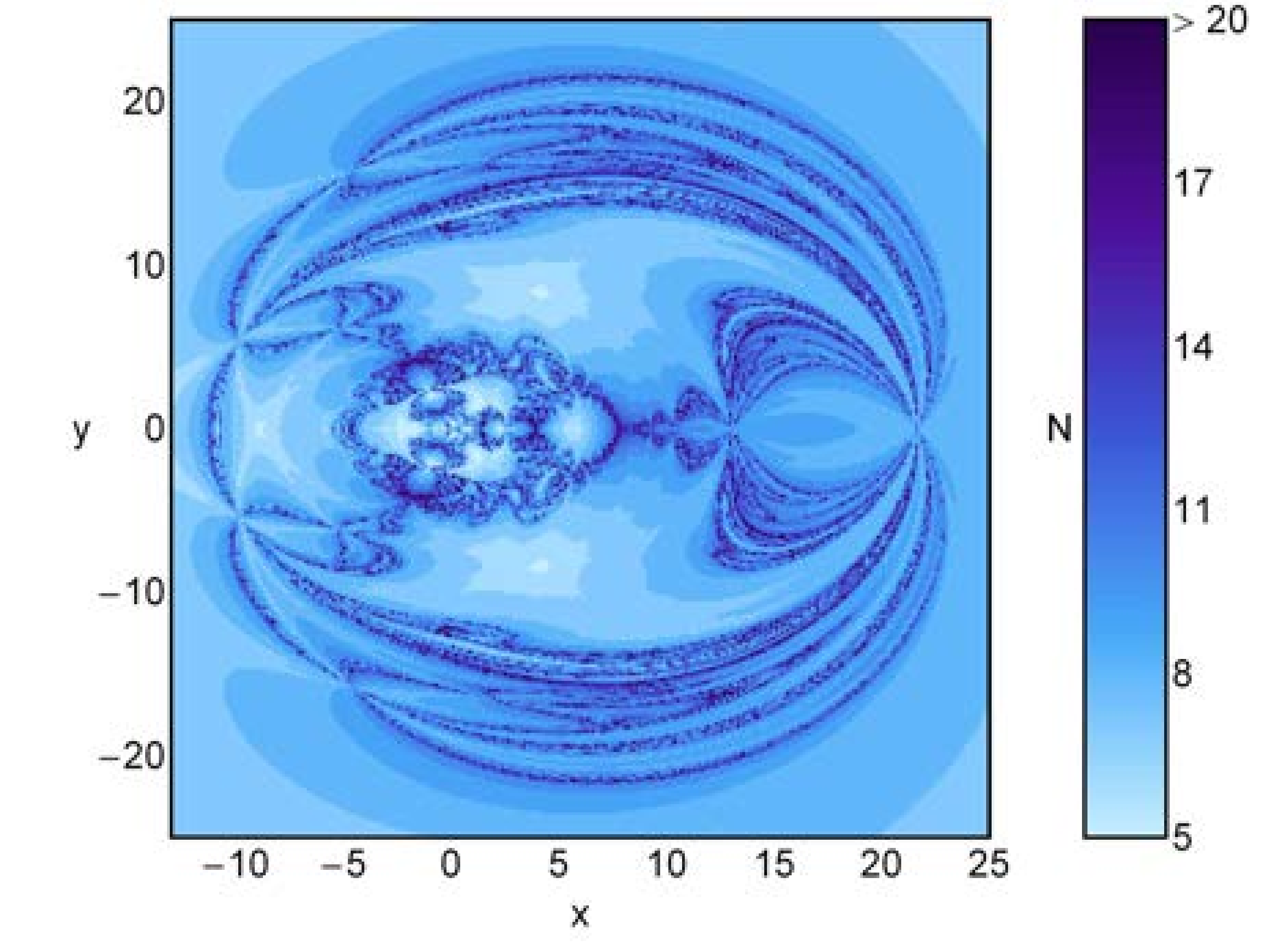}
(c)\includegraphics[scale=.25]{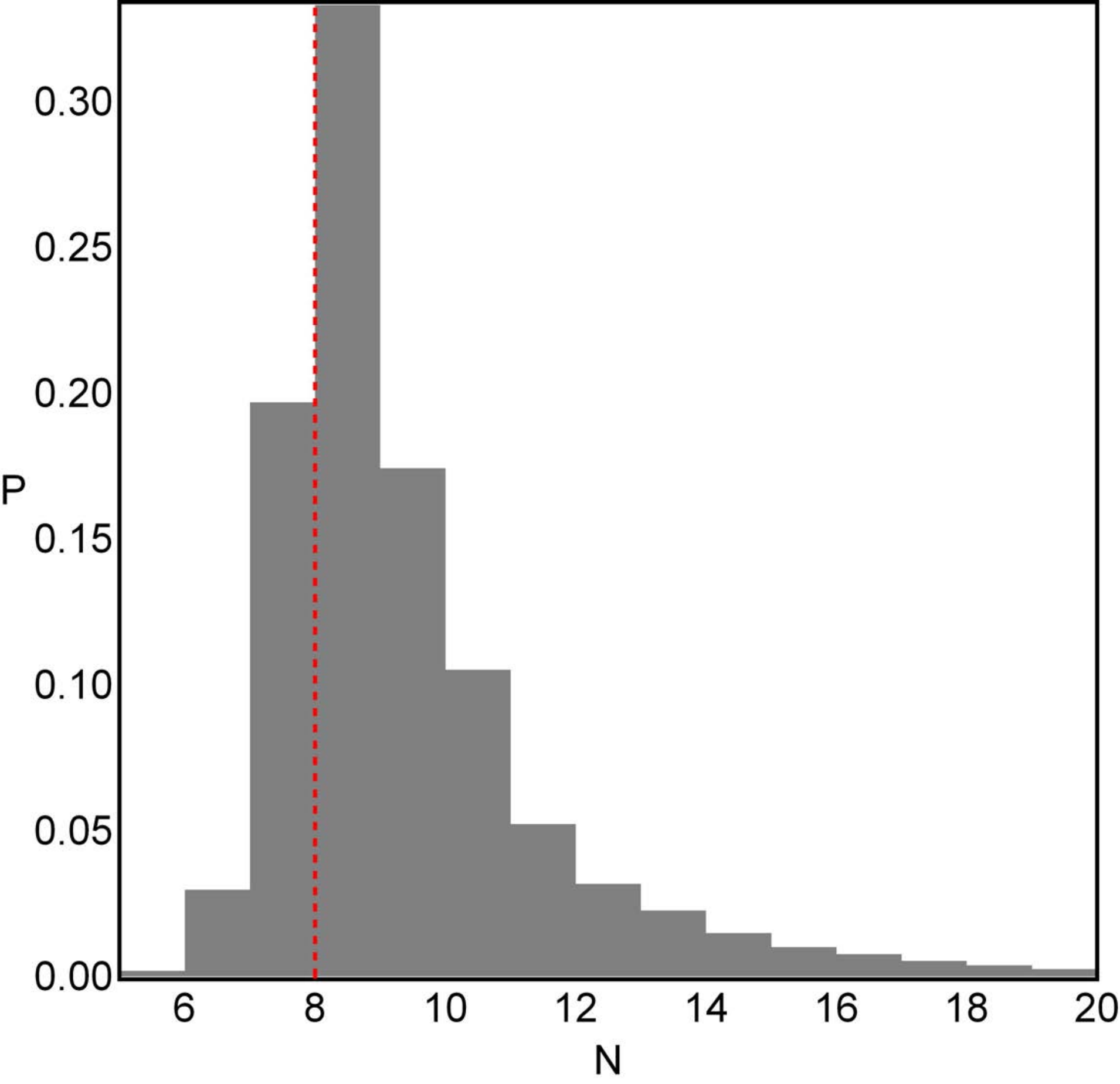}\\
(d)\includegraphics[scale=.27]{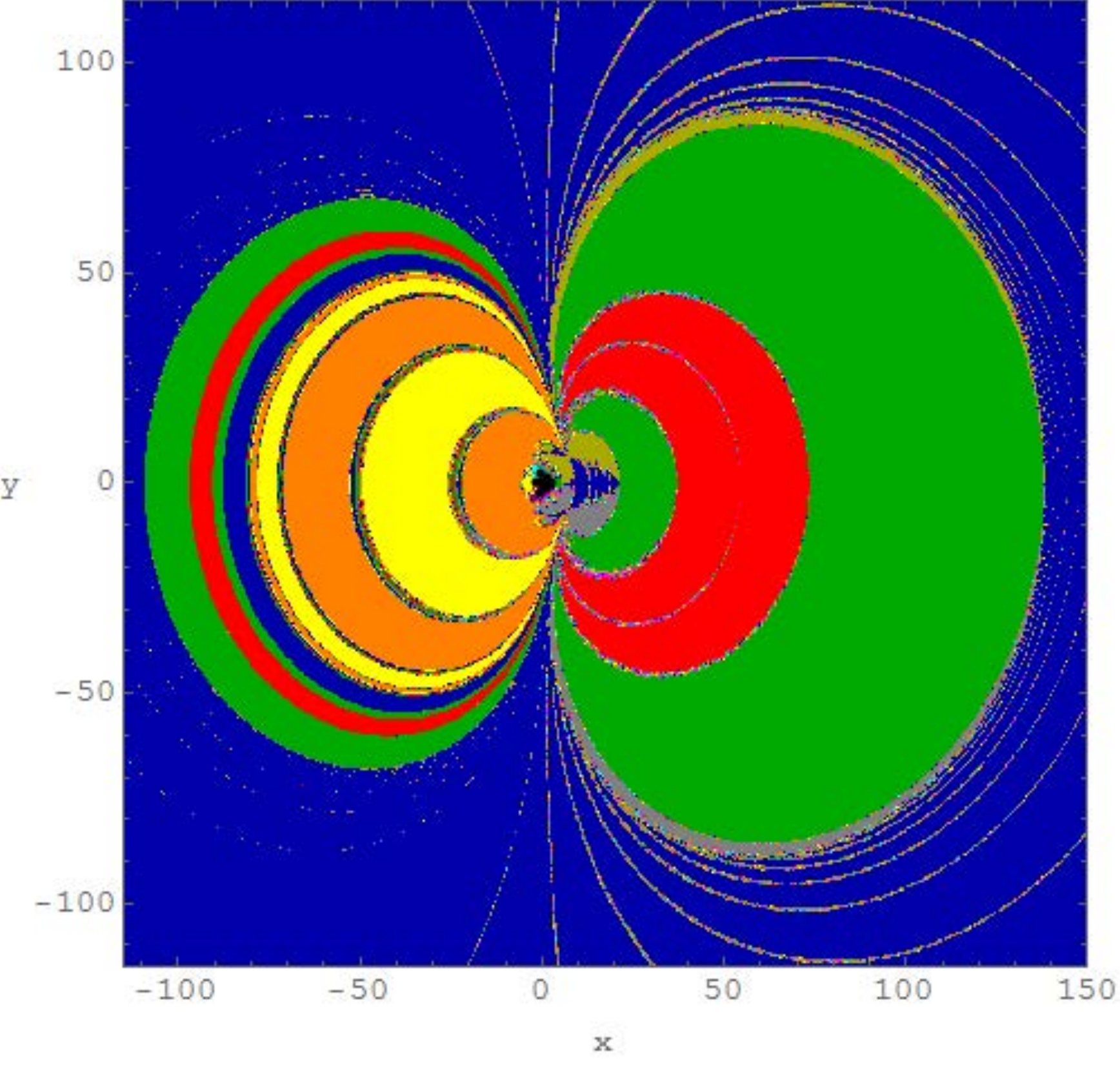}%Fig_42
(e)\includegraphics[scale=.27]{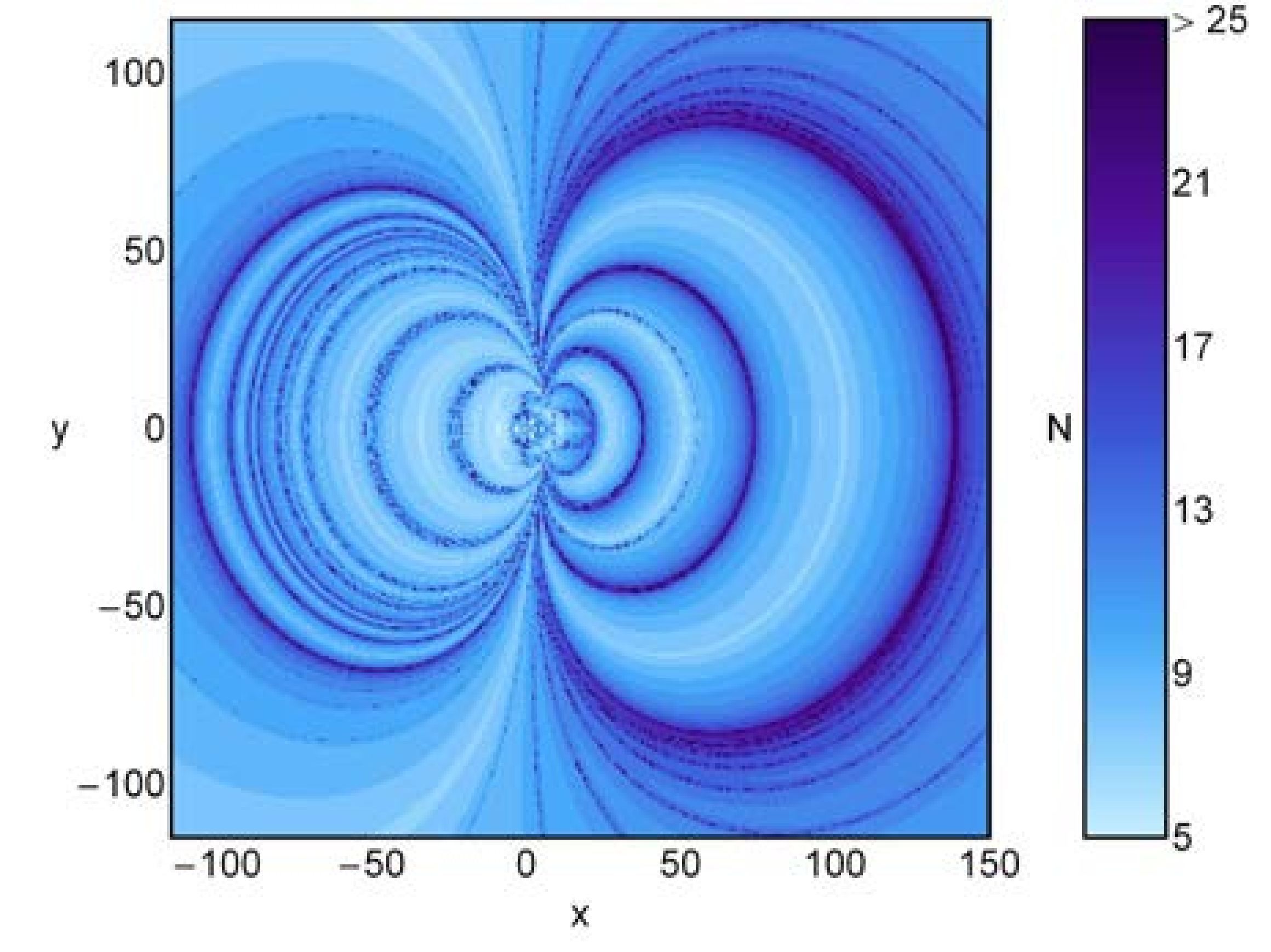}
(f)\includegraphics[scale=.25]{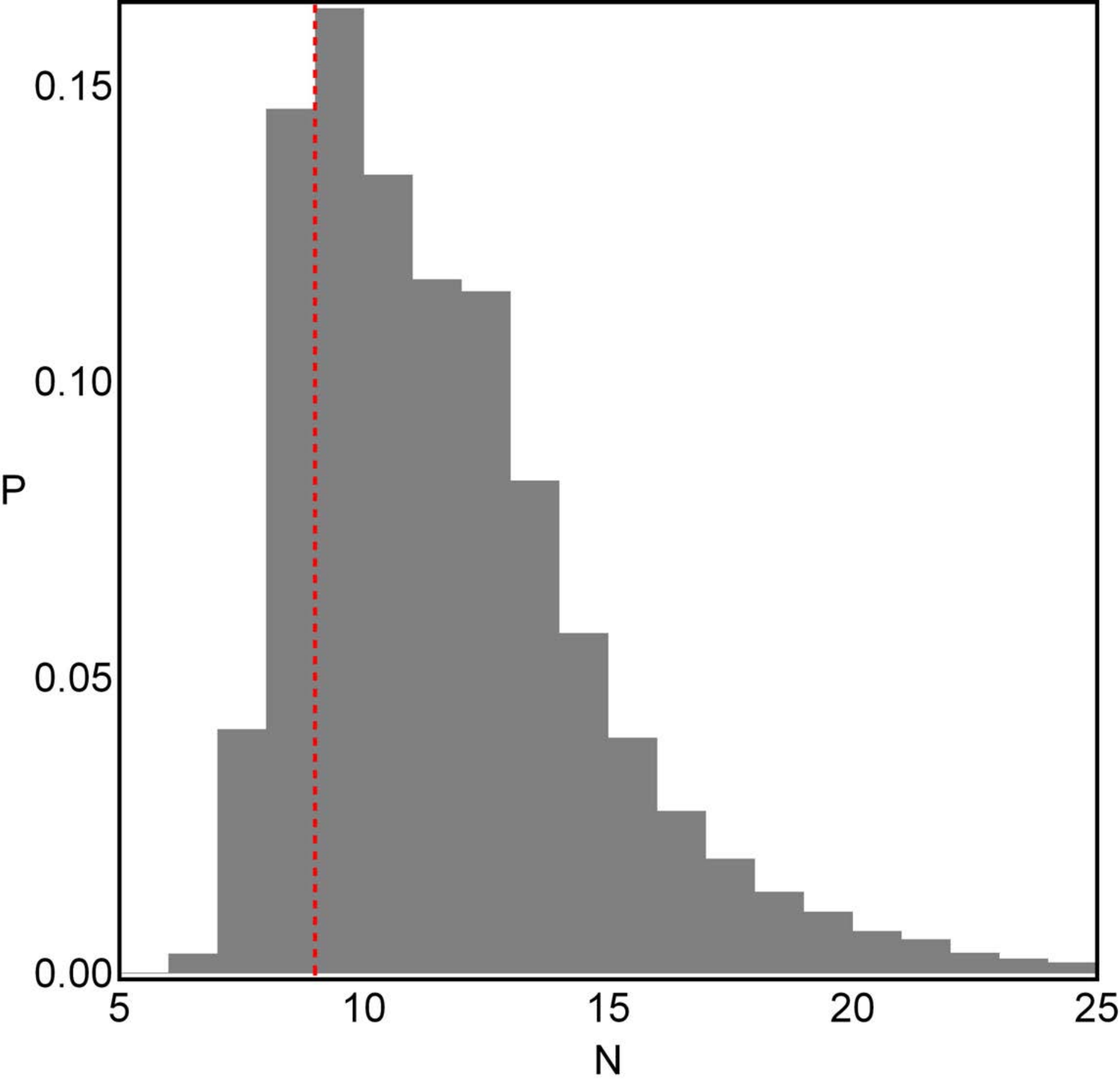}\\
(g)\includegraphics[scale=.27]{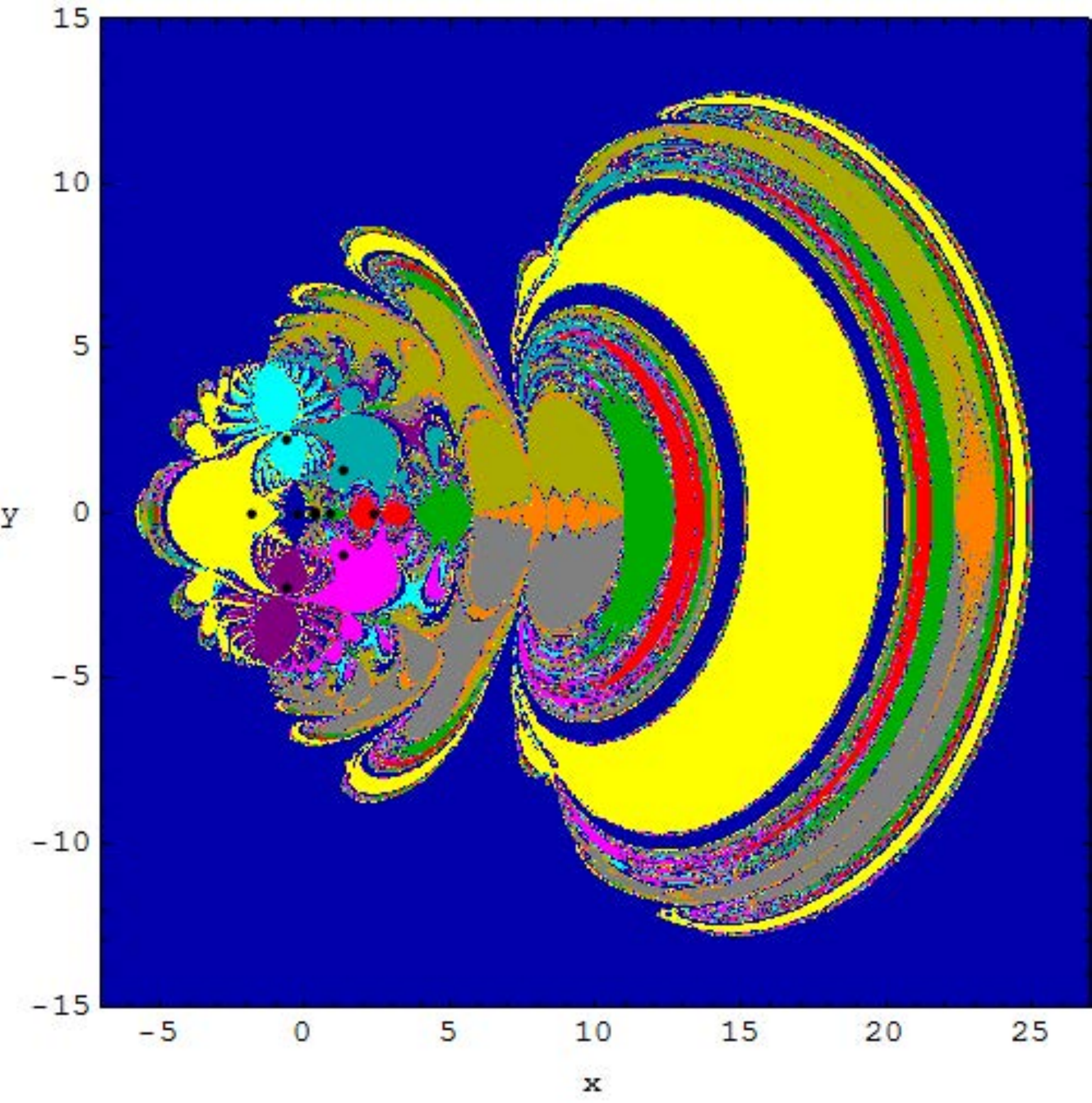}%Fig_44
(h)\includegraphics[scale=.27]{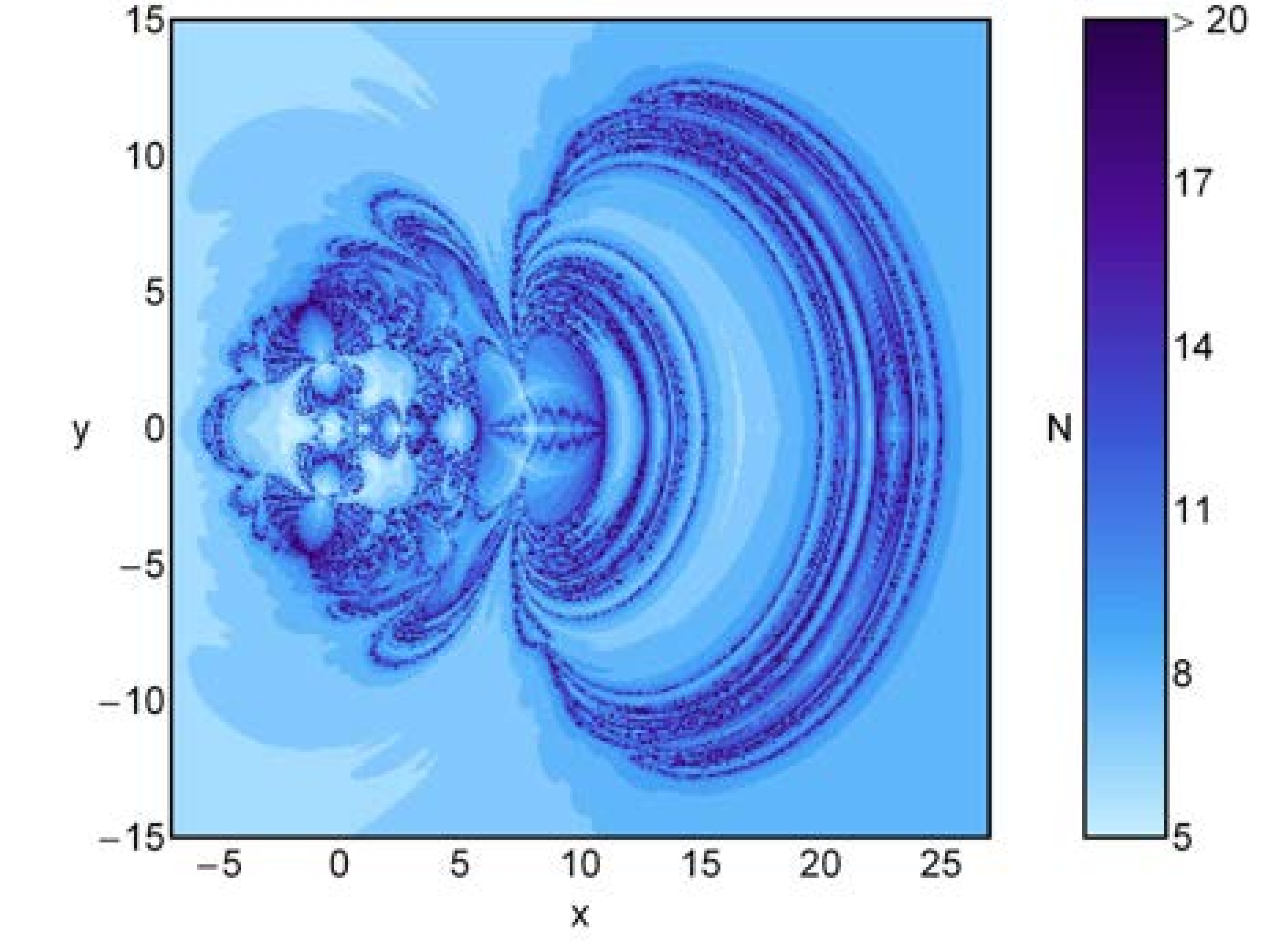}
(i)\includegraphics[scale=.25]{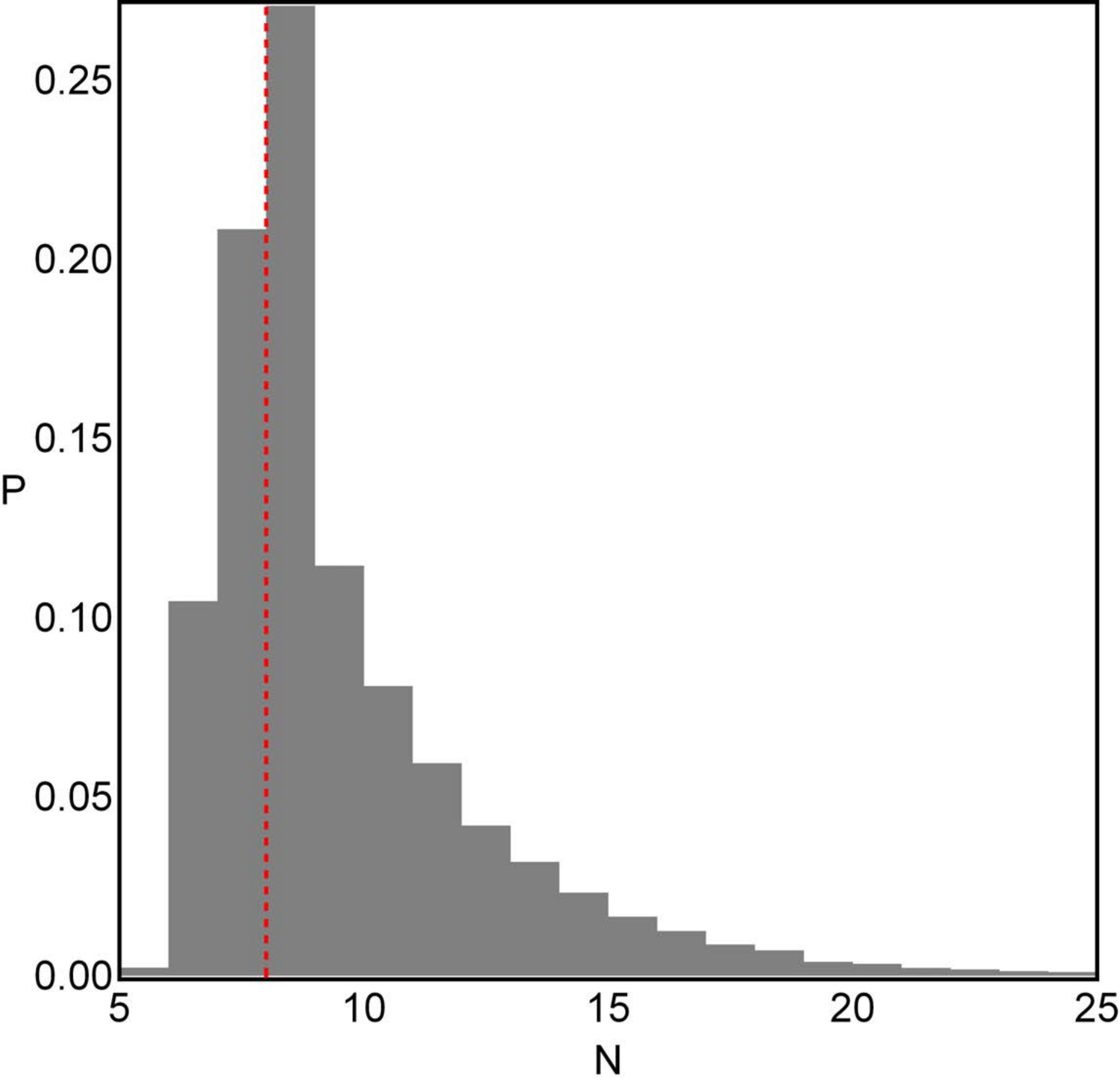}\\
\caption{The Newton-Raphson basins of attraction on the xy-plane for the
case when eleven libration points exist for  fixed value of
$\alpha=61 \degree$  and for:
(a) $\beta=39\degree$; (d) $\beta=42\degree$;
(g) $\beta=44\degree$. The color code for the libration points $L_1$,...,$L_9$,  is same as in Fig \ref{NR_Fig_1} and $L_{10}\emph{(orange)}$; $L_{11}
\emph{(blue)}$; and non-converging points (white);  (b, e,  h) and (c, f, i) are the distribution of the corresponding number $(N)$ and the  probability distributions of required iterations for obtaining the Newton-Raphson basins of attraction shown in (a, d, g), respectively. (Color figure online).}
\label{NR_Fig_C4}
\end{figure*}

In this last subsection, we have discussed the basins of convergence associated with the libration points in the second concave case, for which $11$ libration points exist. The Fig. \ref{NR_Fig_C4} is depicted for the fixed value of angle parameter $\alpha=61\degree$ and three different values of $\beta$. One may observe the drastic change in the topology of the finite domain of the basins of convergence associated with the libration points when $\beta$ increases. For $\beta=39\degree$, Fig. \ref{NR_Fig_C4} panel-a, we have observed that all the non-collinear libration points are symmetrical with respect to $x$-axis. An interesting phenomenon is also observed that the extent of basins of convergence corresponding to the collinear libration point $L_{11}$ is infinite whereas, for all the other libration point, the domains of basins of convergence are finite. Further, it is also observed that most of the area in the configuration plane $(x, y)$ is covered by the finite domain of the basins of convergence corresponding to the libration points $L_{8, 9, 10}$ in which $L_{8, 9}$ are symmetrical with respect to $x-$axis and $L_{10}$ encloses the topology of the basins of convergence associated with finite domains. The basins of convergence linked with the libration points $L_{1, 6, 7}$ alike butterfly with wings and many antennas in  which $L_{6, 7}$ are symmetrical with respect to $x-$axis and cover more area than the area covered by $L_{1}$. The basins of convergence associated with the libration points $L_{3, 4, 5}$ look like the flying birds whose wings are tied with each other mutually. Moreover, $L_{4, 5}$ are symmetrical but $L_{3}$ covers the majority of area than $L_{4, 5}$.

Fig. \ref{NR_Fig_C4} panel-d has been drawn for $\beta=42\degree$, a sudden change in the structure of the basins of convergence associated with the libration points has been observed. From the water melon shape (in previous panel), its shape changes to horizontal teddy-bear shape. In this panel, the majority of the area on the configuration plane is covered by the finite domain of convergence corresponding to the libration point $L_2$ which is green in color. All the basins of convergence are originating from the center of the teddy-bear.\\
In Fig. \ref{NR_Fig_C4} panel-g, we have taken $\beta=44\degree$, a drastic change in the basins of convergence is occurred and most of the area on the configuration plane is covered by the finite domain of convergence corresponding to the libration point $L_3$ (yellow color). We have observed that the domain of basins of convergence is well shaped and surrounded by the chaotic sea composed of the initial conditions.

 In Fig. \ref{NR_Fig_C4} panels-(b, e, h), we have illustrated  the corresponding number $N$ of iterations to obtain the predefined accuracy. It is also noticed that most of the initial conditions which lie in the domain of the basins of attraction have relatively fast convergence ($N<10$) whereas, for $N>10$
 have slow rate of convergence. In Fig. \ref{NR_Fig_C4} panels-(c, f, i), the corresponding probability distribution of the iterations are shown and most probable number $N^*$ of the iterations is found. We have observed that the value of $N^*$ is neither increasing nor decreasing as we increase the values of $\beta$.
%%%%
%%%%
%%%%
%%%%
%%%%%
%%%%
%%%%
\section{Concluding remarks}
\label{Concluding remarks}
In the present paper, we have expanded out the work initiated in our Paper-$1$ \cite{sur18f} by taking the case of concave configuration to discuss the topology of the basins of the convergence associated with the libration points for different values of the angle parameters $\alpha$ and $\beta$. In the presented model of the restricted five-body problem, we have considered two different configurations namely, the first concave case and the second concave case regarding the concave configuration in which we have considered three (in first concave case) and two (in second concave case) different values of the angle parameter $\alpha$ and the corresponding permissible set of the values of the angle parameter $\beta$. In both the cases, it is observed that the total number of libration points as well as their positions strongly depend upon the angle parameters $\alpha$ and $\beta$. However, the numerical calculations revealed that these libration points are linearly unstable for all the combinations of the angle parameters taken.

One of the other aspect of the present work is to study the basins of convergence associated to the libration points of the dynamical system by using the multivariate version of the Newton-Raphson iterative method. These basins of convergence unveil the fact how the libration points (which act as attractors) attract each point on the configuration $(x, y)$ plane. To the best of our knowledge, this is first time where this mathematical model is chosen to perform a systematic and thorough numerical investigation, regarding the basins of convergence and this fact reveals the novelty and importance of the present work.

The main results of the numerical calculations can be summarized in following statements:
\begin{itemize}
  \item[*] \textcolor[rgb]{0.00,0.00,0.00}{The total number of libration points are either nine, eleven or thirteen which are unstable for the considered values of the angle parameters.}
  \item[*] It is observed that the configuration $(x, y)$ plane is composed of the mixture of the basins of attraction and highly fractal regions. The value of the angle parameters is very sensitive as a slight change leads to a completely different topology of the basins of convergence. Moreover, any choice of initial conditions in the fractal regions is completely unpredictable about its final state.
  \item[*] In most of the cases, the several basins of attraction are well-formed and very intricately interwoven. The chaotic regions are on the other hands mainly appear in the vicinity of the basins boundaries.
  \item[*] The extent of the basins of convergence associated with one of the collinear libration points is infinite in all the cases while for the non-collinear libration points it is always finite.
  \item[*] The presented iterative scheme was found to converge extremely fast for those initial conditions which fall in the vicinity of the libration points whereas, it is relatively very slow for the initial conditions lying in the neighbourhood of the basins boundaries.
  \item[*]The required number of iterations to obtain the predefined accuracy varies in almost each combination of the angle parameters.
  \end{itemize}
For all the calculations and the graphical illustrations, we used the latest version 11 of Mathematica$^\circledR$ \cite{wol03}. Moreover, it is  supposed that the present study and the obtained results will be useful in the field of basins of convergence in dynamical systems of the restricted five-body problem.
%%%%
\appendix
\section{}
\label{Appendix}{}
\begin{align}
\Delta&=\omega^2=\frac{m_2/|P_1P_2|^2+2m\sin \alpha /|P_1P_3|^2}{P_1O},\nonumber\\
m_3&=m_4=m= -\frac{m_1 a+m_2 b}{2c},\nonumber\\
a&=(1-m_{1})\tan \alpha-m_{2}\tan \beta, \nonumber\\
b&=-m_{1}\tan \alpha +(1-m_{2})\tan \beta, \nonumber\\
c&=\big(\cos^3\beta-\frac{1}{8}\big)\tan\alpha+\big(\cos^3\alpha-\frac{1}{8}\big)\tan\beta, \nonumber\\
m_{1}&=\frac{(b_{1}+a_{0}-b_{0})b_{0}}{a_{0}b_{1}+a_{1}b_{0}-a_{1}b_{1}},\nonumber
\end{align}
\begin{align}
m_{2}&=\frac{(a_{1}+b_{0}-a_{0})a_{0}}{a_{0}b_{1}+a_{1}b_{0}-a_{1}b_{1}},\nonumber\\
a_0&=\big(\cos^3\alpha-\frac{1}{8}\big)\tan\alpha, \nonumber\\
b_0&=-\big(\cos^3\beta-\frac{1}{8}\big)\tan\beta, \nonumber
\end{align}
\begin{align}
a_1&=-\big(\frac{1}{8}-\cos^3\alpha-\cos^3\beta \big)\tan\beta+\frac{1}{(\tan\alpha-\tan \beta)^2}\nonumber\\
   &-\frac{1}{8}\tan\alpha, \nonumber\\
b_1&=\big(\frac{1}{8}-\cos^3\alpha-\cos^3\beta \big)\tan\alpha+\frac{1}{(\tan\alpha-\tan\beta)^2} \nonumber\\
   &+\frac{1}{8}\tan\beta, \nonumber
\end{align}
\section*{Acknowledgments}
\footnotesize
The authors are thankful to Center for Fundamental Research in Space dynamics and Celestial mechanics (CFRSC), New Delhi, India for providing research facilities.%\par
\section*{Compliance with Ethical Standards}
\begin{description}
  \item[-] Funding: The authors state that they have not received any research
grants.
  \item[-] Conflict of interest: The authors declare that they have no conflict of
interest.
\end{description}
%%%%%%%%%%%%%%%

\end{document}